\documentclass[a4paper,11pt]{article}
\usepackage{amsmath, amsfonts, amscd, amssymb}
\usepackage{amsthm}
\usepackage{latexsym}       \usepackage{amsmath}
\usepackage{amsfonts}       \usepackage{amssymb}
\usepackage{amscd}          \usepackage[greek,english]{babel}
\usepackage[iso-8859-7]{inputenc}
\usepackage[dvipsnames,usenames]{color}
\usepackage{graphics}
\usepackage{graphicx}
\usepackage{longtable}
\usepackage{yhmath}

\numberwithin{equation}{section}

\theoremstyle{plain}

\theoremstyle{definition}

\theoremstyle{proposition}

\theoremstyle{theorem}

\theoremstyle{corollary}

\theoremstyle{lemma}

\catcode`\@=11

\newcount\printerresolution
\global\printerresolution=300

\newcount\headsize
\global\headsize=450

\font\dotfont = lcircle10 at 3pt

\newcount\epihead
\global\epihead=200

\newcount\defaultscale
\def\setdefaultscale#1{\global\defaultscale=#1}
\setdefaultscale{100}

\newcount\actualtextarrowspace
\newcount\actualtextarrowlength

\newcommand{\computetextparameters}%
{\global\actualtextarrowspace=\textarrowlength%
\global\advance\actualtextarrowspace by 3%
\global\actualtextarrowlength=\textarrowlength%
\global\multiply\actualtextarrowlength by 100}

\newcount\textarrowlength
\def\settextarrowlength#1{\global\textarrowlength=#1%
\computetextparameters} \settextarrowlength{20}

\newcount\actualdisplayarrowspace
\newcount\actualdisplayarrowlength

\newcommand{\computedisplayparameters}%
{\global\actualdisplayarrowspace=\displayarrowlength%
\global\advance\actualdisplayarrowspace by 3%
\global\actualdisplayarrowlength=\displayarrowlength%
\global\multiply\actualdisplayarrowlength by 100}

\newcount\displayarrowlength
\def\setdisplayarrowlength#1{\global\displayarrowlength=#1%
\computedisplayparameters} \setdisplayarrowlength{30}

\def\@ifnexttok#1#2#3{\let\@tempe #1\def\@tempa{#2}\def\@tempb{#3}%
\futurelet\@tempc\@ifntok}
\def\@ifntok{\ifx \@tempc \@tempe\let\@tempd\@tempa\else\let\@tempd\@tempb\fi%
\@tempd}

\def\@diagramerror#1#2{%
\edef\@@tempc{#2}\expandafter\errhelp\expandafter{\@@tempc}%
\typeout{Diagram error. \space See User's guide for
explanation.^^J
 \space\@spaces\@spaces\@spaces Type \space H <return> \space for
 immediate help.}\errmessage{#1}}

\newif\ifdiagram

\def\testtextmode{%
\ifdiagram\@diagramerror{Text arrows are not allowed in
diagrams}{Here you should use east or west diagram arrows, not
forward or backward text arrows. Try proceeding now, typeset could
succeed but with unpredictable output.}
\else\ifmmode\relax\else%
\@diagramerror{Missing \string$}{Text arrows should be introduced
in math mode. Try proceeding now, typeset could succeed but output
could not be what you expected.}\fi\fi}

\def\testdiagrammode{\ifdiagram\relax\else
\@diagramerror{Diagram arrows are not allowed in formulas}{Here
you should use forward or backward text arrows, not diagram
arrows. Proceeding could work with unpredictable output, but
overflow arithmetic could also occur.}\fi}

\def\checkmode{\ifmmode\@diagramerror{Wrong mode: no diagrams
allowed in math mode.}{You should leave math mode before
introducing your diagram. All items in the diagram will
automatically be processed in math
mode.}\else\relax\fi\global\diagramtrue}

\global\diagramfalse

\def\DOT{{\dotfont q}}

\newcount\basicstep
\newcount\numberofsteps
\numberofsteps=\headsize \global\divide\numberofsteps by
\basicstep

\newcommand{\makehead}[3]{%
\begin{picture}(0,0)%
\multiput(0,0)(#1,#2){#3}{\DOT}%
\multiput(0,0)(-#2,#1){#3}{\DOT}%
\end{picture}}

\newcount\xstep
\newcount\ystep

\newsavebox{\northhead}
\savebox{\northhead}{%
\xstep=-\basicstep%
\multiply\xstep by 7071%
\divide\xstep by 10000%
\ystep=\xstep%
\makehead{\xstep}{\ystep}{\numberofsteps}}
\newcommand{\nhead}{\usebox{\northhead}}

\newsavebox{\easthead}
\savebox{\easthead}{%
\xstep=-\basicstep%
\multiply\xstep by 7071%
\divide\xstep by 10000%
\ystep=-\xstep%
\makehead{\xstep}{\ystep}{\numberofsteps}}
\newcommand{\ehead}{\usebox{\easthead}}

\newsavebox{\southhead}
\savebox{\southhead}{%
\xstep=\basicstep%
\multiply\xstep by 7071%
\divide\xstep by 10000%
\ystep=\xstep%
\makehead{\xstep}{\ystep}{\numberofsteps}}
\newcommand{\shead}{\usebox{\southhead}}

\newsavebox{\westhead}
\savebox{\westhead}{%
\xstep=\basicstep%
\multiply\xstep by 7071%
\divide\xstep by 10000%
\ystep=-\xstep%
\makehead{\xstep}{\ystep}{\numberofsteps}}
\newcommand{\whead}{\usebox{\westhead}}

\newsavebox{\northwesthead}
\savebox{\northwesthead}{%
\makehead{0}{-\basicstep}{\numberofsteps}}
\newcommand{\nwhead}{\usebox{\northwesthead}}

\newsavebox{\northeasthead}
\savebox{\northeasthead}{%
\makehead{-\basicstep}{0}{\numberofsteps}}
\newcommand{\nehead}{\usebox{\northeasthead}}

\newsavebox{\southwesthead}
\savebox{\southwesthead}{%
\makehead{\basicstep}{0}{\numberofsteps}}
\newcommand{\swhead}{\usebox{\southwesthead}}

\newsavebox{\southeasthead}
\savebox{\southeasthead}{%
\makehead{0}{\basicstep}{\numberofsteps}}
\newcommand{\sehead}{\usebox{\southeasthead}}

\newsavebox{\eastnortheasthead}
\savebox{\eastnortheasthead}{%
\xstep=-\basicstep%
\multiply\xstep by 9486%
\divide\xstep by 10000%
\ystep=\xstep%
\divide\ystep by -3%
\makehead{\xstep}{\ystep}{\numberofsteps}}
\newcommand{\enehead}{\usebox{\eastnortheasthead}}

\newsavebox{\northnortheasthead}
\savebox{\northnortheasthead}{%
\xstep=-\basicstep%
\multiply\xstep by 9486%
\divide\xstep by 10000%
\ystep=\xstep%
\divide\ystep by 3%
\makehead{\xstep}{\ystep}{\numberofsteps}}
\newcommand{\nnehead}{\usebox{\northnortheasthead}}

\newsavebox{\southsouthwesthead}
\savebox{\southsouthwesthead}{%
\xstep=\basicstep%
\multiply\xstep by 9486%
\divide\xstep by 10000%
\ystep=\xstep%
\divide\ystep by 3%
\makehead{\xstep}{\ystep}{\numberofsteps}}
\newcommand{\sswhead}{\usebox{\southsouthwesthead}}

\newsavebox{\westsouthwesthead}
\savebox{\westsouthwesthead}{%
\xstep=\basicstep%
\multiply\xstep by 9486%
\divide\xstep by 10000%
\ystep=\xstep%
\divide\ystep by -3%
\makehead{\xstep}{\ystep}{\numberofsteps}}
\newcommand{\wswhead}{\usebox{\westsouthwesthead}}

\newsavebox{\westnorthwesthead}
\savebox{\westnorthwesthead}{%
\xstep=\basicstep%
\multiply\xstep by 3162%
\divide\xstep by 10000%
\ystep=\xstep%
\multiply\ystep by -3%
\makehead{\xstep}{\ystep}{\numberofsteps}}
\newcommand{\wnwhead}{\usebox{\westnorthwesthead}}

\newsavebox{\eastsoutheasthead}
\savebox{\eastsoutheasthead}{%
\xstep=-\basicstep%
\multiply\xstep by 3162%
\divide\xstep by 10000%
\ystep=\xstep%
\multiply\ystep by -3%
\makehead{\xstep}{\ystep}{\numberofsteps}}
\newcommand{\esehead}{\usebox{\eastsoutheasthead}}

\newsavebox{\northnorthwesthead}
\savebox{\northnorthwesthead}{%
\xstep=-\basicstep%
\multiply\xstep by 3162%
\divide\xstep by 10000%
\ystep=\xstep%
\multiply\ystep by 3%
\makehead{\xstep}{\ystep}{\numberofsteps}}
\newcommand{\nnwhead}{\usebox{\northnorthwesthead}}

\newsavebox{\southsoutheasthead}
\savebox{\southsoutheasthead}{%
\xstep=\basicstep%
\multiply\xstep by 3162%
\divide\xstep by 10000%
\ystep=\xstep%
\multiply\ystep by 3%
\makehead{\xstep}{\ystep}{\numberofsteps}}
\newcommand{\ssehead}{\usebox{\southsoutheasthead}}

\newsavebox{\easteastnortheasthead}
\savebox{\easteastnortheasthead}{%
\xstep=-\basicstep%
\multiply\xstep by 8944%
\divide\xstep by 10000%
\ystep=\xstep%
\divide\ystep by -2%
\makehead{\xstep}{\ystep}{\numberofsteps}}
\newcommand{\eenehead}{\usebox{\easteastnortheasthead}}

\newsavebox{\northnorthnortheasthead}
\savebox{\northnorthnortheasthead}{%
\xstep=-\basicstep%
\multiply\xstep by 8944%
\divide\xstep by 10000%
\ystep=\xstep%
\divide\ystep by 2%
\makehead{\xstep}{\ystep}{\numberofsteps}}
\newcommand{\nnnehead}{\usebox{\northnorthnortheasthead}}

\newsavebox{\southsouthsouthwesthead}
\savebox{\southsouthsouthwesthead}{%
\xstep=\basicstep%
\multiply\xstep by 8944%
\divide\xstep by 10000%
\ystep=\xstep%
\divide\ystep by 2%
\makehead{\xstep}{\ystep}{\numberofsteps}}
\newcommand{\ssswhead}{\usebox{\southsouthsouthwesthead}}

\newsavebox{\westwestsouthwesthead}
\savebox{\westwestsouthwesthead}{%
\xstep=\basicstep%
\multiply\xstep by 8944%
\divide\xstep by 10000%
\ystep=\xstep%
\divide\ystep by -2%
\makehead{\xstep}{\ystep}{\numberofsteps}}
\newcommand{\wwswhead}{\usebox{\westwestsouthwesthead}}

\newsavebox{\westwestnorthwesthead}
\savebox{\westwestnorthwesthead}{%
\xstep=\basicstep%
\multiply\xstep by 4472%
\divide\xstep by 10000%
\ystep=\xstep%
\multiply\ystep by -2%
\makehead{\xstep}{\ystep}{\numberofsteps}}
\newcommand{\wwnwhead}{\usebox{\westwestnorthwesthead}}

\newsavebox{\easteastsoutheasthead}
\savebox{\easteastsoutheasthead}{%
\xstep=-\basicstep%
\multiply\xstep by 4472%
\divide\xstep by 10000%
\ystep=\xstep%
\multiply\ystep by -2%
\makehead{\xstep}{\ystep}{\numberofsteps}}
\newcommand{\eesehead}{\usebox{\easteastsoutheasthead}}

\newsavebox{\northnorthnorthwesthead}
\savebox{\northnorthnorthwesthead}{%
\xstep=-\basicstep%
\multiply\xstep by 4472%
\divide\xstep by 10000%
\ystep=\xstep%
\multiply\ystep by 2%
\makehead{\xstep}{\ystep}{\numberofsteps}}
\newcommand{\nnnwhead}{\usebox{\northnorthnorthwesthead}}

\newsavebox{\southsouthsoutheasthead}
\savebox{\southsouthsoutheasthead}{%
\xstep=\basicstep%
\multiply\xstep by 4472%
\divide\xstep by 10000%
\ystep=\xstep%
\multiply\ystep by 2%
\makehead{\xstep}{\ystep}{\numberofsteps}}
\newcommand{\sssehead}{\usebox{\southsouthsoutheasthead}}

\newsavebox{\northeasteastnortheasthead}
\savebox{\northeasteastnortheasthead}{%
\xstep=-\basicstep%
\multiply\xstep by 9806%
\divide\xstep by 10000%
\ystep=\xstep%
\divide\ystep by -5%
\makehead{\xstep}{\ystep}{\numberofsteps}}
\newcommand{\neenehead}{\usebox{\northeasteastnortheasthead}}

\newsavebox{\northeastnorthnortheasthead}
\savebox{\northeastnorthnortheasthead}{%
\xstep=-\basicstep%
\multiply\xstep by 9806%
\divide\xstep by 10000%
\ystep=\xstep%
\divide\ystep by 5%
\makehead{\xstep}{\ystep}{\numberofsteps}}
\newcommand{\nennehead}{\usebox{\northeastnorthnortheasthead}}

\newsavebox{\southwestsouthsouthwesthead}
\savebox{\southwestsouthsouthwesthead}{%
\xstep=\basicstep%
\multiply\xstep by 9806%
\divide\xstep by 10000%
\ystep=\xstep%
\divide\ystep by 5%
\makehead{\xstep}{\ystep}{\numberofsteps}}
\newcommand{\swsswhead}{\usebox{\southwestsouthsouthwesthead}}

\newsavebox{\southwestwestsouthwesthead}
\savebox{\southwestwestsouthwesthead}{%
\xstep=\basicstep%
\multiply\xstep by 9806%
\divide\xstep by 10000%
\ystep=\xstep%
\divide\ystep by -5%
\makehead{\xstep}{\ystep}{\numberofsteps}}
\newcommand{\swwswhead}{\usebox{\southwestwestsouthwesthead}}

\newsavebox{\northwestwestnorthwesthead}
\savebox{\northwestwestnorthwesthead}{%
\xstep=\basicstep%
\multiply\xstep by 1961%
\divide\xstep by 10000%
\ystep=\xstep%
\multiply\ystep by -5%
\makehead{\xstep}{\ystep}{\numberofsteps}}
\newcommand{\nwwnwhead}{\usebox{\northwestwestnorthwesthead}}

\newsavebox{\southeasteastsoutheasthead}
\savebox{\southeasteastsoutheasthead}{%
\xstep=-\basicstep%
\multiply\xstep by 1961%
\divide\xstep by 10000%
\ystep=\xstep%
\multiply\ystep by -5%
\makehead{\xstep}{\ystep}{\numberofsteps}}
\newcommand{\seesehead}{\usebox{\southeasteastsoutheasthead}}

\newsavebox{\northwestnorthnorthwesthead}
\savebox{\northwestnorthnorthwesthead}{%
\xstep=-\basicstep%
\multiply\xstep by 1961%
\divide\xstep by 10000%
\ystep=\xstep%
\multiply\ystep by 5%
\makehead{\xstep}{\ystep}{\numberofsteps}}
\newcommand{\nwnnwhead}{\usebox{\northwestnorthnorthwesthead}}

\newsavebox{\southeastsouthsoutheasthead}
\savebox{\southeastsouthsoutheasthead}{%
\xstep=\basicstep%
\multiply\xstep by 1961%
\divide\xstep by 10000%
\ystep=\xstep%
\multiply\ystep by 5%
\makehead{\xstep}{\ystep}{\numberofsteps}}
\newcommand{\sessehead}{\usebox{\southeastsouthsoutheasthead}}

\newsavebox{\isomorphismmark}
\newcommand{\isomark}[1]{\savebox{\isomorphismmark}{#1}}
\isomark{$\cong$}

\newif\ifuserdist
\newsavebox{\distributormark}
\newcommand{\distmark}[1]{\ifx#1\distcircle\userdistfalse\else%
\userdisttrue\savebox{\distributormark}{#1}\fi}
\newsavebox{\distributorcircle}
\savebox{\distributorcircle}{\begin{picture}(0,0)%
\put(0,0){\circle{4}}\end{picture}}

\userdistfalse

\newcount\monolength
\newcount\epilength
\newcount\bimolength
\newcount\secondmonolength
\newcount\secondepilength

\newcount\monotail%
\global\monotail=\headsize%
\global\multiply\monotail by 7070%
\global\divide\monotail by 10000%

\newcount\truemonotail%
\newcount\trueepihead%

\def\truetail{\truemonotail=\monotail%
\multiply\truemonotail by 100%
\divide\truemonotail by \SCALE}

\def\truehead{\trueepihead=\epihead%
\multiply\trueepihead by 100%
\divide\trueepihead by \SCALE}

\newcount\Monotail%
\global\Monotail=\headsize%
\global\divide\Monotail by 2%

\newcount\Epihead%
\global\Epihead=\epihead%
\global\multiply\Epihead by 7070%
\global\divide\Epihead by 10000%

\newcount\Truemonotail%
\newcount\Trueepihead%

\def\Truetail{\Truemonotail=\Monotail%
\multiply\Truemonotail by 100%
\divide\Truemonotail by \SCALE}%

\def\Truehead{\Trueepihead=\Epihead%
\multiply\Trueepihead by 100%
\divide\Trueepihead by \SCALE}

\newcount\MonoTail%
\global\MonoTail=\headsize%
\global\multiply\MonoTail by 6324%
\global\divide\MonoTail by 10000%

\newcount\EpiHead%
\global\EpiHead=\epihead%
\global\multiply\EpiHead by 8944%
\global\divide\EpiHead by 10000%

\newcount\TrueMonoTail%
\newcount\TrueEpiHead%

\def\TrueTail{\TrueMonoTail=\MonoTail%
\multiply\TrueMonoTail by 100%
\divide\TrueMonoTail by \SCALE}%

\def\TrueHead{\TrueEpiHead=\EpiHead%
\multiply\TrueEpiHead by 100%
\divide\TrueEpiHead by \SCALE}

\newcount\monotaiL%
\global\monotaiL=\headsize%
\global\multiply\monotaiL by 3162%
\global\divide\monotaiL by 10000%

\newcount\epiheaD%
\global\epiheaD=\epihead%
\global\multiply\epiheaD by 4472%
\global\divide\epiheaD by 10000%

\newcount\truemonotaiL%
\newcount\trueepiheaD%

\def\truetaiL{\truemonotaiL=\monotaiL%
\multiply\truemonotaiL by 100%
\divide\truemonotaiL by \SCALE}%

\def\trueheaD{\trueepiheaD=\epiheaD%
\multiply\trueepiheaD by 100%
\divide\trueepiheaD by \SCALE}

\newcount\SCALE%

\newcount\NUMBER%

\newcount\LINE%

\newcount\COLUMN%

\newcount\WIDTH%

\newcount\SOURCE%

\newcount\ARROW%

\newcount\TARGET%

\newcount\ARROWLENGTH%

\newcount\NUMBEROFDOTS%

\newcounter{x}%

\newcounter{y}%

\newcounter{z}%

\newcounter{horizontal}%

\newcounter{vertical}%

\newskip\itemlength%

\newskip\firstitem%

\newskip\seconditem%

\newcommand{\printarrow}{}%

\newcount\X%

\newcount\Y%

\newcount\Z%

\newcommand{\truex}[1]{%
\NUMBER=#1%
\multiply\NUMBER by 100%
\divide\NUMBER by \SCALE%
\setcounter{x}{\NUMBER}}%

\newcommand{\truey}[1]{%
\NUMBER=#1%
\multiply\NUMBER by 100%
\divide\NUMBER by \SCALE%
\setcounter{y}{\NUMBER}}%

\newcommand{\truez}[1]{%
\NUMBER=#1%
\multiply\NUMBER by 100%
\divide\NUMBER by \SCALE%
\setcounter{z}{\NUMBER}}%

\newcommand{\changecounters}[1]{%
\SOURCE=\ARROW%
\ARROW=\TARGET%
\settowidth{\itemlength}{#1}%
\ifdim \itemlength > 2800\unitlength%
\addtolength{\itemlength}{-2800\unitlength}%
\TARGET=\itemlength%
\divide\TARGET by 1310%
\multiply\TARGET by 100%
\divide\TARGET by \SCALE%
\else%
\TARGET=0%
\fi%
\ARROWLENGTH=5000%
\advance\ARROWLENGTH by -\SOURCE%
\advance\ARROWLENGTH by -\TARGET%
\divide\ARROWLENGTH by 100%
\advance\SOURCE by -\TARGET}%

\newcommand{\initialize}[1]{%
\LINE=0%
\COLUMN=0%
\WIDTH=0%
\ARROW=0%
\TARGET=0%
\changecounters{#1}%
\renewcommand{\printarrow}{#1}%
\begin{center}%
\vspace{2pt}%
\begin{picture}(0,0)}%

\newcommand{\DIAGV}[2]{%
\checkmode%
\SCALE=#1%
\setlength{\unitlength}{655sp}%
\multiply\unitlength by \SCALE%
\divide\unitlength by 100%
\initialize{\mbox{$#2$}}}%

\newcommand{\n}[1]{%
\changecounters{\mbox{$#1$}}%
\put(\COLUMN,\LINE){\makebox(0,0){\printarrow}}%
\thinlines%
\renewcommand{\printarrow}{\mbox{$#1$}}%
\advance\COLUMN by 4000}%

\newcommand{\nn}[1]{%
\put(\COLUMN,\LINE){\makebox(0,0){\printarrow}}%
\thinlines%
\ifnum \WIDTH < \COLUMN%
\WIDTH=\COLUMN%
\else%
\fi%
\advance\LINE by -4000%
\COLUMN=0%
\ARROW=0%
\TARGET=0%
\changecounters{\mbox{$#1$}}%
\renewcommand{\printarrow}{\mbox{$#1$}}}%

\newcommand{\conclude}{%
\put(\COLUMN,\LINE){\makebox(0,0){\printarrow}}%
\thinlines%
\ifnum \WIDTH < \COLUMN%
\WIDTH=\COLUMN%
\else%
\fi%
\setcounter{horizontal}{\WIDTH}%
\setcounter{vertical}{-\LINE}%
\end{picture}}%

\newcommand{\diag}{%
\conclude%
\raisebox{0pt}[0pt][\value{vertical}\unitlength]{}%
\hspace*{\value{horizontal}\unitlength}%
\vspace{12pt}%
\end{center}%
\setlength{\unitlength}{1pt}%
\global\diagramfalse}%

\newcommand{\diagv}[3]{%
\conclude%
\NUMBER=#1%
\rule{0pt}{\NUMBER pt}%
\hspace*{-#2pt}%
\raisebox{0pt}[0pt][\value{vertical}\unitlength]{}%
\hspace*{\value{horizontal}\unitlength}
\NUMBER=#3%
\advance\NUMBER by 12%
\vspace*{\NUMBER pt}%
\end{center}%
\setlength{\unitlength}{1pt}%
\global\diagramfalse}%

\def\movename(#1,#2)#3{%
\hspace{#1pt}%
\raisebox{#2pt}[5pt][2pt]{\raisebox{#2pt}{$#3$}}%
\hspace{-#1pt}}%

\def\movearrow(#1,#2)#3{%
\makebox[0pt]{%
\hspace{#1pt}\hspace{#1pt}%
\raisebox{#2pt}[0pt][0pt]{\raisebox{#2pt}{$#3$}}}}%

\def\movevertex(#1,#2)#3{%
\mbox{\hspace{#1pt}%
\raisebox{#2pt}{\raisebox{#2pt}{$#3$}}%
\hspace{-#1pt}}}%

\newcommand{\crosslength}[2]{%
\settowidth{\firstitem}{#1}%
\settowidth{\seconditem}{#2}%
\ifdim\firstitem < \seconditem%
\itemlength=\seconditem%
\else%
\itemlength=\firstitem%
\fi%
\divide\itemlength by 2%
\hspace{\itemlength}}%

\def\basicDIAG#1€{\DIAGV{\defaultscale}{#1}\@ifnexttok€{\finishline}{\basicn}}
\def\basicDIAGV[#1]#2€{\DIAGV{#1}{#2}\@ifnexttok€{\finishline}{\basicn}}
\def\basicn#1€{\n{#1}\@ifnexttok€{\finishline}{\basicn}}
\def\basicnn#1€{\nn{#1}\@ifnexttok€{\finishline}{\basicn}}
\def\finishline#1{\@ifnextchar\end{\diag}%
{\@ifnextchar\spacing{\relax}{\basicnn}}}
\def\spacing(#1,#2,#3){\diagv{#1}{#2}{#3}}

\newif\ifcaption%

\newenvironment{diagram}{%
\iffloatdiag\relax\else
\global\def\diagramcaption##1{%
\global\captiontrue%
\global\def\@diagcaption{##1}}%
\global\def\@diagcaption{}\fi%
\@ifnextchar[{\basicDIAGV}{\basicDIAG}}%
{\iffloatdiag\relax\else%
\ifcaption
\begin{center}\mbox{}\@diagcaption\end{center}%
\else\relax\fi\fi\global\captionfalse}

\global\captionfalse

\gdef\@diaglabel{Diagram}
\gdef\diagramlabel#1{\gdef\@diaglabel{#1}}

\newcounter{Diagram}
\def\theDiagram{\@arabic\c@Diagram}
\def\fps@Diagram{tbp}
\def\ftype@Diagram{1}
\def\ext@Diagram{lof}
\def\fnum@Diagram{\@diaglabel\ \theDiagram}
\def\Diagram{\@float{Diagram}}
\let\endDiagram\end@float
\@namedef{Diagram*}{\@dblfloat{Diagram}}
\@namedef{endDiagram*}{\end@dblfloat}

\def\setdiagramcounter#1{\@addtoreset{Diagram}{#1}%
\def\theDiagram{\arabic{#1}.\@arabic\c@Diagram}}

\newif\iffloatdiag

{\end{diagram}\caption{\@diagcaption}\end{Diagram}%
\global\floatdiagfalse}

\global\floatdiagfalse

\newcommand{\TUP}[1]{\raisebox{0pt}[0pt][3pt]{}#1}

\newcommand{\TDOWN}[1]{\raisebox{0pt}[6pt][0pt]{}#1}

\newcommand{\tlowername}[2]%
{$\stackrel{\makebox[1pt]{#1}}%
{\begin{picture}(0,0)%
\put(0,0){\makebox(0,6)[t]{\makebox[1pt]{$\scriptstyle#2$}}}%
\end{picture}}$}%

\newcommand{\tcase}[1]{%
\testtextmode%
\setlength{\unitlength}{0.01pt}%
\makebox[\actualtextarrowspace pt]%
{\raisebox{2.5pt}{#1{\actualtextarrowlength}}}%
\setlength{\unitlength}{1pt}}%

\newcommand{\Tcase}[2]{%
\testtextmode%
\setlength{\unitlength}{0.01pt}%
\makebox[\actualtextarrowspace pt]%
{\raisebox{2.5pt}{$\stackrel{\scriptstyle #2}{#1{\actualtextarrowlength}}$}}%
\setlength{\unitlength}{1pt}}%

\newcommand{\tbicase}[1]{%
\testtextmode%
\setlength{\unitlength}{0.01pt}%
\makebox[\actualtextarrowspace pt]%
{\raisebox{1pt}{#1{\actualtextarrowlength}}}%
\setlength{\unitlength}{1pt}}%

\newcommand{\Tbicase}[3]{%
\testtextmode%
\setlength{\unitlength}{0.01pt}%
\makebox[\actualtextarrowspace pt]%
{\raisebox{-1pt}%
{$\stackrel{\scriptstyle #2}%
{\mbox{\tlowername{#1{\actualtextarrowlength}}%
{\scriptstyle #3}}}$}}%
\setlength{\unitlength}{1pt}}%

\newcommand{\DUP}[1]{\raisebox{0pt}[0pt][4pt]{}#1}

\newcommand{\DDOWN}[1]{\raisebox{0pt}[9pt][0pt]{}#1}

\newcommand{\dlowername}[2]%
{$\stackrel{\makebox[1pt]{#1}}%
{\begin{picture}(0,0)%
\put(0,0){\makebox(0,6)[t]{\makebox[1pt]{$\textstyle#2$}}}%
\end{picture}}$}%

\newcommand{\dcase}[1]{%
\testtextmode%
\setlength{\unitlength}{0.01pt}%
\makebox[\actualdisplayarrowspace pt]%
{\raisebox{2.5pt}{#1{\actualdisplayarrowlength}}}%
\setlength{\unitlength}{1pt}}%

\newcommand{\Dcase}[2]{%
\testtextmode%
\setlength{\unitlength}{0.01pt}%
\makebox[\actualdisplayarrowspace pt]%
{\raisebox{2.5pt}{$\stackrel{\textstyle #2}{#1{\actualdisplayarrowlength}}$}}%
\setlength{\unitlength}{1pt}}%

\newcommand{\dbicase}[1]{%
\testtextmode%
\setlength{\unitlength}{0.01pt}%
\makebox[\actualdisplayarrowspace pt]%
{\raisebox{1pt}{#1{\actualdisplayarrowlength}}}%
\setlength{\unitlength}{1pt}}%

\newcommand{\Dbicase}[3]{%
\testtextmode%
\setlength{\unitlength}{0.01pt}%
\makebox[\actualdisplayarrowspace pt]%
{\raisebox{-1pt}%
{$\stackrel{\textstyle #2}%
{\mbox{\tlowername{#1{\actualdisplayarrowlength}}%
{\textstyle #3}}}$}}%
\setlength{\unitlength}{1pt}}%

\newcommand{\AR}[1]%
{\begin{picture}(#1,0)%
\put(0,0){\line(1,0){#1}}%
\put(#1,0){\ehead}%
\end{picture}}%

\newcommand{\DIST}[1]%
{\begin{picture}(#1,0)%
\put(0,0){\line(1,0){#1}}%
\put(#1,0){\ehead}%
\NUMBER=#1%
\divide\NUMBER by 2%
\put(\NUMBER,0){\circle{400}}%
\end{picture}}%

\newcommand{\DOTAR}[1]%
{\NUMBEROFDOTS=#1%
\divide\NUMBEROFDOTS by 300%
\advance\NUMBEROFDOTS by 1%
\begin{picture}(#1,0)%
\multiput(0,0)(300,0){\NUMBEROFDOTS}{\circle*{100}}%
\put(#1,0){\ehead}%
\end{picture}}%

\newcommand{\MONO}[1]%
{\monolength=#1%
\advance\monolength by -\monotail%
\begin{picture}(#1,0)%
\put(\monotail,0){\line(1,0){\monolength}}%
\put(#1,0){\ehead}%
\put(\monotail,0){\ehead}%
\end{picture}}%

\newcommand{\EPI}[1]%
{\epilength=#1%
\advance\epilength by -\epihead%
\begin{picture}(#1,0)(-#1,0)%
\put(-#1,0){\line(1,0){\epilength}}%
\put(-\epihead,0){\ehead}%
\put(0,0){\ehead}%
\end{picture}}%

\newcommand{\BIMO}[1]%
{\monolength=#1%
\advance\monolength by -\monotail%
\epilength=\monolength%
\advance\epilength by -\epihead%
\begin{picture}(#1,0)(-#1,0)%
\put(-\monolength,0){\line(1,0){\epilength}}%
\put(-\monolength,0){\ehead}%
\put(-\epihead,0){\ehead}%
\put(0,0){\ehead}%
\end{picture}}%

\newcommand{\BIAR}[1]%
{\begin{picture}(#1,700)%
\put(0,0){\line(1,0){#1}}%
\put(#1,0){\ehead}%
\put(0,700){\line(1,0){#1}}%
\put(#1,700){\ehead}%
\end{picture}}%

\newcommand{\BIDIST}[1]%
{\begin{picture}(#1,700)%
\put(0,0){\line(1,0){#1}}%
\put(#1,0){\ehead}%
\put(0,700){\line(1,0){#1}}%
\put(#1,700){\ehead}%
\NUMBER=#1%
\divide\NUMBER by 2%
\put(\NUMBER,0){\circle{400}}%
\put(\NUMBER,700){\circle{400}}%
\end{picture}}%

\newcommand{\EQL}[1]%
{\begin{picture}(#1,0)%
\put(0,100){\line(1,0){#1}}%
\put(0,-100){\line(1,0){#1}}%
\end{picture}}%

\newcommand{\ADJAR}[1]%
{\begin{picture}(#1,700)%
\put(0,0){\line(1,0){#1}}%
\put(#1,0){\ehead}%
\put(#1,700){\line(-1,0){#1}}%
\put(0,700){\whead}
\end{picture}}%

\newcommand{\ADJDIST}[1]%
{\begin{picture}(#1,700)%
\put(0,0){\line(1,0){#1}}%
\put(#1,0){\ehead}%
\put(#1,700){\line(-1,0){#1}}%
\put(0,700){\whead}
\NUMBER=#1%
\divide\NUMBER by 2%
\put(\NUMBER,0){\circle{400}}%
\put(\NUMBER,700){\circle{400}}%
\end{picture}}%

\newcommand{\ar}{\ifinner\tcase{\AR}\else\dcase{\AR}\fi}%

\newcommand{\Ar}[1]{\ifinner\Tcase{\AR}{#1}\else\Dcase{\AR}{#1}\fi}%

\newcommand{\dist}{\ifinner\tcase{\DIST}\else\dcase{\DIST}\fi}%

\newcommand{\Dist}[1]{\ifinner\Tcase{\DIST}{\TUP{#1}}%
\else\Dcase{\DIST}{\TUP{#1}}\fi}%

\newcommand{\dotar}{\ifinner\tcase{\DOTAR}\else\dcase{\DOTAR}\fi}%

\newcommand{\Dotar}[1]{\ifinner\Tcase{\DOTAR}{#1}%
\else\Dcase{\DOTAR}{#1}\fi}%

\newcommand{\mono}{\ifinner\tcase{\MONO}\else\dcase{\MONO}\fi}%

\newcommand{\Mono}[1]{\ifinner\Tcase{\MONO}{#1}\else\Dcase{\MONO}{#1}\fi}%

\newcommand{\epi}{\ifinner\tcase{\EPI}\else\dcase{\EPI}\fi}%

\newcommand{\Epi}[1]{\ifinner\Tcase{\EPI}{#1}\else\Dcase{\EPI}{#1}\fi}%

\newcommand{\bimo}{\ifinner\tcase{\BIMO}\else\dcase{\BIMO}\fi}%

\newcommand{\Bimo}[1]{\ifinner\Tcase{\BIMO}{#1}%
\else\Dcase{\BIMO}{#1}\fi}%

\newcommand{\iso}{\ifinner\Tcase{\AR}{\cong}\else\Dcase{\AR}{\cong}\fi}%

\newcommand{\Iso}[1]{\ifinner\Tcase{\AR}{\cong{#1}}%
\else\Dcase{\AR}{\cong{#1}}\fi}%

\newcommand{\biar}{\ifinner\tbicase{\BIAR}\else\dbicase{\BIAR}\fi}%

\newcommand{\Biar}[2]{\ifinner\Tbicase{\BIAR}{#1}{#2}%
\else\Dbicase{\BIAR}{#1}{#2}\fi}%

\newcommand{\bidist}{\ifinner\tbicase{\BIDIST}\else\dbicase{\BIDIST}\fi}%

\newcommand{\Bidist}[2]{\ifinner\Tbicase{\BIDIST}{\TUP{#1}}{\TDOWN{#2}}%
\else\Dbicase{\BIDIST}{\DUP{#1}}{\DDOWN{#2}}\fi}%

\newcommand{\eql}{\ifinner\tcase{\EQL}\else\dcase{\EQL}\fi}%

\newcommand{\Eql}[1]{\ifinner\Tcase{\EQL}{\TUP{#1}}%
\else\Dcase{\EQL}{\DUP{#1}}\fi}%

\newcommand{\adjar}{\ifinner\tbicase{\ADJAR}\else\dbicase{\ADJAR}\fi}%

\newcommand{\Adjar}[2]{\ifinner\Tbicase{\ADJAR}{#1}{#2}%
\else\Dbicase{\ADJAR}{#1}{#2}\fi}%

\newcommand{\adjdist}{\ifinner\tbicase{\ADJDIST}\else\dbicase{\ADJDIST}\fi}%

\newcommand{\Adjdist}[2]{\ifinner\Tbicase{\ADJDIST}{\TUP{#1}}{\TDOWN{#2}}%
\else\Dbicase{\ADJDIST}{\DUP{#1}}{\DDOWN{#2}}\fi}%

\newcommand{\BKAR}[1]%
{\begin{picture}(#1,0)%
\put(#1,0){\line(-1,0){#1}}%
\put(0,0){\whead}%
\end{picture}}%

\newcommand{\BKDIST}[1]%
{\begin{picture}(#1,0)%
\put(#1,0){\line(-1,0){#1}}%
\put(0,0){\whead}%
\NUMBER=#1%
\divide\NUMBER by 2%
\put(\NUMBER,0){\circle{400}}%
\end{picture}}%

\newcommand{\BKDOTAR}[1]%
{\NUMBEROFDOTS=#1%
\divide\NUMBEROFDOTS by 300%
\advance\NUMBEROFDOTS by 1%
\begin{picture}(#1,0)%
\multiput(#1,0)(-300,0){\NUMBEROFDOTS}{\circle*{100}}%
\put(0,0){\whead}%
\end{picture}}%

\newcommand{\BKMONO}[1]%
{\monolength=#1%
\advance\monolength by -\monotail%
\begin{picture}(#1,0)(-#1,0)%
\put(-\monotail,0){\line(-1,0){\monolength}}%
\put(-\monotail,0){\whead}%
\put(-#1,0){\whead}%
\end{picture}}%

\newcommand{\BKEPI}[1]%
{\epilength=#1%
\advance\epilength by -\epihead%
\begin{picture}(#1,0)%
\put(#1,0){\line(-1,0){\epilength}}%
\put(\epihead,0){\whead}%
\put(0,0){\whead}%
\end{picture}}%

\newcommand{\BKBIMO}[1]%
{\monolength=#1%
\advance\monolength by -\monotail%
\epilength=\monolength%
\advance\epilength by -\epihead%
\begin{picture}(#1,0)%
\put(\monolength,0){\line(-1,0){\epilength}}%
\put(\monolength,0){\whead}%
\put(\epihead,0){\whead}%
\put(0,0){\whead}%
\end{picture}}%

\newcommand{\BKBIAR}[1]%
{\begin{picture}(#1,700)%
\put(#1,0){\line(-1,0){#1}}%
\put(0,0){\whead}%
\put(#1,700){\line(-1,0){#1}}%
\put(0,700){\whead}%
\end{picture}}%

\newcommand{\BKBIDIST}[1]%
{\begin{picture}(#1,700)%
\put(#1,0){\line(-1,0){#1}}%
\put(0,0){\whead}%
\put(#1,700){\line(-1,0){#1}}%
\put(0,700){\whead}%
\NUMBER=#1%
\divide\NUMBER by 2%
\put(\NUMBER,0){\circle{400}}%
\put(\NUMBER,700){\circle{400}}%
\end{picture}}%

\newcommand{\BKADJAR}[1]%
{\begin{picture}(#1,700)%
\put(0,700){\line(1,0){#1}}%
\put(#1,700){\ehead}%
\put(#1,0){\line(-1,0){#1}}%
\put(0,0){\whead}%
\end{picture}}%

\newcommand{\BKADJDIST}[1]%
{\begin{picture}(#1,700)%
\put(0,700){\line(1,0){#1}}%
\put(#1,700){\ehead}%
\put(#1,0){\line(-1,0){#1}}%
\put(0,0){\whead}%
\NUMBER=#1%
\divide\NUMBER by 2%
\put(\NUMBER,0){\circle{400}}%
\put(\NUMBER,700){\circle{400}}%
\end{picture}}%

\newcommand{\bkar}{\ifinner\tcase{\BKAR}\else\dcase{\BKAR}\fi}%

\newcommand{\Bkar}[1]{\ifinner\Tcase{\BKAR}{#1}\else\Dcase{\BKAR}{#1}\fi}%

\newcommand{\bkdist}{\ifinner\tcase{\BKDIST}\else\dcase{\BKDIST}\fi}%

\newcommand{\Bkdist}[1]{\ifinner\Tcase{\BKDIST}{\TUP{#1}}%
\else\Dcase{\BKDIST}{\TUP{#1}}\fi}%

\newcommand{\bkdotar}{\ifinner\tcase{\BKDOTAR}\else\dcase{\BKDOTAR}\fi}%

\newcommand{\Bkdotar}[1]{\ifinner\Tcase{\BKDOTAR}{#1}%
\else\Dcase{\BKDOTAR}{#1}\fi}%

\newcommand{\bkmono}{\ifinner\tcase{\BKMONO}\else\dcase{\BKMONO}\fi}%

\newcommand{\Bkmono}[1]{\ifinner\Tcase{\BKMONO}{#1}%
\else\Dcase{\BKMONO}{#1}\fi}%

\newcommand{\bkepi}{\ifinner\tcase{\BKEPI}\else\dcase{\BKEPI}\fi}%

\newcommand{\Bkepi}[1]{\ifinner\Tcase{\BKEPI}{#1}%
\else\Dcase{\BKEPI}{#1}\fi}%

\newcommand{\bkbimo}{\ifinner\tcase{\BKBIMO}\else\dcase{\BKBIMO}\fi}%

\newcommand{\Bkbimo}[1]{\ifinner\Tcase{\BKBIMO}{\hspace{9pt}#1}%
\else\Dcase{\BKBIMO}{\hspace{9pt}#1}\fi}%

\newcommand{\bkiso}{\ifinner\Tcase{\BKAR}{\cong}%
\else\Dcase{\BKAR}{\cong}\fi}%

\newcommand{\Bkiso}[1]{\ifinner\Tcase{\BKAR}{\cong{#1}}%
\else\Dcase{\BKAR}{\cong{#1}}\fi}%

\newcommand{\bkbiar}{\ifinner\tbicase{\BKBIAR}\else\dbicase{\BKBIAR}\fi}%

\newcommand{\Bkbiar}[2]{\ifinner\Tbicase{\BKBIAR}{#1}{#2}%
\else\Dbicase{\BKBIAR}{#1}{#2}\fi}%

\newcommand{\bkbidist}{\ifinner\tbicase{\BKBIDIST}%
\else\dbicase{\BKBIDIST}\fi}%

\newcommand{\Bkbidist}[2]{\ifinner\Tbicase{\BKBIDIST}{\TUP{#1}}{\TDOWN{#2}}%
\else\Tbicase{\BKBIDIST}{\DUP{#1}}{\DDOWN{#2}}\fi}%

\newcommand{\bkadjar}{\ifinner\tbicase{\BKADJAR}%
\else\dbicase{\BKADJAR}\fi}%

\newcommand{\Bkadjar}[2]{\ifinner\Tbicase{\BKADJAR}{#1}{#2}%
\else\Dbicase{\BKADJAR}{#1}{#2}\fi}%

\newcommand{\bkadjdist}{\ifinner\tbicase{\BKADJDIST}%
\else\dbicase{\BKADJDIST}\fi}%

\newcommand{\Bkadjdist}[2]{\ifinner\Tbicase{\BKADJDIST}{\TUP{#1}}{\TDOWN{#2}}%
\else\Dbicase{\BKADJDIST}{\TUP{#1}}{\TDOWN{#2}}\fi}%

\newcommand{\lowername}[2]%
{$\stackrel{\makebox[1pt]{#1}}%
{\begin{picture}(0,0)%
\truex{600}%
\put(0,0){\makebox(0,\value{x})[t]{\makebox[1pt]{$#2$}}}%
\end{picture}}$}%

\newcommand{\hcase}[2]%
{\testdiagrammode\makebox[0pt]%
{\raisebox{0pt}[0pt][0pt]{#1{#2}}}}%

\newcommand{\Hcase}[3]%
{\testdiagrammode\makebox[0pt]
{\raisebox{0pt}[0pt][0pt]%
{$\stackrel{\makebox[0pt]{$\textstyle{#2}$}}{#1{#3}}$}}}%

\newcommand{\hcasE}[3]%
{\testdiagrammode\makebox[0pt]%
{\raisebox{-8pt}[0pt][0pt]%
{\lowername{#1{#3}}{#2}}}}%

\newcommand{\Hisocase}[4]%
{\testdiagrammode\makebox[0pt]
{\raisebox{-8pt}[0pt][0pt]%
{$\stackrel{\makebox[0pt]{$\textstyle{#2}$}}%
{\mbox{\lowername{#1{#4}}{#3}}}$}}}%

\newcommand{\hbicase}[2]%
{\testdiagrammode\makebox[0pt]%
{\raisebox{-2.4pt}[0pt][0pt]{#1{#2}}}}%

\newcommand{\Hbicase}[4]%
{\testdiagrammode\makebox[0pt]
{\raisebox{-10.4pt}[0pt][0pt]%
{$\stackrel{\makebox[0pt]{$\textstyle{#2}$}}%
{\mbox{\lowername{#1{#4}}{#3}}}$}}}%

\newcommand{\EAR}[1]%
{\begin{picture}(#1,0)%
\put(0,0){\line(1,0){#1}}%
\put(#1,0){\ehead}%
\end{picture}}%

\newcommand{\EDIST}[1]%
{\begin{picture}(#1,0)%
\put(0,0){\line(1,0){#1}}%
\put(#1,0){\ehead}%
\truex{400}
\NUMBER=#1%
\divide\NUMBER by 2%
\put(\NUMBER,0){\circle{\value{x}}}
\end{picture}}%

\newcommand{\EDOTAR}[1]%
{\truex{100}\truey{300}%
\NUMBEROFDOTS=#1%
\divide\NUMBEROFDOTS by \value{y}%
\advance\NUMBEROFDOTS by 1%
\begin{picture}(#1,0)%
\multiput(0,0)(\value{y},0){\NUMBEROFDOTS}%
{\circle*{\value{x}}}%
\put(#1,0){\ehead}%
\end{picture}}%

\newcommand{\EMONO}[1]%
{\truetail
\monolength=#1%
\advance\monolength by -\truemonotail%
\begin{picture}(#1,0)%
\put(\truemonotail,0){\line(1,0){\monolength}}%
\put(#1,0){\ehead}%
\put(\truemonotail,0){\ehead}%
\end{picture}}%

\newcommand{\EEPI}[1]%
{\truehead%
\epilength=#1%
\advance\epilength by -\trueepihead%
\begin{picture}(#1,0)(-#1,0)%
\put(-#1,0){\line(1,0){\epilength}}%
\put(-\trueepihead,0){\ehead}%
\put(0,0){\ehead}%
\end{picture}}%

\newcommand{\EBIMO}[1]%
{\truehead\truetail%
\monolength=#1%
\advance\monolength by -\truemonotail%
\epilength=\monolength%
\advance\epilength by -\trueepihead%
\begin{picture}(#1,0)(-#1,0)%
\put(-\monolength,0){\line(1,0){\epilength}}%
\put(-\monolength,0){\ehead}%
\put(-\trueepihead,0){\ehead}%
\put(0,0){\ehead}%
\end{picture}}%

\newcommand{\EBIAR}[1]%
{\truex{700}%
\begin{picture}(#1,\value{x})%
\put(0,0){\line(1,0){#1}}%
\put(#1,0){\ehead}%
\put(0,\value{x}){\line(1,0){#1}}%
\put(#1,\value{x}){\ehead}%
\end{picture}}%

\newcommand{\EBIDIST}[1]%
{\truex{700}%
\begin{picture}(#1,\value{x})%
\put(0,0){\line(1,0){#1}}%
\put(#1,0){\ehead}%
\put(0,\value{x}){\line(1,0){#1}}%
\put(#1,\value{x}){\ehead}%
\truey{400}%
\NUMBER=#1%
\divide\NUMBER by 2%
\put(\NUMBER,0){\circle{\value{y}}}
\put(\NUMBER,\value{x}){\circle{\value{y}}}%
\end{picture}}%

\newcommand{\EEQL}[1]%
{\begin{picture}(#1,0)%
\truex{200}%
\put(0,\value{x}){\line(1,0){#1}}%
\put(0,0){\line(1,0){#1}}%
\end{picture}}%

\newcommand{\EADJAR}[1]%
{\truex{700}%
\begin{picture}(#1,\value{x})%
\put(0,0){\line(1,0){#1}}%
\put(#1,0){\ehead}%
\put(#1,\value{x}){\line(-1,0){#1}}%
\put(0,\value{x}){\whead}%
\end{picture}}%

\newcommand{\EADJDIST}[1]%
{\truex{700}%
\begin{picture}(#1,\value{x})%
\put(0,0){\line(1,0){#1}}%
\put(#1,0){\ehead}%
\put(#1,\value{x}){\line(-1,0){#1}}%
\put(0,\value{x}){\whead}%
\truey{400}%
\NUMBER=#1%
\divide\NUMBER by 2%
\put(\NUMBER,0){\circle{\value{y}}}
\put(\NUMBER,\value{x}){\circle{\value{y}}}%
\end{picture}}%

\def\basicear[#1]{%
\Z=#1%
\multiply \Z by 100%
\hcase{\EAR}{\Z}}%

\newcommand{\ear}{\@ifnextchar[{\basicear}%
{\hspace{\SOURCE\unitlength}\basicear[\ARROWLENGTH]}}%

\def\basicEar[#1]#2{%
\Z=#1%
\multiply \Z by 100%
\Hcase{\EAR}{#2}{\Z}}%

\newcommand{\Ear}{\@ifnextchar[{\basicEar}%
{\hspace{\SOURCE\unitlength}\basicEar[\ARROWLENGTH]}}%

\def\basiceaR[#1]#2{%
\Z=#1%
\multiply \Z by 100%
\hcasE{\EAR}{#2}{\Z}}%

\newcommand{\eaR}{\@ifnextchar[{\basiceaR}%
{\hspace{\SOURCE\unitlength}\basiceaR[\ARROWLENGTH]}}%

\def\basicedist[#1]{%
\Z=#1%
\multiply \Z by 100%
\hcase{\EDIST}{\Z}}%

\newcommand{\edist}{\@ifnextchar[{\basicedist}%
{\hspace{\SOURCE\unitlength}\basicedist[\ARROWLENGTH]}}%

\def\basicEdist[#1]#2{%
\Z=#1%
\multiply \Z by 100%
\Hcase{\EDIST}{\DUP{#2}}{\Z}}%

\newcommand{\Edist}{\@ifnextchar[{\basicEdist}%
{\hspace{\SOURCE\unitlength}\basicEdist[\ARROWLENGTH]}}%

\def\basicedisT[#1]#2{%
\Z=#1%
\multiply \Z by 100%
\hcasE{\EDIST}{\DDOWN{#2}}{\Z}}%

\newcommand{\edisT}{\@ifnextchar[{\basicedisT}%
{\hspace{\SOURCE\unitlength}\basicedisT[\ARROWLENGTH]}}%

\def\basicedotar[#1]{%
\Z=#1%
\multiply \Z by 100%
\hcase{\EDOTAR}{\Z}}%

\newcommand{\edotar}{\@ifnextchar[{\basicedotar}%
{\hspace{\SOURCE\unitlength}\basicedotar[\ARROWLENGTH]}}%

\def\basicEdotar[#1]#2{%
\Z=#1%
\multiply \Z by 100%
\Hcase{\EDOTAR}{#2}{\Z}}%

\newcommand{\Edotar}{\@ifnextchar[{\basicEdotar}%
{\hspace{\SOURCE\unitlength}\basicEdotar[\ARROWLENGTH]}}%

\def\basicedotaR[#1]#2{%
\Z=#1%
\multiply \Z by 100%
\hcasE{\EDOTAR}{#2}{\Z}}%

\newcommand{\edotaR}{\@ifnextchar[{\basicedotaR}%
{\hspace{\SOURCE\unitlength}\basicedotaR[\ARROWLENGTH]}}%

\def\basicemono[#1]{%
\Z=#1%
\multiply \Z by 100%
\hcase{\EMONO}{\Z}}%

\newcommand{\emono}{\@ifnextchar[{\basicemono}%
{\hspace{\SOURCE\unitlength}\basicemono[\ARROWLENGTH]}}%

\def\basicEmono[#1]#2{%
\Z=#1%
\multiply \Z by 100%
\Hcase{\EMONO}{#2}{\Z}}%

\newcommand{\Emono}{\@ifnextchar[{\basicEmono}%
{\hspace{\SOURCE\unitlength}\basicEmono[\ARROWLENGTH]}}%

\def\basicemonO[#1]#2{%
\Z=#1%
\multiply \Z by 100%
\hcasE{\EMONO}{#2}{\Z}}%

\newcommand{\emonO}{\@ifnextchar[{\basicemonO}%
{\hspace{\SOURCE\unitlength}\basicemonO[\ARROWLENGTH]}}%

\def\basiceepi[#1]{%
\Z=#1%
\multiply \Z by 100%
\hcase{\EEPI}{\Z}}%

\newcommand{\eepi}{\@ifnextchar[{\basiceepi}%
{\hspace{\SOURCE\unitlength}\basiceepi[\ARROWLENGTH]}}%

\def\basicEepi[#1]#2{%
\Z=#1%
\multiply \Z by 100%
\Hcase{\EEPI}{#2}{\Z}}%

\newcommand{\Eepi}{\@ifnextchar[{\basicEepi}%
{\hspace{\SOURCE\unitlength}\basicEepi[\ARROWLENGTH]}}%

\def\basiceepI[#1]#2{%
\Z=#1%
\multiply \Z by 100%
\hcasE{\EEPI}{#2}{\Z}}%

\newcommand{\eepI}{\@ifnextchar[{\basiceepI}%
{\hspace{\SOURCE\unitlength}\basiceepI[\ARROWLENGTH]}}%

\def\basicebimo[#1]{%
\Z=#1%
\multiply \Z by 100%
\hcase{\EBIMO}{\Z}}%

\newcommand{\ebimo}{\@ifnextchar[{\basicebimo}%
{\hspace{\SOURCE\unitlength}\basicebimo[\ARROWLENGTH]}}%

\def\basicEbimo[#1]#2{%
\Z=#1%
\multiply \Z by 100%
\Hcase{\EBIMO}{#2}{\Z}}%

\newcommand{\Ebimo}{\@ifnextchar[{\basicEbimo}%
{\hspace{\SOURCE\unitlength}\basicEbimo[\ARROWLENGTH]}}%

\def\basicebimO[#1]#2{%
\Z=#1%
\multiply \Z by 100%
\hcasE{\EBIMO}{#2}{\Z}}%

\newcommand{\ebimO}{\@ifnextchar[{\basicebimO}%
{\hspace{\SOURCE\unitlength}\basicebimO[\ARROWLENGTH]}}%

\def\basiceiso[#1]{%
\Z=#1%
\multiply \Z by 100%
\Hisocase{\EAR}{\cong}{}{\Z}}%

\newcommand{\eiso}{\@ifnextchar[{\basiceiso}%
{\hspace{\SOURCE\unitlength}\basiceiso[\ARROWLENGTH]}}%

\def\basicEiso[#1]#2{%
\Z=#1%
\multiply \Z by 100%
\Hisocase{\EAR}{#2}{\cong}{\Z}}%

\newcommand{\Eiso}{\@ifnextchar[{\basicEiso}%
{\hspace{\SOURCE\unitlength}\basicEiso[\ARROWLENGTH]}}%

\def\basiceisO[#1]#2{%
\Z=#1%
\multiply \Z by 100%
\Hisocase{\EAR}{\cong}{#2}{\Z}}%

\newcommand{\eisO}{\@ifnextchar[{\basiceisO}%
{\hspace{\SOURCE\unitlength}\basiceisO[\ARROWLENGTH]}}%

\def\basiceeql[#1]{%
\Z=#1%
\multiply \Z by 100%
\hcase{\EEQL}{\Z}}%

\newcommand{\eeql}{\@ifnextchar[{\basiceeql}%
{\hspace{\SOURCE\unitlength}\basiceeql[\ARROWLENGTH]}}%

\def\basicEeql[#1]#2{%
\Z=#1%
\multiply \Z by 100%
\Hcase{\EEQL}{\DUP{#2}}{\Z}}%

\newcommand{\Eeql}{\@ifnextchar[{\basicEeql}%
{\hspace{\SOURCE\unitlength}\basicEeql[\ARROWLENGTH]}}%

\def\basiceeqL[#1]#2{%
\Z=#1%
\multiply \Z by 100%
\hcasE{\EEQL}{#2}{\Z}}%

\newcommand{\eeqL}{\@ifnextchar[{\basiceeqL}%
{\hspace{\SOURCE\unitlength}\basiceeqL[\ARROWLENGTH]}}%

\def\basicebiar[#1]{%
\Z=#1%
\multiply \Z by 100%
\hbicase{\EBIAR}{\Z}}%

\newcommand{\ebiar}{\@ifnextchar[{\basicebiar}%
{\hspace{\SOURCE\unitlength}\basicebiar[\ARROWLENGTH]}}%

\def\basicEbiar[#1]#2#3{%
\Z=#1%
\multiply \Z by 100%
\Hbicase{\EBIAR}{#2}{#3}{\Z}}%

\newcommand{\Ebiar}{\@ifnextchar[{\basicEbiar}%
{\hspace{\SOURCE\unitlength}\basicEbiar[\ARROWLENGTH]}}%

\def\basicebidist[#1]{%
\Z=#1%
\multiply \Z by 100%
\hbicase{\EBIDIST}{\Z}}%

\newcommand{\ebidist}{\@ifnextchar[{\basicebidist}%
{\hspace{\SOURCE\unitlength}\basicebidist[\ARROWLENGTH]}}%

\def\basicEbidist[#1]#2#3{%
\Z=#1%
\multiply \Z by 100%
\Hbicase{\EBIDIST}{\DUP{#2}}{\DDOWN{#3}}{\Z}}%

\newcommand{\Ebidist}{\@ifnextchar[{\basicEbidist}%
{\hspace{\SOURCE\unitlength}\basicEbidist[\ARROWLENGTH]}}%

\def\basiceadjar[#1]{%
\Z=#1%
\multiply \Z by 100%
\hbicase{\EADJAR}{\Z}}%

\newcommand{\eadjar}{\@ifnextchar[{\basiceadjar}%
{\hspace{\SOURCE\unitlength}\basiceadjar[\ARROWLENGTH]}}%

\def\basicEadjar[#1]#2#3{%
\Z=#1%
\multiply \Z by 100%
\Hbicase{\EADJAR}{#2}{#3}{\Z}}%

\newcommand{\Eadjar}{\@ifnextchar[{\basicEadjar}%
{\hspace{\SOURCE\unitlength}\basicEadjar[\ARROWLENGTH]}}%

\def\basiceadjdist[#1]{%
\Z=#1%
\multiply \Z by 100%
\hbicase{\EADJDIST}{\Z}}%

\newcommand{\eadjdist}{\@ifnextchar[{\basiceadjdist}%
{\hspace{\SOURCE\unitlength}\basiceadjdist[\ARROWLENGTH]}}%

\def\basicEadjdist[#1]#2#3{%
\Z=#1%
\multiply \Z by 100%
\Hbicase{\EADJDIST}{\DUP{#2}}{\DDOWN{#3}}{\Z}}%

\newcommand{\Eadjdist}{\@ifnextchar[{\basicEadjdist}%
{\hspace{\SOURCE\unitlength}\basicEadjdist[\ARROWLENGTH]}}%

\newcommand{\WAR}[1]%
{\begin{picture}(#1,0)%
\put(#1,0){\line(-1,0){#1}}%
\put(0,0){\whead}%
\end{picture}}%

\newcommand{\WDIST}[1]%
{\begin{picture}(#1,0)%
\put(#1,0){\line(-1,0){#1}}%
\put(0,0){\whead}%
\truex{400}%
\NUMBER=#1%
\divide\NUMBER by 2%
\put(\NUMBER,0){\circle{\value{x}}}%
\end{picture}}%

\newcommand{\WDOTAR}[1]%
{\truex{100}\truey{300}%
\NUMBEROFDOTS=#1%
\divide\NUMBEROFDOTS by \value{y}%
\advance\NUMBEROFDOTS by 1%
\begin{picture}(#1,0)%
\multiput(#1,0)(-\value{y},0){\NUMBEROFDOTS}%
{\circle*{\value{x}}}%
\put(0,0){\whead}%
\end{picture}}%

\newcommand{\WMONO}[1]%
{\truetail%
\monolength=#1%
\advance\monolength by -\truemonotail%
\begin{picture}(#1,0)(-#1,0)%
\put(-\truemonotail,0){\line(-1,0){\monolength}}%
\put(-\truemonotail,0){\whead}%
\put(-#1,0){\whead}%
\end{picture}}%

\newcommand{\WEPI}[1]%
{\truehead%
\epilength=#1%
\advance\epilength by -\trueepihead%
\begin{picture}(#1,0)%
\put(#1,0){\line(-1,0){\epilength}}%
\put(\trueepihead,0){\whead}%
\put(0,0){\whead}%
\end{picture}}%

\newcommand{\WBIMO}[1]%
{\truehead\truetail%
\monolength=#1
\advance\monolength by -\truemonotail%
\epilength=\monolength%
\advance\epilength by -\trueepihead%
\begin{picture}(#1,0)%
\put(\monolength,0){\line(-1,0){\epilength}}%
\put(\monolength,0){\whead}%
\put(\trueepihead,0){\whead}%
\put(0,0){\whead}%
\end{picture}}%

\newcommand{\WBIAR}[1]%
{\truex{700}%
\begin{picture}(#1,\value{x})%
\put(#1,0){\line(-1,0){#1}}%
\put(0,0){\whead}%
\put(#1,\value{x}){\line(-1,0){#1}}%
\put(0,\value{x}){\whead}%
\end{picture}}%

\newcommand{\WBIDIST}[1]%
{\truex{700}%
\begin{picture}(#1,\value{x})%
\put(#1,0){\line(-1,0){#1}}%
\put(0,0){\whead}%
\put(#1,\value{x}){\line(-1,0){#1}}%
\put(0,\value{x}){\whead}%
\truey{400}%
\NUMBER=#1%
\divide\NUMBER by 2%
\put(\NUMBER,0){\circle{\value{y}}}%
\put(\NUMBER,\value{x}){\circle{\value{y}}}%
\end{picture}}%

\newcommand{\WADJAR}[1]%
{\truex{700}%
\begin{picture}(#1,\value{x})%
\put(0,\value{x}){\line(1,0){#1}}%
\put(#1,\value{x}){\ehead}%
\put(#1,0){\line(-1,0){#1}}%
\put(0,0){\whead}%
\end{picture}}%

\newcommand{\WADJDIST}[1]%
{\truex{700}%
\begin{picture}(#1,\value{x})%
\put(0,\value{x}){\line(1,0){#1}}%
\put(#1,\value{x}){\ehead}%
\put(#1,0){\line(-1,0){#1}}%
\put(0,0){\whead}%
\truey{400}%
\NUMBER=#1%
\divide\NUMBER by 2%
\put(\NUMBER,0){\circle{\value{y}}}%
\put(\NUMBER,\value{x}){\circle{\value{y}}}%
\end{picture}}%

\def\basicwar[#1]{%
\Z=#1%
\multiply \Z by 100%
\hcase{\WAR}{\Z}}%

\newcommand{\war}{\@ifnextchar[{\basicwar}%
{\hspace{\SOURCE\unitlength}\basicwar[\ARROWLENGTH]}}%

\def\basicWar[#1]#2{%
\Z=#1%
\multiply \Z by 100%
\Hcase{\WAR}{#2}{\Z}}%

\newcommand{\War}{\@ifnextchar[{\basicWar}%
{\hspace{\SOURCE\unitlength}\basicWar[\ARROWLENGTH]}}%

\def\basicwaR[#1]#2{%
\Z=#1%
\multiply \Z by 100%
\hcasE{\WAR}{#2}{\Z}}%

\newcommand{\waR}{\@ifnextchar[{\basicwaR}%
{\hspace{\SOURCE\unitlength}\basicwaR[\ARROWLENGTH]}}%

\def\basicwdist[#1]{%
\Z=#1%
\multiply \Z by 100%
\hcase{\WDIST}{\Z}}%

\newcommand{\wdist}{\@ifnextchar[{\basicwdist}%
{\hspace{\SOURCE\unitlength}\basicwdist[\ARROWLENGTH]}}%

\def\basicWdist[#1]#2{%
\Z=#1%
\multiply \Z by 100%
\Hcase{\WDIST}{\DUP{#2}}{\Z}}%

\newcommand{\Wdist}{\@ifnextchar[{\basicWdist}%
{\hspace{\SOURCE\unitlength}\basicWdist[\ARROWLENGTH]}}%

\def\basicwdisT[#1]#2{%
\Z=#1%
\multiply \Z by 100%
\hcasE{\WDIST}{\DDOWN{#2}}{\Z}}%

\newcommand{\wdisT}{\@ifnextchar[{\basicwdisT}%
{\hspace{\SOURCE\unitlength}\basicwdisT[\ARROWLENGTH]}}%

\def\basicwdotar[#1]{%
\Z=#1%
\multiply \Z by 100%
\hcase{\WDOTAR}{\Z}}%

\newcommand{\wdotar}{\@ifnextchar[{\basicwdotar}%
{\hspace{\SOURCE\unitlength}\basicwdotar[\ARROWLENGTH]}}%

\def\basicWdotar[#1]#2{%
\Z=#1%
\multiply \Z by 100%
\Hcase{\WDOTAR}{#2}{\Z}}%

\newcommand{\Wdotar}{\@ifnextchar[{\basicWdotar}%
{\hspace{\SOURCE\unitlength}\basicWdotar[\ARROWLENGTH]}}%

\def\basicwdotaR[#1]#2{%
\Z=#1%
\multiply \Z by 100%
\hcasE{\WDOTAR}{#2}{\Z}}%

\newcommand{\wdotaR}{\@ifnextchar[{\basicwdotaR}%
{\hspace{\SOURCE\unitlength}\basicwdotaR[\ARROWLENGTH]}}%

\def\basicwmono[#1]{%
\Z=#1%
\multiply \Z by 100%
\hcase{\WMONO}{\Z}}%

\newcommand{\wmono}{\@ifnextchar[{\basicwmono}%
{\hspace{\SOURCE\unitlength}\basicwmono[\ARROWLENGTH]}}%

\def\basicWmono[#1]#2{%
\Z=#1%
\multiply \Z by 100%
\Hcase{\WMONO}{#2}{\Z}}%

\newcommand{\Wmono}{\@ifnextchar[{\basicWmono}%
{\hspace{\SOURCE\unitlength}\basicWmono[\ARROWLENGTH]}}%

\def\basicwmonO[#1]#2{%
\Z=#1%
\multiply \Z by 100%
\hcasE{\WMONO}{#2}{\Z}}%

\newcommand{\wmonO}{\@ifnextchar[{\basicwmonO}%
{\hspace{\SOURCE\unitlength}\basicwmonO[\ARROWLENGTH]}}%

\def\basicwepi[#1]{%
\Z=#1%
\multiply \Z by 100%
\hcase{\WEPI}{\Z}}%

\newcommand{\wepi}{\@ifnextchar[{\basicwepi}%
{\hspace{\SOURCE\unitlength}\basicwepi[\ARROWLENGTH]}}%

\def\basicWepi[#1]#2{%
\Z=#1%
\multiply \Z by 100%
\Hcase{\WEPI}{#2}{\Z}}%

\newcommand{\Wepi}{\@ifnextchar[{\basicWepi}%
{\hspace{\SOURCE\unitlength}\basicWepi[\ARROWLENGTH]}}%

\def\basicwepI[#1]#2{%
\Z=#1%
\multiply \Z by 100%
\hcasE{\WEPI}{#2}{\Z}}%

\newcommand{\wepI}{\@ifnextchar[{\basicwepI}%
{\hspace{\SOURCE\unitlength}\basicwepI[\ARROWLENGTH]}}%

\def\basicwbimo[#1]{%
\Z=#1%
\multiply \Z by 100%
\hcase{\WBIMO}{\Z}}%

\newcommand{\wbimo}{\@ifnextchar[{\basicwbimo}%
{\hspace{\SOURCE\unitlength}\basicwbimo[\ARROWLENGTH]}}%

\def\basicWbimo[#1]#2{%
\Z=#1%
\multiply \Z by 100%
\Hcase{\WBIMO}{#2}{\Z}}%

\newcommand{\Wbimo}{\@ifnextchar[{\basicWbimo}%
{\hspace{\SOURCE\unitlength}\basicWbimo[\ARROWLENGTH]}}%

\def\basicwbimO[#1]#2{%
\Z=#1%
\multiply \Z by 100%
\hcasE{\WBIMO}{#2}{\Z}}%

\newcommand{\wbimO}{\@ifnextchar[{\basicwbimO}%
{\hspace{\SOURCE\unitlength}\basicwbimO[\ARROWLENGTH]}}%

\def\basicwiso[#1]{%
\Z=#1%
\multiply \Z by 100%
\Hisocase{\WAR}{\cong}{}{\Z}}%

\newcommand{\wiso}{\@ifnextchar[{\basicwiso}%
{\hspace{\SOURCE\unitlength}\basicwiso[\ARROWLENGTH]}}%

\def\basicWiso[#1]#2{%
\Z=#1%
\multiply \Z by 100%
\Hisocase{\WAR}{#2}{\cong}{\Z}}%

\newcommand{\Wiso}{\@ifnextchar[{\basicWiso}%
{\hspace{\SOURCE\unitlength}\basicWiso[\ARROWLENGTH]}}%

\def\basicwisO[#1]#2{%
\Z=#1%
\multiply \Z by 100%
\Hisocase{\WAR}{\cong}{#2}{\Z}}%

\newcommand{\wisO}{\@ifnextchar[{\basicwisO}%
{\hspace{\SOURCE\unitlength}\basicwisO[\ARROWLENGTH]}}%

\def\basicwbiar[#1]{%
\Z=#1%
\multiply \Z by 100%
\hbicase{\WBIAR}{\Z}}%

\newcommand{\wbiar}{\@ifnextchar[{\basicwbiar}%
{\hspace{\SOURCE\unitlength}\basicwbiar[\ARROWLENGTH]}}%

\def\basicWbiar[#1]#2#3{%
\Z=#1%
\multiply \Z by 100%
\Hbicase{\WBIAR}{#2}{#3}{\Z}}%

\newcommand{\Wbiar}{\@ifnextchar[{\basicWbiar}%
{\hspace{\SOURCE\unitlength}\basicWbiar[\ARROWLENGTH]}}%

\def\basicwbidist[#1]{%
\Z=#1%
\multiply \Z by 100%
\hbicase{\WBIDIST}{\Z}}%

\newcommand{\wbidist}{\@ifnextchar[{\basicwbidist}%
{\hspace{\SOURCE\unitlength}\basicwbidist[\ARROWLENGTH]}}%

\def\basicWbidist[#1]#2#3{%
\Z=#1%
\multiply \Z by 100%
\Hbicase{\WBIDIST}{\DUP{#2}}{\DDOWN{#3}}{\Z}}%

\newcommand{\Wbidist}{\@ifnextchar[{\basicWbidist}%
{\hspace{\SOURCE\unitlength}\basicWbidist[\ARROWLENGTH]}}%

\def\basicwadjar[#1]{%
\Z=#1%
\multiply \Z by 100%
\hbicase{\WADJAR}{\Z}}%

\newcommand{\wadjar}{\@ifnextchar[{\basicwadjar}%
{\hspace{\SOURCE\unitlength}\basicwadjar[\ARROWLENGTH]}}%

\def\basicWadjar[#1]#2#3{%
\Z=#1%
\multiply \Z by 100%
\Hbicase{\WADJAR}{#2}{#3}{\Z}}%

\newcommand{\Wadjar}{\@ifnextchar[{\basicWadjar}%
{\hspace{\SOURCE\unitlength}\basicWadjar[\ARROWLENGTH]}}%

\def\basicwadjdist[#1]{%
\Z=#1%
\multiply \Z by 100%
\hbicase{\WADJDIST}{\Z}}%

\newcommand{\wadjdist}{\@ifnextchar[{\basicwadjdist}%
{\hspace{\SOURCE\unitlength}\basicwadjdist[\ARROWLENGTH]}}%

\def\basicWadjdist[#1]#2#3{%
\Z=#1%
\multiply \Z by 100%
\Hbicase{\WADJDIST}{\DUP{#2}}{\DDOWN{#3}}{\Z}}%

\newcommand{\Wadjdist}{\@ifnextchar[{\basicWadjdist}%
{\hspace{\SOURCE\unitlength}\basicWadjdist[\ARROWLENGTH]}}%

\newcommand{\vcase}[2]{\testdiagrammode#1{#2}}%

\newcommand{\Vcase}[3]{\testdiagrammode\makebox[0pt]%
{\makebox[0pt][r]{\raisebox{0pt}[0pt][0pt]{${#2}\hspace{2pt}$}}}#1{#3}}%

\newcommand{\vcasE}[3]{\testdiagrammode\makebox[0pt]%
{#1{#3}\makebox[0pt][l]{\raisebox{0pt}[0pt][0pt]{\hspace{2pt}$#2$}}}}%

\newcommand{\Visocase}[4]{\testdiagrammode\makebox[0pt]%
{\makebox[0pt][r]{\raisebox{0pt}[0pt][0pt]{$#2$\hspace{2pt}}}#1{#4}%
\makebox[0pt][l]{\raisebox{0pt}[0pt][0pt]{\hspace{2pt}$#3$}}}}%
\newcommand{\vbicase}[2]{\testdiagrammode\makebox[0pt]{{#1{#2}}}}%

\newcommand{\Vbicase}[4]{\testdiagrammode\makebox[0pt]%
{\makebox[0pt][r]{\raisebox{0pt}[0pt][0pt]{$#2$\hspace{5.5pt}}}#1{#4}%
\makebox[0pt][l]{\raisebox{0pt}[0pt][0pt]{\hspace{6.5pt}$#3$}}}}%

\newcommand{\SAR}[1]%
{\begin{picture}(0,0)%
\put(0,0){\makebox(0,0)%
{\begin{picture}(0,#1)%
\put(0,#1){\line(0,-1){#1}}%
\put(0,0){\shead}%
\end{picture}}}\end{picture}}%

\newcommand{\SDIST}[1]%
{\begin{picture}(0,0)%
\put(0,0){\makebox(0,0)%
{\begin{picture}(0,#1)%
\put(0,#1){\line(0,-1){#1}}%
\put(0,0){\shead}%
\end{picture}}}%
\truex{400}%
\put(0,0){\circle{\value{x}}}%
\end{picture}}%

\newcommand{\SDOTAR}[1]%
{\truex{100}\truey{300}%
\NUMBEROFDOTS=#1%
\divide\NUMBEROFDOTS by \value{y}%
\advance\NUMBEROFDOTS by 1%
\begin{picture}(0,0)%
\put(0,0){\makebox(0,0)%
{\begin{picture}(0,#1)%
\multiput(0,#1)(0,-\value{y}){\NUMBEROFDOTS}%
{\circle*{\value{x}}}%
\put(0,0){\shead}%
\end{picture}}}\end{picture}}%

\newcommand{\SMONO}[1]%
{\truetail%
\monolength=#1%
\advance\monolength by -\truemonotail%
\begin{picture}(0,0)%
\put(0,0){\makebox(0,0)%
{\begin{picture}(0,#1)%
\put(0,\monolength){\line(0,-1){\monolength}}%
\put(0,\monolength){\shead}%
\put(0,0){\shead}%
\end{picture}}}\end{picture}}%

\newcommand{\SEPI}[1]%
{\truehead%
\epilength=#1%
\advance\epilength by -\trueepihead%
\begin{picture}(0,0)%
\put(0,0){\makebox(0,0)%
{\begin{picture}(0,#1)%
\put(0,#1){\line(0,-1){\epilength}}%
\put(0,\trueepihead){\shead}%
\put(0,0){\shead}%
\end{picture}}}\end{picture}}%

\newcommand{\SBIMO}[1]%
{\truehead\truetail%
\monolength=#1%
\advance\monolength by -\truemonotail%
\epilength=\monolength%
\advance\epilength by -\trueepihead%
\begin{picture}(0,0)%
\put(0,0){\makebox(0,0)%
{\begin{picture}(0,#1)%
\put(0,\monolength){\line(0,-1){\epilength}}%
\put(0,\monolength){\shead}%
\put(0,\trueepihead){\shead}%
\put(0,0){\shead}%
\end{picture}}}\end{picture}}%

\newcommand{\SBIAR}[1]%
{\begin{picture}(0,0)%
\truex{350}%
\put(0,0){\makebox(0,0)%
{\begin{picture}(0,#1)%
\put(-\value{x},#1){\line(0,-1){#1}}%
\put(-\value{x},0){\shead}%
\put(\value{x},#1){\line(0,-1){#1}}%
\put(\value{x},0){\shead}%
\end{picture}}}\end{picture}}%

\newcommand{\SBIDIST}[1]%
{\begin{picture}(0,0)%
\truex{350}%
\put(0,0){\makebox(0,0)%
{\begin{picture}(0,#1)%
\put(-\value{x},#1){\line(0,-1){#1}}%
\put(-\value{x},0){\shead}%
\put(\value{x},#1){\line(0,-1){#1}}%
\put(\value{x},0){\shead}%
\end{picture}}}%
\truey{400}%
\put(-\value{x},0){\circle{\value{y}}}%
\put(\value{x},0){\circle{\value{y}}}%
\end{picture}}%

\newcommand{\SEQL}[1]%
{\begin{picture}(0,0)%
\truex{100}%
\put(0,0){\makebox(0,0)%
{\begin{picture}(0,#1)\put(-\value{x},#1){\line(0,-1){#1}}%
\put(\value{x},#1){\line(0,-1){#1}}%
\end{picture}}}\end{picture}}%

\newcommand{\SADJAR}[1]{\begin{picture}(0,0)%
\truex{350}%
\put(0,0){\makebox(0,0)%
{\begin{picture}(0,#1)%
\put(-\value{x},#1){\line(0,-1){#1}}%
\put(-\value{x},0){\shead}%
\put(\value{x},0){\line(0,1){#1}}%
\put(\value{x},#1){\nhead}%
\end{picture}}}\end{picture}}%

\newcommand{\SADJDIST}[1]{\begin{picture}(0,0)%
\truex{350}%
\put(0,0){\makebox(0,0)%
{\begin{picture}(0,#1)%
\put(-\value{x},#1){\line(0,-1){#1}}%
\put(-\value{x},0){\shead}%
\put(\value{x},0){\line(0,1){#1}}%
\put(\value{x},#1){\nhead}%
\end{picture}}}%
\truey{400}%
\put(-\value{x},0){\circle{\value{y}}}%
\put(\value{x},0){\circle{\value{y}}}%
\end{picture}}%

\def\basicsar[#1]{\vcase{\SAR}{#100}}%

\newcommand{\sar}{\@ifnextchar[{\basicsar}{\basicsar[50]}}%

\def\basicSar[#1]#2{\Vcase{\SAR}{#2}{#100}}%

\newcommand{\Sar}{\@ifnextchar[{\basicSar}{\basicSar[50]}}%

\def\basicsaR[#1]#2{\vcasE{\SAR}{#2}{#100}}%

\newcommand{\saR}{\@ifnextchar[{\basicsaR}{\basicsaR[50]}}%

\def\basicsdist[#1]{\vcase{\SDIST}{#100}}%

\newcommand{\sdist}{\@ifnextchar[{\basicsdist}{\basicsdist[50]}}%

\def\basicSdist[#1]#2{\Vcase{\SDIST}{#2\hspace*{2pt}}{#100}}%

\newcommand{\Sdist}{\@ifnextchar[{\basicSdist}{\basicSdist[50]}}%

\def\basicsdisT[#1]#2{\vcasE{\SDIST}{\hspace*{2pt}#2}{#100}}%

\newcommand{\sdisT}{\@ifnextchar[{\basicsdisT}{\basicsdisT[50]}}%

\def\basicsdotar[#1]{\vcase{\SDOTAR}{#100}}%

\newcommand{\sdotar}{\@ifnextchar[{\basicsdotar}{\basicsdotar[50]}}%

\def\basicSdotar[#1]#2{\Vcase{\SDOTAR}{#2}{#100}}%

\newcommand{\Sdotar}{\@ifnextchar[{\basicSdotar}{\basicSdotar[50]}}%

\def\basicsdotaR[#1]#2{\vcasE{\SDOTAR}{#2}{#100}}%

\newcommand{\sdotaR}{\@ifnextchar[{\basicsdotaR}{\basicsdotaR[50]}}%

\def\basicsmono[#1]{\vcase{\SMONO}{#100}}%

\newcommand{\smono}{\@ifnextchar[{\basicsmono}{\basicsmono[50]}}%

\def\basicSmono[#1]#2{\Vcase{\SMONO}{#2}{#100}}%

\newcommand{\Smono}{\@ifnextchar[{\basicSmono}{\basicSmono[50]}}%

\def\basicsmonO[#1]#2{\vcasE{\SMONO}{#2}{#100}}%

\newcommand{\smonO}{\@ifnextchar[{\basicsmonO}{\basicsmonO[50]}}%

\def\basicsepi[#1]{\vcase{\SEPI}{#100}}%

\newcommand{\sepi}{\@ifnextchar[{\basicsepi}{\basicsepi[50]}}%

\def\basicSepi[#1]#2{\Vcase{\SEPI}{#2}{#100}}%

\newcommand{\Sepi}{\@ifnextchar[{\basicSepi}{\basicSepi[50]}}%

\def\basicsepI[#1]#2{\vcasE{\SEPI}{#2}{#100}}%

\newcommand{\sepI}{\@ifnextchar[{\basicsepI}{\basicsepI[50]}}%

\def\basicsbimo[#1]{\vcase{\SBIMO}{#100}}%

\newcommand{\sbimo}{\@ifnextchar[{\basicsbimo}{\basicsbimo[50]}}%

\def\basicSbimo[#1]#2{\Vcase{\SBIMO}{#2}{#100}}%

\newcommand{\Sbimo}{\@ifnextchar[{\basicSbimo}{\basicSbimo[50]}}%

\def\basicsbimO[#1]#2{\vcasE{\SBIMO}{#2}{#100}}%

\newcommand{\sbimO}{\@ifnextchar[{\basicsbimO}{\basicsbimO[50]}}%

\def\basicsiso[#1]{\Visocase{\SAR}{\cong}{}{#100}}%

\newcommand{\siso}{\@ifnextchar[{\basicsiso}{\basicsiso[50]}}%

\def\basicSiso[#1]#2{\Visocase{\SAR}{#2}{\cong}{#100}}%

\newcommand{\Siso}{\@ifnextchar[{\basicSiso}{\basicSiso[50]}}%

\def\basicsisO[#1]#2{\Visocase{\SAR}{\cong}{#2}{#100}}%

\newcommand{\sisO}{\@ifnextchar[{\basicsisO}{\basicsisO[50]}}%

\def\basicseql[#1]{\vcase{\SEQL}{#100}}%

\newcommand{\seql}{\@ifnextchar[{\basicseql}{\basicseql[50]}}%

\def\basicSeql[#1]#2{\Vcase{\SEQL}{#2\hspace*{2pt}}{#100}}%

\newcommand{\Seql}{\@ifnextchar[{\basicSeql}{\basicSeql[50]}}%

\def\basicseqL[#1]#2{\vcasE{\SEQL}{\hspace*{2pt}#2}{#100}}%

\newcommand{\seqL}{\@ifnextchar[{\basicseqL}{\basicseqL[50]}}%

\def\basicsbiar[#1]{\vbicase{\SBIAR}{#100}}%

\newcommand{\sbiar}{\@ifnextchar[{\basicsbiar}{\basicsbiar[50]}}%

\def\basicSbiar[#1]#2#3{\Vbicase{\SBIAR}{#2}{#3}{#100}}%

\newcommand{\Sbiar}{\@ifnextchar[{\basicSbiar}{\basicSbiar[50]}}%

\def\basicsbidist[#1]{\vbicase{\SBIDIST}{#100}}%

\newcommand{\sbidist}{\@ifnextchar[{\basicsbidist}{\basicsbidist[50]}}%

\def\basicSbidist[#1]#2#3%
{\Vbicase{\SBIDIST}{#2\hspace*{2pt}}{\hspace*{2pt}#3}{#100}}%

\newcommand{\Sbidist}{\@ifnextchar[{\basicSbidist}{\basicSbidist[50]}}%

\def\basicsadjar[#1]{\vbicase{\SADJAR}{#100}}%

\newcommand{\sadjar}{\@ifnextchar[{\basicsadjar}{\basicsadjar[50]}}%

\def\basicSadjar[#1]#2#3{\Vbicase{\SADJAR}{#2}{#3}{#100}}%

\newcommand{\Sadjar}{\@ifnextchar[{\basicSadjar}{\basicSadjar[50]}}%

\def\basicsadjdist[#1]{\vbicase{\SADJDIST}{#100}}%

\newcommand{\sadjdist}{\@ifnextchar[{\basicsadjdist}{\basicsadjdist[50]}}%

\def\basicSadjdist[#1]#2#3%
{\Vbicase{\SADJDIST}{#2\hspace*{2pt}}{\hspace*{2pt}#3}{#100}}%

\newcommand{\Sadjdist}{\@ifnextchar[{\basicSadjdist}{\basicSadjdist[50]}}%

\newcommand{\NAR}[1]%
{\begin{picture}(0,0)%
\put(0,0){\makebox(0,0)%
{\begin{picture}(0,#1)%
\put(0,0){\line(0,1){#1}}%
\put(0,#1){\nhead}%
\end{picture}}}\end{picture}}%

\newcommand{\NDIST}[1]%
{\begin{picture}(0,0)%
\put(0,0){\makebox(0,0)%
{\begin{picture}(0,#1)%
\put(0,0){\line(0,1){#1}}%
\put(0,#1){\nhead}%
\end{picture}}}
\truex{400}%
\put(0,0){\circle{\value{x}}}%
\end{picture}}%

\newcommand{\NDOTAR}[1]%
{\truex{100}\truey{300}%
\NUMBEROFDOTS=#1%
\divide\NUMBEROFDOTS by \value{y}%
\advance\NUMBEROFDOTS by 1%
\begin{picture}(0,0)%
\put(0,0){\makebox(0,0)%
{\begin{picture}(0,#1)%
\multiput(0,0)(0,\value{y}){\NUMBEROFDOTS}%
{\circle*{\value{x}}}%
\put(0,#1){\nhead}%
\end{picture}}}\end{picture}}%

\newcommand{\NMONO}[1]%
{\truetail%
\monolength=#1%
\advance\monolength by -\truemonotail%
\begin{picture}(0,0)%
\put(0,0){\makebox(0,0)%
{\begin{picture}(0,#1)%
\put(0,\truemonotail){\line(0,1){\monolength}}%
\put(0,#1){\nhead}%
\put(0,\truemonotail){\nhead}%
\end{picture}}}\end{picture}}%

\newcommand{\NEPI}[1]%
{\truehead%
\epilength=#1%
\advance\epilength by -\trueepihead%
\begin{picture}(0,0)%
\put(0,0){\makebox(0,0)%
{\begin{picture}(0,#1)%
\put(0,0){\line(0,1){\epilength}}%
\put(0,#1){\nhead}%
\put(0,\epilength){\nhead}%
\end{picture}}}\end{picture}}%

\newcommand{\NBIMO}[1]%
{\truehead\truetail%
\epilength=#1%
\advance\epilength by -\trueepihead%
\monolength=\epilength%
\advance\monolength by -\truemonotail%
\begin{picture}(0,0)%
\put(0,0){\makebox(0,0)%
{\begin{picture}(0,#1)%
\put(0,\truemonotail){\line(0,1){\monolength}}%
\put(0,#1){\nhead}%
\put(0,\truemonotail){\nhead}%
\put(0,\epilength){\nhead}%
\end{picture}}}\end{picture}}%

\newcommand{\NBIAR}[1]%
{\begin{picture}(0,0)%
\truex{350}%
\put(0,0){\makebox(0,0)%
{\begin{picture}(0,#1)%
\put(-\value{x},0){\line(0,1){#1}}%
\put(-\value{x},#1){\nhead}%
\put(\value{x},0){\line(0,1){#1}}%
\put(\value{x},#1){\nhead}%
\end{picture}}}\end{picture}}%

\newcommand{\NBIDIST}[1]%
{\begin{picture}(0,0)%
\truex{350}%
\put(0,0){\makebox(0,0)%
{\begin{picture}(0,#1)%
\put(-\value{x},0){\line(0,1){#1}}%
\put(-\value{x},#1){\nhead}%
\put(\value{x},0){\line(0,1){#1}}%
\put(\value{x},#1){\nhead}%
\end{picture}}}
\truey{400}%
\put(-\value{x},0){\circle{\value{y}}}%
\put(\value{x},0){\circle{\value{y}}}%
\end{picture}}%

\newcommand{\NADJAR}[1]{\begin{picture}(0,0)%
\truex{350}%
\put(0,0){\makebox(0,0)%
{\begin{picture}(0,#1)%
\put(\value{x},#1){\line(0,-1){#1}}%
\put(\value{x},0){\shead}%
\put(-\value{x},0){\line(0,1){#1}}%
\put(-\value{x},#1){\nhead}%
\end{picture}}}\end{picture}}%

\newcommand{\NADJDIST}[1]{\begin{picture}(0,0)%
\truex{350}%
\put(0,0){\makebox(0,0)%
{\begin{picture}(0,#1)%
\put(\value{x},#1){\line(0,-1){#1}}%
\put(\value{x},0){\shead}%
\put(-\value{x},0){\line(0,1){#1}}%
\put(-\value{x},#1){\nhead}%
\end{picture}}}
\truey{400}%
\put(-\value{x},0){\circle{\value{y}}}%
\put(\value{x},0){\circle{\value{y}}}%
\end{picture}}%

\def\basicnar[#1]{\vcase{\NAR}{#100}}%

\newcommand{\nar}{\@ifnextchar[{\basicnar}{\basicnar[50]}}%

\def\basicNar[#1]#2{\Vcase{\NAR}{#2}{#100}}%

\newcommand{\Nar}{\@ifnextchar[{\basicNar}{\basicNar[50]}}%

\def\basicnaR[#1]#2{\vcasE{\NAR}{#2}{#100}}%

\newcommand{\naR}{\@ifnextchar[{\basicnaR}{\basicnaR[50]}}%

\def\basicndist[#1]{\vcase{\NDIST}{#100}}%

\newcommand{\ndist}{\@ifnextchar[{\basicndist}{\basicndist[50]}}%

\def\basicNdist[#1]#2{\Vcase{\NDIST}{#2\hspace*{2pt}}{#100}}%

\newcommand{\Ndist}{\@ifnextchar[{\basicNdist}{\basicNdist[50]}}%

\def\basicndisT[#1]#2{\vcasE{\NDIST}{\hspace*{2pt}#2}{#100}}%

\newcommand{\ndisT}{\@ifnextchar[{\basicndisT}{\basicndisT[50]}}%

\def\basicndotar[#1]{\vcase{\NDOTAR}{#100}}%

\newcommand{\ndotar}{\@ifnextchar[{\basicndotar}{\basicndotar[50]}}%

\def\basicNdotar[#1]#2{\Vcase{\NDOTAR}{#2}{#100}}%

\newcommand{\Ndotar}{\@ifnextchar[{\basicNdotar}{\basicNdotar[50]}}%

\def\basicndotaR[#1]#2{\vcasE{\NDOTAR}{#2}{#100}}%

\newcommand{\ndotaR}{\@ifnextchar[{\basicndotaR}{\basicndotaR[50]}}%

\def\basicnmono[#1]{\vcase{\NMONO}{#100}}%

\newcommand{\nmono}{\@ifnextchar[{\basicnmono}%
{\basicnmono[50]}}%

\def\basicNmono[#1]#2{\Vcase{\NMONO}{#2}{#100}}%

\newcommand{\Nmono}{\@ifnextchar[{\basicNmono}{\basicNmono[50]}}%

\def\basicnmonO[#1]#2{\vcasE{\NMONO}{#2}{#100}}%

\newcommand{\nmonO}{\@ifnextchar[{\basicnmonO}{\basicnmonO[50]}}%

\def\basicnepi[#1]{\vcase{\NEPI}{#100}}%

\newcommand{\nepi}{\@ifnextchar[{\basicnepi}{\basicnepi[50]}}%

\def\basicNepi[#1]#2{\Vcase{\NEPI}{#2}{#100}}%

\newcommand{\Nepi}{\@ifnextchar[{\basicNepi}{\basicNepi[50]}}%

\def\basicnepI[#1]#2{\vcasE{\NEPI}{#2}{#100}}%

\newcommand{\nepI}{\@ifnextchar[{\basicnepI}{\basicnepI[50]}}%

\def\basicnbimo[#1]{\vcase{\NBIMO}{#100}}%

\newcommand{\nbimo}{\@ifnextchar[{\basicnbimo}{\basicnbimo[50]}}%

\def\basicNbimo[#1]#2{\Vcase{\NBIMO}{#2}{#100}}%

\newcommand{\Nbimo}{\@ifnextchar[{\basicNbimo}{\basicNbimo[50]}}%

\def\basicnbimO[#1]#2{\vcasE{\NBIMO}{#2}{#100}}%

\newcommand{\nbimO}{\@ifnextchar[{\basicnbimO}{\basicnbimO[50]}}%

\def\basicniso[#1]{\Visocase{\NAR}{\cong}{}{#100}}%

\newcommand{\niso}{\@ifnextchar[{\basicniso}{\basicniso[50]}}%

\def\basicNiso[#1]#2{\Visocase{\NAR}{#2}{\cong}{#100}}%

\newcommand{\Niso}{\@ifnextchar[{\basicNiso}{\basicNiso[50]}}%

\def\basicnisO[#1]#2{\Visocase{\NAR}{\cong}{#2}{#100}}%

\newcommand{\nisO}{\@ifnextchar[{\basicnisO}{\basicnisO[50]}}%

\def\basicnbiar[#1]{\vbicase{\NBIAR}{#100}}%

\newcommand{\nbiar}{\@ifnextchar[{\basicnbiar}{\basicnbiar[50]}}%

\def\basicNbiar[#1]#2#3{\Vbicase{\NBIAR}{#2}{#3}{#100}}%

\newcommand{\Nbiar}{\@ifnextchar[{\basicNbiar}{\basicNbiar[50]}}%

\def\basicnbidist[#1]{\vbicase{\NBIDIST}{#100}}%

\newcommand{\nbidist}{\@ifnextchar[{\basicnbidist}{\basicnbidist[50]}}%

\def\basicNbidist[#1]#2#3%
{\Vbicase{\NBIDIST}{#2\hspace*{2pt}}{\hspace*{2pt}#3}{#100}}%

\newcommand{\Nbidist}{\@ifnextchar[{\basicNbidist}{\basicNbidist[50]}}%

\def\basicnadjar[#1]{\vbicase{\NADJAR}{#100}}%

\newcommand{\nadjar}{\@ifnextchar[{\basicnadjar}{\basicnadjar[50]}}%

\def\basicNadjar[#1]#2#3{\Vbicase{\NADJAR}{#2}{#3}{#100}}%

\newcommand{\Nadjar}{\@ifnextchar[{\basicNadjar}{\basicNadjar[50]}}%

\def\basicnadjdist[#1]{\vbicase{\NADJDIST}{#100}}%

\newcommand{\nadjdist}{\@ifnextchar[{\basicnadjdist}{\basicnadjdist[50]}}%

\def\basicNadjdist[#1]#2#3%
{\Vbicase{\NADJDIST}{#2\hspace*{2pt}}{\hspace*{2pt}#3}{#100}}%

\newcommand{\Nadjdist}{\@ifnextchar[{\basicNadjdist}{\basicNadjdist[50]}}%

\newcommand{\fdcase}[4]{\testdiagrammode\begin{picture}(0,0)%
\put(0,0){#1{#4}}%
\truex{200}\truey{600}\truez{600}%
\put(-\value{x},-\value{x}){\makebox(0,\value{z})[r]{${#2}$}}%
\put(\value{x},-\value{y}){\makebox(0,\value{z})[l]{${#3}$}}%
\end{picture}}%

\newcommand{\fdbicase}[4]{\testdiagrammode\begin{picture}(0,0)%
\put(0,0){#1{#4}}%
\truex{900}\truey{150}%
\put(-\value{x},\value{y}){${#2}$}%
\truex{300}\truey{1050}%
\put(\value{x},-\value{y}){${#3}$}%
\end{picture}}%

\newcommand{\NEAR}[1]{%
\Y=#1%
\divide\Y by 2%
\begin{picture}(0,0)%
\put(-\Y,-\Y){\line(1,1){#1}}%
\put(\Y,\Y){\nehead}%
\end{picture}}%

\newcommand{\NEDIST}[1]{%
\Y=#1%
\divide\Y by 2%
\begin{picture}(0,0)%
\put(-\Y,-\Y){\line(1,1){#1}}%
\put(\Y,\Y){\nehead}%
\truex{400}%
\put(0,0){\circle{\value{x}}}%
\end{picture}}%

\newcommand{\NEDOTAR}[1]%
{\truex{100}\truey{212}%
\Y=#1%
\divide\Y by 2%
\NUMBEROFDOTS=#1%
\divide\NUMBEROFDOTS by \value{y}%
\advance\NUMBEROFDOTS by 1%
\begin{picture}(0,0)%
\multiput(-\Y,-\Y)(\value{y},\value{y}){\NUMBEROFDOTS}%
{\circle*{\value{x}}}%
\put(\Y,\Y){\nehead}%
\end{picture}}%

\newcommand{\NEMONO}[1]{%
\Y=#1%
\divide \Y by 2%
\Truetail%
\bimolength=#1%
\advance\bimolength by -\Truemonotail%
\monolength=\bimolength%
\advance\monolength by -\Y%
\begin{picture}(0,0)%
\put(-\monolength,-\monolength){\line(1,1){\bimolength}}%
\put(-\monolength,-\monolength){\nehead}%
\put(\Y,\Y){\nehead}%
\end{picture}}%

\newcommand{\NEEPI}[1]{%
\Y=#1%
\divide\Y by 2%
\Truehead%
\bimolength=#1%
\advance\bimolength by -\Trueepihead%
\epilength=\bimolength%
\advance\epilength by -\Y%
\begin{picture}(0,0)%
\put(-\Y,-\Y){\line(1,1){\bimolength}}%
\put(\epilength,\epilength){\nehead}%
\put(\Y,\Y){\nehead}%
\end{picture}}%

\newcommand{\NEBIMO}[1]{%
\Y=#1%
\divide\Y by 2%
\Truetail\Truehead%
\bimolength=#1%
\advance\bimolength by -\Truemonotail%
\monolength=\bimolength%
\advance\monolength by -\Y%
\advance\bimolength by -\Trueepihead%
\epilength=\bimolength%
\advance\epilength by -\monolength%
\begin{picture}(0,0)%
\put(-\monolength,-\monolength){\line(1,1){\bimolength}}%
\put(-\monolength,-\monolength){\nehead}%
\put(\epilength,\epilength){\nehead}%
\put(\Y,\Y){\nehead}%
\end{picture}}%

\newcommand{\NEBIAR}[1]{%
\Y=#1%
\divide\Y by 2%
\begin{picture}(0,0)%
\put(-\Y,-\Y){\begin{picture}(0,0)%
\truex{247}%
\put(-\value{x},\value{x}){\line(1,1){#1}}%
\put(\value{x},-\value{x}){\line(1,1){#1}}%
\monolength=#1%
\advance\monolength by -\value{x}%
\epilength=#1%
\advance\epilength by \value{x}%
\put(\monolength,\epilength){\nehead}%
\put(\epilength,\monolength){\nehead}%
\end{picture}}\end{picture}}%

\newcommand{\NEBIDIST}[1]{%
\Y=#1%
\divide\Y by 2%
\truey{400}%
\begin{picture}(0,0)%
\put(-\Y,-\Y){\begin{picture}(0,0)%
\truex{247}%
\monolength=#1%
\advance\monolength by -\value{x}%
\epilength=#1%
\advance\epilength by \value{x}%
\put(\value{x},-\value{x}){\line(1,1){#1}}%
\put(\epilength,\monolength){\nehead}%
\end{picture}}%
\put(-\Y,-\Y){\begin{picture}(0,0)%
\truex{247}%
\monolength=#1%
\advance\monolength by \value{x}%
\epilength=#1%
\advance\epilength by -\value{x}%
\put(-\value{x},\value{x}){\line(1,1){#1}}%
\put(\epilength,\monolength){\nehead}%
\end{picture}}%
\put(-\value{x},\value{x}){\circle{\value{y}}}%
\put(\value{x},-\value{x}){\circle{\value{y}}}%
\end{picture}}%

\newcommand{\NEEQL}[1]{%
\Y=#1%
\divide\Y by 2%
\begin{picture}(0,0)%
\put(-\Y,-\Y){\begin{picture}(0,0)%
\truex{70}%
\put(-\value{x},\value{x}){\line(1,1){#1}}%
\put(\value{x},-\value{x}){\line(1,1){#1}}%
\end{picture}}\end{picture}}%

\newcommand{\NEADJAR}[1]{%
\Y=#1%
\divide\Y by 2%
\begin{picture}(0,0)%
\put(-\Y,-\Y){\begin{picture}(0,0)%
\truex{247}%
\monolength=#1%
\advance\monolength by -\value{x}%
\epilength=#1%
\advance\epilength by \value{x}%
\put(\value{x},-\value{x}){\line(1,1){#1}}%
\put(\epilength,\monolength){\nehead}%
\end{picture}}%
\put(\Y,\Y){\begin{picture}(0,0)%
\truex{247}%
\monolength=#1%
\advance\monolength by -\value{x}%
\epilength=#1%
\advance\epilength by \value{x}%
\put(-\value{x},\value{x}){\line(-1,-1){#1}}%
\put(-\epilength,-\monolength){\swhead}%
\end{picture}}\end{picture}}%

\newcommand{\NEADJDIST}[1]{%
\Y=#1%
\divide\Y by 2%
\truey{400}%
\begin{picture}(0,0)%
\put(-\Y,-\Y){\begin{picture}(0,0)%
\truex{247}%
\monolength=#1%
\advance\monolength by -\value{x}%
\epilength=#1%
\advance\epilength by \value{x}%
\put(\value{x},-\value{x}){\line(1,1){#1}}%
\put(\epilength,\monolength){\nehead}%
\end{picture}}%
\put(\Y,\Y){\begin{picture}(0,0)%
\truex{247}%
\monolength=#1%
\advance\monolength by -\value{x}%
\epilength=#1%
\advance\epilength by \value{x}%
\put(-\value{x},\value{x}){\line(-1,-1){#1}}%
\put(-\epilength,-\monolength){\swhead}%
\end{picture}}%
\put(-\value{x},\value{x}){\circle{\value{y}}}%
\put(\value{x},-\value{x}){\circle{\value{y}}}%
\end{picture}}%

\def\basicnear[#1]{\fdcase{\NEAR}{}{}{#100}}%

\newcommand{\near}{\@ifnextchar[{\basicnear}{\basicnear[59]}}%

\def\basicNear[#1]#2{\fdcase{\NEAR}{#2}{}{#100}}%

\newcommand{\Near}{\@ifnextchar[{\basicNear}{\basicNear[59]}}%

\def\basicneaR[#1]#2{\fdcase{\NEAR}{}{#2}{#100}}%

\newcommand{\neaR}{\@ifnextchar[{\basicneaR}{\basicneaR[59]}}%

\def\basicnedist[#1]{\fdcase{\NEDIST}{}{}{#100}}%

\newcommand{\nedist}{\@ifnextchar[{\basicnedist}{\basicnedist[59]}}%

\def\basicNedist[#1]#2{\fdcase{\NEDIST}{#2}{}{#100}}%

\newcommand{\Nedist}{\@ifnextchar[{\basicNedist}{\basicNedist[59]}}%

\def\basicnedisT[#1]#2{\fdcase{\NEDIST}{}{#2}{#100}}%

\newcommand{\nedisT}{\@ifnextchar[{\basicnedisT}{\basicnedisT[59]}}%

\def\basicnedotar[#1]{\fdcase{\NEDOTAR}{}{}{#100}}%

\newcommand{\nedotar}{\@ifnextchar[{\basicnedotar}{\basicnedotar[59]}}%

\def\basicNedotar[#1]#2{\fdcase{\NEDOTAR}{#2}{}{#100}}%

\newcommand{\Nedotar}{\@ifnextchar[{\basicNedotar}{\basicNedotar[59]}}%

\def\basicnedotaR[#1]#2{\fdcase{\NEDOTAR}{}{#2}{#100}}%

\newcommand{\nedotaR}{\@ifnextchar[{\basicnedotaR}{\basicnedotaR[59]}}%

\def\basicnemono[#1]{\fdcase{\NEMONO}{}{}{#100}}%

\newcommand{\nemono}{\@ifnextchar[{\basicnemono}{\basicnemono[59]}}%

\def\basicNemono[#1]#2{\fdcase{\NEMONO}{#2}{}{#100}}%

\newcommand{\Nemono}{\@ifnextchar[{\basicNemono}{\basicNemono[59]}}%

\def\basicnemonO[#1]#2{\fdcase{\NEMONO}{}{#2}{#100}}%

\newcommand{\nemonO}{\@ifnextchar[{\basicnemonO}{\basicnemonO[59]}}%

\def\basicneepi[#1]{\fdcase{\NEEPI}{}{}{#100}}%

\newcommand{\neepi}{\@ifnextchar[{\basicneepi}{\basicneepi[59]}}%

\def\basicNeepi[#1]#2{\fdcase{\NEEPI}{#2}{}{#100}}%

\newcommand{\Neepi}{\@ifnextchar[{\basicNeepi}{\basicNeepi[59]}}%

\def\basicneepI[#1]#2{\fdcase{\NEEPI}{}{#2}{#100}}%

\newcommand{\neepI}{\@ifnextchar[{\basicneepI}{\basicneepI[59]}}%

\def\basicnebimo[#1]{\fdcase{\NEBIMO}{}{}{#100}}%

\newcommand{\nebimo}{\@ifnextchar[{\basicnebimo}{\basicnebimo[59]}}%

\def\basicNebimo[#1]#2{\fdcase{\NEBIMO}{#2}{}{#100}}%

\newcommand{\Nebimo}{\@ifnextchar[{\basicNebimo}{\basicNebimo[59]}}%

\def\basicnebimO[#1]#2{\fdcase{\NEBIMO}{}{#2}{#100}}%

\newcommand{\nebimO}{\@ifnextchar[{\basicnebimO}{\basicnebimO[59]}}%

\def\basicneiso[#1]{\fdcase{\NEAR}{\hspace{-2pt}\cong}{}{#100}}%

\newcommand{\neiso}{\@ifnextchar[{\basicneiso}{\basicneiso[59]}}%

\def\basicNeiso[#1]#2{\fdcase{\NEAR}{#2}{\cong}{#100}}%

\newcommand{\Neiso}{\@ifnextchar[{\basicNeiso}{\basicNeiso[59]}}%

\def\basicneisO[#1]#2{\fdcase{\NEAR}{\hspace{-2pt}\cong}{#2}{#100}}%

\newcommand{\neisO}{\@ifnextchar[{\basicneisO}{\basicneisO[59]}}%

\def\basicneeql[#1]{\fdcase{\NEEQL}{}{}{#100}}%

\newcommand{\neeql}{\@ifnextchar[{\basicneeql}{\basicneeql[59]}}%

\def\basicNeeql[#1]#2{\fdcase{\NEEQL}{#2}{}{#100}}%

\newcommand{\Neeql}{\@ifnextchar[{\basicNeeql}{\basicNeeql[59]}}%

\def\basicneeqL[#1]#2{\fdcase{\NEEQL}{}{#2}{#100}}%

\newcommand{\neeqL}{\@ifnextchar[{\basicneeqL}{\basicneeqL[59]}}%

\def\basicnebiar[#1]{\fdbicase{\NEBIAR}{}{}{#100}}%

\newcommand{\nebiar}{\@ifnextchar[{\basicnebiar}{\basicnebiar[59]}}%

\def\basicNebiar[#1]#2#3{\fdbicase{\NEBIAR}{#2}{#3}{#100}}%

\newcommand{\Nebiar}{\@ifnextchar[{\basicNebiar}{\basicNebiar[59]}}%

\def\basicneadjar[#1]{\fdbicase{\NEADJAR}{}{}{#100}}%

\newcommand{\neadjar}{\@ifnextchar[{\basicneadjar}{\basicneadjar[59]}}%

\def\basicNeadjar[#1]#2#3{\fdbicase{\NEADJAR}{#2}{#3}{#100}}%

\newcommand{\Neadjar}{\@ifnextchar[{\basicNeadjar}{\basicNeadjar[59]}}%

\def\basicnebidist[#1]{\fdbicase{\NEBIDIST}{}{}{#100}}%

\newcommand{\nebidist}{\@ifnextchar[{\basicnebidist}{\basicnebidist[59]}}%

\def\basicNebidist[#1]#2#3{\fdbicase{\NEBIDIST}{#2}{#3}{#100}}%

\newcommand{\Nebidist}{\@ifnextchar[{\basicNebidist}{\basicNebidist[59]}}%

\def\basicneadjdist[#1]{\fdbicase{\NEADJDIST}{}{}{#100}}%

\newcommand{\neadjdist}{\@ifnextchar[{\basicneadjdist}{\basicneadjdist[59]}}%

\def\basicNeadjdist[#1]#2#3{\fdbicase{\NEADJDIST}{#2}{#3}{#100}}%

\newcommand{\Neadjdist}{\@ifnextchar[{\basicNeadjdist}{\basicNeadjdist[59]}}%

\newcommand{\SWAR}[1]{%
\Y=#1%
\divide\Y by 2%
\begin{picture}(0,0)%
\put(\Y,\Y){\line(-1,-1){#1}}%
\put(-\Y,-\Y){\swhead}%
\end{picture}}%

\newcommand{\SWDIST}[1]{%
\Y=#1%
\divide\Y by 2%
\begin{picture}(0,0)%
\put(\Y,\Y){\line(-1,-1){#1}}%
\put(-\Y,-\Y){\swhead}%
\truex{400}%
\put(0,0){\circle{\value{x}}}%
\end{picture}}%

\newcommand{\SWDOTAR}[1]%
{\truex{100}\truey{212}%
\Y=#1%
\divide\Y by 2%
\NUMBEROFDOTS=#1%
\divide\NUMBEROFDOTS by \value{y}%
\advance\NUMBEROFDOTS by 1%
\begin{picture}(0,0)%
\multiput(\Y,\Y)(-\value{y},-\value{y}){\NUMBEROFDOTS}%
{\circle*{\value{x}}}%
\put(-\Y,-\Y){\swhead}%
\end{picture}}%

\newcommand{\SWMONO}[1]{%
\Y=#1%
\divide \Y by 2%
\Truetail%
\bimolength=#1%
\advance\bimolength by -\Truemonotail%
\monolength=\bimolength%
\advance\monolength by -\Y%
\begin{picture}(0,0)%
\put(\monolength,\monolength){\line(-1,-1){\bimolength}}%
\put(\monolength,\monolength){\swhead}%
\put(-\Y,-\Y){\swhead}%
\end{picture}}%

\newcommand{\SWEPI}[1]{%
\Y=#1%
\divide\Y by 2%
\Truehead%
\bimolength=#1%
\advance\bimolength by -\Trueepihead%
\epilength=\bimolength%
\advance\epilength by -\Y%
\begin{picture}(0,0)%
\put(\Y,\Y){\line(-1,-1){\bimolength}}%
\put(-\epilength,-\epilength){\swhead}%
\put(-\Y,-\Y){\swhead}%
\end{picture}}%

\newcommand{\SWBIMO}[1]{%
\Y=#1%
\divide\Y by 2%
\Truetail\Truehead%
\bimolength=#1%
\advance\bimolength by -\Truemonotail%
\monolength=\bimolength%
\advance\monolength by -\Y%
\advance\bimolength by -\Trueepihead%
\epilength=\bimolength%
\advance\epilength by -\monolength%
\begin{picture}(0,0)%
\put(\monolength,\monolength){\line(-1,-1){\bimolength}}%
\put(\monolength,\monolength){\swhead}%
\put(-\epilength,-\epilength){\swhead}%
\put(-\Y,-\Y){\swhead}%
\end{picture}}%

\newcommand{\SWBIAR}[1]{%
\Y=#1%
\divide\Y by 2%
\begin{picture}(0,0)%
\put(\Y,\Y){\begin{picture}(0,0)%
\truex{247}%
\put(\value{x},-\value{x}){\line(-1,-1){#1}}%
\put(-\value{x},\value{x}){\line(-1,-1){#1}}%
\monolength=#1%
\advance\monolength by -\value{x}%
\epilength=#1%
\advance\epilength by \value{x}%
\put(-\monolength,-\epilength){\swhead}%
\put(-\epilength,-\monolength){\swhead}%
\end{picture}}\end{picture}}%

\newcommand{\SWBIDIST}[1]{%
\Y=#1%
\divide\Y by 2%
\truey{400}%
\begin{picture}(0,0)%
\put(\Y,\Y){\begin{picture}(0,0)%
\truex{247}%
\monolength=#1%
\advance\monolength by -\value{x}%
\epilength=#1%
\advance\epilength by \value{x}%
\put(-\value{x},\value{x}){\line(-1,-1){#1}}%
\put(-\epilength,-\monolength){\swhead}%
\end{picture}}%
\put(\Y,\Y){\begin{picture}(0,0)%
\truex{247}%
\monolength=#1%
\advance\monolength by \value{x}%
\epilength=#1%
\advance\epilength by -\value{x}%
\put(\value{x},-\value{x}){\line(-1,-1){#1}}%
\put(-\epilength,-\monolength){\swhead}%
\end{picture}}%
\put(\value{x},-\value{x}){\circle{\value{y}}}%
\put(-\value{x},\value{x}){\circle{\value{y}}}%
\end{picture}}%

\newcommand{\SWADJAR}[1]{%
\Y=#1%
\divide\Y by 2%
\begin{picture}(0,0)%
\put(\Y,\Y){\begin{picture}(0,0)%
\truex{247}%
\monolength=#1%
\advance\monolength by -\value{x}%
\epilength=#1%
\advance\epilength by \value{x}%
\put(\value{x},-\value{x}){\line(-1,-1){#1}}%
\put(-\monolength,-\epilength){\swhead}%
\end{picture}}%
\put(-\Y,-\Y){\begin{picture}(0,0)%
\truex{247}%
\monolength=#1%
\advance\monolength by -\value{x}%
\epilength=#1%
\advance\epilength by \value{x}%
\put(-\value{x},\value{x}){\line(1,1){#1}}%
\put(\monolength,\epilength){\nehead}%
\end{picture}}\end{picture}}%

\newcommand{\SWADJDIST}[1]{%
\Y=#1%
\divide\Y by 2%
\truey{400}%
\begin{picture}(0,0)%
\put(\Y,\Y){\begin{picture}(0,0)%
\truex{247}%
\monolength=#1%
\advance\monolength by -\value{x}%
\epilength=#1%
\advance\epilength by \value{x}%
\put(\value{x},-\value{x}){\line(-1,-1){#1}}%
\put(-\monolength,-\epilength){\swhead}%
\end{picture}}%
\put(-\Y,-\Y){\begin{picture}(0,0)%
\truex{247}%
\monolength=#1%
\advance\monolength by -\value{x}%
\epilength=#1%
\advance\epilength by \value{x}%
\put(-\value{x},\value{x}){\line(1,1){#1}}%
\put(\monolength,\epilength){\nehead}%
\end{picture}}%
\put(-\value{x},\value{x}){\circle{\value{y}}}%
\put(\value{x},-\value{x}){\circle{\value{y}}}%
\end{picture}}%

\def\basicswar[#1]{\fdcase{\SWAR}{}{}{#100}}%

\newcommand{\swar}{\@ifnextchar[{\basicswar}{\basicswar[59]}}%

\def\basicSwar[#1]#2{\fdcase{\SWAR}{#2}{}{#100}}%

\newcommand{\Swar}{\@ifnextchar[{\basicSwar}{\basicSwar[59]}}%

\def\basicswaR[#1]#2{\fdcase{\SWAR}{}{#2}{#100}}%

\newcommand{\swaR}{\@ifnextchar[{\basicswaR}{\basicswaR[59]}}%

\def\basicswdist[#1]{\fdcase{\SWDIST}{}{}{#100}}%

\newcommand{\swdist}{\@ifnextchar[{\basicswdist}{\basicswdist[59]}}%

\def\basicSwdist[#1]#2{\fdcase{\SWDIST}{#2}{}{#100}}%

\newcommand{\Swdist}{\@ifnextchar[{\basicSwdist}{\basicSwdist[59]}}%

\def\basicswdisT[#1]#2{\fdcase{\SWDIST}{}{#2}{#100}}%

\newcommand{\swdisT}{\@ifnextchar[{\basicswdisT}{\basicswdisT[59]}}%

\def\basicswdotar[#1]{\fdcase{\SWDOTAR}{}{}{#100}}%

\newcommand{\swdotar}{\@ifnextchar[{\basicswdotar}{\basicswdotar[59]}}%

\def\basicSwdotar[#1]#2{\fdcase{\SWDOTAR}{#2}{}{#100}}%

\newcommand{\Swdotar}{\@ifnextchar[{\basicSwdotar}{\basicSwdotar[59]}}%

\def\basicswdotaR[#1]#2{\fdcase{\SWDOTAR}{}{#2}{#100}}%

\newcommand{\swdotaR}{\@ifnextchar[{\basicswdotaR}{\basicswdotaR[59]}}%

\def\basicswmono[#1]{\fdcase{\SWMONO}{}{}{#100}}%

\newcommand{\swmono}{\@ifnextchar[{\basicswmono}{\basicswmono[59]}}%

\def\basicSwmono[#1]#2{\fdcase{\SWMONO}{#2}{}{#100}}%

\newcommand{\Swmono}{\@ifnextchar[{\basicSwmono}{\basicSwmono[59]}}%

\def\basicswmonO[#1]#2{\fdcase{\SWMONO}{}{#2}{#100}}%

\newcommand{\swmonO}{\@ifnextchar[{\basicswmonO}{\basicswmonO[59]}}%

\def\basicswepi[#1]{\fdcase{\SWEPI}{}{}{#100}}%

\newcommand{\swepi}{\@ifnextchar[{\basicswepi}{\basicswepi[59]}}%

\def\basicSwepi[#1]#2{\fdcase{\SWEPI}{#2}{}{#100}}%

\newcommand{\Swepi}{\@ifnextchar[{\basicSwepi}{\basicSwepi[59]}}%

\def\basicswepI[#1]#2{\fdcase{\SWEPI}{}{#2}{#100}}%

\newcommand{\swepI}{\@ifnextchar[{\basicswepI}{\basicswepI[59]}}%

\def\basicswbimo[#1]{\fdcase{\SWBIMO}{}{}{#100}}%

\newcommand{\swbimo}{\@ifnextchar[{\basicswbimo}{\basicswbimo[59]}}%

\def\basicSwbimo[#1]#2{\fdcase{\SWBIMO}{#2}{}{#100}}%

\newcommand{\Swbimo}{\@ifnextchar[{\basicSwbimo}{\basicSwbimo[59]}}%

\def\basicswbimO[#1]#2{\fdcase{\SWBIMO}{}{#2}{#100}}%

\newcommand{\swbimO}{\@ifnextchar[{\basicswbimO}{\basicswbimO[59]}}%

\def\basicswiso[#1]{\fdcase{\SWAR}{\hspace{-2pt}\cong}{}{#100}}%

\newcommand{\swiso}{\@ifnextchar[{\basicswiso}{\basicswiso[59]}}%

\def\basicSwiso[#1]#2{\fdcase{\SWAR}{#2}{\cong}{#100}}%

\newcommand{\Swiso}{\@ifnextchar[{\basicSwiso}{\basicSwiso[59]}}%

\def\basicswisO[#1]#2{\fdcase{\SWAR}{\hspace{-2pt}\cong}{#2}{#100}}%

\newcommand{\swisO}{\@ifnextchar[{\basicswisO}{\basicswisO[59]}}%

\def\basicswbiar[#1]{\fdbicase{\SWBIAR}{}{}{#100}}%

\newcommand{\swbiar}{\@ifnextchar[{\basicswbiar}{\basicswbiar[59]}}%

\def\basicSwbiar[#1]#2#3{\fdbicase{\SWBIAR}{#2}{#3}{#100}}%

\newcommand{\Swbiar}{\@ifnextchar[{\basicSwbiar}{\basicSwbiar[59]}}%

\def\basicswadjar[#1]{\fdbicase{\SWADJAR}{}{}{#100}}%

\newcommand{\swadjar}{\@ifnextchar[{\basicswadjar}{\basicswadjar[59]}}%

\def\basicSwadjar[#1]#2#3{\fdbicase{\SWADJAR}{#2}{#3}{#100}}%

\newcommand{\Swadjar}{\@ifnextchar[{\basicSwadjar}{\basicSwadjar[59]}}%

\def\basicswbidist[#1]{\fdbicase{\SWBIDIST}{}{}{#100}}%

\newcommand{\swbidist}{\@ifnextchar[{\basicswbidist}{\basicswbidist[59]}}%

\def\basicSwbidist[#1]#2#3{\fdbicase{\SWBIDIST}{#2}{#3}{#100}}%

\newcommand{\Swbidist}{\@ifnextchar[{\basicSwbidist}{\basicSwbidist[59]}}%

\def\basicswadjdist[#1]{\fdbicase{\SWADJDIST}{}{}{#100}}%

\newcommand{\swadjdist}{\@ifnextchar[{\basicswadjdist}{\basicswadjdist[59]}}%

\def\basicSwadjdist[#1]#2#3{\fdbicase{\SWADJDIST}{#2}{#3}{#100}}%

\newcommand{\Swadjdist}{\@ifnextchar[{\basicSwadjdist}{\basicSwadjdist[59]}}%

\newcommand{\sdcase}[4]{\testdiagrammode\begin{picture}(0,0)%
\put(0,0){#1{#4}}%
\truex{100}\truez{600}%
\put(\value{x},\value{x}){\makebox(0,\value{z})[l]{${#2}$}}%
\truex{300}\truey{800}%
\put(-\value{x},-\value{y}){\makebox(0,\value{z})[r]{${#3}$}}%
\end{picture}}%

\newcommand{\sdbicase}[4]{\testdiagrammode\begin{picture}(0,0)%
\put(0,0){#1{#4}}%
\truex{350}\truey{600}\truez{950}%
\put(\value{x},\value{x}){\makebox(0,\value{y})[l]{${#2}$}}%
\truex{450}\truey{600}\truez{1050}%
\put(-\value{x},-\value{z}){\makebox(0,\value{y})[r]{${#3}$}}%
\end{picture}}%

\newcommand{\SEAR}[1]{%
\Y=#1%
\divide\Y by 2%
\begin{picture}(0,0)%
\put(-\Y,\Y){\line(1,-1){#1}}%
\put(\Y,-\Y){\sehead}%
\end{picture}}%

\newcommand{\SEDIST}[1]{%
\Y=#1%
\divide\Y by 2%
\begin{picture}(0,0)%
\put(-\Y,\Y){\line(1,-1){#1}}%
\put(\Y,-\Y){\sehead}%
\truex{400}%
\put(0,0){\circle{\value{x}}}%
\end{picture}}%

\newcommand{\SEDOTAR}[1]%
{\truex{100}\truey{212}%
\Y=#1%
\divide\Y by 2%
\NUMBEROFDOTS=#1%
\divide\NUMBEROFDOTS by \value{y}%
\advance\NUMBEROFDOTS by 1%
\begin{picture}(0,0)%
\multiput(-\Y,\Y)(\value{y},-\value{y}){\NUMBEROFDOTS}%
{\circle*{\value{x}}}%
\put(\Y,-\Y){\sehead}%
\end{picture}}%

\newcommand{\SEMONO}[1]{%
\Y=#1%
\divide \Y by 2%
\Truetail%
\bimolength=#1%
\advance\bimolength by -\Truemonotail%
\monolength=\bimolength%
\advance\monolength by -\Y%
\begin{picture}(0,0)%
\put(-\monolength,\monolength){\line(1,-1){\bimolength}}%
\put(-\monolength,\monolength){\sehead}%
\put(\Y,-\Y){\sehead}%
\end{picture}}%

\newcommand{\SEEPI}[1]{%
\Y=#1%
\divide\Y by 2%
\Truehead%
\bimolength=#1%
\advance\bimolength by -\Trueepihead%
\epilength=\bimolength%
\advance\epilength by -\Y%
\begin{picture}(0,0)%
\put(-\Y,\Y){\line(1,-1){\bimolength}}%
\put(\epilength,-\epilength){\sehead}%
\put(\Y,-\Y){\sehead}%
\end{picture}}%

\newcommand{\SEBIMO}[1]{%
\Y=#1%
\divide\Y by 2%
\Truetail\Truehead%
\bimolength=#1%
\advance\bimolength by -\Truemonotail%
\monolength=\bimolength%
\advance\monolength by -\Y%
\advance\bimolength by -\Trueepihead%
\epilength=\bimolength%
\advance\epilength by -\monolength%
\begin{picture}(0,0)%
\put(-\monolength,\monolength){\line(1,-1){\bimolength}}%
\put(-\monolength,\monolength){\sehead}%
\put(\epilength,-\epilength){\sehead}%
\put(\Y,-\Y){\sehead}%
\end{picture}}%

\newcommand{\SEBIAR}[1]{%
\Y=#1%
\divide\Y by 2%
\begin{picture}(0,0)%
\put(-\Y,\Y){\begin{picture}(0,0)%
\truex{247}%
\put(-\value{x},-\value{x}){\line(1,-1){#1}}%
\put(\value{x},\value{x}){\line(1,-1){#1}}%
\monolength=#1%
\advance\monolength by -\value{x}%
\epilength=#1%
\advance\epilength by \value{x}%
\put(\monolength,-\epilength){\sehead}%
\put(\epilength,-\monolength){\sehead}%
\end{picture}}\end{picture}}%

\newcommand{\SEBIDIST}[1]{%
\Y=#1%
\divide\Y by 2%
\truey{400}%
\begin{picture}(0,0)%
\put(-\Y,\Y){\begin{picture}(0,0)%
\truex{247}%
\monolength=#1%
\advance\monolength by -\value{x}%
\epilength=#1%
\advance\epilength by \value{x}%
\put(\value{x},\value{x}){\line(1,-1){#1}}%
\put(\epilength,-\monolength){\sehead}%
\end{picture}}%
\put(-\Y,\Y){\begin{picture}(0,0)%
\truex{247}%
\monolength=#1%
\advance\monolength by \value{x}%
\epilength=#1%
\advance\epilength by -\value{x}%
\put(-\value{x},-\value{x}){\line(1,-1){#1}}%
\put(\epilength,-\monolength){\sehead}%
\end{picture}}%
\put(-\value{x},-\value{x}){\circle{\value{y}}}%
\put(\value{x},\value{x}){\circle{\value{y}}}%
\end{picture}}%

\newcommand{\SEEQL}[1]{%
\Y=#1%
\divide\Y by 2%
\begin{picture}(0,0)%
\put(-\Y,\Y){\begin{picture}(0,0)%
\truex{70}%
\put(-\value{x},-\value{x}){\line(1,-1){#1}}%
\put(\value{x},\value{x}){\line(1,-1){#1}}%
\end{picture}}\end{picture}}%

\newcommand{\SEADJAR}[1]{%
\Y=#1%
\divide\Y by 2%
\begin{picture}(0,0)%
\put(-\Y,\Y){\begin{picture}(0,0)%
\truex{247}%
\monolength=#1%
\advance\monolength by -\value{x}%
\epilength=#1%
\advance\epilength by \value{x}%
\put(-\value{x},-\value{x}){\line(1,-1){#1}}%
\put(\monolength,-\epilength){\sehead}%
\end{picture}}%
\put(\Y,-\Y){\begin{picture}(0,0)%
\truex{247}%
\monolength=#1%
\advance\monolength by -\value{x}%
\epilength=#1%
\advance\epilength by \value{x}%
\put(\value{x},\value{x}){\line(-1,1){#1}}%
\put(-\monolength,\epilength){\nwhead}%
\end{picture}}\end{picture}}%

\newcommand{\SEADJDIST}[1]{%
\Y=#1%
\divide\Y by 2%
\truey{400}%
\begin{picture}(0,0)%
\put(-\Y,\Y){\begin{picture}(0,0)%
\truex{247}%
\monolength=#1%
\advance\monolength by -\value{x}%
\epilength=#1%
\advance\epilength by \value{x}%
\put(-\value{x},-\value{x}){\line(1,-1){#1}}%
\put(\monolength,-\epilength){\sehead}%
\end{picture}}%
\put(\Y,-\Y){\begin{picture}(0,0)%
\truex{247}%
\monolength=#1%
\advance\monolength by -\value{x}%
\epilength=#1%
\advance\epilength by \value{x}%
\put(\value{x},\value{x}){\line(-1,1){#1}}%
\put(-\monolength,\epilength){\nwhead}%
\end{picture}}%
\put(-\value{x},-\value{x}){\circle{\value{y}}}%
\put(\value{x},\value{x}){\circle{\value{y}}}%
\end{picture}}%

\def\basicsear[#1]{\sdcase{\SEAR}{}{}{#100}}%

\newcommand{\sear}{\@ifnextchar[{\basicsear}{\basicsear[59]}}%

\def\basicSear[#1]#2{\sdcase{\SEAR}{#2}{}{#100}}%

\newcommand{\Sear}{\@ifnextchar[{\basicSear}{\basicSear[59]}}%

\def\basicseaR[#1]#2{\sdcase{\SEAR}{}{#2}{#100}}%

\newcommand{\seaR}{\@ifnextchar[{\basicseaR}{\basicseaR[59]}}%

\def\basicsedist[#1]{\sdcase{\SEDIST}{}{}{#100}}%

\newcommand{\sedist}{\@ifnextchar[{\basicsedist}{\basicsedist[59]}}%

\def\basicSedist[#1]#2{\sdcase{\SEDIST}{#2}{}{#100}}%

\newcommand{\Sedist}{\@ifnextchar[{\basicSedist}{\basicSedist[59]}}%

\def\basicsedisT[#1]#2{\sdcase{\SEDIST}{}{#2}{#100}}%

\newcommand{\sedisT}{\@ifnextchar[{\basicsedisT}{\basicsedisT[59]}}%

\def\basicsedotar[#1]{\sdcase{\SEDOTAR}{}{}{#100}}%

\newcommand{\sedotar}{\@ifnextchar[{\basicsedotar}{\basicsedotar[59]}}%

\def\basicSedotar[#1]#2{\sdcase{\SEDOTAR}{#2}{}{#100}}%

\newcommand{\Sedotar}{\@ifnextchar[{\basicSedotar}{\basicSedotar[59]}}%

\def\basicsedotaR[#1]#2{\sdcase{\SEDOTAR}{}{#2}{#100}}%

\newcommand{\sedotaR}{\@ifnextchar[{\basicsedotaR}{\basicsedotaR[59]}}%

\def\basicsemono[#1]{\sdcase{\SEMONO}{}{}{#100}}%

\newcommand{\semono}{\@ifnextchar[{\basicsemono}{\basicsemono[59]}}%

\def\basicSemono[#1]#2{\sdcase{\SEMONO}{#2}{}{#100}}%

\newcommand{\Semono}{\@ifnextchar[{\basicSemono}{\basicSemono[59]}}%

\def\basicsemonO[#1]#2{\sdcase{\SEMONO}{}{#2}{#100}}%

\newcommand{\semonO}{\@ifnextchar[{\basicsemonO}{\basicsemonO[59]}}%

\def\basicseepi[#1]{\sdcase{\SEEPI}{}{}{#100}}%

\newcommand{\seepi}{\@ifnextchar[{\basicseepi}{\basicseepi[59]}}%

\def\basicSeepi[#1]#2{\sdcase{\SEEPI}{#2}{}{#100}}%

\newcommand{\Seepi}{\@ifnextchar[{\basicSeepi}{\basicSeepi[59]}}%

\def\basicseepI[#1]#2{\sdcase{\SEEPI}{}{#2}{#100}}%

\newcommand{\seepI}{\@ifnextchar[{\basicseepI}{\basicseepI[59]}}%

\def\basicsebimo[#1]{\sdcase{\SEBIMO}{}{}{#100}}%

\newcommand{\sebimo}{\@ifnextchar[{\basicsebimo}{\basicsebimo[59]}}%

\def\basicSebimo[#1]#2{\sdcase{\SEBIMO}{#2}{}{#100}}%

\newcommand{\Sebimo}{\@ifnextchar[{\basicSebimo}{\basicSebimo[59]}}%

\def\basicsebimO[#1]#2{\sdcase{\SEBIMO}{}{#2}{#100}}%

\newcommand{\sebimO}{\@ifnextchar[{\basicsebimO}{\basicsebimO[59]}}%

\def\basicseiso[#1]{\sdcase{\SEAR}{\hspace{-2pt}\cong}{}{#100}}%

\newcommand{\seiso}{\@ifnextchar[{\basicseiso}{\basicseiso[59]}}%

\def\basicSeiso[#1]#2{\sdcase{\SEAR}{#2}{\cong}{#100}}%

\newcommand{\Seiso}{\@ifnextchar[{\basicSeiso}{\basicSeiso[59]}}%

\def\basicseisO[#1]#2{\sdcase{\SEAR}{\hspace{-2pt}\cong}{#2}{#100}}%

\newcommand{\seisO}{\@ifnextchar[{\basicseisO}{\basicseisO[59]}}%

\def\basicseeql[#1]{\sdcase{\SEEQL}{}{}{#100}}%

\newcommand{\seeql}{\@ifnextchar[{\basicseeql}{\basicseeql[59]}}%

\def\basicSeeql[#1]#2{\sdcase{\SEEQL}{#2}{}{#100}}%

\newcommand{\Seeql}{\@ifnextchar[{\basicSeeql}{\basicSeeql[59]}}%

\def\basicseeqL[#1]#2{\sdcase{\SEEQL}{}{#2}{#100}}%

\newcommand{\seeqL}{\@ifnextchar[{\basicseeqL}{\basicseeqL[59]}}%

\def\basicsebiar[#1]{\sdbicase{\SEBIAR}{}{}{#100}}%

\newcommand{\sebiar}{\@ifnextchar[{\basicsebiar}{\basicsebiar[59]}}%

\def\basicSebiar[#1]#2#3{\sdbicase{\SEBIAR}{#2}{#3}{#100}}%

\newcommand{\Sebiar}{\@ifnextchar[{\basicSebiar}{\basicSebiar[59]}}%

\def\basicseadjar[#1]{\sdbicase{\SEADJAR}{}{}{#100}}%

\newcommand{\seadjar}{\@ifnextchar[{\basicseadjar}{\basicseadjar[59]}}%

\def\basicSeadjar[#1]#2#3{\sdbicase{\SEADJAR}{#2}{#3}{#100}}%

\newcommand{\Seadjar}{\@ifnextchar[{\basicSeadjar}{\basicSeadjar[59]}}%

\def\basicsebidist[#1]{\sdbicase{\SEBIDIST}{}{}{#100}}%

\newcommand{\sebidist}{\@ifnextchar[{\basicsebidist}{\basicsebidist[59]}}%

\def\basicSebidist[#1]#2#3{\sdbicase{\SEBIDIST}{#2}{#3}{#100}}%

\newcommand{\Sebidist}{\@ifnextchar[{\basicSebidist}{\basicSebidist[59]}}%

\def\basicseadjdist[#1]{\sdbicase{\SEADJDIST}{}{}{#100}}%

\newcommand{\seadjdist}{\@ifnextchar[{\basicseadjdist}{\basicseadjdist[59]}}%

\def\basicSeadjdist[#1]#2#3{\sdbicase{\SEADJDIST}{#2}{#3}{#100}}%

\newcommand{\Seadjdist}{\@ifnextchar[{\basicSeadjdist}{\basicSeadjdist[59]}}%

\newcommand{\NWAR}[1]{%
\Y=#1%
\divide\Y by 2%
\begin{picture}(0,0)%
\put(\Y,-\Y){\line(-1,1){#1}}%
\put(-\Y,\Y){\nwhead}%
\end{picture}}%

\newcommand{\NWDIST}[1]{%
\Y=#1%
\divide\Y by 2%
\begin{picture}(0,0)%
\put(\Y,-\Y){\line(-1,1){#1}}%
\put(-\Y,\Y){\nwhead}%
\truex{400}%
\put(0,0){\circle{\value{x}}}%
\end{picture}}%

\newcommand{\NWDOTAR}[1]%
{\truex{100}\truey{212}%
\Y=#1%
\divide\Y by 2%
\NUMBEROFDOTS=#1%
\divide\NUMBEROFDOTS by \value{y}%
\advance\NUMBEROFDOTS by 1%
\begin{picture}(0,0)%
\multiput(\Y,-\Y)(-\value{y},\value{y}){\NUMBEROFDOTS}%
{\circle*{\value{x}}}%
\put(-\Y,\Y){\nwhead}%
\end{picture}}%

\newcommand{\NWMONO}[1]{%
\Y=#1%
\divide \Y by 2%
\Truetail%
\bimolength=#1%
\advance\bimolength by -\Truemonotail%
\monolength=\bimolength%
\advance\monolength by -\Y%
\begin{picture}(0,0)%
\put(\monolength,-\monolength){\line(-1,1){\bimolength}}%
\put(\monolength,-\monolength){\nwhead}%
\put(-\Y,\Y){\nwhead}%
\end{picture}}%

\newcommand{\NWEPI}[1]{%
\Y=#1%
\divide\Y by 2%
\Truehead%
\bimolength=#1%
\advance\bimolength by -\Trueepihead%
\epilength=\bimolength%
\advance\epilength by -\Y%
\begin{picture}(0,0)%
\put(\Y,-\Y){\line(-1,1){\bimolength}}%
\put(-\epilength,\epilength){\nwhead}%
\put(-\Y,\Y){\nwhead}%
\end{picture}}%

\newcommand{\NWBIMO}[1]{%
\Y=#1%
\divide\Y by 2%
\Truetail\Truehead%
\bimolength=#1%
\advance\bimolength by -\Truemonotail%
\monolength=\bimolength%
\advance\monolength by -\Y%
\advance\bimolength by -\Trueepihead%
\epilength=\bimolength%
\advance\epilength by -\monolength%
\begin{picture}(0,0)%
\put(\monolength,-\monolength){\line(-1,1){\bimolength}}%
\put(\monolength,-\monolength){\nwhead}%
\put(-\epilength,\epilength){\nwhead}%
\put(-\Y,\Y){\nwhead}%
\end{picture}}%

\newcommand{\NWBIAR}[1]{%
\Y=#1%
\divide\Y by 2%
\begin{picture}(0,0)%
\put(\Y,-\Y){\begin{picture}(0,0)%
\truex{247}%
\put(-\value{x},-\value{x}){\line(-1,1){#1}}%
\put(\value{x},\value{x}){\line(-1,1){#1}}%
\monolength=#1%
\advance\monolength by -\value{x}%
\epilength=#1%
\advance\epilength by \value{x}%
\put(-\monolength,\epilength){\nwhead}%
\put(-\epilength,\monolength){\nwhead}%
\end{picture}}\end{picture}}%

\newcommand{\NWBIDIST}[1]{%
\Y=#1%
\divide\Y by 2%
\truey{400}%
\begin{picture}(0,0)%
\put(\Y,-\Y){\begin{picture}(0,0)%
\truex{247}%
\monolength=#1%
\advance\monolength by -\value{x}%
\epilength=#1%
\advance\epilength by \value{x}%
\put(-\value{x},-\value{x}){\line(-1,1){#1}}%
\put(-\epilength,\monolength){\nwhead}%
\end{picture}}%
\put(\Y,-\Y){\begin{picture}(0,0)%
\truex{247}%
\monolength=#1%
\advance\monolength by \value{x}%
\epilength=#1%
\advance\epilength by -\value{x}%
\put(\value{x},\value{x}){\line(-1,1){#1}}%
\put(-\epilength,\monolength){\nwhead}%
\end{picture}}%
\put(-\value{x},-\value{x}){\circle{\value{y}}}%
\put(\value{x},\value{x}){\circle{\value{y}}}%
\end{picture}}%

\newcommand{\NWADJAR}[1]{%
\Y=#1%
\divide\Y by 2%
\begin{picture}(0,0)%
\put(\Y,-\Y){\begin{picture}(0,0)%
\truex{247}%
\monolength=#1%
\advance\monolength by -\value{x}%
\epilength=#1%
\advance\epilength by \value{x}%
\put(-\value{x},-\value{x}){\line(-1,1){#1}}%
\put(-\epilength,\monolength){\nwhead}%
\end{picture}}%
\put(-\Y,\Y){\begin{picture}(0,0)%
\truex{247}%
\monolength=#1%
\advance\monolength by -\value{x}%
\epilength=#1%
\advance\epilength by \value{x}%
\put(\value{x},\value{x}){\line(1,-1){#1}}%
\put(\epilength,-\monolength){\sehead}%
\end{picture}}\end{picture}}%

\newcommand{\NWADJDIST}[1]{%
\Y=#1%
\divide\Y by 2%
\truey{400}%
\begin{picture}(0,0)%
\put(\Y,-\Y){\begin{picture}(0,0)%
\truex{247}%
\monolength=#1%
\advance\monolength by -\value{x}%
\epilength=#1%
\advance\epilength by \value{x}%
\put(-\value{x},-\value{x}){\line(-1,1){#1}}%
\put(-\epilength,\monolength){\nwhead}%
\end{picture}}%
\put(-\Y,\Y){\begin{picture}(0,0)%
\truex{247}%
\monolength=#1%
\advance\monolength by -\value{x}%
\epilength=#1%
\advance\epilength by \value{x}%
\put(\value{x},\value{x}){\line(1,-1){#1}}%
\put(\epilength,-\monolength){\sehead}%
\end{picture}}%
\put(-\value{x},-\value{x}){\circle{\value{y}}}%
\put(\value{x},\value{x}){\circle{\value{y}}}%
\end{picture}}%

\def\basicnwar[#1]{\sdcase{\NWAR}{}{}{#100}}%

\newcommand{\nwar}{\@ifnextchar[{\basicnwar}{\basicnwar[59]}}%

\def\basicNwar[#1]#2{\sdcase{\NWAR}{#2}{}{#100}}%

\newcommand{\Nwar}{\@ifnextchar[{\basicNwar}{\basicNwar[59]}}%

\def\basicnwaR[#1]#2{\sdcase{\NWAR}{}{#2}{#100}}%

\newcommand{\nwaR}{\@ifnextchar[{\basicnwaR}{\basicnwaR[59]}}%

\def\basicnwdist[#1]{\sdcase{\NWDIST}{}{}{#100}}%

\newcommand{\nwdist}{\@ifnextchar[{\basicnwdist}{\basicnwdist[59]}}%

\def\basicNwdist[#1]#2{\sdcase{\NWDIST}{#2}{}{#100}}%

\newcommand{\Nwdist}{\@ifnextchar[{\basicNwdist}{\basicNwdist[59]}}%

\def\basicnwdisT[#1]#2{\sdcase{\NWDIST}{}{#2}{#100}}%

\newcommand{\nwdisT}{\@ifnextchar[{\basicnwdisT}{\basicnwdisT[59]}}%

\def\basicnwdotar[#1]{\sdcase{\NWDOTAR}{}{}{#100}}%

\newcommand{\nwdotar}{\@ifnextchar[{\basicnwdotar}{\basicnwdotar[59]}}%

\def\basicNwdotar[#1]#2{\sdcase{\NWDOTAR}{#2}{}{#100}}%

\newcommand{\Nwdotar}{\@ifnextchar[{\basicNwdotar}{\basicNwdotar[59]}}%

\def\basicnwdotaR[#1]#2{\sdcase{\NWDOTAR}{}{#2}{#100}}%

\newcommand{\nwdotaR}{\@ifnextchar[{\basicnwdotaR}{\basicnwdotaR[59]}}%

\def\basicnwmono[#1]{\sdcase{\NWMONO}{}{}{#100}}%

\newcommand{\nwmono}{\@ifnextchar[{\basicnwmono}{\basicnwmono[59]}}%

\def\basicNwmono[#1]#2{\sdcase{\NWMONO}{#2}{}{#100}}%

\newcommand{\Nwmono}{\@ifnextchar[{\basicNwmono}{\basicNwmono[59]}}%

\def\basicnwmonO[#1]#2{\sdcase{\NWMONO}{}{#2}{#100}}%

\newcommand{\nwmonO}{\@ifnextchar[{\basicnwmonO}{\basicnwmonO[59]}}%

\def\basicnwepi[#1]{\sdcase{\NWEPI}{}{}{#100}}%

\newcommand{\nwepi}{\@ifnextchar[{\basicnwepi}{\basicnwepi[59]}}%

\def\basicNwepi[#1]#2{\sdcase{\NWEPI}{#2}{}{#100}}%

\newcommand{\Nwepi}{\@ifnextchar[{\basicNwepi}{\basicNwepi[59]}}%

\def\basicnwepI[#1]#2{\sdcase{\NWEPI}{}{#2}{#100}}%

\newcommand{\nwepI}{\@ifnextchar[{\basicnwepI}{\basicnwepI[59]}}%

\def\basicnwbimo[#1]{\sdcase{\NWBIMO}{}{}{#100}}%

\newcommand{\nwbimo}{\@ifnextchar[{\basicnwbimo}{\basicnwbimo[59]}}%

\def\basicNwbimo[#1]#2{\sdcase{\NWBIMO}{#2}{}{#100}}%

\newcommand{\Nwbimo}{\@ifnextchar[{\basicNwbimo}{\basicNwbimo[59]}}%

\def\basicnwbimO[#1]#2{\sdcase{\NWBIMO}{}{#2}{#100}}%

\newcommand{\nwbimO}{\@ifnextchar[{\basicnwbimO}{\basicnwbimO[59]}}%

\def\basicnwiso[#1]{\sdcase{\NWAR}{\hspace{-2pt}\cong}{}{#100}}%

\newcommand{\nwiso}{\@ifnextchar[{\basicnwiso}{\basicnwiso[59]}}%

\def\basicNwiso[#1]#2{\sdcase{\NWAR}{#2}{\cong}{#100}}%

\newcommand{\Nwiso}{\@ifnextchar[{\basicNwiso}{\basicNwiso[59]}}%

\def\basicnwisO[#1]#2{\sdcase{\NWAR}{\hspace{-2pt}\cong}{#2}{#100}}%

\newcommand{\nwisO}{\@ifnextchar[{\basicnwisO}{\basicnwisO[59]}}%

\def\basicnwbiar[#1]{\sdbicase{\NWBIAR}{}{}{#100}}%

\newcommand{\nwbiar}{\@ifnextchar[{\basicnwbiar}{\basicnwbiar[59]}}%

\def\basicNwbiar[#1]#2#3{\sdbicase{\NWBIAR}{#2}{#3}{#100}}%

\newcommand{\Nwbiar}{\@ifnextchar[{\basicNwbiar}{\basicNwbiar[59]}}%

\def\basicnwadjar[#1]{\sdbicase{\NWADJAR}{}{}{#100}}%

\newcommand{\nwadjar}{\@ifnextchar[{\basicnwadjar}{\basicnwadjar[59]}}%

\def\basicNwadjar[#1]#2#3{\sdbicase{\NWADJAR}{#2}{#3}{#100}}%

\newcommand{\Nwadjar}{\@ifnextchar[{\basicNwadjar}{\basicNwadjar[59]}}%

\def\basicnwbidist[#1]{\sdbicase{\NWBIDIST}{}{}{#100}}%

\newcommand{\nwbidist}{\@ifnextchar[{\basicnwbidist}{\basicnwbidist[59]}}%

\def\basicNwbidist[#1]#2#3{\sdbicase{\NWBIDIST}{#2}{#3}{#100}}%

\newcommand{\Nwbidist}{\@ifnextchar[{\basicNwbidist}{\basicNwbidist[59]}}%

\def\basicnwadjdist[#1]{\sdbicase{\NWADJDIST}{}{}{#100}}%

\newcommand{\nwadjdist}{\@ifnextchar[{\basicnwadjdist}{\basicnwadjdist[59]}}%

\def\basicNwadjdist[#1]#2#3{\sdbicase{\NWADJDIST}{#2}{#3}{#100}}%

\newcommand{\Nwadjdist}{\@ifnextchar[{\basicNwadjdist}{\basicNwadjdist[59]}}%

\newcommand{\ENEAR}[3]{\testdiagrammode%
\Y=#3%
\divide\Y by 2%
\Z=\Y%
\divide\Z by 2%
\begin{picture}(0,0)%
\put(-\Y,-\Z){\line(2,1){#3}}%
\put(\Y,\Z){\enehead}%
\truex{200}\truey{800}\truez{600}%
\put(-\value{x},\value{x}){\makebox(0,\value{z})[r]{${#1}$}}%
\put(\value{x},-\value{y}){\makebox(0,\value{z})[l]{${#2}$}}%
\end{picture}}%

\newcommand{\ENEDIST}[3]{\testdiagrammode%
\Y=#3%
\divide\Y by 2%
\Z=\Y%
\divide\Z by 2%
\begin{picture}(0,0)%
\put(-\Y,-\Z){\line(2,1){#3}}%
\put(\Y,\Z){\enehead}%
\truex{400}%
\put(0,0){\circle{\value{x}}}%
\truex{200}\truey{800}\truez{600}%
\put(-\value{x},\value{x}){\makebox(0,\value{z})[r]{${#1}$}}%
\put(\value{x},-\value{y}){\makebox(0,\value{z})[l]{${#2}$}}%
\end{picture}}%

\newcommand{\ENEDOTAR}[3]{\testdiagrammode%
\truex{100}\truey{268}\truez{134}%
\Y=#3%
\divide\Y by 2%
\Z=\Y%
\divide\Z by 2%
\NUMBEROFDOTS=#3%
\divide\NUMBEROFDOTS by \value{y}%
\advance\NUMBEROFDOTS by 1%
\begin{picture}(0,0)%
\multiput(-\Y,-\Z)(\value{y},\value{z}){\NUMBEROFDOTS}%
{\circle*{\value{x}}}%
\put(\Y,\Z){\enehead}%
\truex{200}\truey{800}\truez{600}%
\put(-\value{x},\value{x}){\makebox(0,\value{z})[r]{${#1}$}}%
\put(\value{x},-\value{y}){\makebox(0,\value{z})[l]{${#2}$}}%
\end{picture}}%

\newcommand{\ENEMONO}[3]{\testdiagrammode%
\Y=#3%
\divide\Y by 2%
\Z=\Y%
\divide\Z by 2%
\TrueTail%
\bimolength=#3%
\advance\bimolength by -\TrueMonoTail%
\monolength=\bimolength%
\advance\monolength by -\Y%
\secondmonolength=\monolength%
\divide\secondmonolength by 2%
\begin{picture}(0,0)%
\put(-\monolength,-\secondmonolength){\line(2,1){\bimolength}}%
\put(-\monolength,-\secondmonolength){\enehead}%
\put(\Y,\Z){\enehead}%
\truex{200}\truey{800}\truez{600}%
\put(-\value{x},\value{x}){\makebox(0,\value{z})[r]{${#1}$}}%
\put(\value{x},-\value{y}){\makebox(0,\value{z})[l]{${#2}$}}%
\end{picture}}%

\newcommand{\ENEEPI}[3]{\testdiagrammode%
\Y=#3%
\divide\Y by 2%
\Z=\Y%
\divide\Z by 2%
\TrueHead%
\bimolength=#3%
\advance\bimolength by -\TrueEpiHead%
\epilength=\bimolength%
\advance\epilength by -\Y%
\secondepilength=\epilength%
\divide\secondepilength by 2%
\begin{picture}(0,0)%
\put(-\Y,-\Z){\line(2,1){\bimolength}}%
\put(\epilength,\secondepilength){\enehead}%
\put(\Y,\Z){\enehead}%
\truex{200}\truey{800}\truez{600}%
\put(-\value{x},\value{x}){\makebox(0,\value{z})[r]{${#1}$}}%
\put(\value{x},-\value{y}){\makebox(0,\value{z})[l]{${#2}$}}%
\end{picture}}%

\newcommand{\ENEBIMO}[3]{\testdiagrammode%
\Y=#3%
\divide\Y by 2%
\Z=\Y%
\divide\Z by 2%
\TrueTail\TrueHead%
\bimolength=#3%
\advance\bimolength by -\TrueMonoTail%
\monolength=\bimolength%
\advance\monolength by -\Y%
\advance\bimolength by -\TrueEpiHead%
\epilength=\bimolength%
\advance\epilength by -\monolength%
\secondmonolength=\monolength%
\divide\secondmonolength by 2%
\secondepilength=\epilength%
\divide\secondepilength by 2%
\begin{picture}(0,0)%
\put(-\monolength,-\secondmonolength){\line(2,1){\bimolength}}%
\put(-\monolength,-\secondmonolength){\enehead}%
\put(\epilength,\secondepilength){\enehead}%
\put(\Y,\Z){\enehead}%
\truex{200}\truey{800}\truez{600}%
\put(-\value{x},\value{x}){\makebox(0,\value{z})[r]{${#1}$}}%
\put(\value{x},-\value{y}){\makebox(0,\value{z})[l]{${#2}$}}%
\end{picture}}%

\newcommand{\ENEEQL}[3]{\testdiagrammode%
\Y=#3%
\divide\Y by 2%
\Z=\Y%
\divide\Z by 2%
\begin{picture}(0,0)%
\put(-\Y,-\Z){\begin{picture}(0,0)%
\truex{44}\truey{89}%
\put(-\value{x},\value{y}){\line(2,1){#3}}%
\put(\value{x},-\value{y}){\line(2,1){#3}}%
\end{picture}}%
\truex{200}\truey{800}\truez{600}%
\put(-\value{x},\value{x}){\makebox(0,\value{z})[r]{${#1}$}}%
\put(\value{x},-\value{y}){\makebox(0,\value{z})[l]{${#2}$}}%
\end{picture}}%

\newcommand{\ENEBIAR}[3]{\testdiagrammode%
\Y=#3%
\divide\Y by 2%
\Z=\Y%
\divide\Z by 2%
\begin{picture}(0,0)%
\put(-\Y,-\Z){\begin{picture}(0,0)%
\truex{156}\truey{313}%
\put(-\value{x},\value{y}){\line(2,1){#3}}%
\put(\value{x},-\value{y}){\line(2,1){#3}}%
\monolength=#3%
\advance\monolength by -\value{x}%
\epilength=#3%
\advance\epilength by \value{x}%
\secondmonolength=\Y%
\advance\secondmonolength by -\value{y}%
\secondepilength=\Y%
\advance\secondepilength by \value{y}%
\put(\monolength,\secondepilength){\enehead}%
\put(\epilength,\secondmonolength){\enehead}%
\end{picture}}
\truex{300}\truey{1000}\truez{600}%
\put(-\value{x},\value{x}){\makebox(0,\value{z})[r]{${#1}$}}%
\put(\value{x},-\value{y}){\makebox(0,\value{z})[l]{${#2}$}}%
\end{picture}}%

\newcommand{\ENEBIDIST}[3]{\testdiagrammode%
\Y=#3%
\divide\Y by 2%
\Z=\Y%
\divide\Z by 2%
\begin{picture}(0,0)%
\truex{156}\truey{313}\truez{400}%
\put(-\Y,-\Z){\begin{picture}(0,0)%
\put(-\value{x},\value{y}){\line(2,1){#3}}%
\put(\value{x},-\value{y}){\line(2,1){#3}}%
\monolength=#3%
\advance\monolength by -\value{x}%
\epilength=#3%
\advance\epilength by \value{x}%
\secondmonolength=\Y%
\advance\secondmonolength by -\value{y}%
\secondepilength=\Y%
\advance\secondepilength by \value{y}%
\put(\monolength,\secondepilength){\enehead}%
\put(\epilength,\secondmonolength){\enehead}%
\end{picture}}
\put(-\value{x},\value{y}){\circle{\value{z}}}%
\put(\value{x},-\value{y}){\circle{\value{z}}}%
\truex{300}\truey{1000}\truez{600}%
\put(-\value{x},\value{x}){\makebox(0,\value{z})[r]{${#1}$}}%
\put(\value{x},-\value{y}){\makebox(0,\value{z})[l]{${#2}$}}%
\end{picture}}%

\newcommand{\ENEADJAR}[3]{\testdiagrammode%
\Y=#3%
\divide\Y by 2%
\Z=\Y%
\divide\Z by 2%
\begin{picture}(0,0)%
\put(-\Y,-\Z){\begin{picture}(0,0)%
\truex{156}\truey{313}%
\monolength=#3%
\advance\monolength by -\value{x}%
\epilength=#3%
\advance\epilength by \value{x}%
\secondmonolength=\Y%
\advance\secondmonolength by -\value{y}%
\secondepilength=\Y%
\advance\secondepilength by \value{y}%
\put(\value{x},-\value{y}){\line(2,1){#3}}%
\put(\epilength,\secondmonolength){\enehead}%
\put(\monolength,\secondepilength){\line(-2,-1){#3}}%
\put(-\value{x},\value{y}){\wswhead}%
\end{picture}}
\truex{300}\truey{1000}\truez{600}%
\put(-\value{x},\value{x}){\makebox(0,\value{z})[r]{${#1}$}}%
\put(\value{x},-\value{y}){\makebox(0,\value{z})[l]{${#2}$}}%
\end{picture}}%

\newcommand{\ENEADJDIST}[3]{\testdiagrammode%
\Y=#3%
\divide\Y by 2%
\Z=\Y%
\divide\Z by 2%
\begin{picture}(0,0)%
\truex{156}\truey{313}\truez{400}%
\put(-\Y,-\Z){\begin{picture}(0,0)%
\monolength=#3%
\advance\monolength by -\value{x}%
\epilength=#3%
\advance\epilength by \value{x}%
\secondmonolength=\Y%
\advance\secondmonolength by -\value{y}%
\secondepilength=\Y%
\advance\secondepilength by \value{y}%
\put(\value{x},-\value{y}){\line(2,1){#3}}%
\put(\epilength,\secondmonolength){\enehead}%
\put(\monolength,\secondepilength){\line(-2,-1){#3}}%
\put(-\value{x},\value{y}){\wswhead}%
\end{picture}}
\put(-\value{x},\value{y}){\circle{\value{z}}}%
\put(\value{x},-\value{y}){\circle{\value{z}}}%
\truex{300}\truey{1000}\truez{600}%
\put(-\value{x},\value{x}){\makebox(0,\value{z})[r]{${#1}$}}%
\put(\value{x},-\value{y}){\makebox(0,\value{z})[l]{${#2}$}}%
\end{picture}}%

\def\basicenear[#1]{\ENEAR{}{}{#100}}%

\newcommand{\enear}{\@ifnextchar[{\basicenear}{\basicenear[133]}}%

\def\basicEnear[#1]#2{\ENEAR{#2}{}{#100}}%

\newcommand{\Enear}{\@ifnextchar[{\basicEnear}{\basicEnear[133]}}%

\def\basiceneaR[#1]#2{\ENEAR{}{#2}{#100}}%

\newcommand{\eneaR}{\@ifnextchar[{\basiceneaR}{\basiceneaR[133]}}%

\def\basicenedist[#1]{\ENEDIST{}{}{#100}}%

\newcommand{\enedist}{\@ifnextchar[{\basicenedist}{\basicenedist[133]}}%

\def\basicEnedist[#1]#2{\ENEDIST{#2}{}{#100}}%

\newcommand{\Enedist}{\@ifnextchar[{\basicEnedist}{\basicEnedist[133]}}%

\def\basicenedisT[#1]#2{\ENEDIST{}{#2}{#100}}%

\newcommand{\enedisT}{\@ifnextchar[{\basicenedisT}{\basicenedisT[133]}}%

\def\basicenedotar[#1]{\ENEDOTAR{}{}{#100}}%

\newcommand{\enedotar}{\@ifnextchar[{\basicenedotar}{\basicenedotar[133]}}%

\def\basicEnedotar[#1]#2{\ENEDOTAR{#2}{}{#100}}%

\newcommand{\Enedotar}{\@ifnextchar[{\basicEnedotar}{\basicEnedotar[133]}}%

\def\basicenedotaR[#1]#2{\ENEDOTAR{}{#2}{#100}}%

\newcommand{\enedotaR}{\@ifnextchar[{\basicenedotaR}{\basicenedotaR[133]}}%

\def\basicenemono[#1]{\ENEMONO{}{}{#100}}%

\newcommand{\enemono}{\@ifnextchar[{\basicenemono}{\basicenemono[133]}}%

\def\basicEnemono[#1]#2{\ENEMONO{#2}{}{#100}}%

\newcommand{\Enemono}{\@ifnextchar[{\basicEnemono}{\basicEnemono[133]}}%

\def\basicenemonO[#1]#2{\ENEMONO{}{#2}{#100}}%

\newcommand{\enemonO}{\@ifnextchar[{\basicenemonO}{\basicenemonO[133]}}%

\def\basiceneepi[#1]{\ENEEPI{}{}{#100}}%

\newcommand{\eneepi}{\@ifnextchar[{\basiceneepi}{\basiceneepi[133]}}%

\def\basicEneepi[#1]#2{\ENEEPI{#2}{}{#100}}%

\newcommand{\Eneepi}{\@ifnextchar[{\basicEneepi}{\basicEneepi[133]}}%

\def\basiceneepI[#1]#2{\ENEEPI{}{#2}{#100}}%

\newcommand{\eneepI}{\@ifnextchar[{\basiceneepI}{\basiceneepI[133]}}%

\def\basicenebimo[#1]{\ENEBIMO{}{}{#100}}%

\newcommand{\enebimo}{\@ifnextchar[{\basicenebimo}{\basicenebimo[133]}}%

\def\basicEnebimo[#1]#2{\ENEBIMO{#2}{}{#100}}%

\newcommand{\Enebimo}{\@ifnextchar[{\basicEnebimo}{\basicEnebimo[133]}}%

\def\basicenebimO[#1]#2{\ENEBIMO{}{#2}{#100}}%

\newcommand{\enebimO}{\@ifnextchar[{\basicenebimO}{\basicenebimO[133]}}%

\def\basiceneiso[#1]{\ENEAR{\cong}{}{#100}}%

\newcommand{\eneiso}{\@ifnextchar[{\basiceneiso}{\basiceneiso[133]}}%

\def\basicEneiso[#1]#2{\ENEAR{#2}{\cong}{#100}}%

\newcommand{\Eneiso}{\@ifnextchar[{\basicEneiso}{\basicEneiso[133]}}%

\def\basiceneisO[#1]#2{\ENEAR{\cong}{#2}{#100}}%

\newcommand{\eneisO}{\@ifnextchar[{\basiceneisO}{\basiceneisO[133]}}%

\def\basiceneeql[#1]{\ENEEQL{}{}{#100}}%

\newcommand{\eneeql}{\@ifnextchar[{\basiceneeql}{\basiceneeql[133]}}%

\def\basicEneeql[#1]#2{\ENEEQL{#2}{}{#100}}%

\newcommand{\Eneeql}{\@ifnextchar[{\basicEneeql}{\basicEneeql[133]}}%

\def\basiceneeqL[#1]#2{\ENEEQL{}{#2}{#100}}%

\newcommand{\eneeqL}{\@ifnextchar[{\basiceneeqL}{\basiceneeqL[133]}}%

\def\basicenebiar[#1]{\ENEBIAR{}{}{#100}}%

\newcommand{\enebiar}{\@ifnextchar[{\basicenebiar}{\basicenebiar[133]}}%

\def\basicEnebiar[#1]#2#3{\ENEBIAR{#2}{#3}{#100}}%

\newcommand{\Enebiar}{\@ifnextchar[{\basicEnebiar}{\basicEnebiar[133]}}%

\def\basicenebidist[#1]{\ENEBIDIST{}{}{#100}}%

\newcommand{\enebidist}{\@ifnextchar[{\basicenebidist}{\basicenebidist[133]}}%

\def\basicEnebidist[#1]#2#3{\ENEBIDIST{#2}{#3}{#100}}%

\newcommand{\Enebidist}{\@ifnextchar[{\basicEnebidist}{\basicEnebidist[133]}}%

\def\basiceneadjar[#1]{\ENEADJAR{}{}{#100}}%

\newcommand{\eneadjar}{\@ifnextchar[{\basiceneadjar}{\basiceneadjar[133]}}%

\def\basicEneadjar[#1]#2#3{\ENEADJAR{#2}{#3}{#100}}%

\newcommand{\Eneadjar}{\@ifnextchar[{\basicEneadjar}{\basicEneadjar[133]}}%

\def\basiceneadjdist[#1]{\ENEADJDIST{}{}{#100}}%

\newcommand{\eneadjdist}{\@ifnextchar[{\basiceneadjdist}{\basiceneadjdist[133]}}%

\def\basicEneadjdist[#1]#2#3{\ENEADJDIST{#2}{#3}{#100}}%

\newcommand{\Eneadjdist}{\@ifnextchar[{\basicEneadjdist}{\basicEneadjdist[133]}}%

\newcommand{\ESEAR}[3]{\testdiagrammode%
\Y=#3%
\divide\Y by 2%
\Z=\Y%
\divide\Z by 2%
\begin{picture}(0,0)%
\put(-\Y,\Z){\line(2,-1){#3}}%
\put(\Y,-\Z){\esehead}%
\truex{200}\truey{800}\truez{600}%
\put(\value{x},\value{x}){\makebox(0,\value{z})[l]{${#1}$}}%
\put(-\value{x},-\value{y}){\makebox(0,\value{z})[r]{${#2}$}}%
\end{picture}}%

\newcommand{\ESEDIST}[3]{\testdiagrammode%
\Y=#3%
\divide\Y by 2%
\Z=\Y%
\divide\Z by 2%
\begin{picture}(0,0)%
\put(-\Y,\Z){\line(2,-1){#3}}%
\put(\Y,-\Z){\esehead}%
\truex{400}%
\put(0,0){\circle{\value{x}}}%
\truex{200}\truey{800}\truez{600}%
\put(\value{x},\value{x}){\makebox(0,\value{z})[l]{${#1}$}}%
\put(-\value{x},-\value{y}){\makebox(0,\value{z})[r]{${#2}$}}%
\end{picture}}%

\newcommand{\ESEDOTAR}[3]{\testdiagrammode%
\truex{100}\truey{268}\truez{134}%
\Y=#3%
\divide\Y by 2%
\Z=\Y%
\divide\Z by 2%
\NUMBEROFDOTS=#3%
\divide\NUMBEROFDOTS by \value{y}%
\advance\NUMBEROFDOTS by 1%
\begin{picture}(0,0)%
\multiput(-\Y,\Z)(\value{y},-\value{z}){\NUMBEROFDOTS}%
{\circle*{\value{x}}}%
\put(\Y,-\Z){\esehead}%
\truex{200}\truey{800}\truez{600}%
\put(\value{x},\value{x}){\makebox(0,\value{z})[l]{${#1}$}}%
\put(-\value{x},-\value{y}){\makebox(0,\value{z})[r]{${#2}$}}%
\end{picture}}%

\newcommand{\ESEMONO}[3]{\testdiagrammode%
\Y=#3%
\divide\Y by 2%
\Z=\Y%
\divide\Z by 2%
\TrueTail%
\bimolength=#3%
\advance\bimolength by -\TrueMonoTail%
\monolength=\bimolength%
\advance\monolength by -\Y%
\secondmonolength=\monolength%
\divide\secondmonolength by 2%
\begin{picture}(0,0)%
\put(-\monolength,\secondmonolength){\line(2,-1){\bimolength}}%
\put(-\monolength,\secondmonolength){\esehead}%
\put(\Y,-\Z){\esehead}%
\truex{200}\truey{800}\truez{600}%
\put(\value{x},\value{x}){\makebox(0,\value{z})[l]{${#1}$}}%
\put(-\value{x},-\value{y}){\makebox(0,\value{z})[r]{${#2}$}}%
\end{picture}}%

\newcommand{\ESEEPI}[3]{\testdiagrammode%
\Y=#3%
\divide\Y by 2%
\Z=\Y%
\divide\Z by 2%
\TrueHead%
\bimolength=#3%
\advance\bimolength by -\TrueEpiHead%
\epilength=\bimolength%
\advance\epilength by -\Y%
\secondepilength=\epilength%
\divide\secondepilength by 2%
\begin{picture}(0,0)%
\put(-\Y,\Z){\line(2,-1){\bimolength}}%
\put(\epilength,-\secondepilength){\esehead}%
\put(\Y,-\Z){\esehead}%
\truex{200}\truey{800}\truez{600}%
\put(\value{x},\value{x}){\makebox(0,\value{z})[l]{${#1}$}}%
\put(-\value{x},-\value{y}){\makebox(0,\value{z})[r]{${#2}$}}%
\end{picture}}%

\newcommand{\ESEBIMO}[3]{\testdiagrammode%
\Y=#3%
\divide\Y by 2%
\Z=\Y%
\divide\Z by 2%
\TrueTail\TrueHead%
\bimolength=#3%
\advance\bimolength by -\TrueMonoTail%
\monolength=\bimolength%
\advance\monolength by -\Y%
\advance\bimolength by -\TrueEpiHead%
\epilength=\bimolength%
\advance\epilength by -\monolength%
\secondmonolength=\monolength%
\divide\secondmonolength by 2%
\secondepilength=\epilength%
\divide\secondepilength by 2%
\begin{picture}(0,0)%
\put(-\monolength,\secondmonolength){\line(2,-1){\bimolength}}%
\put(-\monolength,\secondmonolength){\esehead}%
\put(\epilength,-\secondepilength){\esehead}%
\put(\Y,-\Z){\esehead}%
\truex{200}\truey{800}\truez{600}%
\put(\value{x},\value{x}){\makebox(0,\value{z})[l]{${#1}$}}%
\put(-\value{x},-\value{y}){\makebox(0,\value{z})[r]{${#2}$}}%
\end{picture}}%

\newcommand{\ESEEQL}[3]{\testdiagrammode%
\Y=#3%
\divide\Y by 2%
\Z=\Y%
\divide\Z by 2%
\begin{picture}(0,0)%
\put(-\Y,\Z){\begin{picture}(0,0)%
\truex{44}\truey{89}%
\put(-\value{x},-\value{y}){\line(2,-1){#3}}%
\put(\value{x},\value{y}){\line(2,-1){#3}}%
\end{picture}}%
\truex{200}\truey{800}\truez{600}%
\put(\value{x},\value{x}){\makebox(0,\value{z})[l]{${#1}$}}%
\put(-\value{x},-\value{y}){\makebox(0,\value{z})[r]{${#2}$}}%
\end{picture}}%

\newcommand{\ESEBIAR}[3]{%
\Y=#3%
\divide\Y by 2%
\Z=\Y%
\divide\Z by 2%
\begin{picture}(0,0)%
\put(-\Y,\Z){\begin{picture}(0,0)%
\truex{156}\truey{313}%
\put(-\value{x},-\value{y}){\line(2,-1){#3}}%
\put(\value{x},\value{y}){\line(2,-1){#3}}%
\monolength=#3%
\advance\monolength by -\value{x}%
\epilength=#3%
\advance\epilength by \value{x}%
\secondmonolength=\Y%
\advance\secondmonolength by -\value{y}%
\secondepilength=\Y%
\advance\secondepilength by \value{y}%
\put(\monolength,-\secondepilength){\esehead}%
\put(\epilength,-\secondmonolength){\esehead}%
\end{picture}}
\truex{400}\truey{1000}\truez{600}%
\put(\value{x},\value{x}){\makebox(0,\value{z})[l]{${#1}$}}%
\put(-\value{x},-\value{y}){\makebox(0,\value{z})[r]{${#2}$}}%
\end{picture}}%

\newcommand{\ESEBIDIST}[3]{\testdiagrammode%
\Y=#3%
\divide\Y by 2%
\Z=\Y%
\divide\Z by 2%
\begin{picture}(0,0)%
\truex{156}\truey{313}\truez{400}%
\put(-\Y,\Z){\begin{picture}(0,0)%
\put(-\value{x},-\value{y}){\line(2,-1){#3}}%
\put(\value{x},\value{y}){\line(2,-1){#3}}%
\monolength=#3%
\advance\monolength by -\value{x}%
\epilength=#3%
\advance\epilength by \value{x}%
\secondmonolength=\Y%
\advance\secondmonolength by -\value{y}%
\secondepilength=\Y%
\advance\secondepilength by \value{y}%
\put(\monolength,-\secondepilength){\esehead}%
\put(\epilength,-\secondmonolength){\esehead}%
\end{picture}}
\put(\value{x},\value{y}){\circle{\value{z}}}%
\put(-\value{x},-\value{y}){\circle{\value{z}}}%
\truex{400}\truey{1000}\truez{600}%
\put(\value{x},\value{x}){\makebox(0,\value{z})[l]{${#1}$}}%
\put(-\value{x},-\value{y}){\makebox(0,\value{z})[r]{${#2}$}}%
\end{picture}}%

\newcommand{\ESEADJAR}[3]{\testdiagrammode%
\Y=#3%
\divide\Y by 2%
\Z=\Y%
\divide\Z by 2%
\begin{picture}(0,0)%
\put(-\Y,\Z){\begin{picture}(0,0)%
\truex{156}\truey{313}%
\monolength=#3%
\advance\monolength by -\value{x}%
\epilength=#3%
\advance\epilength by \value{x}%
\secondmonolength=\Y%
\advance\secondmonolength by -\value{y}%
\secondepilength=\Y%
\advance\secondepilength by \value{y}%
\put(-\value{x},-\value{y}){\line(2,-1){#3}}%
\put(\monolength,-\secondepilength){\esehead}%
\put(\epilength,-\secondmonolength){\line(-2,1){#3}}%
\put(\value{x},\value{y}){\wnwhead}%
\end{picture}}
\truex{400}\truey{1000}\truez{600}%
\put(\value{x},\value{x}){\makebox(0,\value{z})[l]{${#1}$}}%
\put(-\value{x},-\value{y}){\makebox(0,\value{z})[r]{${#2}$}}%
\end{picture}}%

\newcommand{\ESEADJDIST}[3]{\testdiagrammode%
\Y=#3%
\divide\Y by 2%
\Z=\Y%
\divide\Z by 2%
\begin{picture}(0,0)%
\truex{156}\truey{313}\truez{400}%
\put(-\Y,\Z){\begin{picture}(0,0)%
\monolength=#3%
\advance\monolength by -\value{x}%
\epilength=#3%
\advance\epilength by \value{x}%
\secondmonolength=\Y%
\advance\secondmonolength by -\value{y}%
\secondepilength=\Y%
\advance\secondepilength by \value{y}%
\put(-\value{x},-\value{y}){\line(2,-1){#3}}%
\put(\monolength,-\secondepilength){\esehead}%
\put(\epilength,-\secondmonolength){\line(-2,1){#3}}%
\put(\value{x},\value{y}){\wnwhead}%
\end{picture}}
\put(\value{x},\value{y}){\circle{\value{z}}}%
\put(-\value{x},-\value{y}){\circle{\value{z}}}%
\truex{400}\truey{1000}\truez{600}%
\put(\value{x},\value{x}){\makebox(0,\value{z})[l]{${#1}$}}%
\put(-\value{x},-\value{y}){\makebox(0,\value{z})[r]{${#2}$}}%
\end{picture}}%

\def\basicesear[#1]{\ESEAR{}{}{#100}}%

\newcommand{\esear}{\@ifnextchar[{\basicesear}{\basicesear[133]}}%

\def\basicEsear[#1]#2{\ESEAR{#2}{}{#100}}%

\newcommand{\Esear}{\@ifnextchar[{\basicEsear}{\basicEsear[133]}}%

\def\basiceseaR[#1]#2{\ESEAR{}{#2}{#100}}%

\newcommand{\eseaR}{\@ifnextchar[{\basiceseaR}{\basiceseaR[133]}}%

\def\basicesedist[#1]{\ESEDIST{}{}{#100}}%

\newcommand{\esedist}{\@ifnextchar[{\basicesedist}{\basicesedist[133]}}%

\def\basicEsedist[#1]#2{\ESEDIST{#2}{}{#100}}%

\newcommand{\Esedist}{\@ifnextchar[{\basicEsedist}{\basicEsedist[133]}}%

\def\basicesedisT[#1]#2{\ESEDIST{}{#2}{#100}}%

\newcommand{\esedisT}{\@ifnextchar[{\basicesedisT}{\basicesedisT[133]}}%

\def\basicesedotar[#1]{\ESEDOTAR{}{}{#100}}%

\newcommand{\esedotar}{\@ifnextchar[{\basicesedotar}{\basicesedotar[133]}}%

\def\basicEsedotar[#1]#2{\ESEDOTAR{#2}{}{#100}}%

\newcommand{\Esedotar}{\@ifnextchar[{\basicEsedotar}{\basicEsedotar[133]}}%

\def\basicesedotaR[#1]#2{\ESEDOTAR{}{#2}{#100}}%

\newcommand{\esedotaR}{\@ifnextchar[{\basicesedotaR}{\basicesedotaR[133]}}%

\def\basicesemono[#1]{\ESEMONO{}{}{#100}}%

\newcommand{\esemono}{\@ifnextchar[{\basicesemono}{\basicesemono[133]}}%

\def\basicEsemono[#1]#2{\ESEMONO{#2}{}{#100}}%

\newcommand{\Esemono}{\@ifnextchar[{\basicEsemono}{\basicEsemono[133]}}%

\def\basicesemonO[#1]#2{\ESEMONO{}{#2}{#100}}%

\newcommand{\esemonO}{\@ifnextchar[{\basicesemonO}{\basicesemonO[133]}}%

\def\basiceseepi[#1]{\ESEEPI{}{}{#100}}%

\newcommand{\eseepi}{\@ifnextchar[{\basiceseepi}{\basiceseepi[133]}}%

\def\basicEseepi[#1]#2{\ESEEPI{#2}{}{#100}}%

\newcommand{\Eseepi}{\@ifnextchar[{\basicEseepi}{\basicEseepi[133]}}%

\def\basiceseepI[#1]#2{\ESEEPI{}{#2}{#100}}%

\newcommand{\eseepI}{\@ifnextchar[{\basiceseepI}{\basiceseepI[133]}}%

\def\basicesebimo[#1]{\ESEBIMO{}{}{#100}}%

\newcommand{\esebimo}{\@ifnextchar[{\basicesebimo}{\basicesebimo[133]}}%

\def\basicEsebimo[#1]#2{\ESEBIMO{#2}{}{#100}}%

\newcommand{\Esebimo}{\@ifnextchar[{\basicEsebimo}{\basicEsebimo[133]}}%

\def\basicesebimO[#1]#2{\ESEBIMO{}{#2}{#100}}%

\newcommand{\esebimO}{\@ifnextchar[{\basicesebimO}{\basicesebimO[133]}}%

\def\basiceseiso[#1]{\ESEAR{\cong}{}{#100}}%

\newcommand{\eseiso}{\@ifnextchar[{\basiceseiso}{\basiceseiso[133]}}%

\def\basicEseiso[#1]#2{\ESEAR{#2}{\cong}{#100}}%

\newcommand{\Eseiso}{\@ifnextchar[{\basicEseiso}{\basicEseiso[133]}}%

\def\basiceseisO[#1]#2{\ESEAR{\cong}{#2}{#100}}%

\newcommand{\eseisO}{\@ifnextchar[{\basiceseisO}{\basiceseisO[133]}}%

\def\basiceseeql[#1]{\ESEEQL{}{}{#100}}%

\newcommand{\eseeql}{\@ifnextchar[{\basiceseeql}{\basiceseeql[133]}}%

\def\basicEseeql[#1]#2{\ESEEQL{#2}{}{#100}}%

\newcommand{\Eseeql}{\@ifnextchar[{\basicEseeql}{\basicEseeql[133]}}%

\def\basiceseeqL[#1]#2{\ESEEQL{}{#2}{#100}}%

\newcommand{\eseeqL}{\@ifnextchar[{\basiceseeqL}{\basiceseeqL[133]}}%

\def\basicesebiar[#1]{\ESEBIAR{}{}{#100}}%

\newcommand{\esebiar}{\@ifnextchar[{\basicesebiar}{\basicesebiar[133]}}%

\def\basicEsebiar[#1]#2#3{\ESEBIAR{#2}{#3}{#100}}%

\newcommand{\Esebiar}{\@ifnextchar[{\basicEsebiar}{\basicEsebiar[133]}}%

\def\basicesebidist[#1]{\ESEBIDIST{}{}{#100}}%

\newcommand{\esebidist}{\@ifnextchar[{\basicesebidist}{\basicesebidist[133]}}%

\def\basicEsebidist[#1]#2#3{\ESEBIDIST{#2}{#3}{#100}}%

\newcommand{\Esebidist}{\@ifnextchar[{\basicEsebidist}{\basicEsebidist[133]}}%

\def\basiceseadjar[#1]{\ESEADJAR{}{}{#100}}%

\newcommand{\eseadjar}{\@ifnextchar[{\basiceseadjar}{\basiceseadjar[133]}}%

\def\basicEseadjar[#1]#2#3{\ESEADJAR{#2}{#3}{#100}}%

\newcommand{\Eseadjar}{\@ifnextchar[{\basicEseadjar}{\basicEseadjar[133]}}%

\def\basiceseadjdist[#1]{\ESEADJDIST{}{}{#100}}%

\newcommand{\eseadjdist}{\@ifnextchar[{\basiceseadjdist}{\basiceseadjdist[133]}}%

\def\basicEseadjdist[#1]#2#3{\ESEADJDIST{#2}{#3}{#100}}%

\newcommand{\Eseadjdist}{\@ifnextchar[{\basicEseadjdist}{\basicEseadjdist[133]}}%

\newcommand{\WSWAR}[3]{\testdiagrammode%
\Y=#3%
\divide\Y by 2%
\Z=\Y%
\divide\Z by 2%
\begin{picture}(0,0)%
\put(\Y,\Z){\line(-2,-1){#3}}%
\put(-\Y,-\Z){\wswhead}%
\truex{200}\truey{800}\truez{600}%
\put(-\value{x},\value{x}){\makebox(0,\value{z})[r]{${#1}$}}%
\put(\value{x},-\value{y}){\makebox(0,\value{z})[l]{${#2}$}}%
\end{picture}}%

\newcommand{\WSWDIST}[3]{\testdiagrammode%
\Y=#3%
\divide\Y by 2%
\Z=\Y%
\divide\Z by 2%
\begin{picture}(0,0)%
\put(\Y,\Z){\line(-2,-1){#3}}%
\put(-\Y,-\Z){\wswhead}%
\truex{400}%
\put(0,0){\circle{\value{x}}}%
\truex{200}\truey{800}\truez{600}%
\put(-\value{x},\value{x}){\makebox(0,\value{z})[r]{${#1}$}}%
\put(\value{x},-\value{y}){\makebox(0,\value{z})[l]{${#2}$}}%
\end{picture}}%

\newcommand{\WSWDOTAR}[3]{\testdiagrammode%
\truex{100}\truey{268}\truez{134}%
\Y=#3%
\divide\Y by 2%
\Z=\Y%
\divide\Z by 2%
\NUMBEROFDOTS=#3%
\divide\NUMBEROFDOTS by \value{y}%
\advance\NUMBEROFDOTS by 1%
\begin{picture}(0,0)%
\multiput(\Y,\Z)(-\value{y},-\value{z}){\NUMBEROFDOTS}%
{\circle*{\value{x}}}%
\put(-\Y,-\Z){\wswhead}%
\truex{200}\truey{800}\truez{600}%
\put(-\value{x},\value{x}){\makebox(0,\value{z})[r]{${#1}$}}%
\put(\value{x},-\value{y}){\makebox(0,\value{z})[l]{${#2}$}}%
\end{picture}}%

\newcommand{\WSWMONO}[3]{\testdiagrammode%
\Y=#3%
\divide\Y by 2%
\Z=\Y%
\divide\Z by 2%
\TrueTail%
\bimolength=#3%
\advance\bimolength by -\TrueMonoTail%
\monolength=\bimolength%
\advance\monolength by -\Y%
\secondmonolength=\monolength%
\divide\secondmonolength by 2%
\begin{picture}(0,0)%
\put(\monolength,\secondmonolength){\line(-2,-1){\bimolength}}%
\put(\monolength,\secondmonolength){\wswhead}%
\put(-\Y,-\Z){\wswhead}%
\truex{200}\truey{800}\truez{600}%
\put(-\value{x},\value{x}){\makebox(0,\value{z})[r]{${#1}$}}%
\put(\value{x},-\value{y}){\makebox(0,\value{z})[l]{${#2}$}}%
\end{picture}}%

\newcommand{\WSWEPI}[3]{\testdiagrammode%
\Y=#3%
\divide\Y by 2%
\Z=\Y%
\divide\Z by 2%
\TrueHead%
\bimolength=#3%
\advance\bimolength by -\TrueEpiHead%
\epilength=\bimolength%
\advance\epilength by -\Y%
\secondepilength=\epilength%
\divide\secondepilength by 2%
\begin{picture}(0,0)%
\put(\Y,\Z){\line(-2,-1){\bimolength}}%
\put(-\epilength,-\secondepilength){\wswhead}%
\put(-\Y,-\Z){\wswhead}%
\truex{200}\truey{800}\truez{600}%
\put(-\value{x},\value{x}){\makebox(0,\value{z})[r]{${#1}$}}%
\put(\value{x},-\value{y}){\makebox(0,\value{z})[l]{${#2}$}}%
\end{picture}}%

\newcommand{\WSWBIMO}[3]{\testdiagrammode%
\Y=#3%
\divide\Y by 2%
\Z=\Y%
\divide\Z by 2%
\TrueTail\TrueHead%
\bimolength=#3%
\advance\bimolength by -\TrueMonoTail%
\monolength=\bimolength%
\advance\monolength by -\Y%
\advance\bimolength by -\TrueEpiHead%
\epilength=\bimolength%
\advance\epilength by -\monolength%
\secondmonolength=\monolength%
\divide\secondmonolength by 2%
\secondepilength=\epilength%
\divide\secondepilength by 2%
\begin{picture}(0,0)%
\put(\monolength,\secondmonolength){\line(-2,-1){\bimolength}}%
\put(\monolength,\secondmonolength){\wswhead}%
\put(-\epilength,-\secondepilength){\wswhead}%
\put(-\Y,-\Z){\wswhead}%
\truex{200}\truey{800}\truez{600}%
\put(-\value{x},\value{x}){\makebox(0,\value{z})[r]{${#1}$}}%
\put(\value{x},-\value{y}){\makebox(0,\value{z})[l]{${#2}$}}%
\end{picture}}%

\newcommand{\WSWBIAR}[3]{\testdiagrammode%
\Y=#3%
\divide\Y by 2%
\Z=\Y%
\divide\Z by 2%
\begin{picture}(0,0)%
\put(\Y,\Z){\begin{picture}(0,0)%
\truex{156}\truey{313}%
\put(-\value{x},\value{y}){\line(-2,-1){#3}}%
\put(\value{x},-\value{y}){\line(-2,-1){#3}}%
\monolength=#3%
\advance\monolength by -\value{x}%
\epilength=#3%
\advance\epilength by \value{x}%
\secondmonolength=\Y%
\advance\secondmonolength by -\value{y}%
\secondepilength=\Y%
\advance\secondepilength by \value{y}%
\put(-\monolength,-\secondepilength){\wswhead}%
\put(-\epilength,-\secondmonolength){\wswhead}%
\end{picture}}
\truex{300}\truey{1000}\truez{600}%
\put(-\value{x},\value{x}){\makebox(0,\value{z})[r]{${#1}$}}%
\put(\value{x},-\value{y}){\makebox(0,\value{z})[l]{${#2}$}}%
\end{picture}}%

\newcommand{\WSWBIDIST}[3]{\testdiagrammode%
\Y=#3%
\divide\Y by 2%
\Z=\Y%
\divide\Z by 2%
\begin{picture}(0,0)%
\truex{156}\truey{313}\truez{400}%
\put(\Y,\Z){\begin{picture}(0,0)%
\put(-\value{x},\value{y}){\line(-2,-1){#3}}%
\put(\value{x},-\value{y}){\line(-2,-1){#3}}%
\monolength=#3%
\advance\monolength by -\value{x}%
\epilength=#3%
\advance\epilength by \value{x}%
\secondmonolength=\Y%
\advance\secondmonolength by -\value{y}%
\secondepilength=\Y%
\advance\secondepilength by \value{y}%
\put(-\monolength,-\secondepilength){\wswhead}%
\put(-\epilength,-\secondmonolength){\wswhead}%
\end{picture}}
\put(-\value{x},\value{y}){\circle{\value{z}}}%
\put(\value{x},-\value{y}){\circle{\value{z}}}%
\truex{300}\truey{1000}\truez{600}%
\put(-\value{x},\value{x}){\makebox(0,\value{z})[r]{${#1}$}}%
\put(\value{x},-\value{y}){\makebox(0,\value{z})[l]{${#2}$}}%
\end{picture}}%

\newcommand{\WSWADJAR}[3]{\testdiagrammode%
\Y=#3%
\divide\Y by 2%
\Z=\Y%
\divide\Z by 2%
\begin{picture}(0,0)%
\put(\Y,\Z){\begin{picture}(0,0)%
\truex{156}\truey{313}%
\monolength=#3%
\advance\monolength by -\value{x}%
\epilength=#3%
\advance\epilength by \value{x}%
\secondmonolength=\Y%
\advance\secondmonolength by -\value{y}%
\secondepilength=\Y%
\advance\secondepilength by \value{y}%
\put(\value{x},-\value{y}){\line(-2,-1){#3}}%
\put(-\monolength,-\secondepilength){\wswhead}%
\put(-\epilength,-\secondmonolength){\line(2,1){#3}}%
\put(-\value{x},\value{y}){\enehead}%
\end{picture}}
\truex{300}\truey{1000}\truez{600}%
\put(-\value{x},\value{x}){\makebox(0,\value{z})[r]{${#1}$}}%
\put(\value{x},-\value{y}){\makebox(0,\value{z})[l]{${#2}$}}%
\end{picture}}%

\newcommand{\WSWADJDIST}[3]{\testdiagrammode%
\Y=#3%
\divide\Y by 2%
\Z=\Y%
\divide\Z by 2%
\begin{picture}(0,0)%
\truex{156}\truey{313}\truez{400}%
\put(\Y,\Z){\begin{picture}(0,0)%
\monolength=#3%
\advance\monolength by -\value{x}%
\epilength=#3%
\advance\epilength by \value{x}%
\secondmonolength=\Y%
\advance\secondmonolength by -\value{y}%
\secondepilength=\Y%
\advance\secondepilength by \value{y}%
\put(\value{x},-\value{y}){\line(-2,-1){#3}}%
\put(-\monolength,-\secondepilength){\wswhead}%
\put(-\epilength,-\secondmonolength){\line(2,1){#3}}%
\put(-\value{x},\value{y}){\enehead}%
\end{picture}}
\put(-\value{x},\value{y}){\circle{\value{z}}}%
\put(\value{x},-\value{y}){\circle{\value{z}}}%
\truex{300}\truey{1000}\truez{600}%
\put(-\value{x},\value{x}){\makebox(0,\value{z})[r]{${#1}$}}%
\put(\value{x},-\value{y}){\makebox(0,\value{z})[l]{${#2}$}}%
\end{picture}}%

\def\basicwswar[#1]{\WSWAR{}{}{#100}}%

\newcommand{\wswar}{\@ifnextchar[{\basicwswar}{\basicwswar[133]}}%

\def\basicWswar[#1]#2{\WSWAR{#2}{}{#100}}%

\newcommand{\Wswar}{\@ifnextchar[{\basicWswar}{\basicWswar[133]}}%

\def\basicwswaR[#1]#2{\WSWAR{}{#2}{#100}}%

\newcommand{\wswaR}{\@ifnextchar[{\basicwswaR}{\basicwswaR[133]}}%

\def\basicwswdist[#1]{\WSWDIST{}{}{#100}}%

\newcommand{\wswdist}{\@ifnextchar[{\basicwswdist}{\basicwswdist[133]}}%

\def\basicWswdist[#1]#2{\WSWDIST{#2}{}{#100}}%

\newcommand{\Wswdist}{\@ifnextchar[{\basicWswdist}{\basicWswdist[133]}}%

\def\basicwswdisT[#1]#2{\WSWDIST{}{#2}{#100}}%

\newcommand{\wswdisT}{\@ifnextchar[{\basicwswdisT}{\basicwswdisT[133]}}%

\def\basicwswdotar[#1]{\WSWDOTAR{}{}{#100}}%

\newcommand{\wswdotar}{\@ifnextchar[{\basicwswdotar}{\basicwswdotar[133]}}%

\def\basicWswdotar[#1]#2{\WSWDOTAR{#2}{}{#100}}%

\newcommand{\Wswdotar}{\@ifnextchar[{\basicWswdotar}{\basicWswdotar[133]}}%

\def\basicwswdotaR[#1]#2{\WSWDOTAR{}{#2}{#100}}%

\newcommand{\wswdotaR}{\@ifnextchar[{\basicwswdotaR}{\basicwswdotaR[133]}}%

\def\basicwswmono[#1]{\WSWMONO{}{}{#100}}%

\newcommand{\wswmono}{\@ifnextchar[{\basicwswmono}{\basicwswmono[133]}}%

\def\basicWswmono[#1]#2{\WSWMONO{#2}{}{#100}}%

\newcommand{\Wswmono}{\@ifnextchar[{\basicWswmono}{\basicWswmono[133]}}%

\def\basicwswmonO[#1]#2{\WSWMONO{}{#2}{#100}}%

\newcommand{\wswmonO}{\@ifnextchar[{\basicwswmonO}{\basicwswmonO[133]}}%

\def\basicwswepi[#1]{\WSWEPI{}{}{#100}}%

\newcommand{\wswepi}{\@ifnextchar[{\basicwswepi}{\basicwswepi[133]}}%

\def\basicWswepi[#1]#2{\WSWEPI{#2}{}{#100}}%

\newcommand{\Wswepi}{\@ifnextchar[{\basicWswepi}{\basicWswepi[133]}}%

\def\basicwswepI[#1]#2{\WSWEPI{}{#2}{#100}}%

\newcommand{\wswepI}{\@ifnextchar[{\basicwswepI}{\basicwswepI[133]}}%

\def\basicwswbimo[#1]{\WSWBIMO{}{}{#100}}%

\newcommand{\wswbimo}{\@ifnextchar[{\basicwswbimo}{\basicwswbimo[133]}}%

\def\basicWswbimo[#1]#2{\WSWBIMO{#2}{}{#100}}%

\newcommand{\Wswbimo}{\@ifnextchar[{\basicWswbimo}{\basicWswbimo[133]}}%

\def\basicwswbimO[#1]#2{\WSWBIMO{}{#2}{#100}}%

\newcommand{\wswbimO}{\@ifnextchar[{\basicwswbimO}{\basicwswbimO[133]}}%

\def\basicwswiso[#1]{\WSWAR{\cong}{}{#100}}%

\newcommand{\wswiso}{\@ifnextchar[{\basicwswiso}{\basicwswiso[133]}}%

\def\basicWswiso[#1]#2{\WSWAR{#2}{\cong}{#100}}%

\newcommand{\Wswiso}{\@ifnextchar[{\basicWswiso}{\basicWswiso[133]}}%

\def\basicwswisO[#1]#2{\WSWAR{\cong}{#2}{#100}}%

\newcommand{\wswisO}{\@ifnextchar[{\basicwswisO}{\basicwswisO[133]}}%

\def\basicwswbiar[#1]{\WSWBIAR{}{}{#100}}%

\newcommand{\wswbiar}{\@ifnextchar[{\basicwswbiar}{\basicwswbiar[133]}}%

\def\basicWswbiar[#1]#2#3{\WSWBIAR{#2}{#3}{#100}}%

\newcommand{\Wswbiar}{\@ifnextchar[{\basicWswbiar}{\basicWswbiar[133]}}%

\def\basicwswbidist[#1]{\WSWBIDIST{}{}{#100}}%

\newcommand{\wswbidist}{\@ifnextchar[{\basicwswbidist}{\basicwswbidist[133]}}%

\def\basicWswbidist[#1]#2#3{\WSWBIDIST{#2}{#3}{#100}}%

\newcommand{\Wswbidist}{\@ifnextchar[{\basicWswbidist}{\basicWswbidist[133]}}%

\def\basicwswadjar[#1]{\WSWADJAR{}{}{#100}}%

\newcommand{\wswadjar}{\@ifnextchar[{\basicwswadjar}{\basicwswadjar[133]}}%

\def\basicWswadjar[#1]#2#3{\WSWADJAR{#2}{#3}{#100}}%

\newcommand{\Wswadjar}{\@ifnextchar[{\basicWswadjar}{\basicWswadjar[133]}}%

\def\basicwswadjdist[#1]{\WSWADJDIST{}{}{#100}}%

\newcommand{\wswadjdist}{\@ifnextchar[{\basicwswadjdist}{\basicwswadjdist[133]}}%

\def\basicWswadjdist[#1]#2#3{\WSWADJDIST{#2}{#3}{#100}}%

\newcommand{\Wswadjdist}{\@ifnextchar[{\basicWswadjdist}{\basicWswadjdist[133]}}%

\newcommand{\WNWAR}[3]{\testdiagrammode%
\Y=#3%
\divide\Y by 2%
\Z=\Y%
\divide\Z by 2%
\begin{picture}(0,0)%
\put(\Y,-\Z){\line(-2,1){#3}}%
\put(-\Y,\Z){\wnwhead}%
\truex{200}\truey{800}\truez{600}%
\put(\value{x},\value{x}){\makebox(0,\value{z})[l]{${#1}$}}%
\put(-\value{x},-\value{y}){\makebox(0,\value{z})[r]{${#2}$}}%
\end{picture}}%

\newcommand{\WNWDIST}[3]{\testdiagrammode%
\Y=#3%
\divide\Y by 2%
\Z=\Y%
\divide\Z by 2%
\begin{picture}(0,0)%
\put(\Y,-\Z){\line(-2,1){#3}}%
\put(-\Y,\Z){\wnwhead}%
\truex{400}%
\put(0,0){\circle{\value{x}}}%
\truex{200}\truey{800}\truez{600}%
\put(\value{x},\value{x}){\makebox(0,\value{z})[l]{${#1}$}}%
\put(-\value{x},-\value{y}){\makebox(0,\value{z})[r]{${#2}$}}%
\end{picture}}%

\newcommand{\WNWDOTAR}[3]{\testdiagrammode%
\truex{100}\truey{268}\truez{134}%
\Y=#3%
\divide\Y by 2%
\Z=\Y%
\divide\Z by 2%
\NUMBEROFDOTS=#3%
\divide\NUMBEROFDOTS by \value{y}%
\advance\NUMBEROFDOTS by 1%
\begin{picture}(0,0)%
\multiput(\Y,-\Z)(-\value{y},\value{z}){\NUMBEROFDOTS}%
{\circle*{\value{x}}}%
\put(-\Y,\Z){\wnwhead}%
\truex{200}\truey{800}\truez{600}%
\put(\value{x},\value{x}){\makebox(0,\value{z})[l]{${#1}$}}%
\put(-\value{x},-\value{y}){\makebox(0,\value{z})[r]{${#2}$}}%
\end{picture}}%

\newcommand{\WNWMONO}[3]{\testdiagrammode%
\Y=#3%
\divide\Y by 2%
\Z=\Y%
\divide\Z by 2%
\TrueTail%
\bimolength=#3%
\advance\bimolength by -\TrueMonoTail%
\monolength=\bimolength%
\advance\monolength by -\Y%
\secondmonolength=\monolength%
\divide\secondmonolength by 2%
\begin{picture}(0,0)%
\put(\monolength,-\secondmonolength){\line(-2,1){\bimolength}}%
\put(\monolength,-\secondmonolength){\wnwhead}%
\put(-\Y,\Z){\wnwhead}%
\truex{200}\truey{800}\truez{600}%
\put(\value{x},\value{x}){\makebox(0,\value{z})[l]{${#1}$}}%
\put(-\value{x},-\value{y}){\makebox(0,\value{z})[r]{${#2}$}}%
\end{picture}}%

\newcommand{\WNWEPI}[3]{\testdiagrammode%
\Y=#3%
\divide\Y by 2%
\Z=\Y%
\divide\Z by 2%
\TrueHead%
\bimolength=#3%
\advance\bimolength by -\TrueEpiHead%
\epilength=\bimolength%
\advance\epilength by -\Y%
\secondepilength=\epilength%
\divide\secondepilength by 2%
\begin{picture}(0,0)%
\put(\Y,-\Z){\line(-2,1){\bimolength}}%
\put(-\epilength,\secondepilength){\wnwhead}%
\put(-\Y,\Z){\wnwhead}%
\truex{200}\truey{800}\truez{600}%
\put(\value{x},\value{x}){\makebox(0,\value{z})[l]{${#1}$}}%
\put(-\value{x},-\value{y}){\makebox(0,\value{z})[r]{${#2}$}}%
\end{picture}}%

\newcommand{\WNWBIMO}[3]{\testdiagrammode%
\Y=#3%
\divide\Y by 2%
\Z=\Y%
\divide\Z by 2%
\TrueTail\TrueHead%
\bimolength=#3%
\advance\bimolength by -\TrueMonoTail%
\monolength=\bimolength%
\advance\monolength by -\Y%
\advance\bimolength by -\TrueEpiHead%
\epilength=\bimolength%
\advance\epilength by -\monolength%
\secondmonolength=\monolength%
\divide\secondmonolength by 2%
\secondepilength=\epilength%
\divide\secondepilength by 2%
\begin{picture}(0,0)%
\put(\monolength,-\secondmonolength){\line(-2,1){\bimolength}}%
\put(\monolength,-\secondmonolength){\wnwhead}%
\put(-\epilength,\secondepilength){\wnwhead}%
\put(-\Y,\Z){\wnwhead}%
\truex{200}\truey{800}\truez{600}%
\put(\value{x},\value{x}){\makebox(0,\value{z})[l]{${#1}$}}%
\put(-\value{x},-\value{y}){\makebox(0,\value{z})[r]{${#2}$}}%
\end{picture}}%

\newcommand{\WNWBIAR}[3]{\testdiagrammode%
\Y=#3%
\divide\Y by 2%
\Z=\Y%
\divide\Z by 2%
\begin{picture}(0,0)%
\put(\Y,-\Z){\begin{picture}(0,0)%
\truex{156}\truey{313}%
\put(-\value{x},-\value{y}){\line(-2,1){#3}}%
\put(\value{x},\value{y}){\line(-2,1){#3}}%
\monolength=#3%
\advance\monolength by -\value{x}%
\epilength=#3%
\advance\epilength by \value{x}%
\secondmonolength=\Y%
\advance\secondmonolength by -\value{y}%
\secondepilength=\Y%
\advance\secondepilength by \value{y}%
\put(-\monolength,\secondepilength){\wnwhead}%
\put(-\epilength,\secondmonolength){\wnwhead}%
\end{picture}}
\truex{400}\truey{1000}\truez{600}%
\put(\value{x},\value{x}){\makebox(0,\value{z})[l]{${#1}$}}%
\put(-\value{x},-\value{y}){\makebox(0,\value{z})[r]{${#2}$}}%
\end{picture}}%

\newcommand{\WNWBIDIST}[3]{\testdiagrammode%
\Y=#3%
\divide\Y by 2%
\Z=\Y%
\divide\Z by 2%
\begin{picture}(0,0)%
\truex{156}\truey{313}\truez{400}%
\put(\Y,-\Z){\begin{picture}(0,0)%
\put(-\value{x},-\value{y}){\line(-2,1){#3}}%
\put(\value{x},\value{y}){\line(-2,1){#3}}%
\monolength=#3%
\advance\monolength by -\value{x}%
\epilength=#3%
\advance\epilength by \value{x}%
\secondmonolength=\Y%
\advance\secondmonolength by -\value{y}%
\secondepilength=\Y%
\advance\secondepilength by \value{y}%
\put(-\monolength,\secondepilength){\wnwhead}%
\put(-\epilength,\secondmonolength){\wnwhead}%
\end{picture}}
\put(\value{x},\value{y}){\circle{\value{z}}}%
\put(-\value{x},-\value{y}){\circle{\value{z}}}%
\truex{400}\truey{1000}\truez{600}%
\put(\value{x},\value{x}){\makebox(0,\value{z})[l]{${#1}$}}%
\put(-\value{x},-\value{y}){\makebox(0,\value{z})[r]{${#2}$}}%
\end{picture}}%

\newcommand{\WNWADJAR}[3]{\testdiagrammode%
\Y=#3%
\divide\Y by 2%
\Z=\Y%
\divide\Z by 2%
\begin{picture}(0,0)%
\put(\Y,-\Z){\begin{picture}(0,0)%
\truex{156}\truey{313}%
\monolength=#3%
\advance\monolength by -\value{x}%
\epilength=#3%
\advance\epilength by \value{x}%
\secondmonolength=\Y%
\advance\secondmonolength by -\value{y}%
\secondepilength=\Y%
\advance\secondepilength by \value{y}%
\put(-\value{x},-\value{y}){\line(-2,1){#3}}%
\put(-\epilength,\secondmonolength){\wnwhead}%
\put(-\monolength,\secondepilength){\line(2,-1){#3}}%
\put(\value{x},\value{y}){\esehead}%
\end{picture}}
\truex{400}\truey{1000}\truez{600}%
\put(\value{x},\value{x}){\makebox(0,\value{z})[l]{${#1}$}}%
\put(-\value{x},-\value{y}){\makebox(0,\value{z})[r]{${#2}$}}%
\end{picture}}%

\newcommand{\WNWADJDIST}[3]{\testdiagrammode%
\Y=#3%
\divide\Y by 2%
\Z=\Y%
\divide\Z by 2%
\begin{picture}(0,0)%
\truex{156}\truey{313}\truez{400}%
\put(\Y,-\Z){\begin{picture}(0,0)%
\monolength=#3%
\advance\monolength by -\value{x}%
\epilength=#3%
\advance\epilength by \value{x}%
\secondmonolength=\Y%
\advance\secondmonolength by -\value{y}%
\secondepilength=\Y%
\advance\secondepilength by \value{y}%
\put(-\value{x},-\value{y}){\line(-2,1){#3}}%
\put(-\epilength,\secondmonolength){\wnwhead}%
\put(-\monolength,\secondepilength){\line(2,-1){#3}}%
\put(\value{x},\value{y}){\esehead}%
\end{picture}}
\put(\value{x},\value{y}){\circle{\value{z}}}%
\put(-\value{x},-\value{y}){\circle{\value{z}}}%
\truex{400}\truey{1000}\truez{600}%
\put(\value{x},\value{x}){\makebox(0,\value{z})[l]{${#1}$}}%
\put(-\value{x},-\value{y}){\makebox(0,\value{z})[r]{${#2}$}}%
\end{picture}}%

\def\basicwnwar[#1]{\WNWAR{}{}{#100}}%

\newcommand{\wnwar}{\@ifnextchar[{\basicwnwar}{\basicwnwar[133]}}%

\def\basicWnwar[#1]#2{\WNWAR{#2}{}{#100}}%

\newcommand{\Wnwar}{\@ifnextchar[{\basicWnwar}{\basicWnwar[133]}}%

\def\basicwnwaR[#1]#2{\WNWAR{}{#2}{#100}}%

\newcommand{\wnwaR}{\@ifnextchar[{\basicwnwaR}{\basicwnwaR[133]}}%

\def\basicwnwdist[#1]{\WNWDIST{}{}{#100}}%

\newcommand{\wnwdist}{\@ifnextchar[{\basicwnwdist}{\basicwnwdist[133]}}%

\def\basicWnwdist[#1]#2{\WNWDIST{#2}{}{#100}}%

\newcommand{\Wnwdist}{\@ifnextchar[{\basicWnwdist}{\basicWnwdist[133]}}%

\def\basicwnwdisT[#1]#2{\WNWDIST{}{#2}{#100}}%

\newcommand{\wnwdisT}{\@ifnextchar[{\basicwnwdisT}{\basicwnwdisT[133]}}%

\def\basicwnwdotar[#1]{\WNWDOTAR{}{}{#100}}%

\newcommand{\wnwdotar}{\@ifnextchar[{\basicwnwdotar}{\basicwnwdotar[133]}}%

\def\basicWnwdotar[#1]#2{\WNWDOTAR{#2}{}{#100}}%

\newcommand{\Wnwdotar}{\@ifnextchar[{\basicWnwdotar}{\basicWnwdotar[133]}}%

\def\basicwnwdotaR[#1]#2{\WNWDOTAR{}{#2}{#100}}%

\newcommand{\wnwdotaR}{\@ifnextchar[{\basicwnwdotaR}{\basicwnwdotaR[133]}}%

\def\basicwnwmono[#1]{\WNWMONO{}{}{#100}}%

\newcommand{\wnwmono}{\@ifnextchar[{\basicwnwmono}{\basicwnwmono[133]}}%

\def\basicWnwmono[#1]#2{\WNWMONO{#2}{}{#100}}%

\newcommand{\Wnwmono}{\@ifnextchar[{\basicWnwmono}{\basicWnwmono[133]}}%

\def\basicwnwmonO[#1]#2{\WNWMONO{}{#2}{#100}}%

\newcommand{\wnwmonO}{\@ifnextchar[{\basicwnwmonO}{\basicwnwmonO[133]}}%

\def\basicwnwepi[#1]{\WNWEPI{}{}{#100}}%

\newcommand{\wnwepi}{\@ifnextchar[{\basicwnwepi}{\basicwnwepi[133]}}%

\def\basicWnwepi[#1]#2{\WNWEPI{#2}{}{#100}}%

\newcommand{\Wnwepi}{\@ifnextchar[{\basicWnwepi}{\basicWnwepi[133]}}%

\def\basicwnwepI[#1]#2{\WNWEPI{}{#2}{#100}}%

\newcommand{\wnwepI}{\@ifnextchar[{\basicwnwepI}{\basicwnwepI[133]}}%

\def\basicwnwbimo[#1]{\WNWBIMO{}{}{#100}}%

\newcommand{\wnwbimo}{\@ifnextchar[{\basicwnwbimo}{\basicwnwbimo[133]}}%

\def\basicWnwbimo[#1]#2{\WNWBIMO{#2}{}{#100}}%

\newcommand{\Wnwbimo}{\@ifnextchar[{\basicWnwbimo}{\basicWnwbimo[133]}}%

\def\basicwnwbimO[#1]#2{\WNWBIMO{}{#2}{#100}}%

\newcommand{\wnwbimO}{\@ifnextchar[{\basicwnwbimO}{\basicwnwbimO[133]}}%

\def\basicwnwiso[#1]{\WNWAR{\cong}{}{#100}}%

\newcommand{\wnwiso}{\@ifnextchar[{\basicwnwiso}{\basicwnwiso[133]}}%

\def\basicWnwiso[#1]#2{\WNWAR{#2}{\cong}{#100}}%

\newcommand{\Wnwiso}{\@ifnextchar[{\basicWnwiso}{\basicWnwiso[133]}}%

\def\basicwnwisO[#1]#2{\WNWAR{\cong}{#2}{#100}}%

\newcommand{\wnwisO}{\@ifnextchar[{\basicwnwisO}{\basicwnwisO[133]}}%

\def\basicwnwbiar[#1]{\WNWBIAR{}{}{#100}}%

\newcommand{\wnwbiar}{\@ifnextchar[{\basicwnwbiar}{\basicwnwbiar[133]}}%

\def\basicWnwbiar[#1]#2#3{\WNWBIAR{#2}{#3}{#100}}%

\newcommand{\Wnwbiar}{\@ifnextchar[{\basicWnwbiar}{\basicWnwbiar[133]}}%

\def\basicwnwbidist[#1]{\WNWBIDIST{}{}{#100}}%

\newcommand{\wnwbidist}{\@ifnextchar[{\basicwnwbidist}{\basicwnwbidist[133]}}%

\def\basicWnwbidist[#1]#2#3{\WNWBIDIST{#2}{#3}{#100}}%

\newcommand{\Wnwbidist}{\@ifnextchar[{\basicWnwbidist}{\basicWnwbidist[133]}}%

\def\basicwnwadjar[#1]{\WNWADJAR{}{}{#100}}%

\newcommand{\wnwadjar}{\@ifnextchar[{\basicwnwadjar}{\basicwnwadjar[133]}}%

\def\basicWnwadjar[#1]#2#3{\WNWADJAR{#2}{#3}{#100}}%

\newcommand{\Wnwadjar}{\@ifnextchar[{\basicWnwadjar}{\basicWnwadjar[133]}}%

\def\basicwnwadjdist[#1]{\WNWADJDIST{}{}{#100}}%

\newcommand{\wnwadjdist}{\@ifnextchar[{\basicwnwadjdist}{\basicwnwadjdist[133]}}%

\def\basicWnwadjdist[#1]#2#3{\WNWADJDIST{#2}{#3}{#100}}%

\newcommand{\Wnwadjdist}{\@ifnextchar[{\basicWnwadjdist}{\basicWnwadjdist[133]}}%

\newcommand{\NNEAR}[3]{\testdiagrammode%
\Z=#3%
\divide\Z by 2%
\begin{picture}(0,0)%
\put(-\Z,-#3){\line(1,2){#3}}%
\put(\Z,#3){\nnehead}%
\truex{200}\truey{800}\truez{600}%
\put(-\value{x},\value{x}){\makebox(0,\value{z})[r]{${#1}$}}%
\put(\value{x},-\value{y}){\makebox(0,\value{z})[l]{${#2}$}}%
\end{picture}}%

\newcommand{\NNEDIST}[3]{\testdiagrammode%
\Z=#3%
\divide\Z by 2%
\begin{picture}(0,0)%
\put(-\Z,-#3){\line(1,2){#3}}%
\put(\Z,#3){\nnehead}%
\truex{400}%
\put(0,0){\circle{\value{x}}}%
\truex{200}\truey{800}\truez{600}%
\put(-\value{x},\value{x}){\makebox(0,\value{z})[r]{${#1}$}}%
\put(\value{x},-\value{y}){\makebox(0,\value{z})[l]{${#2}$}}%
\end{picture}}%

\newcommand{\NNEDOTAR}[3]{\testdiagrammode%
\truex{100}\truey{268}\truez{134}%
\Z=#3%
\divide\Z by 2%
\NUMBEROFDOTS=#3%
\divide\NUMBEROFDOTS by \value{z}%
\advance\NUMBEROFDOTS by 1%
\begin{picture}(0,0)%
\multiput(-\Z,-#3)(\value{z},\value{y}){\NUMBEROFDOTS}%
{\circle*{\value{x}}}%
\put(\Z,#3){\nnehead}%
\truex{200}\truey{800}\truez{600}%
\put(-\value{x},\value{x}){\makebox(0,\value{z})[r]{${#1}$}}%
\put(\value{x},-\value{y}){\makebox(0,\value{z})[l]{${#2}$}}%
\end{picture}}%

\newcommand{\NNEMONO}[3]{\testdiagrammode%
\Z=#3%
\divide\Z by 2%
\truetaiL%
\bimolength=#3%
\advance\bimolength by -\truemonotaiL%
\monolength=\bimolength%
\advance\monolength by -\Z%
\secondmonolength=\monolength%
\multiply\secondmonolength by 2%
\begin{picture}(0,0)%
\put(-\monolength,-\secondmonolength){\line(1,2){\bimolength}}%
\put(-\monolength,-\secondmonolength){\nnehead}%
\put(\Z,#3){\nnehead}%
\truex{200}\truey{800}\truez{600}%
\put(-\value{x},\value{x}){\makebox(0,\value{z})[r]{${#1}$}}%
\put(\value{x},-\value{y}){\makebox(0,\value{z})[l]{${#2}$}}%
\end{picture}}%

\newcommand{\NNEEPI}[3]{\testdiagrammode%
\Z=#3%
\divide\Z by 2%
\trueheaD%
\bimolength=#3%
\advance\bimolength by -\trueepiheaD%
\epilength=\bimolength%
\advance\epilength by -\Z%
\secondepilength=\epilength%
\multiply\secondepilength by 2%
\begin{picture}(0,0)%
\put(-\Z,-#3){\line(1,2){\bimolength}}%
\put(\epilength,\secondepilength){\nnehead}%
\put(\Z,#3){\nnehead}%
\truex{200}\truey{800}\truez{600}%
\put(-\value{x},\value{x}){\makebox(0,\value{z})[r]{${#1}$}}%
\put(\value{x},-\value{y}){\makebox(0,\value{z})[l]{${#2}$}}%
\end{picture}}%

\newcommand{\NNEBIMO}[3]{\testdiagrammode%
\Z=#3%
\divide\Z by 2%
\truetaiL\trueheaD%
\bimolength=#3%
\advance\bimolength by -\truemonotaiL%
\monolength=\bimolength%
\advance\monolength by -\Z%
\advance\bimolength by -\trueepiheaD%
\epilength=\bimolength%
\advance\epilength by -\monolength%
\secondmonolength=\monolength%
\multiply\secondmonolength by 2%
\secondepilength=\epilength%
\multiply\secondepilength by 2%
\begin{picture}(0,0)%
\put(-\monolength,-\secondmonolength){\line(1,2){\bimolength}}%
\put(-\monolength,-\secondmonolength){\nnehead}%
\put(\epilength,\secondepilength){\nnehead}%
\put(\Z,#3){\nnehead}%
\truex{200}\truey{800}\truez{600}%
\put(-\value{x},\value{x}){\makebox(0,\value{z})[r]{${#1}$}}%
\put(\value{x},-\value{y}){\makebox(0,\value{z})[l]{${#2}$}}%
\end{picture}}%

\newcommand{\NNEEQL}[3]{\testdiagrammode%
\Z=#3%
\divide\Z by 2%
\begin{picture}(0,0)%
\put(-\Z,-#3){\begin{picture}(0,0)%
\truex{44}\truey{89}%
\put(-\value{y},\value{x}){\line(1,2){#3}}%
\put(\value{y},-\value{x}){\line(1,2){#3}}%
\end{picture}}%
\truex{200}\truey{800}\truez{600}%
\put(-\value{x},\value{x}){\makebox(0,\value{z})[r]{${#1}$}}%
\put(\value{x},-\value{y}){\makebox(0,\value{z})[l]{${#2}$}}%
\end{picture}}%

\newcommand{\NNEBIAR}[3]{\testdiagrammode%
\Y=#3%
\divide\Y by 2%
\Z=#3%
\multiply \Z by 2%
\begin{picture}(0,0)%
\put(-\Y,-#3){\begin{picture}(0,0)%
\truex{313}\truey{156}%
\put(-\value{x},\value{y}){\line(1,2){#3}}%
\put(\value{x},-\value{y}){\line(1,2){#3}}%
\monolength=#3%
\advance\monolength by -\value{x}%
\epilength=#3%
\advance\epilength by \value{x}%
\secondmonolength=\Z%
\advance\secondmonolength by -\value{y}%
\secondepilength=\Z%
\advance\secondepilength by \value{y}%
\put(\monolength,\secondepilength){\nnehead}%
\put(\epilength,\secondmonolength){\nnehead}%
\end{picture}}
\truex{300}\truey{1000}\truez{600}%
\put(-\value{x},\value{x}){\makebox(0,\value{z})[r]{${#1}$}}%
\put(\value{x},-\value{y}){\makebox(0,\value{z})[l]{${#2}$}}%
\end{picture}}%

\newcommand{\NNEBIDIST}[3]{\testdiagrammode%
\Y=#3%
\divide\Y by 2%
\Z=#3%
\multiply \Z by 2%
\begin{picture}(0,0)%
\truex{313}\truey{156}\truez{400}%
\put(-\Y,-#3){\begin{picture}(0,0)%
\put(-\value{x},\value{y}){\line(1,2){#3}}%
\put(\value{x},-\value{y}){\line(1,2){#3}}%
\monolength=#3%
\advance\monolength by -\value{x}%
\epilength=#3%
\advance\epilength by \value{x}%
\secondmonolength=\Z%
\advance\secondmonolength by -\value{y}%
\secondepilength=\Z%
\advance\secondepilength by \value{y}%
\put(\monolength,\secondepilength){\nnehead}%
\put(\epilength,\secondmonolength){\nnehead}%
\end{picture}}
\put(-\value{x},\value{y}){\circle{\value{z}}}%
\put(\value{x},-\value{y}){\circle{\value{z}}}%
\truex{300}\truey{1000}\truez{600}%
\put(-\value{x},\value{x}){\makebox(0,\value{z})[r]{${#1}$}}%
\put(\value{x},-\value{y}){\makebox(0,\value{z})[l]{${#2}$}}%
\end{picture}}%

\newcommand{\NNEADJAR}[3]{\testdiagrammode%
\Y=#3%
\divide\Y by 2%
\Z=#3%
\multiply \Z by 2%
\begin{picture}(0,0)%
\put(-\Y,-#3){\begin{picture}(0,0)%
\truex{313}\truey{156}%
\monolength=#3%
\advance\monolength by -\value{x}%
\epilength=#3%
\advance\epilength by \value{x}%
\secondmonolength=\Z%
\advance\secondmonolength by -\value{y}%
\secondepilength=\Z%
\advance\secondepilength by \value{y}%
\put(\value{x},-\value{y}){\line(1,2){#3}}%
\put(\epilength,\secondmonolength){\nnehead}%
\put(\monolength,\secondepilength){\line(-1,-2){#3}}%
\put(-\value{x},\value{y}){\sswhead}%
\end{picture}}
\truex{300}\truey{1000}\truez{600}%
\put(-\value{x},\value{x}){\makebox(0,\value{z})[r]{${#1}$}}%
\put(\value{x},-\value{y}){\makebox(0,\value{z})[l]{${#2}$}}%
\end{picture}}%

\newcommand{\NNEADJDIST}[3]{\testdiagrammode%
\Y=#3%
\divide\Y by 2%
\Z=#3%
\multiply \Z by 2%
\begin{picture}(0,0)%
\truex{313}\truey{156}\truez{400}%
\put(-\Y,-#3){\begin{picture}(0,0)%
\monolength=#3%
\advance\monolength by -\value{x}%
\epilength=#3%
\advance\epilength by \value{x}%
\secondmonolength=\Z%
\advance\secondmonolength by -\value{y}%
\secondepilength=\Z%
\advance\secondepilength by \value{y}%
\put(\value{x},-\value{y}){\line(1,2){#3}}%
\put(\epilength,\secondmonolength){\nnehead}%
\put(\monolength,\secondepilength){\line(-1,-2){#3}}%
\put(-\value{x},\value{y}){\sswhead}%
\end{picture}}
\put(-\value{x},\value{y}){\circle{\value{z}}}%
\put(\value{x},-\value{y}){\circle{\value{z}}}%
\truex{300}\truey{1000}\truez{600}%
\put(-\value{x},\value{x}){\makebox(0,\value{z})[r]{${#1}$}}%
\put(\value{x},-\value{y}){\makebox(0,\value{z})[l]{${#2}$}}%
\end{picture}}%

\def\basicnnear[#1]{\NNEAR{}{}{#100}}%

\newcommand{\nnear}{\@ifnextchar[{\basicnnear}{\basicnnear[67]}}%

\def\basicNnear[#1]#2{\NNEAR{#2}{}{#100}}%

\newcommand{\Nnear}{\@ifnextchar[{\basicNnear}{\basicNnear[67]}}%

\def\basicnneaR[#1]#2{\NNEAR{}{#2}{#100}}%

\newcommand{\nneaR}{\@ifnextchar[{\basicnneaR}{\basicnneaR[67]}}%

\def\basicnnedist[#1]{\NNEDIST{}{}{#100}}%

\newcommand{\nnedist}{\@ifnextchar[{\basicnnedist}{\basicnnedist[67]}}%

\def\basicNnedist[#1]#2{\NNEDIST{#2}{}{#100}}%

\newcommand{\Nnedist}{\@ifnextchar[{\basicNnedist}{\basicNnedist[67]}}%

\def\basicnnedisT[#1]#2{\NNEDIST{}{#2}{#100}}%

\newcommand{\nnedisT}{\@ifnextchar[{\basicnnedisT}{\basicnnedisT[67]}}%

\def\basicnnedotar[#1]{\NNEDOTAR{}{}{#100}}%

\newcommand{\nnedotar}{\@ifnextchar[{\basicnnedotar}{\basicnnedotar[67]}}%

\def\basicNnedotar[#1]#2{\NNEDOTAR{#2}{}{#100}}%

\newcommand{\Nnedotar}{\@ifnextchar[{\basicNnedotar}{\basicNnedotar[67]}}%

\def\basicnnedotaR[#1]#2{\NNEDOTAR{}{#2}{#100}}%

\newcommand{\nnedotaR}{\@ifnextchar[{\basicnnedotaR}{\basicnnedotaR[67]}}%

\def\basicnnemono[#1]{\NNEMONO{}{}{#100}}%

\newcommand{\nnemono}{\@ifnextchar[{\basicnnemono}{\basicnnemono[67]}}%

\def\basicNnemono[#1]#2{\NNEMONO{#2}{}{#100}}%

\newcommand{\Nnemono}{\@ifnextchar[{\basicNnemono}{\basicNnemono[67]}}%

\def\basicnnemonO[#1]#2{\NNEMONO{}{#2}{#100}}%

\newcommand{\nnemonO}{\@ifnextchar[{\basicnnemonO}{\basicnnemonO[67]}}%

\def\basicnneepi[#1]{\NNEEPI{}{}{#100}}%

\newcommand{\nneepi}{\@ifnextchar[{\basicnneepi}{\basicnneepi[67]}}%

\def\basicNneepi[#1]#2{\NNEEPI{#2}{}{#100}}%

\newcommand{\Nneepi}{\@ifnextchar[{\basicNneepi}{\basicNneepi[67]}}%

\def\basicnneepI[#1]#2{\NNEEPI{}{#2}{#100}}%

\newcommand{\nneepI}{\@ifnextchar[{\basicnneepI}{\basicnneepI[67]}}%

\def\basicnnebimo[#1]{\NNEBIMO{}{}{#100}}%

\newcommand{\nnebimo}{\@ifnextchar[{\basicnnebimo}{\basicnnebimo[67]}}%

\def\basicNnebimo[#1]#2{\NNEBIMO{#2}{}{#100}}%

\newcommand{\Nnebimo}{\@ifnextchar[{\basicNnebimo}{\basicNnebimo[67]}}%

\def\basicnnebimO[#1]#2{\NNEBIMO{}{#2}{#100}}%

\newcommand{\nnebimO}{\@ifnextchar[{\basicnnebimO}{\basicnnebimO[67]}}%

\def\basicnneiso[#1]{\NNEAR{\cong}{}{#100}}%

\newcommand{\nneiso}{\@ifnextchar[{\basicnneiso}{\basicnneiso[67]}}%

\def\basicNneiso[#1]#2{\NNEAR{#2}{\cong}{#100}}%

\newcommand{\Nneiso}{\@ifnextchar[{\basicNneiso}{\basicNneiso[67]}}%

\def\basicnneisO[#1]#2{\NNEAR{\cong}{#2}{#100}}%

\newcommand{\nneisO}{\@ifnextchar[{\basicnneisO}{\basicnneisO[67]}}%

\def\basicnneeql[#1]{\NNEEQL{}{}{#100}}%

\newcommand{\nneeql}{\@ifnextchar[{\basicnneeql}{\basicnneeql[67]}}%

\def\basicNneeql[#1]#2{\NNEEQL{#2}{}{#100}}%

\newcommand{\Nneeql}{\@ifnextchar[{\basicNneeql}{\basicNneeql[67]}}%

\def\basicnneeqL[#1]#2{\NNEEQL{}{#2}{#100}}%

\newcommand{\nneeqL}{\@ifnextchar[{\basicnneeqL}{\basicnneeqL[67]}}%

\def\basicnnebiar[#1]{\NNEBIAR{}{}{#100}}%

\newcommand{\nnebiar}{\@ifnextchar[{\basicnnebiar}{\basicnnebiar[67]}}%

\def\basicNnebiar[#1]#2#3{\NNEBIAR{#2}{#3}{#100}}%

\newcommand{\Nnebiar}{\@ifnextchar[{\basicNnebiar}{\basicNnebiar[67]}}%

\def\basicnnebidist[#1]{\NNEBIDIST{}{}{#100}}%

\newcommand{\nnebidist}{\@ifnextchar[{\basicnnebidist}{\basicnnebidist[67]}}%

\def\basicNnebidist[#1]#2#3{\NNEBIDIST{#2}{#3}{#100}}%

\newcommand{\Nnebidist}{\@ifnextchar[{\basicNnebidist}{\basicNnebidist[67]}}%

\def\basicnneadjar[#1]{\NNEADJAR{}{}{#100}}%

\newcommand{\nneadjar}{\@ifnextchar[{\basicnneadjar}{\basicnneadjar[67]}}%

\def\basicNneadjar[#1]#2#3{\NNEADJAR{#2}{#3}{#100}}%

\newcommand{\Nneadjar}{\@ifnextchar[{\basicNneadjar}{\basicNneadjar[67]}}%

\def\basicnneadjdist[#1]{\NNEADJDIST{}{}{#100}}%

\newcommand{\nneadjdist}{\@ifnextchar[{\basicnneadjdist}{\basicnneadjdist[67]}}%

\def\basicNneadjdist[#1]#2#3{\NNEADJDIST{#2}{#3}{#100}}%

\newcommand{\Nneadjdist}{\@ifnextchar[{\basicNneadjdist}{\basicNneadjdist[67]}}%

\newcommand{\SSEAR}[3]{\testdiagrammode%
\Z=#3%
\divide\Z by 2%
\begin{picture}(0,0)%
\put(-\Z,#3){\line(1,-2){#3}}%
\put(\Z,-#3){\ssehead}%
\truex{200}\truey{800}\truez{600}%
\put(\value{x},\value{x}){\makebox(0,\value{z})[l]{${#1}$}}%
\put(-\value{x},-\value{y}){\makebox(0,\value{z})[r]{${#2}$}}%
\end{picture}}%

\newcommand{\SSEDIST}[3]{\testdiagrammode%
\Z=#3%
\divide\Z by 2%
\begin{picture}(0,0)%
\put(-\Z,#3){\line(1,-2){#3}}%
\put(\Z,-#3){\ssehead}%
\truex{400}%
\put(0,0){\circle{\value{x}}}%
\truex{200}\truey{800}\truez{600}%
\put(\value{x},\value{x}){\makebox(0,\value{z})[l]{${#1}$}}%
\put(-\value{x},-\value{y}){\makebox(0,\value{z})[r]{${#2}$}}%
\end{picture}}%

\newcommand{\SSEDOTAR}[3]{\testdiagrammode%
\truex{100}\truey{268}\truez{134}%
\Z=#3%
\divide\Z by 2%
\NUMBEROFDOTS=#3%
\divide\NUMBEROFDOTS by \value{z}%
\advance\NUMBEROFDOTS by 1%
\begin{picture}(0,0)%
\multiput(-\Z,#3)(\value{z},-\value{y}){\NUMBEROFDOTS}%
{\circle*{\value{x}}}%
\put(\Z,-#3){\ssehead}%
\truex{200}\truey{800}\truez{600}%
\put(\value{x},\value{x}){\makebox(0,\value{z})[l]{${#1}$}}%
\put(-\value{x},-\value{y}){\makebox(0,\value{z})[r]{${#2}$}}%
\end{picture}}%

\newcommand{\SSEMONO}[3]{\testdiagrammode%
\Z=#3%
\divide\Z by 2%
\truetaiL%
\bimolength=#3%
\advance\bimolength by -\truemonotaiL%
\monolength=\bimolength%
\advance\monolength by -\Z%
\secondmonolength=\monolength%
\multiply\secondmonolength by 2%
\begin{picture}(0,0)%
\put(-\monolength,\secondmonolength){\line(1,-2){\bimolength}}%
\put(-\monolength,\secondmonolength){\ssehead}%
\put(\Z,-#3){\ssehead}%
\truex{200}\truey{800}\truez{600}%
\put(\value{x},\value{x}){\makebox(0,\value{z})[l]{${#1}$}}%
\put(-\value{x},-\value{y}){\makebox(0,\value{z})[r]{${#2}$}}%
\end{picture}}%

\newcommand{\SSEEPI}[3]{\testdiagrammode%
\Z=#3%
\divide\Z by 2%
\trueheaD%
\bimolength=#3%
\advance\bimolength by -\trueepiheaD%
\epilength=\bimolength%
\advance\epilength by -\Z%
\secondepilength=\epilength%
\multiply\secondepilength by 2%
\begin{picture}(0,0)%
\put(-\Z,#3){\line(1,-2){\bimolength}}%
\put(\epilength,-\secondepilength){\ssehead}%
\put(\Z,-#3){\ssehead}%
\truex{200}\truey{800}\truez{600}%
\put(\value{x},\value{x}){\makebox(0,\value{z})[l]{${#1}$}}%
\put(-\value{x},-\value{y}){\makebox(0,\value{z})[r]{${#2}$}}%
\end{picture}}%

\newcommand{\SSEBIMO}[3]{\testdiagrammode%
\Z=#3%
\divide\Z by 2%
\truetaiL\trueheaD%
\bimolength=#3%
\advance\bimolength by -\truemonotaiL%
\monolength=\bimolength%
\advance\monolength by -\Z%
\advance\bimolength by -\trueepiheaD%
\epilength=\bimolength%
\advance\epilength by -\monolength%
\secondmonolength=\monolength%
\multiply\secondmonolength by 2%
\secondepilength=\epilength%
\multiply\secondepilength by 2%
\begin{picture}(0,0)%
\put(-\monolength,\secondmonolength){\line(1,-2){\bimolength}}%
\put(-\monolength,\secondmonolength){\ssehead}%
\put(\epilength,-\secondepilength){\ssehead}%
\put(\Z,-#3){\ssehead}%
\truex{200}\truey{800}\truez{600}%
\put(\value{x},\value{x}){\makebox(0,\value{z})[l]{${#1}$}}%
\put(-\value{x},-\value{y}){\makebox(0,\value{z})[r]{${#2}$}}%
\end{picture}}%

\newcommand{\SSEEQL}[3]{\testdiagrammode%
\Z=#3%
\divide\Z by 2%
\begin{picture}(0,0)%
\put(-\Z,#3){\begin{picture}(0,0)%
\truex{44}\truey{89}%
\put(-\value{y},-\value{x}){\line(1,-2){#3}}%
\put(\value{y},\value{x}){\line(1,-2){#3}}%
\end{picture}}%
\truex{200}\truey{800}\truez{600}%
\put(\value{x},\value{x}){\makebox(0,\value{z})[l]{${#1}$}}%
\put(-\value{x},-\value{y}){\makebox(0,\value{z})[r]{${#2}$}}%
\end{picture}}%

\newcommand{\SSEBIAR}[3]{\testdiagrammode%
\Y=#3%
\divide\Y by 2%
\Z=#3%
\multiply \Z by 2%
\begin{picture}(0,0)%
\put(-\Y,#3){\begin{picture}(0,0)%
\truex{313}\truey{156}%
\put(-\value{x},-\value{y}){\line(1,-2){#3}}%
\put(\value{x},\value{y}){\line(1,-2){#3}}%
\monolength=#3%
\advance\monolength by -\value{x}%
\epilength=#3%
\advance\epilength by \value{x}%
\secondmonolength=\Z%
\advance\secondmonolength by -\value{y}%
\secondepilength=\Z%
\advance\secondepilength by \value{y}%
\put(\monolength,-\secondepilength){\ssehead}%
\put(\epilength,-\secondmonolength){\ssehead}%
\end{picture}}
\truex{400}\truey{1000}\truez{600}%
\put(\value{x},\value{x}){\makebox(0,\value{z})[l]{${#1}$}}%
\put(-\value{x},-\value{y}){\makebox(0,\value{z})[r]{${#2}$}}%
\end{picture}}%

\newcommand{\SSEBIDIST}[3]{\testdiagrammode%
\Y=#3%
\divide\Y by 2%
\Z=#3%
\multiply \Z by 2%
\begin{picture}(0,0)%
\truex{313}\truey{156}\truez{400}%
\put(-\Y,#3){\begin{picture}(0,0)%
\put(-\value{x},-\value{y}){\line(1,-2){#3}}%
\put(\value{x},\value{y}){\line(1,-2){#3}}%
\monolength=#3%
\advance\monolength by -\value{x}%
\epilength=#3%
\advance\epilength by \value{x}%
\secondmonolength=\Z%
\advance\secondmonolength by -\value{y}%
\secondepilength=\Z%
\advance\secondepilength by \value{y}%
\put(\monolength,-\secondepilength){\ssehead}%
\put(\epilength,-\secondmonolength){\ssehead}%
\end{picture}}
\put(-\value{x},-\value{y}){\circle{\value{z}}}%
\put(\value{x},\value{y}){\circle{\value{z}}}%
\truex{500}\truey{1000}\truez{600}%
\put(\value{x},\value{x}){\makebox(0,\value{z})[l]{${#1}$}}%
\put(-\value{x},-\value{y}){\makebox(0,\value{z})[r]{${#2}$}}%
\end{picture}}%

\newcommand{\SSEADJAR}[3]{\testdiagrammode%
\Y=#3%
\divide\Y by 2%
\Z=#3%
\multiply \Z by 2%
\begin{picture}(0,0)%
\put(-\Y,#3){\begin{picture}(0,0)%
\truex{313}\truey{156}%
\monolength=#3%
\advance\monolength by -\value{x}%
\epilength=#3%
\advance\epilength by \value{x}%
\secondmonolength=\Z%
\advance\secondmonolength by -\value{y}%
\secondepilength=\Z%
\advance\secondepilength by \value{y}%
\put(-\value{x},-\value{y}){\line(1,-2){#3}}%
\put(\monolength,-\secondepilength){\ssehead}%
\put(\epilength,-\secondmonolength){\line(-1,2){#3}}%
\put(\value{x},\value{y}){\nnwhead}%
\end{picture}}
\truex{400}\truey{1000}\truez{600}%
\put(\value{x},\value{x}){\makebox(0,\value{z})[l]{${#1}$}}%
\put(-\value{x},-\value{y}){\makebox(0,\value{z})[r]{${#2}$}}%
\end{picture}}%

\newcommand{\SSEADJDIST}[3]{\testdiagrammode%
\Y=#3%
\divide\Y by 2%
\Z=#3%
\multiply \Z by 2%
\begin{picture}(0,0)%
\truex{313}\truey{156}\truez{400}%
\put(-\Y,#3){\begin{picture}(0,0)%
\monolength=#3%
\advance\monolength by -\value{x}%
\epilength=#3%
\advance\epilength by \value{x}%
\secondmonolength=\Z%
\advance\secondmonolength by -\value{y}%
\secondepilength=\Z%
\advance\secondepilength by \value{y}%
\put(-\value{x},-\value{y}){\line(1,-2){#3}}%
\put(\monolength,-\secondepilength){\ssehead}%
\put(\epilength,-\secondmonolength){\line(-1,2){#3}}%
\put(\value{x},\value{y}){\nnwhead}%
\end{picture}}
\put(\value{x},\value{y}){\circle{\value{z}}}%
\put(-\value{x},-\value{y}){\circle{\value{z}}}%
\truex{500}\truey{1000}\truez{600}%
\put(\value{x},\value{x}){\makebox(0,\value{z})[l]{${#1}$}}%
\put(-\value{x},-\value{y}){\makebox(0,\value{z})[r]{${#2}$}}%
\end{picture}}%

\def\basicssear[#1]{\SSEAR{}{}{#100}}%

\newcommand{\ssear}{\@ifnextchar[{\basicssear}{\basicssear[67]}}%

\def\basicSsear[#1]#2{\SSEAR{#2}{}{#100}}%

\newcommand{\Ssear}{\@ifnextchar[{\basicSsear}{\basicSsear[67]}}%

\def\basicsseaR[#1]#2{\SSEAR{}{#2}{#100}}%

\newcommand{\sseaR}{\@ifnextchar[{\basicsseaR}{\basicsseaR[67]}}%

\def\basicssedist[#1]{\SSEDIST{}{}{#100}}%

\newcommand{\ssedist}{\@ifnextchar[{\basicssedist}{\basicssedist[67]}}%

\def\basicSsedist[#1]#2{\SSEDIST{#2}{}{#100}}%

\newcommand{\Ssedist}{\@ifnextchar[{\basicSsedist}{\basicSsedist[67]}}%

\def\basicssedisT[#1]#2{\SSEDIST{}{#2}{#100}}%

\newcommand{\ssedisT}{\@ifnextchar[{\basicssedisT}{\basicssedisT[67]}}%

\def\basicssedotar[#1]{\SSEDOTAR{}{}{#100}}%

\newcommand{\ssedotar}{\@ifnextchar[{\basicssedotar}{\basicssedotar[67]}}%

\def\basicSsedotar[#1]#2{\SSEDOTAR{#2}{}{#100}}%

\newcommand{\Ssedotar}{\@ifnextchar[{\basicSsedotar}{\basicSsedotar[67]}}%

\def\basicssedotaR[#1]#2{\SSEDOTAR{}{#2}{#100}}%

\newcommand{\ssedotaR}{\@ifnextchar[{\basicssedotaR}{\basicssedotaR[67]}}%

\def\basicssemono[#1]{\SSEMONO{}{}{#100}}%

\newcommand{\ssemono}{\@ifnextchar[{\basicssemono}{\basicssemono[67]}}%

\def\basicSsemono[#1]#2{\SSEMONO{#2}{}{#100}}%

\newcommand{\Ssemono}{\@ifnextchar[{\basicSsemono}{\basicSsemono[67]}}%

\def\basicssemonO[#1]#2{\SSEMONO{}{#2}{#100}}%

\newcommand{\ssemonO}{\@ifnextchar[{\basicssemonO}{\basicssemonO[67]}}%

\def\basicsseepi[#1]{\SSEEPI{}{}{#100}}%

\newcommand{\sseepi}{\@ifnextchar[{\basicsseepi}{\basicsseepi[67]}}%

\def\basicSseepi[#1]#2{\SSEEPI{#2}{}{#100}}%

\newcommand{\Sseepi}{\@ifnextchar[{\basicSseepi}{\basicSseepi[67]}}%

\def\basicsseepI[#1]#2{\SSEEPI{}{#2}{#100}}%

\newcommand{\sseepI}{\@ifnextchar[{\basicsseepI}{\basicsseepI[67]}}%

\def\basicssebimo[#1]{\SSEBIMO{}{}{#100}}%

\newcommand{\ssebimo}{\@ifnextchar[{\basicssebimo}{\basicssebimo[67]}}%

\def\basicSsebimo[#1]#2{\SSEBIMO{#2}{}{#100}}%

\newcommand{\Ssebimo}{\@ifnextchar[{\basicSsebimo}{\basicSsebimo[67]}}%

\def\basicssebimO[#1]#2{\SSEBIMO{}{#2}{#100}}%

\newcommand{\ssebimO}{\@ifnextchar[{\basicssebimO}{\basicssebimO[67]}}%

\def\basicsseiso[#1]{\SSEAR{\cong}{}{#100}}%

\newcommand{\sseiso}{\@ifnextchar[{\basicsseiso}{\basicsseiso[67]}}%

\def\basicSseiso[#1]#2{\SSEAR{#2}{\cong}{#100}}%

\newcommand{\Sseiso}{\@ifnextchar[{\basicSseiso}{\basicSseiso[67]}}%

\def\basicsseisO[#1]#2{\SSEAR{\cong}{#2}{#100}}%

\newcommand{\sseisO}{\@ifnextchar[{\basicsseisO}{\basicsseisO[67]}}%

\def\basicsseeql[#1]{\SSEEQL{}{}{#100}}%

\newcommand{\sseeql}{\@ifnextchar[{\basicsseeql}{\basicsseeql[67]}}%

\def\basicSseeql[#1]#2{\SSEEQL{#2}{}{#100}}%

\newcommand{\Sseeql}{\@ifnextchar[{\basicSseeql}{\basicSseeql[67]}}%

\def\basicsseeqL[#1]#2{\SSEEQL{}{#2}{#100}}%

\newcommand{\sseeqL}{\@ifnextchar[{\basicsseeqL}{\basicsseeqL[67]}}%

\def\basicssebiar[#1]{\SSEBIAR{}{}{#100}}%

\newcommand{\ssebiar}{\@ifnextchar[{\basicssebiar}{\basicssebiar[67]}}%

\def\basicSsebiar[#1]#2#3{\SSEBIAR{#2}{#3}{#100}}%

\newcommand{\Ssebiar}{\@ifnextchar[{\basicSsebiar}{\basicSsebiar[67]}}%

\def\basicssebidist[#1]{\SSEBIDIST{}{}{#100}}%

\newcommand{\ssebidist}{\@ifnextchar[{\basicssebidist}{\basicssebidist[67]}}%

\def\basicSsebidist[#1]#2#3{\SSEBIDIST{#2}{#3}{#100}}%

\newcommand{\Ssebidist}{\@ifnextchar[{\basicSsebidist}{\basicSsebidist[67]}}%

\def\basicsseadjar[#1]{\SSEADJAR{}{}{#100}}%

\newcommand{\sseadjar}{\@ifnextchar[{\basicsseadjar}{\basicsseadjar[67]}}%

\def\basicSseadjar[#1]#2#3{\SSEADJAR{#2}{#3}{#100}}%

\newcommand{\Sseadjar}{\@ifnextchar[{\basicSseadjar}{\basicSseadjar[67]}}%

\def\basicsseadjdist[#1]{\SSEADJDIST{}{}{#100}}%

\newcommand{\sseadjdist}{\@ifnextchar[{\basicsseadjdist}{\basicsseadjdist[67]}}%

\def\basicSseadjdist[#1]#2#3{\SSEADJDIST{#2}{#3}{#100}}%

\newcommand{\Sseadjdist}{\@ifnextchar[{\basicSseadjdist}{\basicSseadjdist[67]}}%

\newcommand{\SSWAR}[3]{\testdiagrammode%
\Z=#3%
\divide\Z by 2%
\begin{picture}(0,0)%
\put(\Z,#3){\line(-1,-2){#3}}%
\put(-\Z,-#3){\sswhead}%
\truex{200}\truey{800}\truez{600}%
\put(-\value{x},\value{x}){\makebox(0,\value{z})[r]{${#1}$}}%
\put(\value{x},-\value{y}){\makebox(0,\value{z})[l]{${#2}$}}%
\end{picture}}%

\newcommand{\SSWDIST}[3]{\testdiagrammode%
\Z=#3%
\divide\Z by 2%
\begin{picture}(0,0)%
\put(\Z,#3){\line(-1,-2){#3}}%
\put(-\Z,-#3){\sswhead}%
\truex{400}%
\put(0,0){\circle{\value{x}}}%
\truex{200}\truey{800}\truez{600}%
\put(-\value{x},\value{x}){\makebox(0,\value{z})[r]{${#1}$}}%
\put(\value{x},-\value{y}){\makebox(0,\value{z})[l]{${#2}$}}%
\end{picture}}%

\newcommand{\SSWDOTAR}[3]{\testdiagrammode%
\truex{100}\truey{268}\truez{134}%
\Z=#3%
\divide\Z by 2%
\NUMBEROFDOTS=#3%
\divide\NUMBEROFDOTS by \value{z}%
\advance\NUMBEROFDOTS by 1%
\begin{picture}(0,0)%
\multiput(\Z,#3)(-\value{z},-\value{y}){\NUMBEROFDOTS}%
{\circle*{\value{x}}}%
\put(-\Z,-#3){\sswhead}%
\truex{200}\truey{800}\truez{600}%
\put(-\value{x},\value{x}){\makebox(0,\value{z})[r]{${#1}$}}%
\put(\value{x},-\value{y}){\makebox(0,\value{z})[l]{${#2}$}}%
\end{picture}}%

\newcommand{\SSWMONO}[3]{\testdiagrammode%
\Z=#3%
\divide\Z by 2%
\truetaiL%
\bimolength=#3%
\advance\bimolength by -\truemonotaiL%
\monolength=\bimolength%
\advance\monolength by -\Z%
\secondmonolength=\monolength%
\multiply\secondmonolength by 2%
\begin{picture}(0,0)%
\put(\monolength,\secondmonolength){\line(-1,-2){\bimolength}}%
\put(\monolength,\secondmonolength){\sswhead}%
\put(-\Z,-#3){\sswhead}%
\truex{200}\truey{800}\truez{600}%
\put(-\value{x},\value{x}){\makebox(0,\value{z})[r]{${#1}$}}%
\put(\value{x},-\value{y}){\makebox(0,\value{z})[l]{${#2}$}}%
\end{picture}}%

\newcommand{\SSWEPI}[3]{\testdiagrammode%
\Z=#3%
\divide\Z by 2%
\trueheaD%
\bimolength=#3%
\advance\bimolength by -\trueepiheaD%
\epilength=\bimolength%
\advance\epilength by -\Z%
\secondepilength=\epilength%
\multiply\secondepilength by 2%
\begin{picture}(0,0)%
\put(\Z,#3){\line(-1,-2){\bimolength}}%
\put(-\epilength,-\secondepilength){\sswhead}%
\put(-\Z,-#3){\sswhead}%
\truex{200}\truey{800}\truez{600}%
\put(-\value{x},\value{x}){\makebox(0,\value{z})[r]{${#1}$}}%
\put(\value{x},-\value{y}){\makebox(0,\value{z})[l]{${#2}$}}%
\end{picture}}%

\newcommand{\SSWBIMO}[3]{\testdiagrammode%
\Z=#3%
\divide\Z by 2%
\truetaiL\trueheaD%
\bimolength=#3%
\advance\bimolength by -\truemonotaiL%
\monolength=\bimolength%
\advance\monolength by -\Z%
\advance\bimolength by -\trueepiheaD%
\epilength=\bimolength%
\advance\epilength by -\monolength%
\secondmonolength=\monolength%
\multiply\secondmonolength by 2%
\secondepilength=\epilength%
\multiply\secondepilength by 2%
\begin{picture}(0,0)%
\put(\monolength,\secondmonolength){\line(-1,-2){\bimolength}}%
\put(\monolength,\secondmonolength){\sswhead}%
\put(-\epilength,-\secondepilength){\sswhead}%
\put(-\Z,-#3){\sswhead}%
\truex{200}\truey{800}\truez{600}%
\put(-\value{x},\value{x}){\makebox(0,\value{z})[r]{${#1}$}}%
\put(\value{x},-\value{y}){\makebox(0,\value{z})[l]{${#2}$}}%
\end{picture}}%

\newcommand{\SSWBIAR}[3]{\testdiagrammode%
\Y=#3%
\divide\Y by 2%
\Z=#3%
\multiply \Z by 2%
\begin{picture}(0,0)%
\put(\Y,#3){\begin{picture}(0,0)%
\truex{313}\truey{156}%
\put(-\value{x},\value{y}){\line(-1,-2){#3}}%
\put(\value{x},-\value{y}){\line(-1,-2){#3}}%
\monolength=#3%
\advance\monolength by -\value{x}%
\epilength=#3%
\advance\epilength by \value{x}%
\secondmonolength=\Z%
\advance\secondmonolength by -\value{y}%
\secondepilength=\Z%
\advance\secondepilength by \value{y}%
\put(-\monolength,-\secondepilength){\sswhead}%
\put(-\epilength,-\secondmonolength){\sswhead}%
\end{picture}}
\truex{300}\truey{1000}\truez{600}%
\put(-\value{x},\value{x}){\makebox(0,\value{z})[r]{${#1}$}}%
\put(\value{x},-\value{y}){\makebox(0,\value{z})[l]{${#2}$}}%
\end{picture}}%

\newcommand{\SSWBIDIST}[3]{\testdiagrammode%
\Y=#3%
\divide\Y by 2%
\Z=#3%
\multiply \Z by 2%
\begin{picture}(0,0)%
\truex{313}\truey{156}\truez{400}%
\put(\Y,#3){\begin{picture}(0,0)%
\put(-\value{x},\value{y}){\line(-1,-2){#3}}%
\put(\value{x},-\value{y}){\line(-1,-2){#3}}%
\monolength=#3%
\advance\monolength by -\value{x}%
\epilength=#3%
\advance\epilength by \value{x}%
\secondmonolength=\Z%
\advance\secondmonolength by -\value{y}%
\secondepilength=\Z%
\advance\secondepilength by \value{y}%
\put(-\monolength,-\secondepilength){\sswhead}%
\put(-\epilength,-\secondmonolength){\sswhead}%
\end{picture}}
\put(-\value{x},\value{y}){\circle{\value{z}}}%
\put(\value{x},-\value{y}){\circle{\value{z}}}%
\truex{300}\truey{1000}\truez{600}%
\put(-\value{x},\value{x}){\makebox(0,\value{z})[r]{${#1}$}}%
\put(\value{x},-\value{y}){\makebox(0,\value{z})[l]{${#2}$}}%
\end{picture}}%

\newcommand{\SSWADJAR}[3]{\testdiagrammode%
\Y=#3%
\divide\Y by 2%
\Z=#3%
\multiply \Z by 2%
\begin{picture}(0,0)%
\put(\Y,#3){\begin{picture}(0,0)%
\truex{313}\truey{156}%
\monolength=#3%
\advance\monolength by -\value{x}%
\epilength=#3%
\advance\epilength by \value{x}%
\secondmonolength=\Z%
\advance\secondmonolength by -\value{y}%
\secondepilength=\Z%
\advance\secondepilength by \value{y}%
\put(\value{x},-\value{y}){\line(-1,-2){#3}}%
\put(-\monolength,-\secondepilength){\sswhead}%
\put(-\epilength,-\secondmonolength){\line(1,2){#3}}%
\put(-\value{x},\value{y}){\nnehead}%
\end{picture}}
\truex{300}\truey{1000}\truez{600}%
\put(-\value{x},\value{x}){\makebox(0,\value{z})[r]{${#1}$}}%
\put(\value{x},-\value{y}){\makebox(0,\value{z})[l]{${#2}$}}%
\end{picture}}%

\newcommand{\SSWADJDIST}[3]{\testdiagrammode%
\Y=#3%
\divide\Y by 2%
\Z=#3%
\multiply \Z by 2%
\begin{picture}(0,0)%
\truex{313}\truey{156}\truez{400}%
\put(\Y,#3){\begin{picture}(0,0)%
\monolength=#3%
\advance\monolength by -\value{x}%
\epilength=#3%
\advance\epilength by \value{x}%
\secondmonolength=\Z%
\advance\secondmonolength by -\value{y}%
\secondepilength=\Z%
\advance\secondepilength by \value{y}%
\put(\value{x},-\value{y}){\line(-1,-2){#3}}%
\put(-\monolength,-\secondepilength){\sswhead}%
\put(-\epilength,-\secondmonolength){\line(1,2){#3}}%
\put(-\value{x},\value{y}){\nnehead}%
\end{picture}}
\put(-\value{x},\value{y}){\circle{\value{z}}}%
\put(\value{x},-\value{y}){\circle{\value{z}}}%
\truex{300}\truey{1000}\truez{600}%
\put(-\value{x},\value{x}){\makebox(0,\value{z})[r]{${#1}$}}%
\put(\value{x},-\value{y}){\makebox(0,\value{z})[l]{${#2}$}}%
\end{picture}}%

\def\basicsswar[#1]{\SSWAR{}{}{#100}}%

\newcommand{\sswar}{\@ifnextchar[{\basicsswar}{\basicsswar[67]}}%

\def\basicSswar[#1]#2{\SSWAR{#2}{}{#100}}%

\newcommand{\Sswar}{\@ifnextchar[{\basicSswar}{\basicSswar[67]}}%

\def\basicsswaR[#1]#2{\SSWAR{}{#2}{#100}}%

\newcommand{\sswaR}{\@ifnextchar[{\basicsswaR}{\basicsswaR[67]}}%

\def\basicsswdist[#1]{\SSWDIST{}{}{#100}}%

\newcommand{\sswdist}{\@ifnextchar[{\basicsswdist}{\basicsswdist[67]}}%

\def\basicSswdist[#1]#2{\SSWDIST{#2}{}{#100}}%

\newcommand{\Sswdist}{\@ifnextchar[{\basicSswdist}{\basicSswdist[67]}}%

\def\basicsswdisT[#1]#2{\SSWDIST{}{#2}{#100}}%

\newcommand{\sswdisT}{\@ifnextchar[{\basicsswdisT}{\basicsswdisT[67]}}%

\def\basicsswdotar[#1]{\SSWDOTAR{}{}{#100}}%

\newcommand{\sswdotar}{\@ifnextchar[{\basicsswdotar}{\basicsswdotar[67]}}%

\def\basicSswdotar[#1]#2{\SSWDOTAR{#2}{}{#100}}%

\newcommand{\Sswdotar}{\@ifnextchar[{\basicSswdotar}{\basicSswdotar[67]}}%

\def\basicsswdotaR[#1]#2{\SSWDOTAR{}{#2}{#100}}%

\newcommand{\sswdotaR}{\@ifnextchar[{\basicsswdotaR}{\basicsswdotaR[67]}}%

\def\basicsswmono[#1]{\SSWMONO{}{}{#100}}%

\newcommand{\sswmono}{\@ifnextchar[{\basicsswmono}{\basicsswmono[67]}}%

\def\basicSswmono[#1]#2{\SSWMONO{#2}{}{#100}}%

\newcommand{\Sswmono}{\@ifnextchar[{\basicSswmono}{\basicSswmono[67]}}%

\def\basicsswmonO[#1]#2{\SSWMONO{}{#2}{#100}}%

\newcommand{\sswmonO}{\@ifnextchar[{\basicsswmonO}{\basicsswmonO[67]}}%

\def\basicsswepi[#1]{\SSWEPI{}{}{#100}}%

\newcommand{\sswepi}{\@ifnextchar[{\basicsswepi}{\basicsswepi[67]}}%

\def\basicSswepi[#1]#2{\SSWEPI{#2}{}{#100}}%

\newcommand{\Sswepi}{\@ifnextchar[{\basicSswepi}{\basicSswepi[67]}}%

\def\basicsswepI[#1]#2{\SSWEPI{}{#2}{#100}}%

\newcommand{\sswepI}{\@ifnextchar[{\basicsswepI}{\basicsswepI[67]}}%

\def\basicsswbimo[#1]{\SSWBIMO{}{}{#100}}%

\newcommand{\sswbimo}{\@ifnextchar[{\basicsswbimo}{\basicsswbimo[67]}}%

\def\basicSswbimo[#1]#2{\SSWBIMO{#2}{}{#100}}%

\newcommand{\Sswbimo}{\@ifnextchar[{\basicSswbimo}{\basicSswbimo[67]}}%

\def\basicsswbimO[#1]#2{\SSWBIMO{}{#2}{#100}}%

\newcommand{\sswbimO}{\@ifnextchar[{\basicsswbimO}{\basicsswbimO[67]}}%

\def\basicsswiso[#1]{\SSWAR{\cong}{}{#100}}%

\newcommand{\sswiso}{\@ifnextchar[{\basicsswiso}{\basicsswiso[67]}}%

\def\basicSswiso[#1]#2{\SSWAR{#2}{\cong}{#100}}%

\newcommand{\Sswiso}{\@ifnextchar[{\basicSswiso}{\basicSswiso[67]}}%

\def\basicsswisO[#1]#2{\SSWAR{\cong}{#2}{#100}}%

\newcommand{\sswisO}{\@ifnextchar[{\basicsswisO}{\basicsswisO[67]}}%

\def\basicsswbiar[#1]{\SSWBIAR{}{}{#100}}%

\newcommand{\sswbiar}{\@ifnextchar[{\basicsswbiar}{\basicsswbiar[67]}}%

\def\basicSswbiar[#1]#2#3{\SSWBIAR{#2}{#3}{#100}}%

\newcommand{\Sswbiar}{\@ifnextchar[{\basicSswbiar}{\basicSswbiar[67]}}%

\def\basicsswbidist[#1]{\SSWBIDIST{}{}{#100}}%

\newcommand{\sswbidist}{\@ifnextchar[{\basicsswbidist}{\basicsswbidist[67]}}%

\def\basicSswbidist[#1]#2#3{\SSWBIDIST{#2}{#3}{#100}}%

\newcommand{\Sswbidist}{\@ifnextchar[{\basicSswbidist}{\basicSswbidist[67]}}%

\def\basicsswadjar[#1]{\SSWADJAR{}{}{#100}}%

\newcommand{\sswadjar}{\@ifnextchar[{\basicsswadjar}{\basicsswadjar[67]}}%

\def\basicSswadjar[#1]#2#3{\SSWADJAR{#2}{#3}{#100}}%

\newcommand{\Sswadjar}{\@ifnextchar[{\basicSswadjar}{\basicSswadjar[67]}}%

\def\basicsswadjdist[#1]{\SSWADJDIST{}{}{#100}}%

\newcommand{\sswadjdist}{\@ifnextchar[{\basicsswadjdist}{\basicsswadjdist[67]}}%

\def\basicSswadjdist[#1]#2#3{\SSWADJDIST{#2}{#3}{#100}}%

\newcommand{\Sswadjdist}{\@ifnextchar[{\basicSswadjdist}{\basicSswadjdist[67]}}%

\newcommand{\NNWAR}[3]{\testdiagrammode%
\Z=#3%
\divide\Z by 2%
\begin{picture}(0,0)%
\put(\Z,-#3){\line(-1,2){#3}}%
\put(-\Z,#3){\nnwhead}%
\truex{200}\truey{800}\truez{600}%
\put(\value{x},\value{x}){\makebox(0,\value{z})[l]{${#1}$}}%
\put(-\value{x},-\value{y}){\makebox(0,\value{z})[r]{${#2}$}}%
\end{picture}}%

\newcommand{\NNWDIST}[3]{\testdiagrammode%
\Z=#3%
\divide\Z by 2%
\begin{picture}(0,0)%
\put(\Z,-#3){\line(-1,2){#3}}%
\put(-\Z,#3){\nnwhead}%
\truex{400}%
\put(0,0){\circle{\value{x}}}%
\truex{200}\truey{800}\truez{600}%
\put(\value{x},\value{x}){\makebox(0,\value{z})[l]{${#1}$}}%
\put(-\value{x},-\value{y}){\makebox(0,\value{z})[r]{${#2}$}}%
\end{picture}}%

\newcommand{\NNWDOTAR}[3]{\testdiagrammode%
\truex{100}\truey{268}\truez{134}%
\Z=#3%
\divide\Z by 2%
\NUMBEROFDOTS=#3%
\divide\NUMBEROFDOTS by \value{z}%
\advance\NUMBEROFDOTS by 1%
\begin{picture}(0,0)%
\multiput(\Z,-#3)(-\value{z},\value{y}){\NUMBEROFDOTS}%
{\circle*{\value{x}}}%
\put(-\Z,#3){\nnwhead}%
\truex{200}\truey{800}\truez{600}%
\put(\value{x},\value{x}){\makebox(0,\value{z})[l]{${#1}$}}%
\put(-\value{x},-\value{y}){\makebox(0,\value{z})[r]{${#2}$}}%
\end{picture}}%

\newcommand{\NNWMONO}[3]{\testdiagrammode%
\Z=#3%
\divide\Z by 2%
\truetaiL%
\bimolength=#3%
\advance\bimolength by -\truemonotaiL%
\monolength=\bimolength%
\advance\monolength by -\Z%
\secondmonolength=\monolength%
\multiply\secondmonolength by 2%
\begin{picture}(0,0)%
\put(\monolength,-\secondmonolength){\line(-1,2){\bimolength}}%
\put(\monolength,-\secondmonolength){\nnwhead}%
\put(-\Z,#3){\nnwhead}%
\truex{200}\truey{800}\truez{600}%
\put(\value{x},\value{x}){\makebox(0,\value{z})[l]{${#1}$}}%
\put(-\value{x},-\value{y}){\makebox(0,\value{z})[r]{${#2}$}}%
\end{picture}}%

\newcommand{\NNWEPI}[3]{\testdiagrammode%
\Z=#3%
\divide\Z by 2%
\trueheaD%
\bimolength=#3%
\advance\bimolength by -\trueepiheaD%
\epilength=\bimolength%
\advance\epilength by -\Z%
\secondepilength=\epilength%
\multiply\secondepilength by 2%
\begin{picture}(0,0)%
\put(\Z,-#3){\line(-1,2){\bimolength}}%
\put(-\epilength,\secondepilength){\nnwhead}%
\put(-\Z,#3){\nnwhead}%
\truex{200}\truey{800}\truez{600}%
\put(\value{x},\value{x}){\makebox(0,\value{z})[l]{${#1}$}}%
\put(-\value{x},-\value{y}){\makebox(0,\value{z})[r]{${#2}$}}%
\end{picture}}%

\newcommand{\NNWBIMO}[3]{\testdiagrammode%
\Z=#3%
\divide\Z by 2%
\truetaiL\trueheaD%
\bimolength=#3%
\advance\bimolength by -\truemonotaiL%
\monolength=\bimolength%
\advance\monolength by -\Z%
\advance\bimolength by -\trueepiheaD%
\epilength=\bimolength%
\advance\epilength by -\monolength%
\secondmonolength=\monolength%
\multiply\secondmonolength by 2%
\secondepilength=\epilength%
\multiply\secondepilength by 2%
\begin{picture}(0,0)%
\put(\monolength,-\secondmonolength){\line(-1,2){\bimolength}}%
\put(\monolength,-\secondmonolength){\nnwhead}%
\put(-\epilength,\secondepilength){\nnwhead}%
\put(-\Z,#3){\nnwhead}%
\truex{200}\truey{800}\truez{600}%
\put(\value{x},\value{x}){\makebox(0,\value{z})[l]{${#1}$}}%
\put(-\value{x},-\value{y}){\makebox(0,\value{z})[r]{${#2}$}}%
\end{picture}}%

\newcommand{\NNWBIAR}[3]{\testdiagrammode%
\Y=#3%
\divide\Y by 2%
\Z=#3%
\multiply \Z by 2%
\begin{picture}(0,0)%
\put(\Y,-#3){\begin{picture}(0,0)%
\truex{313}\truey{156}%
\put(-\value{x},-\value{y}){\line(-1,2){#3}}%
\put(\value{x},\value{y}){\line(-1,2){#3}}%
\monolength=#3%
\advance\monolength by -\value{x}%
\epilength=#3%
\advance\epilength by \value{x}%
\secondmonolength=\Z%
\advance\secondmonolength by -\value{y}%
\secondepilength=\Z%
\advance\secondepilength by \value{y}%
\put(-\monolength,\secondepilength){\nnwhead}%
\put(-\epilength,\secondmonolength){\nnwhead}%
\end{picture}}
\truex{400}\truey{1000}\truez{600}%
\put(\value{x},\value{x}){\makebox(0,\value{z})[l]{${#1}$}}%
\put(-\value{x},-\value{y}){\makebox(0,\value{z})[r]{${#2}$}}%
\end{picture}}%

\newcommand{\NNWBIDIST}[3]{\testdiagrammode%
\Y=#3%
\divide\Y by 2%
\Z=#3%
\multiply \Z by 2%
\begin{picture}(0,0)%
\truex{313}\truey{156}\truez{400}%
\put(\Y,-#3){\begin{picture}(0,0)%
\put(-\value{x},-\value{y}){\line(-1,2){#3}}%
\put(\value{x},\value{y}){\line(-1,2){#3}}%
\monolength=#3%
\advance\monolength by -\value{x}%
\epilength=#3%
\advance\epilength by \value{x}%
\secondmonolength=\Z%
\advance\secondmonolength by -\value{y}%
\secondepilength=\Z%
\advance\secondepilength by \value{y}%
\put(-\monolength,\secondepilength){\nnwhead}%
\put(-\epilength,\secondmonolength){\nnwhead}%
\end{picture}}
\put(-\value{x},-\value{y}){\circle{\value{z}}}%
\put(\value{x},\value{y}){\circle{\value{z}}}%
\truex{500}\truey{1000}\truez{600}%
\put(\value{x},\value{x}){\makebox(0,\value{z})[l]{${#1}$}}%
\put(-\value{x},-\value{y}){\makebox(0,\value{z})[r]{${#2}$}}%
\end{picture}}%

\newcommand{\NNWADJAR}[3]{\testdiagrammode%
\Y=#3%
\divide\Y by 2%
\Z=#3%
\multiply \Z by 2%
\begin{picture}(0,0)%
\put(\Y,-#3){\begin{picture}(0,0)%
\truex{313}\truey{156}%
\monolength=#3%
\advance\monolength by -\value{x}%
\epilength=#3%
\advance\epilength by \value{x}%
\secondmonolength=\Z%
\advance\secondmonolength by -\value{y}%
\secondepilength=\Z%
\advance\secondepilength by \value{y}%
\put(-\value{x},-\value{y}){\line(-1,2){#3}}%
\put(-\epilength,\secondmonolength){\nnwhead}%
\put(-\monolength,\secondepilength){\line(1,-2){#3}}%
\put(\value{x},\value{y}){\ssehead}%
\end{picture}}
\truex{400}\truey{1000}\truez{600}%
\put(\value{x},\value{x}){\makebox(0,\value{z})[l]{${#1}$}}%
\put(-\value{x},-\value{y}){\makebox(0,\value{z})[r]{${#2}$}}%
\end{picture}}%

\newcommand{\NNWADJDIST}[3]{\testdiagrammode%
\Y=#3%
\divide\Y by 2%
\Z=#3%
\multiply \Z by 2%
\begin{picture}(0,0)%
\truex{313}\truey{156}\truez{400}%
\put(\Y,-#3){\begin{picture}(0,0)%
\monolength=#3%
\advance\monolength by -\value{x}%
\epilength=#3%
\advance\epilength by \value{x}%
\secondmonolength=\Z%
\advance\secondmonolength by -\value{y}%
\secondepilength=\Z%
\advance\secondepilength by \value{y}%
\put(-\value{x},-\value{y}){\line(-1,2){#3}}%
\put(-\epilength,\secondmonolength){\nnwhead}%
\put(-\monolength,\secondepilength){\line(1,-2){#3}}%
\put(\value{x},\value{y}){\ssehead}%
\end{picture}}
\put(\value{x},\value{y}){\circle{\value{z}}}%
\put(-\value{x},-\value{y}){\circle{\value{z}}}%
\truex{500}\truey{1000}\truez{600}%
\put(\value{x},\value{x}){\makebox(0,\value{z})[l]{${#1}$}}%
\put(-\value{x},-\value{y}){\makebox(0,\value{z})[r]{${#2}$}}%
\end{picture}}%

\def\basicnnwar[#1]{\NNWAR{}{}{#100}}%

\newcommand{\nnwar}{\@ifnextchar[{\basicnnwar}{\basicnnwar[67]}}%

\def\basicNnwar[#1]#2{\NNWAR{#2}{}{#100}}%

\newcommand{\Nnwar}{\@ifnextchar[{\basicNnwar}{\basicNnwar[67]}}%

\def\basicnnwaR[#1]#2{\NNWAR{}{#2}{#100}}%

\newcommand{\nnwaR}{\@ifnextchar[{\basicnnwaR}{\basicnnwaR[67]}}%

\def\basicnnwdist[#1]{\NNWDIST{}{}{#100}}%

\newcommand{\nnwdist}{\@ifnextchar[{\basicnnwdist}{\basicnnwdist[67]}}%

\def\basicNnwdist[#1]#2{\NNWDIST{#2}{}{#100}}%

\newcommand{\Nnwdist}{\@ifnextchar[{\basicNnwdist}{\basicNnwdist[67]}}%

\def\basicnnwdisT[#1]#2{\NNWDIST{}{#2}{#100}}%

\newcommand{\nnwdisT}{\@ifnextchar[{\basicnnwdisT}{\basicnnwdisT[67]}}%

\def\basicnnwdotar[#1]{\NNWDOTAR{}{}{#100}}%

\newcommand{\nnwdotar}{\@ifnextchar[{\basicnnwdotar}{\basicnnwdotar[67]}}%

\def\basicNnwdotar[#1]#2{\NNWDOTAR{#2}{}{#100}}%

\newcommand{\Nnwdotar}{\@ifnextchar[{\basicNnwdotar}{\basicNnwdotar[67]}}%

\def\basicnnwdotaR[#1]#2{\NNWDOTAR{}{#2}{#100}}%

\newcommand{\nnwdotaR}{\@ifnextchar[{\basicnnwdotaR}{\basicnnwdotaR[67]}}%

\def\basicnnwmono[#1]{\NNWMONO{}{}{#100}}%

\newcommand{\nnwmono}{\@ifnextchar[{\basicnnwmono}{\basicnnwmono[67]}}%

\def\basicNnwmono[#1]#2{\NNWMONO{#2}{}{#100}}%

\newcommand{\Nnwmono}{\@ifnextchar[{\basicNnwmono}{\basicNnwmono[67]}}%

\def\basicnnwmonO[#1]#2{\NNWMONO{}{#2}{#100}}%

\newcommand{\nnwmonO}{\@ifnextchar[{\basicnnwmonO}{\basicnnwmonO[67]}}%

\def\basicnnwepi[#1]{\NNWEPI{}{}{#100}}%

\newcommand{\nnwepi}{\@ifnextchar[{\basicnnwepi}{\basicnnwepi[67]}}%

\def\basicNnwepi[#1]#2{\NNWEPI{#2}{}{#100}}%

\newcommand{\Nnwepi}{\@ifnextchar[{\basicNnwepi}{\basicNnwepi[67]}}%

\def\basicnnwepI[#1]#2{\NNWEPI{}{#2}{#100}}%

\newcommand{\nnwepI}{\@ifnextchar[{\basicnnwepI}{\basicnnwepI[67]}}%

\def\basicnnwbimo[#1]{\NNWBIMO{}{}{#100}}%

\newcommand{\nnwbimo}{\@ifnextchar[{\basicnnwbimo}{\basicnnwbimo[67]}}%

\def\basicNnwbimo[#1]#2{\NNWBIMO{#2}{}{#100}}%

\newcommand{\Nnwbimo}{\@ifnextchar[{\basicNnwbimo}{\basicNnwbimo[67]}}%

\def\basicnnwbimO[#1]#2{\NNWBIMO{}{#2}{#100}}%

\newcommand{\nnwbimO}{\@ifnextchar[{\basicnnwbimO}{\basicnnwbimO[67]}}%

\def\basicnnwiso[#1]{\NNWAR{\cong}{}{#100}}%

\newcommand{\nnwiso}{\@ifnextchar[{\basicnnwiso}{\basicnnwiso[67]}}%

\def\basicNnwiso[#1]#2{\NNWAR{#2}{\cong}{#100}}%

\newcommand{\Nnwiso}{\@ifnextchar[{\basicNnwiso}{\basicNnwiso[67]}}%

\def\basicnnwisO[#1]#2{\NNWAR{\cong}{#2}{#100}}%

\newcommand{\nnwisO}{\@ifnextchar[{\basicnnwisO}{\basicnnwisO[67]}}%

\def\basicnnwbiar[#1]{\NNWBIAR{}{}{#100}}%

\newcommand{\nnwbiar}{\@ifnextchar[{\basicnnwbiar}{\basicnnwbiar[67]}}%

\def\basicNnwbiar[#1]#2#3{\NNWBIAR{#2}{#3}{#100}}%

\newcommand{\Nnwbiar}{\@ifnextchar[{\basicNnwbiar}{\basicNnwbiar[67]}}%

\def\basicnnwbidist[#1]{\NNWBIDIST{}{}{#100}}%

\newcommand{\nnwbidist}{\@ifnextchar[{\basicnnwbidist}{\basicnnwbidist[67]}}%

\def\basicNnwbidist[#1]#2#3{\NNWBIDIST{#2}{#3}{#100}}%

\newcommand{\Nnwbidist}{\@ifnextchar[{\basicNnwbidist}{\basicNnwbidist[67]}}%

\def\basicnnwadjar[#1]{\NNWADJAR{}{}{#100}}%

\newcommand{\nnwadjar}{\@ifnextchar[{\basicnnwadjar}{\basicnnwadjar[67]}}%

\def\basicNnwadjar[#1]#2#3{\NNWADJAR{#2}{#3}{#100}}%

\newcommand{\Nnwadjar}{\@ifnextchar[{\basicNnwadjar}{\basicNnwadjar[67]}}%

\def\basicnnwadjdist[#1]{\NNWADJDIST{}{}{#100}}%

\newcommand{\nnwadjdist}{\@ifnextchar[{\basicnnwadjdist}{\basicnnwadjdist[67]}}%

\def\basicNnwadjdist[#1]#2#3{\NNWADJDIST{#2}{#3}{#100}}%

\newcommand{\Nnwadjdist}{\@ifnextchar[{\basicNnwadjdist}{\basicNnwadjdist[67]}}%

\newcommand{\EENEAR}[3]{\testdiagrammode%
\Y=#3%
\divide \Y by 2%
\Z=\Y%
\divide \Z by 3%
\begin{picture}(0,0)%
\put(-\Y,-\Z){\line(3,1){#3}}%
\put(\Y,\Z){\eenehead}%
\truex{200}\truey{800}\truez{600}%
\put(-\value{x},\value{x}){\makebox(0,\value{z})[r]{${#1}$}}%
\put(\value{x},-\value{y}){\makebox(0,\value{z})[l]{${#2}$}}%
\end{picture}}%

\def\basiceenear[#1]{\EENEAR{}{}{#100}}%

\newcommand{\eenear}{\@ifnextchar[{\basiceenear}{\basiceenear[211]}}%

\def\basicEenear[#1]#2{\EENEAR{#2}{}{#100}}%

\newcommand{\Eenear}{\@ifnextchar[{\basicEenear}{\basicEenear[211]}}%

\def\basiceeneaR[#1]#2{\EENEAR{}{#2}{#100}}%

\newcommand{\eeneaR}{\@ifnextchar[{\basiceeneaR}{\basiceeneaR[211]}}%

\newcommand{\EESEAR}[3]{\testdiagrammode%
\Y=#3%
\divide \Y by 2%
\Z=\Y%
\divide \Z by 3%
\begin{picture}(0,0)%
\put(-\Y,\Z){\line(3,-1){#3}}%
\put(\Y,-\Z){\eesehead}%
\truex{200}\truey{800}\truez{600}%
\put(\value{x},\value{x}){\makebox(0,\value{z})[l]{${#1}$}}%
\put(-\value{x},-\value{y}){\makebox(0,\value{z})[r]{${#2}$}}%
\end{picture}}%

\def\basiceesear[#1]{\EESEAR{}{}{#100}}%

\newcommand{\eesear}{\@ifnextchar[{\basiceesear}{\basiceesear[211]}}%

\def\basicEesear[#1]#2{\EESEAR{#2}{}{#100}}%

\newcommand{\Eesear}{\@ifnextchar[{\basicEesear}{\basicEesear[211]}}%

\def\basiceeseaR[#1]#2{\EESEAR{}{#2}{#100}}%

\newcommand{\eeseaR}{\@ifnextchar[{\basiceeseaR}{\basiceeseaR[211]}}%

\newcommand{\WWNWAR}[3]{\testdiagrammode%
\Y=#3%
\divide \Y by 2%
\Z=\Y%
\divide \Z by 3%
\begin{picture}(0,0)%
\put(\Y,-\Z){\line(-3,1){#3}}%
\put(-\Y,\Z){\wwnwhead}%
\truex{200}\truey{800}\truez{600}%
\put(\value{x},\value{x}){\makebox(0,\value{z})[l]{${#1}$}}%
\put(-\value{x},-\value{y}){\makebox(0,\value{z})[r]{${#2}$}}%
\end{picture}}%

\def\basicwwnwar[#1]{\WWNWAR{}{}{#100}}%

\newcommand{\wwnwar}{\@ifnextchar[{\basicwwnwar}{\basicwwnwar[211]}}%

\def\basicWwnwar[#1]#2{\WWNWAR{#2}{}{#100}}%

\newcommand{\Wwnwar}{\@ifnextchar[{\basicWwnwar}{\basicWwnwar[211]}}%

\def\basicwwnwaR[#1]#2{\WWNWAR{}{#2}{#100}}%

\newcommand{\wwnwaR}{\@ifnextchar[{\basicwwnwaR}{\basicwwnwaR[211]}}%

\newcommand{\WWSWAR}[3]{\testdiagrammode%
\Y=#3%
\divide \Y by 2%
\Z=\Y%
\divide \Z by 3%
\begin{picture}(0,0)%
\put(\Y,\Z){\line(-3,-1){#3}}%
\put(-\Y,-\Z){\wwswhead}%
\truex{200}\truey{800}\truez{600}%
\put(-\value{x},\value{x}){\makebox(0,\value{z})[r]{${#1}$}}%
\put(\value{x},-\value{y}){\makebox(0,\value{z})[l]{${#2}$}}%
\end{picture}}%

\def\basicwwswar[#1]{\WWSWAR{}{}{#100}}%

\newcommand{\wwswar}{\@ifnextchar[{\basicwwswar}{\basicwwswar[211]}}%

\def\basicWwswar[#1]#2{\WWSWAR{#2}{}{#100}}%

\newcommand{\Wwswar}{\@ifnextchar[{\basicWwswar}{\basicWwswar[211]}}%

\def\basicwwswaR[#1]#2{\WWSWAR{}{#2}{#100}}%

\newcommand{\wwswaR}{\@ifnextchar[{\basicwwswaR}{\basicwwswaR[211]}}%

\newcommand{\NNNEAR}[3]{\testdiagrammode%
\Y=#3%
\divide \Y by 2%
\Z=\Y%
\multiply \Z by 3%
\begin{picture}(0,0)%
\put(-\Y,-\Z){\line(1,3){#3}}%
\put(\Y,\Z){\nnnehead}%
\truex{100}\truez{600}%
\put(-\value{x},\value{x}){\makebox(0,\value{z})[r]{${#1}$}}%
\put(\value{x},-\value{z}){\makebox(0,\value{z})[l]{${#2}$}}%
\end{picture}}%

\def\basicnnnear[#1]{\NNNEAR{}{}{#100}}%

\newcommand{\nnnear}{\@ifnextchar[{\basicnnnear}{\basicnnnear[71]}}%

\def\basicNnnear[#1]#2{\NNNEAR{#2}{}{#100}}%

\newcommand{\Nnnear}{\@ifnextchar[{\basicNnnear}{\basicNnnear[71]}}%

\def\basicnnneaR[#1]#2{\NNNEAR{}{#2}{#100}}%

\newcommand{\nnneaR}{\@ifnextchar[{\basicnnneaR}{\basicnnneaR[71]}}%

\newcommand{\SSSWAR}[3]{\testdiagrammode%
\Y=#3%
\divide \Y by 2%
\Z=\Y%
\multiply \Z by 3%
\begin{picture}(0,0)%
\put(\Y,\Z){\line(-1,-3){#3}}%
\put(-\Y,-\Z){\ssswhead}%
\truex{100}\truez{600}%
\put(-\value{x},\value{x}){\makebox(0,\value{z})[r]{${#1}$}}%
\put(\value{x},-\value{z}){\makebox(0,\value{z})[l]{${#2}$}}%
\end{picture}}%

\def\basicssswar[#1]{\SSSWAR{}{}{#100}}%

\newcommand{\ssswar}{\@ifnextchar[{\basicssswar}{\basicssswar[71]}}%

\def\basicSsswar[#1]#2{\SSSWAR{#2}{}{#100}}%

\newcommand{\Ssswar}{\@ifnextchar[{\basicSsswar}{\basicSsswar[71]}}%

\def\basicssswaR[#1]#2{\SSSWAR{}{#2}{#100}}%

\newcommand{\ssswaR}{\@ifnextchar[{\basicssswaR}{\basicssswaR[71]}}%

\newcommand{\SSSEAR}[3]{\testdiagrammode%
\Y=#3%
\divide \Y by 2%
\Z=\Y%
\multiply \Z by 3%
\begin{picture}(0,0)%
\put(-\Y,\Z){\line(1,-3){#3}}%
\put(\Y,-\Z){\sssehead}%
\truex{200}\truez{600}%
\put(\value{x},\value{x}){\makebox(0,\value{z})[l]{${#1}$}}%
\put(-\value{x},-\value{z}){\makebox(0,\value{z})[r]{${#2}$}}%
\end{picture}}%

\def\basicsssear[#1]{\SSSEAR{}{}{#100}}%

\newcommand{\sssear}{\@ifnextchar[{\basicsssear}{\basicsssear[71]}}%

\def\basicSssear[#1]#2{\SSSEAR{#2}{}{#100}}%

\newcommand{\Sssear}{\@ifnextchar[{\basicSssear}{\basicSssear[71]}}%

\def\basicssseaR[#1]#2{\SSSEAR{}{#2}{#100}}%

\newcommand{\ssseaR}{\@ifnextchar[{\basicssseaR}{\basicssseaR[71]}}%

\newcommand{\NNNWAR}[3]{\testdiagrammode%
\Y=#3%
\divide \Y by 2%
\Z=\Y%
\multiply \Z by 3%
\begin{picture}(0,0)%
\put(\Y,-\Z){\line(-1,3){#3}}%
\put(-\Y,\Z){\nnnwhead}%
\truex{200}\truez{600}%
\put(\value{x},\value{x}){\makebox(0,\value{z})[l]{${#1}$}}%
\put(-\value{x},-\value{z}){\makebox(0,\value{z})[r]{${#2}$}}%
\end{picture}}%

\def\basicnnnwar[#1]{\NNNWAR{}{}{#100}}%

\newcommand{\nnnwar}{\@ifnextchar[{\basicnnnwar}{\basicnnnwar[71]}}%

\def\basicNnnwar[#1]#2{\NNNWAR{#2}{}{#100}}%

\newcommand{\Nnnwar}{\@ifnextchar[{\basicNnnwar}{\basicNnnwar[71]}}%

\def\basicnnnwaR[#1]#2{\NNNWAR{}{#2}{#100}}%

\newcommand{\nnnwaR}{\@ifnextchar[{\basicnnnwaR}{\basicnnnwaR[71]}}%

\newcommand{\NEENEAR}[3]{\testdiagrammode%
\Y=#3%
\divide \Y by 2%
\Z=#3%
\divide \Z by 3%
\begin{picture}(0,0)%
\put(-\Y,-\Z){\line(3,2){#3}}%
\put(\Y,\Z){\neenehead}%
\truex{200}\truey{800}\truez{600}%
\put(-\value{x},\value{x}){\makebox(0,\value{z})[r]{${#1}$}}%
\put(\value{x},-\value{y}){\makebox(0,\value{z})[l]{${#2}$}}%
\end{picture}}%

\def\basicneenear[#1]{\NEENEAR{}{}{#100}}%

\newcommand{\neenear}{\@ifnextchar[{\basicneenear}{\basicneenear[215]}}%

\def\basicNeenear[#1]#2{\NEENEAR{#2}{}{#100}}%

\newcommand{\Neenear}{\@ifnextchar[{\basicNeenear}{\basicNeenear[215]}}%

\def\basicneeneaR[#1]#2{\NEENEAR{}{#2}{#100}}%

\newcommand{\neeneaR}{\@ifnextchar[{\basicneeneaR}{\basicneeneaR[215]}}%

\newcommand{\SEESEAR}[3]{\testdiagrammode%
\Y=#3%
\divide \Y by 2%
\Z=#3%
\divide \Z by 3%
\begin{picture}(0,0)%
\put(-\Y,\Z){\line(3,-2){#3}}%
\put(\Y,-\Z){\seesehead}%
\truex{200}\truey{800}\truez{600}%
\put(\value{x},\value{x}){\makebox(0,\value{z})[l]{${#1}$}}%
\put(-\value{x},-\value{y}){\makebox(0,\value{z})[r]{${#2}$}}%
\end{picture}}%

\def\basicseesear[#1]{\SEESEAR{}{}{#100}}%

\newcommand{\seesear}{\@ifnextchar[{\basicseesear}{\basicseesear[215]}}%

\def\basicSeesear[#1]#2{\SEESEAR{#2}{}{#100}}%

\newcommand{\Seesear}{\@ifnextchar[{\basicSeesear}{\basicSeesear[215]}}%

\def\basicseeseaR[#1]#2{\SEESEAR{}{#2}{#100}}%

\newcommand{\seeseaR}{\@ifnextchar[{\basicseeseaR}{\basicseeseaR[215]}}%

\newcommand{\NWWNWAR}[3]{\testdiagrammode%
\Y=#3%
\divide \Y by 2%
\Z=#3%
\divide \Z by 3%
\begin{picture}(0,0)%
\put(\Y,-\Z){\line(-3,2){#3}}%
\put(-\Y,\Z){\nwwnwhead}%
\truex{200}\truey{800}\truez{600}%
\put(\value{x},\value{x}){\makebox(0,\value{z})[l]{${#1}$}}%
\put(-\value{x},-\value{y}){\makebox(0,\value{z})[r]{${#2}$}}%
\end{picture}}%

\def\basicnwwnwar[#1]{\NWWNWAR{}{}{#100}}%

\newcommand{\nwwnwar}{\@ifnextchar[{\basicnwwnwar}{\basicnwwnwar[215]}}%

\def\basicNwwnwar[#1]#2{\NWWNWAR{#2}{}{#100}}%

\newcommand{\Nwwnwar}{\@ifnextchar[{\basicNwwnwar}{\basicNwwnwar[215]}}%

\def\basicnwwnwaR[#1]#2{\NWWNWAR{}{#2}{#100}}%

\newcommand{\nwwnwaR}{\@ifnextchar[{\basicnwwnwaR}{\basicnwwnwaR[215]}}%

\newcommand{\SWWSWAR}[3]{\testdiagrammode%
\Y=#3%
\divide \Y by 2%
\Z=#3%
\divide \Z by 3%
\begin{picture}(0,0)%
\put(\Y,\Z){\line(-3,-2){#3}}%
\put(-\Y,-\Z){\swwswhead}%
\truex{200}\truey{800}\truez{600}%
\put(-\value{x},\value{x}){\makebox(0,\value{z})[r]{${#1}$}}%
\put(\value{x},-\value{y}){\makebox(0,\value{z})[l]{${#2}$}}%
\end{picture}}%

\def\basicswwswar[#1]{\SWWSWAR{}{}{#100}}%

\newcommand{\swwswar}{\@ifnextchar[{\basicswwswar}{\basicswwswar[215]}}%

\def\basicSwwswar[#1]#2{\SWWSWAR{#2}{}{#100}}%

\newcommand{\Swwswar}{\@ifnextchar[{\basicSwwswar}{\basicSwwswar[215]}}%

\def\basicswwswaR[#1]#2{\SWWSWAR{}{#2}{#100}}%

\newcommand{\swwswaR}{\@ifnextchar[{\basicswwswaR}{\basicswwswaR[215]}}%

\newcommand{\NENNEAR}[3]{\testdiagrammode%
\Y=#3%
\divide \Y by 2%
\Z=#3%
\multiply \Z by 3%
\divide \Z by 4%
\begin{picture}(0,0)%
\put(-\Y,-\Z){\line(2,3){#3}}%
\put(\Y,\Z){\nennehead}%
\truex{100}\truez{600}%
\put(-\value{x},\value{x}){\makebox(0,\value{z})[r]{${#1}$}}%
\put(\value{x},-\value{z}){\makebox(0,\value{z})[l]{${#2}$}}%
\end{picture}}%

\def\basicnennear[#1]{\NENNEAR{}{}{#100}}%

\newcommand{\nennear}{\@ifnextchar[{\basicnennear}{\basicnennear[143]}}%

\def\basicNennear[#1]#2{\NENNEAR{#2}{}{#100}}%

\newcommand{\Nennear}{\@ifnextchar[{\basicNennear}{\basicNennear[143]}}%

\def\basicnenneaR[#1]#2{\NENNEAR{}{#2}{#100}}%

\newcommand{\nenneaR}{\@ifnextchar[{\basicnenneaR}{\basicnenneaR[143]}}%

\newcommand{\SWSSWAR}[3]{\testdiagrammode%
\Y=#3%
\divide \Y by 2%
\Z=#3%
\multiply \Z by 3%
\divide \Z by 4%
\begin{picture}(0,0)%
\put(\Y,\Z){\line(-2,-3){#3}}%
\put(-\Y,-\Z){\swsswhead}%
\truex{100}\truez{600}%
\put(-\value{x},\value{x}){\makebox(0,\value{z})[r]{${#1}$}}%
\put(\value{x},-\value{z}){\makebox(0,\value{z})[l]{${#2}$}}%
\end{picture}}%

\def\basicswsswar[#1]{\SWSSWAR{}{}{#100}}%

\newcommand{\swsswar}{\@ifnextchar[{\basicswsswar}{\basicswsswar[143]}}%

\def\basicSwsswar[#1]#2{\SWSSWAR{#2}{}{#100}}%

\newcommand{\Swsswar}{\@ifnextchar[{\basicSwsswar}{\basicSwsswar[143]}}%

\def\basicswsswaR[#1]#2{\SWSSWAR{}{#2}{#100}}%

\newcommand{\swsswaR}{\@ifnextchar[{\basicswsswaR}{\basicswsswaR[143]}}%

\newcommand{\SESSEAR}[3]{\testdiagrammode%
\Y=#3%
\divide \Y by 2%
\Z=#3%
\multiply \Z by 3%
\divide \Z by 4%
\begin{picture}(0,0)%
\put(-\Y,\Z){\line(2,-3){#3}}%
\put(\Y,-\Z){\sessehead}%
\truex{200}\truez{600}%
\put(\value{x},\value{x}){\makebox(0,\value{z})[l]{${#1}$}}%
\put(-\value{x},-\value{z}){\makebox(0,\value{z})[r]{${#2}$}}%
\end{picture}}%

\def\basicsessear[#1]{\SESSEAR{}{}{#100}}%

\newcommand{\sessear}{\@ifnextchar[{\basicsessear}{\basicsessear[143]}}%

\def\basicSessear[#1]#2{\SESSEAR{#2}{}{#100}}%

\newcommand{\Sessear}{\@ifnextchar[{\basicSessear}{\basicSessear[143]}}%

\def\basicsesseaR[#1]#2{\SESSEAR{}{#2}{#100}}%

\newcommand{\sesseaR}{\@ifnextchar[{\basicsesseaR}{\basicsesseaR[143]}}%

\newcommand{\NWNNWAR}[3]{\testdiagrammode%
\Y=#3%
\divide \Y by 2%
\Z=#3%
\multiply \Z by 3%
\divide \Z by 4%
\begin{picture}(0,0)%
\put(\Y,-\Z){\line(-2,3){#3}}%
\put(-\Y,\Z){\nwnnwhead}%
\truex{200}\truez{600}%
\put(\value{x},\value{x}){\makebox(0,\value{z})[l]{${#1}$}}%
\put(-\value{x},-\value{z}){\makebox(0,\value{z})[r]{${#2}$}}%
\end{picture}}%

\def\basicnwnnwar[#1]{\NWNNWAR{}{}{#100}}%

\newcommand{\nwnnwar}{\@ifnextchar[{\basicnwnnwar}{\basicnwnnwar[143]}}%

\def\basicNwnnwar[#1]#2{\NWNNWAR{#2}{}{#100}}%

\newcommand{\Nwnnwar}{\@ifnextchar[{\basicNwnnwar}{\basicNwnnwar[143]}}%

\def\basicnwnnwaR[#1]#2{\NWNNWAR{}{#2}{#100}}%

\newcommand{\nwnnwaR}{\@ifnextchar[{\basicnwnnwaR}{\basicnwnnwaR[143]}}%

\newcommand{\Necurve}[2]%
{\testdiagrammode\begin{picture}(0,0)%
\truex{1300}\truey{2000}\truez{200}%
\put(0,\value{x}){\oval(#200,\value{y})[t]}%
\put(0,\value{x}){\makebox(0,0){\begin{picture}(#200,0)%
\put(#200,0){\line(0,-1){\value{z}}}%
\put(#200,-\value{z}){\shead}%
\put(0,0){\line(0,-1){\value{z}}}\end{picture}}}%
\truex{2500}%
\put(0,\value{x}){\makebox(0,0)[b]{${#1}$}}%
\end{picture}}%

\def\basicnecurvar[#1]{\Necurve{}{#1}}

\newcommand{\necurvar}{\@ifnextchar[{\basicnecurvar}{\basicnecurvar[160]}}%

\def\basicNecurvar[#1]#2{\Necurve{#2}{#1}}%

\newcommand{\Necurvar}{\@ifnextchar[{\basicNecurvar}{\basicNecurvar[160]}}%

\newcommand{\Nwcurve}[2]%
{\testdiagrammode\begin{picture}(0,0)%
\truex{1300}\truey{2000}\truez{200}%
\put(0,\value{x}){\oval(#200,\value{y})[t]}%
\put(0,\value{x}){\makebox(0,0){\begin{picture}(#200,0)%
\put(#200,0){\line(0,-1){\value{z}}}%
\put(0,0){\line(0,-1){\value{z}}}%
\put(0,-\value{z}){\shead}%
\end{picture}}}%
\truex{2500}%
\put(0,\value{x}){\makebox(0,0)[b]{${#1}$}}%
\end{picture}}%

\def\basicnwcurvar[#1]{\Nwcurve{}{#1}}

\newcommand{\nwcurvar}{\@ifnextchar[{\basicnwcurvar}{\basicnwcurvar[160]}}%

\def\basicNwcurvar[#1]#2{\Nwcurve{#2}{#1}}%

\newcommand{\Nwcurvar}{\@ifnextchar[{\basicNwcurvar}{\basicNwcurvar[160]}}%

\newcommand{\Securve}[2]%
{\testdiagrammode\begin{picture}(0,0)%
\truex{1300}\truey{2000}\truez{200}%
\put(0,-\value{x}){\oval(#200,\value{y})[b]}%
\put(0,-\value{x}){\makebox(0,0){\begin{picture}(#200,0)%
\put(#200,0){\line(0,1){\value{z}}}%
\put(0,0){\line(0,1){\value{z}}}%
\put(#200,\value{z}){\nhead}%
\end{picture}}}%
\truex{2500}%
\put(0,-\value{x}){\makebox(0,0)[t]{${#1}$}}%
\end{picture}}%

\def\basicsecurvar[#1]{\Securve{}{#1}}

\newcommand{\securvar}{\@ifnextchar[{\basicsecurvar}{\basicsecurvar[160]}}%

\def\basicSecurvar[#1]#2{\Securve{#2}{#1}}%

\newcommand{\Securvar}{\@ifnextchar[{\basicSecurvar}{\basicSecurvar[160]}}%

\newcommand{\Swcurve}[2]%
{\testdiagrammode\begin{picture}(0,0)%
\truex{1300}\truey{2000}\truez{200}%
\put(0,-\value{x}){\oval(#200,\value{y})[b]}%
\put(0,-\value{x}){\makebox(0,0){\begin{picture}(#200,0)%
\put(#200,0){\line(0,1){\value{z}}}%
\put(0,0){\line(0,1){\value{z}}}%
\put(0,\value{z}){\nhead}%
\end{picture}}}%
\truex{2500}%
\put(0,-\value{x}){\makebox(0,0)[t]{${#1}$}}%
\end{picture}}%

\def\basicswcurvar[#1]{\Swcurve{}{#1}}

\newcommand{\swcurvar}{\@ifnextchar[{\basicswcurvar}{\basicswcurvar[160]}}%

\def\basicSwcurvar[#1]#2{\Swcurve{#2}{#1}}%

\newcommand{\Swcurvar}{\@ifnextchar[{\basicSwcurvar}{\basicSwcurvar[160]}}%

\newcommand{\Escurve}[2]%
{\testdiagrammode\begin{picture}(0,0)%
\truex{1400}\truey{2000}\truez{200}%
\put(\value{x},0){\oval(\value{y},#200)[r]}%
\put(\value{x},0){\makebox(0,0){\begin{picture}(0,#200)%
\put(0,0){\line(-1,0){\value{z}}}%
\put(0,#200){\line(-1,0){\value{z}}}%
\put(-\value{z},0){\whead}%
\end{picture}}}%
\truex{2500}%
\put(\value{x},0){\makebox(0,0)[l]{${#1}$}}%
\end{picture}}%

\def\basicescurvar[#1]{\Escurve{}{#1}}

\newcommand{\escurvar}{\@ifnextchar[{\basicescurvar}{\basicescurvar[160]}}%

\def\basicEscurvar[#1]#2{\Escurve{#2}{#1}}%

\newcommand{\Escurvar}{\@ifnextchar[{\basicEscurvar}{\basicEscurvar[160]}}%

\newcommand{\Encurve}[2]%
{\testdiagrammode\begin{picture}(0,0)%
\truex{1400}\truey{2000}\truez{200}%
\put(\value{x},0){\oval(\value{y},#200)[r]}%
\put(\value{x},0){\makebox(0,0){\begin{picture}(0,#200)%
\put(0,0){\line(-1,0){\value{z}}}%
\put(0,#200){\line(-1,0){\value{z}}}%
\put(-\value{z},#200){\whead}%
\end{picture}}}%
\truex{2500}%
\put(\value{x},0){\makebox(0,0)[l]{${#1}$}}%
\end{picture}}%

\def\basicencurvar[#1]{\Encurve{}{#1}}

\newcommand{\encurvar}{\@ifnextchar[{\basicencurvar}{\basicencurvar[160]}}%

\def\basicEncurvar[#1]#2{\Encurve{#2}{#1}}%

\newcommand{\Encurvar}{\@ifnextchar[{\basicEncurvar}{\basicEncurvar[160]}}%

\newcommand{\Wscurve}[2]%
{\testdiagrammode\begin{picture}(0,0)%
\truex{1300}\truey{2000}\truez{200}%
\put(-\value{x},0){\oval(\value{y},#200)[l]}%
\put(-\value{x},0){\makebox(0,0){\begin{picture}(0,#200)%
\put(0,0){\line(1,0){\value{z}}}%
\put(0,#200){\line(1,0){\value{z}}}%
\put(\value{z},0){\ehead}%
\end{picture}}}%
\truex{2400}%
\put(-\value{x},0){\makebox(0,0)[r]{${#1}$}}%
\end{picture}}%

\def\basicwscurvar[#1]{\Wscurve{}{#1}}

\newcommand{\wscurvar}{\@ifnextchar[{\basicwscurvar}{\basicwscurvar[160]}}%

\def\basicWscurvar[#1]#2{\Wscurve{#2}{#1}}%

\newcommand{\Wscurvar}{\@ifnextchar[{\basicWscurvar}{\basicWscurvar[160]}}%

\newcommand{\Wncurve}[2]%
{\testdiagrammode\begin{picture}(0,0)%
\truex{1300}\truey{2000}\truez{200}%
\put(-\value{x},0){\oval(\value{y},#200)[l]}%
\put(-\value{x},0){\makebox(0,0){\begin{picture}(0,#200)%
\put(0,0){\line(1,0){\value{z}}}%
\put(\value{z},#200){\ehead}%
\put(0,#200){\line(1,0){\value{z}}}%
\end{picture}}}%
\truex{2400}%
\put(-\value{x},0){\makebox(0,0)[r]{${#1}$}}%
\end{picture}}%

\def\basicwncurvar[#1]{\Wncurve{}{#1}}

\newcommand{\wncurvar}{\@ifnextchar[{\basicwncurvar}{\basicwncurvar[160]}}%

\def\basicWncurvar[#1]#2{\Wncurve{#2}{#1}}%

\newcommand{\Wncurvar}{\@ifnextchar[{\basicWncurvar}{\basicWncurvar[160]}}%

\catcode`\@=12

\begin{document}

\begin{flushright}
{\large \bf Published in Foundations of Physics 2025, 55:76 \\ https://doi.org/10.1007/s10701-025-00888-5 \\ shared link https://rdcu.be/eKsx0}
\end{flushright}

\vspace{10mm}

\begin{center}
{\LARGE \bf Contemporary Perspectivism as a Framework \\ 
of Scientific Inquiry in Quantum Mechanics and Beyond} \\
\end{center}

\vspace{.2mm}

\begin{center}
{\large \bf  Vassilios Karakostas and Elias Zafiris}
\end{center}

\footnotetext{To contact the authors, please write to: Vassilios Karakostas, {\em Department of History and Philosophy of Science, Faculty of Science, National and Kapodistrian University of Athens, Athens 157 71, Greece; e-mail: karakost@phs.uoa.gr}, or to Elias Zafiris, {\em Parmenides Foundation, Center for the Conceptual Foundations of Science, Kirchplatz 1, Pullach, 82049 Munich, Germany; e-mail: ezafiris@math.uoa.gr}.}

\begin{abstract}
\hspace{-45.2pt}
\begin{minipage}{1\textwidth}

\noindent
Contemporary perspectivism is viewed as a framework of scientific inquiry concerning the origin, generation and systematization of scientific knowledge of nature by focusing on the conditions under which such knowledge may arise in perspectivist terms and investigating the essential ramifications of these conditions. To this end, we develop the conceptual, methodological and semantic framework of contemporary perspectivism according to the norms of the proposed endo-theoretic approach. Implementation of the preceding three-fold scheme in quantum mechanics implies that the global structure of a quantum algebra of events can be consistently comprehended through a multilevel structure of locally variable Boolean perspectives, interconnected in a category-theoretic environment, yielding jointly all the information encoded in the former. In this respect, the proposed approach validates the perspectivist/contextual nature of quantum mechanics at a fundamental level of discourse. Furthermore, due to its general character, it may acquire the form of a theoretical pattern of scientific inquiry in the natural sciences, especially when dealing with complex trans-perspectival phenomena, the analysis of which requires the use of information resulting from more than one perspective. Finally, in the appendix, we provide a concise comparative assessment between our perspectivist framework of quantum theory and Rovelli's relational interpretation of quantum mechanics.

\end{minipage}
\end{abstract}

\vspace{.2cm}

\renewcommand{\baselinestretch}{1.15}

\section{On the Notion of Perspective in the Natural Sciences}
Contemporary scientific perspectivism has gradually emerged in twenty-first century as a collection of positions in philosophy of science that in various ways place emphasis on the situated character of knowledge. The latter feature is understood in the sense that our scientific knowledge is the product of the historical period and the prevailing cultural tradition in which working hypotheses, modeling practices, data gathering, and scientific theories belong.
In this context of discourse, as a central position of perspectivism is advanced the reliance of scientific knowledge on perspective, such as a conceptual scheme or a theoretical lexicon [1], scientific hypotheses and practices in the form of models for classifying phenomena and observations [2], or invoking epistemic norms of justification with respect to specific scientific communities at any particular historical time [3].

Accordingly, in general philosophy of science, scientific perspectivism is often subsumed under the thesis that knowledge of nature is possible only within the boundaries of historically well-defined scientific perspectives composed of data analysis, theoretical models and principles relative to the perspective adopted, so that one may refer, for instance, to the Newtonian perspective, the Einsteinian perspective, and so on [1]. Alternatively, in a weaker sense, scientific perspectivism is viewed as a means for assessing or evaluating, at the same time, rival modeling practices or incompatible research programmes which may give rise to perspectival knowledge (e.g. [4]). In both cases---characterising, correspondingly, the diachronic and synchronic version of perspectivism---the usual sense attributed to the notion of a ``perspective" refers to ``the actual, historically and culturally situated, scientific practice of a real scientific community at a given historical time" [3, p. 5]. Thus, the notion of a perspective is broadly taken to include, firstly, the body of knowledge claims advanced by the scientific community at a given time, secondly, the community's resources (theoretical, experimental, technological) available at the time for reliably generating such claims, and, thirdly, second-order epistemic assertions concerning the justification of the knowledge claims so advanced [3, p. 6].
Admittedly, defined that way, a perspective echoes Kuhn's historicism, specifically, his concept of a ``disciplinary matrix" (see also [5]).

In the present study, contemporary perspectivism is viewed as a framework of scientific inquiry concerning the origin, generation and systematization of scientific knowledge of nature by focusing on the conditions under which such knowledge may arise in perspectivist terms, and investigating the essential ramifications of these conditions.
In view of the fact that the customary position strictly associates a scientific perspective to a given epistemic community, we introduce alternatively, a concrete spatiotemporal concept of ``perspective" that aims to serve as the basis of scientific inquiry in the natural sciences. 
The proposed notion of ``perspective" is theorized, in relation to its ordinary meaning, and is conceived as the primary vehicle of tracing and investigating the world, as the principal structural unit of probing the natural world.
According to the proposed framework of scientific perspectivism:
\begin{itemize}
\item[-] A perspective is characterised {\it endo-theoretically}, namely, within a specific discipline, by a set of variables that are used to describe systems or to partition objects into parts, which together give a systematic account of a domain of phenomena.
\item[-] A perspective, therefore, in its functionality to act as a probe of an object, should be {\it structurally adaptable} to the species of the object of inquiry [6]. The condition of structural adaptability of a potential perspective is satisfied if and only if it encodes a structurally invariant context of resolution, grouping together all information that can be delineated from the investigated object in terms of the probe's resolving power.
\item[-] The proposed conception of perspective is of an interactive nature and, therefore, should also be differentiated from the common pictorial understanding of a perspective as a visual metaphor involving a viewing projection, which, by itself, depicts the process of knowing as a passive activity.
\end{itemize}

Due to the aforementioned characteristics of the notion of a perspective, our approach to scientific perspectivism may naturally be called ``the endo-theoretic approach".
Such a kind of perspective provides observational accessibility on the investigated system, resolves a targeted object of inquiry, functions locally as a probing frame for the individuation of events, in general, probes consistently the natural world, thus serving as a ``window'' on physical reality.
The indicated conception of perspective, broadly defined at this stage, is rigorously formulated in Sect. 3 in the case of Hilbert space quantum mechanics.

\section{The Architecture of Scientific Perspectivism According to the Norms of the Endo-Theoretic Approach}
Since our objective is to actively engage scientific perspectivism with the science of today, as a means of analyzing the origin, generation and systematization of knowledge, we initially outline the conceptual-philosophical framework, secondly, its methodological component and, thirdly, the semantic stance of contemporary perspectivism according to the norms of the proposed endo-theoretic approach.

\subsection{The Conceptual-Philosophical Framework}
\begin{itemize}
\item[-] Scientific perspectivism in our approach amply recognizes the existence of an autonomous mind-independent world as being logically prior to experience and knowledge, constituting the overarching condition for the possibility of knowledge.
\item[-] Scientific knowledge of the world, however, can never be pure, direct, or unmediated since it requires pre-conceptualization or structural organization; it requires the adoption of a perspective. Scientific knowledge independent of all perspectives is beyond the human condition; it is unattainable.
\item[-] For perspectivism, the separation between the knowing subject and the object to be known, the partition between the observer and the observed, required for an objective description of phenomena, is neither absolute nor catholic, as Cartesian-like epistemological approaches advocate, thus promoting an allegedly context-free account of the world. The subject-object partition is accomplished upon the condition of the adopted perspective.
\item[-] The choice to adopt a particular perspective also signifies the approval of a conceptual scheme on the basis of which one may isolate, which of the many available properties do, and which do not count for the purposes of description, since the world does not come with one preferred system of description.
\item[-] Consequently, scientific observation may be regarded as perspectival in the sense that claims about what is observed cannot be detached, in all circumstances, from the context of observation. The significance of this point is particularly pertinent to quantum theory due to the existence of incompatible quantities pertaining to any non-trivial quantum system. Experimental or measurement contexts of such quantities are mutually exclusive in quantum mechanics; they cannot be held simultaneously. Thus, each mode of observation of incompatible quantum mechanical quantities gives rise in an ineliminable way to a particular kind of representation or description of the system. Accordingly, the representation of all aspects, at the same time, of a quantum entity is in principle unreachable. The essentially probabilistic structure of the theory and the subsequent phenomenon of quantum contextuality pose a fundamental limit on objectification (e.g. [7]).
\item[-] In this respect, a context-free and interpretation-free access to reality as such seems an illusion. The `book of nature' proves too subtle and complex to be determined by just reading off reality.
    The philosophical conception of perspectivism necessitates, therefore, a critique of traditional realism and, thus, requires a shift towards a new kind of scientific realism that is suitable to today's actual theorizing and practice.
\item[-] The traditional doctrine of scientific realism adopts a view of the world as constituted by autonomous, self-contained physical entities that exist wholly externally---that is, independently of the synthesizing power of the mind, the application of theoretical and conceptual frameworks, the consideration of contextual factors, the performance of experimental interventions or measurement procedures, our epistemic capacities, and so on (e.g. [8]). Subsequently, the metaphysical stance of traditional realism concerns a world that is pre-determined, pre-structured---a world expected to be reflected in physical theories and represented in the ideal limit as the `way it truly is'. The primary implicit assumption pertaining to this view is a presumed absolute kinematic (pre-dynamic) independence between the knowing subject and the object of knowledge (the world), or, at a local level of physical analysis, between the measuring system (as an extension of the knowing subject) and the system under measurement [9]. Such an idealized assumption concerning the kinematically independent behavior of a physical system is possible in classical physics due to the Cartesian product structure of phase space, namely, the state space of classical theories, and the absence of genuine indeterminism in the course of events or of an element of authentic chance in the measurement process. This reflects in a rather formal manner Pauli's portrayal of the conceptual description of nature in classical physics as ``the ideal of the detached observer". As Pauli [10, p. 43] himself put it: ``the observer has according to this ideal to disappear entirely in a discrete manner as hidden spectator, never as actor, nature being left alone in a predetermined course of events, independent of the way in which the phenomena are observed". It becomes then apparent that the traditional realist conception of the ``kinematically independent or detached observer", designates a worldview that has been left behind once and for all in the territory of contemporary physics, especially in view of the ubiquitous phenomenon of quantum contextuality. As established in Sect. 3, various no-go results in the foundations of quantum mechanics, predominantly the Kochen-Specker theorem, show formally that such a detached spectator of worldly states of affairs cannot possibly exist without contradiction.
\item[-] In contradistinction to the traditional doctrine of scientific realism, perspectival realism in our approach is distinguished by the premise that the physical reality is not considered as a ready-made existence, a predetermined truth governed by an external, universally fixed reference point. Instead, it is regarded as a dynamic existence that requires the establishment of a particular perspective for its investigation.
    Perspectival realism forms a novel philosophical framework in which metaphysical issues of science are contextualized within scientific theories and involved epistemic processes, underlying the intertwinement between ontic and epistemic aspects, kept completely apart in traditional thinking. The focal point of investigation lies now upon the world-theory relationship in accordance with scientific principles and practices, rather than contemplating the world in isolation.
\item[-] Hence, scientific perspectivism opposes a traditional substantivalist conception of nature, which is founded exclusively on the category of the object as a stand-alone, self-contained entity---possessing intrinsic properties, regardless of its environment or the context of conditions in which may be embedded---thus being unapt to contemporary scientific thought, especially in relation to quantum mechanics.
\item[-] Accordingly, the simplified assumption that knowledge of an object is achieved by forming a representation of that object as an immutable substance is rejected and, subsequently, replaced by the contemporary realist view of formulating local or partial contextual theoretical structures enabling different (even mutually incompatible, as in the quantum case) physical descriptions, grounded on the same actually existing object.
\item[-] The aforementioned considerations call further into question the traditional realist appeal that ``the world consists of some fixed totality of mind-independent objects", giving rise eventually to the thesis that ``there is exactly one true and complete description of `the way the world is'" ([11, p. 49]). In contradistinction to such a God's eye view of the world, there is simply no global Archimedean standpoint on the basis of which one may justify the traditional realist's belief that reality is definite and determinate, once and for all.
    Nature, especially in its fundamental microphysical dimension, is not a passive, fixed, unchangeable entity, waiting to be copied in our scientific representations [9]. As a generic example, quantum phenomena are not particularly stable across series of measurement of non-commuting incompatible quantities in order to be treated as direct reflections of predetermined intrinsic properties. Quantum mechanics refers to a value-indefinite reality; it is impossible, in principle, to assign simultaneously determined values to all the properties of a system.
    Be sure, no unique perspective can capture the entirety of a quantum object, let alone nature in its totality (Sect. 3).
    Consequently, a supposed absolute conception of the world cannot even be considered as an ideal goal, since it is not attainable.
\item[-] In our view, therefore, scientific perspectivism provides a promising middle path at the philosophical spectrum that keeps a firm grasp on reality, while rejecting the fallacious assumption that the human mind mirrors nature or that access to the world, particularly for scientific purposes, is possible independently of any prior conceptualization. Perspectival realism is the kind of realistic realism that science can provide.
    It gives rise to a comprehensive understanding of scientific knowledge as being genuinely informative about reality;
    its foundation is related, not to some unworkable obsolete metaphysical dictum, but to the process of obtaining knowledge about the world.
\end{itemize}

\subsection{The Methodological Component}
In consonance with the preceding conceptual scheme, the proposed methodological framework of scientific perspectivism in the light of the endo-theoretic approach consists of the following broad desiderata, to be further specified when applied to particular scientific domains of physical systems, concrete theoretical models and actual scientific practices.
\begin{itemize}
\item[-] It is instructive to note initially that the suggested concept of ``perspective'' in Sect. 1 constitutes a fundamental notion of ``point of view" or of ``probing frame", one that grounds facts about the world, and serves, due to its local variability, as the basis for the different ways the world could be.
\item[-] Henceforth, measurement values of physical quantities can, in general, be regarded as pertaining to an object under investigation within a local probing frame or perspective from which the object-system is considered. For instance, in quantum theory, the dependence of values of physical quantities upon the experimental or measurement context originates in Kochen-Specker's [12] celebrated theorem, leading to the contextual character of fundamental microphysical reality. As argued in Sect. 6, in the quantum domain, the consideration of perspective is of an ontic nature; it concerns the identity and definiteness of being at a given time.
\item[-] In the whole realm of physics, be it classical or quantum, measurement outcomes are registered as events within a macroscopic environment. The acknowledgement of the definite character of an event as something that actually occurs or not presupposes a suitable structure of alternative distinct possibilities corresponding to a Boolean algebra, characterized by Boole [13] as capturing the ``conditions of possible experience". A perspective, therefore, conceptualized as the principal structural unit of probing the natural world, and with respect to which measurement results are coordinatized, is ipso facto of a Boolean character.
\item[-] In relation to the empirical investigation of nature, theoretical and methodological considerations of contemporary physical theories specify the perspective from which we articulate the elementary yes-no experimental propositions or questions associated with properties of physical systems, in the sense that, on the one hand, they supply with a well-defined meaning the question that is put to nature, and, on the other hand, they specify the kind of the operations to be performed in order to ascertain particular answers to them.\renewcommand{\baselinestretch}{1}\footnote{The question-answer dialectic scheme in investigating reality is reminiscent of Kant's [14, p. 109] reading that the experimental method stands to nature ``not like a pupil, who has recited to him whatever the teacher wants to say, but like an appointed judge who compels witnesses to answer the questions he puts to them. [...] This is how natural science was first brought to the secure course of a science after groping about for so many centuries".}\renewcommand{\baselinestretch}{1.15} An experimenter, for instance, may choose to pass a beam of photons through an analyzer that authorizes propositions about their circular polarization (clockwise or counterclockwise) or through an analyzer that allows propositions about their plane polarization (vertical or horizontal). According to the formal structure of quantum theory, these two choices materially exclude each other. 
    It is legitimate, therefore, to say that the perspectival nature of experimental/empirical knowledge is an essential characteristic of acquiring scientific knowledge.
\item[-] The perspectivist representation of an object-system implies, in general, that a single perspective provides a partial/local and, thus, an incomplete description of the system to which it applies.
\item[-] Yet, the systematization of knowledge requires that perspectives associated with all aspects of a system can be correlated forming a synthesized unity, but they cannot be simply combined as independent integral parts of a third perspective. Hence, a perspective of all perspectives or, equivalently, a panoptical perspective from nowhere does not exist.
\item[-] It is crucial, however, that a full-fledged analysis of a successful framework of perspectivism in science ought to provide a syntax of perspectives, illustrating how locally shared perspectives can (or cannot) be meaningfully combined at a higher theoretical level.
\item[-] Consequently, nature can be grasped scientifically, through structured multitudes of local variable perspectives, forming a coherent multilevel theoretical structure, exemplified by experimental procedures that render possible specific access to specific aspects of physical reality.
\end{itemize}

Precisely this demanding task is accomplished by the category-theoretic perspectivist approach to quantum mechanics of Karakostas and Zafiris [6], outlined in Sect. 5, thus establishing a perspectival aspect to microphysical reality.
On this account, it is explicitly shown that the global structure of a quantum algebra of events can be consistently comprehended through a multilevel structure of locally variable Boolean perspectives, interconnected in a category-theoretic environment, yielding jointly all the information encoded in the former.

\subsection{The Semantic Stance}
The endo-theoretic approach to scientific perspectivism effectively localizes and subsequently contextualizes truth valuation with respect to distinct Boolean probing frames or perspectives targeting the physical system of interest, thus, inducing a perspectivist/contex\-tual account of truth that bears the following broad characteristics.
\begin{itemize}
\item[-] Perspectivist truth conforms to perspective or context-bound correspondence of a de re nature, designating an objectively existing state of affairs. Thus, it essentially opposes a non-perspectival, metaphysically fixed point of reference or a panoptical standpoint from which to logically evaluate all facts of nature.
\item[-] In the quantum domain of discourse, such an account derives by virtue of the microphysical nature of physical reality in displaying a context-dependence of facts. Furthermore, due to the inherently probabilistic structure of quantum theory and the resulting semantic indeterminacy, the specification of the perspective or the local context of reference provides the necessary conditions whereby bivalent assignment of truth values to quantum mechanical propositions is in principle applicable. In the quantum realm, perspectivist/contextual truth is an empirical fact.
\item[-] The proposed account of truth of perspectivist/contextual correspondence does not subscribe to relational truth. Given the specification of a perspective, the attribution of truth values to propositions does not depend upon the epistemic system of a knower's cognizance or her personal experiences. Agents sharing the same perspective agree on the determinateness and repeatability of empirical facts, established by measurement processes.
\item[-] Importantly, the proposed alethic scheme of perspectivist/contextual correspondence is applicable to both the propositional structure of fundamental theories of physics and also to natural ordinary discourse. In fact, it remedies intrinsic weaknesses of the traditional conception of correspondence truth by providing, in particular, a concrete explanation of the nature of the correspondence truthmaking relation and, thus, enlightens further the very notion of truth.
\item[-] On this account, the traditional conception of correspondence truth may be viewed as a limit case of the more generic alethic scheme of perspectivist/contextual correspondence, when the latter is applied in straightforward unproblematic circumstances where the non-explicit specification of a perspective or of a context of discourse poses no further implications.
\end{itemize}

The semantic stance of scientific perspectivism, addressing all preceding aspects, is developed in Sect. 6 in relation to truth-theoretic semantics in quantum mechanics.
In the following section we provide succinct argumentation in revealing the compelling association of quantum theory to perspectivist reasoning through the endemic feature of quantum contextuality.

\section{The Perspectivist/Contextual Nature of Quantum Mechanics}
In the standard Hilbert space formulation of quantum theory, each physical system is associated with a separable complex Hilbert space $\mathcal H$, the unit vectors of which correspond to possible states of the system.
Every physical quantity or observable $A$ pertaining to the system is represented by a self-adjoint operator $\hat{A}$ acting on $\mathcal H$, the spectrum of which is the set of possible values of $A$.
In marked contrast to classical mechanics, quantum mechanical observables are, in general, non-commuting under the pointwise multiplication of operators.
From a physical point of view, the non-commutativity condition means that incompatible operators are not co-measurable and thus their corresponding potential eigenvalues cannot qualify as measurement values prior to and independently of the actual measurement process.

Within this framework, quantum events or elementary propositions---that is, true/false questions concerning values of observables pertaining to a quantum system---are represented by orthogonal projection operators $\{\hat {P}_i\}$ on the system's Hilbert space $\mathcal H$ or, equivalently, by the closed linear subspace ${\mathcal H}_{\hat {P}_i}$ of $\mathcal H$ upon which the projection operator $\hat {P}_i$ projects.
The isomorphism between the set of all closed linear subspaces of $\mathcal H$ and the set of all projection operators, denoted by ${L}_{\mathcal H}$, allows us to translate the lattice structure of the subspaces of Hilbert space into the algebra of projections with the appropriate lattice theoretic characterizations [15].
Then, a quantum algebra of events is identified with the algebraic structure of all projection operators on Hilbert space, ordered by inclusion and carrying an orthocomplementation operation, thus forming a complete, atomic, orthomodular lattice. In effect, a non-classical, non-Boolean logical structure is induced which has its origin in quantum theory.

An efficient way for revealing the association of the perspectivist/contextual reasoning to quantum mechanics is provided by Kochen-Specker's celebrated theorem and its recent ramifications (e.g. [16]-[18]).
According to the Kochen and Specker result, for any quantum system associated to a Hilbert space of dimension greater than two, there does not exist a two-valued homomorphism, or, equivalently, a truth-functional assignment $h: {L}_{\mathcal H} \rightarrow \{0, 1\}$ on the set of projection operators, ${L}_{\mathcal H}$, interpretable as quantum mechanical propositions, preserving the lattice operations and the orthocomplement, even if these lattice operations are carried out among commuting elements only.
The essence of the theorem, when interpreted semantically, asserts the impossibility of assigning definite truth values to all propositions pertaining to a physical system at any one time, for any of its quantum states, without generating a contradiction.
In view of Gleason's [19] fundamental theorem, Kochen-Specker's theorem reaffirms the fact that on a Hilbert space of dimension greater than two, no two-valued probability measure exists globally and, thus, no representation of probability measures as a convex sum of \{0, 1\}-valuations is possible.

From  a physical point of view, the Kochen-Specker result shows that in a system represented by a Hilbert space of dimension $d>2$, there exist projection operators $\{\hat {P}_i\}$ such that it is not always possible to assign truth values 0 and 1 to all corresponding propositions pertaining to the system, so that the following conditions are fulfilled:
\begin{enumerate}
\item[(i)] For any orthogonal $i-$tuple of projection operators, $\{\hat {P}_i\}$, the assignment satisfies $\sum_{i}h(\hat {P}_{i}) = h(\hat {\mathbb I}) = 1$, that is, one projection operator is mapped onto 1 (``true") and the remaining $i-1$ projection operators are mapped onto 0 (``false") (completeness of the basis condition).
\item[(ii)] If a projection operator, $\hat {P}_k$, belongs to multiple complete orthogonal bases, then, it is consistently assigned the same value in all bases (non-contextuality condition).
\end{enumerate}

The initial proof of Kochen and Specker establishes that no such assignment of $\{0, 1\}$-valuation is possible for a special case restricted to a finite sublattice of projection operators on a three-dimensional Hilbert space, associated to a spin-1 quantum system, in a way that preserves the non-contextuality condition.
In their original proof, Kochen and Specker used a three-dimensional real Hilbert space and a set of 117 rays (rank-1 projectors) for which the preceding two conditions are shown to be contradictory.
Hence, the established contradiction entails the existence of a contradiction in higher dimensional complex Hilbert spaces.
Recent developments have remarkably reduced the size and complexity of proofs of the Kochen-Specker theorem [20] and transformed such proofs into experimentally testable inequalities, providing also loophole-free tests of quantum contextuality [21], thus establishing the contextual character of the theory as a structural feature of the quantum mechanical formalism itself.
Furthermore, quantum contextuality has been established at the inception of twenty-first century as a fundamental resource in quantum information processing, ranging from measurement-based quantum computation to communication complexity scenarios in an endeavor to significantly surpass classical computing (e.g. [22]).

A failure of the non-contextuality condition means that the value assigned to a quantum mechanical observable $A$ depends on the context in which it is considered.
An equivalent way of expressing the above is to say that the value of $A$ depends on what other compatible observables are assigned values at the same time; i.e., it depends on a choice that concerns operators that commute with $\hat A$.
That is, the value of the observable $A$ cannot be regarded as pre-fixed, as being independent of the measurement context actually chosen, as specified by the set of mutually compatible observables one may consider it with.
This dependence captures the endemic feature of quantum contextuality.
As opposed to the classical case, in quantum mechanics there is no a global context of measurement, a single privileged perspective, relative to which all conceivable observables of the system are co-measurable.
One and the same quantum system does exhibit several possible contextual manifestations in the sense that it can be assigned several definite incommensurable properties with respect to distinct incompatible quantum observables corresponding to different aspects of reality which, in principle, cannot be considered simultaneously.

Henceforth, the notion of context giving rise to a perspective that is applied on a quantum system is formalized by a maximal set of mutually compatible observables, or, equivalently, by a complete Boolean algebra of commuting projection operators generated by this set.
It should be underlined that such a complete Boolean algebra of projection operators represents the maximal information that is encodable into a quantum system at any given time by a state preparation procedure (e.g. [23]).
It also bears the status of a local structural invariant characterizing a whole commutative algebra of observables that can be simultaneously spectrally resolved and hence be co-measurable.
Of course, such a local structural invariant is not uniquely defined in the lattice of quantum events since it depends on the chosen set of mutually compatible observables, allowing thereby local variability in targeting any particular aspect of interest of the quantum system considered.
It satisfies, therefore, the general desiderata put forward in Sect. 1 for a structural unit to be qualified as a suitable perspective.

It is probably one of the deepest insights of modern quantum theory that although a quantum event structure is globally non-Boolean, it can be qualified spectrally, and hence be accessed experimentally, only in terms of Boolean event structures operating locally as structural invariants of co-measurable families of physical observables. Consequently, a complete Boolean algebra of projection operators in the lattice of quantum events picked by an observable to be measured functions locally as a {\it Boolean probing frame} or {\it perspective} relative to which results of measurement are being coordinatized. 
Due to the existence of incompatible observables and the subsequent absence of a globally defined Boolean frame over a quantum event structure, it is necessary to consider all possible local ones together with their interrelations.
This naturally leads, by extension, to a horizon of perspectives on the structure of quantum events with respect to a multiplicity of variable overlapping Boolean frames, realized as experimental contexts for measuring physical quantities.  
Of course, no single context or perspective is adequate for delivering a complete picture of the quantum system. However, within a category-theoretical framework, it is possible to combine them suitably in an overall multilevel structure that will capture the entire system. 
It has been recently shown, by applying category-theoretic reasoning to quantum mechanics, that the global structure of a quantum algebra of events can be represented in terms of structured interconnected families of Boolean probing frames, realized as locally variable perspectives on a quantum system, being capable of carrying jointly all the information encoded in the former (Sect. 5).
To this end, we resort to the powerful methods of category theory, which directly captures the idea of structures varying over contexts, thus providing a natural setting for investigating multilevel structures and studying perspectivist/contextual phenomena.\renewcommand{\baselinestretch}{1}\footnote{Among the various interpretations of the quantum formalism, perspectival aspects have been associated with Rovelli's relational account of quantum mechanics [24, 25], Bene and Dieks' [26] perspectival version of the modal interpretation, French's [27] phenomenological reading of the quantum measurement process in the spirit of London and Bauer, and the agent-centered approach of QBism [28], its relation to pragmatist views [29], and its possible recasting within a phenomenological context [30]. 
The distinguishing feature of our study is that it acknowledges the strong affinity between scientific perspectivism as analyzed in Sect. 2 and structural features of Hilbert space quantum mechanics, being modeled and extended within a category-theoretical framework, by linking consistently the variable and local Boolean with the global quantum level in perspectivist terms (Sect. 5.3). In the appendix we provide a concise comparison between our proposed endo-theoretic perspectivist framework of quantum mechanics and Rovelli's relational proposal, especially in its recent development 
involving the postulate of ``cross-perspective links".}\renewcommand{\baselinestretch}{1.15}

\section{Category Theory in a Nutshell}
Category theory provides a general theoretical framework for the study of structured systems in terms of their mutual relations and admissible transformations. Contrary to the atomistic approach of set theory, which crucially depends on the concept of elements-points and the membership relationship of a variable $x$ in a set $X$, $x \in X$, in category theory the notion of {\it morphism} or {\it arrow} undertakes primary role.
A morphism, for instance, $f: A \rightarrow B$ in a category $\mathcal C$ expresses one of the many possible ways in which the object $A$ relates to the object $B$ within the context of category $\mathcal C$. Thus, an incoming morphism to $B$ from any other object $A$ in category $\mathcal C$ may be considered as a {\it perspective} targeting $B$ whose source is $A$ in the same category.
The category theoretic mode of thinking incorporates internally the very nature of ``pointing at" or ``viewing from".
In this vein, the notion of structure does not exclusively refer to a fixed universe of sets of pre-determined elements, but acquires a variable reference [31].

However counterintuitive it may initially appear, in category theory the nature of the objects is a derivative aspect of the patterns described by the morphisms or mappings that connect the objects. In fact, an object can be completely classified and uniquely derived up to an isomorphism by the network of all morphisms---the structure preserving relations---targeting this object within the same category. Most significantly, this does not exclude the inter-level relational determination of objects belonging to other categorical species of structure, under the condition that there exists a bi-directional functorial correlation between them, formulated in the language of {\it adjunctions}. It is precisely the latter development that gradually introduced into category theory a paradigm change in understanding structures of general types and paved the way for forming bridges between seemingly unrelated mathematical disciplines.
As summarised by Goldblatt [32, p. 438]: ``The isolation and explication of the notion of adjointness [or adjunction] is perhaps the most profound contribution that category theory has made to the history of general mathematical ideas''.

The basic categorical principles that we adopt in the subsequent analysis are summarized as follows:
\begin{enumerate}
\item[(i)] To each kind of mathematical structure used to represent a system, there corresponds a {\it category} whose objects have that structure and whose morphisms or arrows preserve it.
\item[(ii)] To any canonical construction on structures of one kind, yielding structures of another kind, there corresponds a {\it functor} from the category of the first specified kind to the category of the second. The implementation of this principle is associated with the fact that a categorical functorial construction is not merely a function from objects of one kind to objects of another kind, but must preserve the essential structural relationships among objects, that is, identity morphisms and composition of morphisms.
\item[(iii)] To each natural translation between two functors having identical domains and co-domains, there corresponds a {\it natural transformation}. The specification ``natural", particularly, in the notion of natural transformations refers to the comparison of two functorial processes, sharing the same source and target categories, in a way that captures the {\it shared structure} or {\it generic common properties} existing in different categorical contexts. The key concept of natural transformations acquires in category theory the status of a principle, analogous to general covariance in physics, that penetrates deeper than is initially discernible.
\item[(iv)] To any natural {\it bi-directional functorial correlation} between two kinds of mathematical structures, there corresponds an {\it adjunction} consisting of a pair of {\it adjoint functors} between the corresponding categories. This means that the objects of the involved categories are related to one another through natural transformations. Then, the pair of adjoint functors constitutes a categorical adjunction. As already pointed out, the latter concept is of fundamental logical and mathematical importance contributed to mathematics by category theory.\renewcommand{\baselinestretch}{1}\footnote{A concise account of the underpinnings of category theory, including an analysis of the preceding notions, can be found in Sect. 3.1 of [6].}
    \renewcommand{\baselinestretch}{1.15}
\end{enumerate}

\section{Perspectivist Representation of Quantum Event Structures via Boolean Probing Frames}
\subsection{Methodological Considerations}
The conceptual basis of the proposed perspectivist representation of a quantum structure of events $L$ in terms of interconnected families of Boolean probing frames, realized as suitable perspectives on $L$, relies on the physically significant fact that it is possible to analyze or `coordinatize' the information contained in a quantum event algebra by means of structure preserving morphisms $B \rightarrow L$, having as their domains locally defined Boolean event algebras $B$. As indicated in Sect. 3, any single map from a Boolean domain to a quantum event algebra does not suffice for a complete determination of the latter's information content.
In fact, it contains only the amount of information related to a particular Boolean frame, prepared for a specific kind of measurement.
This problem is confronted, therefore, by employing a sufficient number of maps, organized in terms of covering sieves, from the coordinatizing Boolean domains to a quantum event algebra so as to cover it completely. These maps furnish the role of local Boolean covers for the filtration of the information associated with a quantum structure of events, in that, their domains $B$ provide Boolean coefficients associated with typical measurement situations of quantum observables.
The local Boolean covers capture, in essence, separate complementary features of the quantum system under investigation, thus providing a structural decomposition of a quantum event algebra in the descriptive terms of Boolean probing frames.
In turn, the incomplete and complementary local Boolean descriptions, arising from a multiplicity of locally variable perspectives, can be smoothly pasted or glued together, by demanding the satisfaction of partial compatibility between overlapping local Boolean covers, so that one may arrive at the synthesis and actual determination of the global quantum event algebra itself.
The methodology involved in the realization of the suggested approach necessitates the application of categorical sheaf theory to quantum structures.\renewcommand{\baselinestretch}{1}\footnote{For a systematic introduction to category-theoretic sheaf theory, the reader may consult [33-35].}
\renewcommand{\baselinestretch}{1.15}

\subsection{Category-Theoretic Modeling of Quantum Event Structures}
As a natural initial step we note that the perspectivist representation of a quantum event algebra in terms of structured multitudes of interconnected Boolean probing frames requires distinct notions of Boolean and quantum categorical event structures, respectively.

A {\it Boolean categorical event structure} is a small category, denoted by
$\mathcal B$, which is called the category of Boolean event
algebras. The objects of $\mathcal B$ are complete Boolean
algebras of events and the morphisms are the corresponding Boolean
algebraic homomorphisms.

A {\it quantum categorical event structure} is a locally small
co-complete category, denoted by $\mathcal L$, which is called the
category of quantum event algebras. The objects of $\mathcal L$
are quantum event algebras and the morphisms are quantum algebraic
homomorphisms.

A quantum event algebra $L$ in $\mathcal L$ is defined as an orthomodular $\sigma$-orthoposet [35]; that is, as a partially ordered set of quantum events, endowed with a maximal element 1 and with an operation of orthocomplementation $[-]^{\ast} : L \rightarrow L$, which satisfy, for all $l \in L$, the following conditions:
[a] $l \leq 1$, [b] $l^{\ast \ast}=l$, [c] $l \vee l^{\ast}=1$, [d] $l
\leq {l}^{\prime} \Rightarrow {{{l}^{\prime}}^{\ast}}   \leq l^{\ast}$,
[e] $ l \bot {l}^{\prime} \Rightarrow l \vee {l}^{\prime} \in L$, [f]
for $l, {l}^{\prime} \in L, \; l \leq {l}^{\prime}$ implies that $l$ and
${l}^{\prime}$ are compatible, where $0:=1^{\ast}$, $l \bot {l}^{\prime} := l \leq {{l}^{\prime}}^{\ast}$, and the operations of meet
$\wedge$ and join $\vee$ represent, respectively, the intersection and linear span of closed subspaces.

Henceforth, a {\it quantum algebraic homomorphism} is a morphism $H: K \rightarrow L$, which satisfies, for all $k \in K$, the following conditions:
[a] $H(1)=1$, [b] $H(k^{\ast})={[H(k)]}^{\ast}$, [c] $k \leq {k}^{\prime} \Rightarrow
H(k) \leq H({k}^{\prime})$, [d] $k \bot {k}^{\prime} \Rightarrow H(k
\vee {k}^{\prime}) \leq H(k) \vee H({k^{\prime}})$, [e] $H({\bigvee}_n
k_n)={\bigvee}_n H(k_n)$, where $k_1, k_2, \ldots$ countable family of mutually orthogonal events.

It is worthy to underline that any arbitrary pair of events $l$ and ${l}^{\prime}$,  belonging to a quantum event algebra $L$ in $\mathcal L$ are compatible, if the sublattice generated by
$\{{l,l^{\ast},{l}^{\prime},{{{l}^{\prime}}^{\ast}}}\}$ is a Boolean algebra, namely, if it is a Boolean sublattice of $L$.
In perspectivist terms, this implies that a Boolean event algebra is {\it structurally adaptable} to a quantum event algebra, since it encodes a structurally invariant context of co-measurable observables by means of their joint compatible spectral resolution.
Furthermore, as indicated in Sect. 3, the global non-Boolean structure of a quantum event algebra is rendered empirically inert without the adjunction of such local spectral invariants to it.
The role of these invariants is to induce partial or local structural congruences with Boolean event structures pertaining to all typical contexts of measurement, depending on the qualification of the corresponding spectral projection operators.

Accordingly, the objective is to derive the non-directly accessible quantum kind of event structure in terms of all possible partial structural congruences with the directly accessible Boolean kind of event structure, via the adjunction of local spectral invariants as probing frames to the former.
In this setting, it is of great importance the consideration of all possible structural relations allowed among the probing Boolean frames, the spectra of which may be disjoint or nested or overlapping and interlocking together non-trivially.
The solution to the posed problem requires a category-theoretic interpretative framework in functorial terms, that is, non-dependent on the artificial choice of particular Boolean frames adjoined to a quantum event algebra.

It is essential, therefore, to consider a Boolean categorical event structure $\mathcal B$ as a category of probes for the quantum categorical event structure $\mathcal L$. The structural adaptability of the category of Boolean event algebras $\mathcal B$ to the category of quantum event algebras $\mathcal L$ gives rise to the Boolean probing or shaping functor of $\mathcal L$ by $\mathcal B$.

A {\it Boolean probing or shaping functor} of a quantum categorical event structure ${\mathcal L}$,  ${\mathbb M}:{\mathcal B} \rightarrow {\mathcal L}$, assigns to each Boolean event algebra in ${\mathcal B}$ the underlying quantum event algebra from ${\mathcal L}$, and to each Boolean homomorphism the underlying quantum algebraic homomorphism.
The shaping functor  ${\mathbb M}:{\mathcal B} \rightarrow {\mathcal L}$ is technically a forgetful functor, not taking into account the extra Boolean structure of $\mathcal B$.
We note, in this respect, that any Boolean event algebra may be thought of as a quantum event algebra of commuting projection operators.
Thus, the functor $\mathbb M$ qualifies functorially the requirement of structural adaptability of the category ${\mathcal B}$ of Boolean probes to the category of quantum event algebras $\mathcal L$.

Then, the quantum algebraic homomorphism, $\psi_B: {\mathbb M} (B) \rightarrow L$, constitutes a Boolean frame of a quantum event algebra $L$, or, equivalently, a Boolean perspective on $L$ whose source is the Boolean probe $B$.
The foregoing proposition is a consequence of the existence of the shaping functor $\mathbb M$ and the consideration of essential features of the quantum mechanical formalism, for instance, the factor that a normalized unit vector in Hilbert space, representing the physical state of a quantum system, is completely analysed and thus specified by the determination of a basis on the state Hilbert space of the system.
Let us note that the consideration of a basis on a system's Hilbert space---be it an orthonormal basis of eigenvectors of a selected observable to be measured, or, more generally, an orthonormal basis of common eigenvectors of a set of mutually compatible observables---always gives rise to a Boolean event algebra $B$, viewed as a special quantum event algebra ${\mathbb M} (B)$ within the quantum categorical event structure $\mathcal L$.
The special structures of the type ${\mathbb M} (B)$ function as variable local Boolean frames of a quantum event algebra $L$ in $\mathcal L$, under their intended physical interpretation as probing frames or perspectives on $L$ for the manifestation and subsequent classification (or contextualization) of quantum events.

We underline the fact that within the considered categorical scheme the variation of the local Boolean probing frames in the category $\mathcal B$ of probes, for each probe $B$ in $\mathcal B$, is actually arising from any experimental praxis aiming to fix or prepare the state of a quantum system and corresponds, in this sense, to the variation of all possible Boolean preparatory contexts pertaining to the system for extracting information about it.
In this respect, Boolean probing frames or instances of concrete experimental arrangements in quantum mechanics play a role analogous to the reference frames of rods and clocks in relativity theory in establishing a perspectival aspect to microphysical reality.

Observe now that the Boolean probing functor of a quantum categorical event structure $\mathcal L$,  ${\mathbb M}:{\mathcal B} \rightarrow {\mathcal L}$, is not invertible. That is, there is no
opposite-directing functor from ${\mathcal L}$ to ${\mathcal B}$, since a quantum event algebra cannot be realized within any Boolean event algebra. Thus, we seek for an extension of ${\mathcal B}$ into a larger categorical environment where such a realization becomes possible.
This extension should conform with the intended perspectivist semantics of adjoining a multiplicity of Boolean probing frames to a quantum event algebra, understood equivalently as Boolean perspectives on the latter.
For this reason, it is necessary to extend the categorical level of $\mathcal B$ to the categorical level of diagrams in $\mathcal B$, such that the global information encoded in a quantum event algebra is expected to be recovered in a structure preserving way by an appropriate sheaf-theoretic construction gluing together categorical diagrams of locally variable
Boolean frames.\renewcommand{\baselinestretch}{1}\footnote{In general, a diagram
$\mathbb X=(\{X_i \}_{i \in I}, \{F_{ij}\}_{i,j \in I})$ in a category $\mathcal C$ is defined as an indexed family of objects $\{X_i \}_{i \in I}$ and a family of
morphisms sets $\{F_{ij}\}_{i,j \in I} \subseteq {Hom_{\mathcal C}}({X_i},{X_j})$.}
\renewcommand{\baselinestretch}{1.15}Technically, this is accomplished by means of the categorical method of Yoneda's embedding ${\mathbb y} : \mathcal B \rightarrow {{\mathcal S}et^{{\mathcal B}^{op}}}$, which is a full and faithful functor, where ${\mathcal B}^{op}$ is the opposite category of ${\mathcal B}$ ([37, pp. 187-189]; [33, p. 26]).

It is apparent, therefore, that the realization of this extension process requires initially the construction of the functor category, ${{\mathcal S}et^{{\mathcal B}^{op}}}$, called the
{\it category of presheaves of sets on Boolean event algebras}.
The objects of ${{\mathcal S}et^{{\mathcal B}^{op}}}$ are all contravariant functors $ {\mathbb P}: {\mathcal B}^{op} \rightarrow {{\mathcal S}et}$ with morphisms all natural transformations between such functors.
Each object ${\mathbb P}$ in the category of presheaves ${{\mathcal S}et^{{\mathcal B}^{op}}}$ is a contravariant set-valued functor on ${\mathcal B}$,  called a {\it presheaf of sets} on ${\mathcal B}$ [34, p. 195].
The functor category of presheaves on Boolean event algebras, ${{\mathcal S}et^{{\mathcal B}^{op}}}$, provides the instantiation of a structure known as {\it topos}. A topos exemplifies a well-defined notion of a universe of variable sets. It can be conceived as a local mathematical framework corresponding to a generalized model of set theory or as a generalized topological space (e.g. [35]).

In order to obtain an unambiguous understanding of the structure of the functor category ${{\mathcal S}et^{{\mathcal B}^{op}}}$, it is useful to think of a presheaf of sets ${\mathbb P}$ in ${{\mathcal S}et^{{\mathcal B}^{op}}}$ as a right action of the category $\mathcal B$ on a set of observables, which is partitioned into a variety of Boolean spectral kinds parameterized  by the Boolean event algebras $B$ in $\mathcal B$.
Such an action $\mathbb P$ is equivalent to the specification of a {\it diagram} in $\mathcal B$, to be thought of as a $\mathcal B$-variable set forming a presheaf ${\mathbb P}(B)$ on ${\mathcal B}$.
For each Boolean algebra $B$ of ${\mathcal B}$,  ${\mathbb P}$(B) is
a set, and for each Boolean homomorphism $f: C \rightarrow B$, ${\mathbb P} (f) :
{\mathbb P}(B) \rightarrow   {\mathbb P}$(C) is a set-theoretic function, such that, if $p \in {\mathbb
P}$(B), the value ${\mathbb P}(f) (p)$ for an arrow $f : C \rightarrow B$
in ${\mathcal B}$ is called the restriction of $p$ along $f$ and
is denoted by ${\mathbb P}(f) (p)=p \cdot f$.
From a physical viewpoint, the purpose of introducing the notion of a presheaf ${\mathbb P}$ on $\mathcal B$, in the environment of the functor category ${{\mathcal S}et^{{\mathcal B}^{op}}}$, is to identify an element of ${\mathbb P}$(B), that is, $p \in {\mathbb P}$(B), with an event observed by means of a physical procedure over a Boolean domain cover for a quantum event algebra.

Henceforth, the {\it Boolean realization functor} of a quantum categorical event structure ${\mathcal L}$ in ${{\mathcal {S} et}^{{\mathcal B}^{op}}}$---namely, the functor of generalized elements of $\mathcal L$ in the environment of the category of presheaves on Boolean event algebras---is defined as follows:
\begin{equation}
{\mathbb {R}}: { {\mathcal L}} \rightarrow {{\mathcal {S} et}}^{{{\mathcal{B}}}^{op}} \\,
\end{equation}
where the action on a Boolean algebra $B$ in $\mathcal B$ is given by
\begin{equation}
{\mathbb {R}}(L)(B)={Hom_{\mathcal L}}({\mathbb M}(B),L) \\.
\end{equation}

The presheaf functor ${\mathbb {R}}(L)(-)={Hom_{\mathcal L}}({\mathbb M}(-),L)$
constitutes the image of ${\mathbb {R}}$ in ${{\mathcal {S} et}}^{{{\mathcal{B}}}^{op}}$ and is called the {\it functor of Boolean frames} or {\it functor of Boolean perspectives} on a quantum event algebra $L$ in ${\mathcal L}$.
Since the physical interpretation of the presheaf functor ${\mathbb {R}}(L)(-)$ refers to the realization of a quantum event algebra $L$ in $\mathcal L$ in terms of structured multitudes of local Boolean frames adjoined to it, intuitively, it is natural to comprehend ${\mathbb {R}}(L)(-)$ as comprising the network of relationships that $L$ has with all admissible Boolean frames-perspectives on $L$.

Specifically, the functor of Boolean frames ${\mathbb {R}}(L)$, being an object in the category of presheaves ${{\mathcal {S}et}}^{{{\mathcal{B}}}^{op}}$, represents a quantum event algebra $L$ in the environment of the topos of presheaves ${{\mathcal {S}et}}^{{{\mathcal{B}}}^{op}}$ over the base category of probes $\mathcal B$. Thus, for a fixed quantum event algebra $L$, the evaluation of the functor
${\mathbb {R}}(L)(-)$, at each Boolean event algebra $B$ in $\mathcal B$, instantiates locally a Boolean probing frame of $L$, or, equivalently, a ($B$-shaped) Boolean perspective on $L$, denoted by ${\psi_B}: {\mathbb M}(B) \rightarrow L$. Importantly, since ${\mathbb {R}}(L)(-)$ is a presheaf functor in ${{\mathcal {S} et}}^{{{\mathcal{B}}}^{op}}$, all resultant structured multitudes of local Boolean frames, adjoined to $L$, are interrelated by the functorial operation of presheaf restriction. In relation to the latter,
it is sufficient to observe that for each Boolean homomorphism $f : C \rightarrow B$, ${{\mathbb {R}}(L)} (f) :
{{\mathbb {R}}(L)}(B) \rightarrow   {{\mathbb {R}}(L)}(C)$ is a function between sets of Boolean frames of $L$ in the opposite direction, such that, if $\psi_B
\in {{\mathbb {R}}(L)}(B)$ is a Boolean frame of $L$, the value of ${{\mathbb {R}}(L)}(f) (\psi_B)$, or, equivalently, the corresponding Boolean frame $\psi_C: {\mathbb M}(C) \rightarrow L$ is given by the restriction or pullback of $\psi_B$ along $f$, denoted by ${{\mathbb {R}}(L)}(f) (\psi_B)=\psi_B \cdot f=\psi_C.$

It is evident that the semantics of the functor ${\mathbb R}(L)$ and, therefore, of any of its subfunctors ${\mathbb T}(L) \triangleright {\mathbb R}(L)$  amounts to the notion of a spectral sieve---comprised by an appropriately interconnected family of, generally, partially compatible local Boolean frames---adjoined to $L$ through processes of measurement of quantum observables.
In this connection, let us draw attention to the following characteristic assertion of Arthur Eddington [38, p. 1566], according to which, in ``Einstein's theory of relativity the observer is a man who sets out in quest of truth armed with a measuring rod. In quantum theory he sets out armed with a sieve".
A spectral sieve adjoined to a quantum event algebra $L$ can be intuitively conceived of as consisting of `filtering holes' ${\mathbb M}(B)$, specified structurally by variable Boolean probing frames targeting $L$, permitting separation of several resolution sizes of observable grain, as well as their compatibility relations.\renewcommand{\baselinestretch}{1}\footnote{Technically, a {\it sieve} on an object $A$ of a category ${\mathcal C}$ is a collection $S$ of morphisms with codomain $A$ in ${\mathcal C}$ such that,
if $f: B \rightarrow A$ belongs to $S$ and $g: C \rightarrow B$ is any morphism, then $f \circ g: C \rightarrow A$ belongs also to $S$.} \renewcommand{\baselinestretch}{1.15}
For instance, if $\psi_B: {\mathbb M}(B) \rightarrow L$ denotes a Boolean frame of $L$  belonging to a sieve ${\mathbb T}(L)$ adjoined to $L$, then any other Boolean frame of $L$ whose spectral resolving power is coarser than $\psi_B$ belongs to this sieve as well.
Thus, spectral sieves adjoined to $L$ effectuate a multilayered classification  of the informational content of $L$ via the `filtering holes' ${\mathbb M}(B)$, for each probe $B$ in $\mathcal B$. 
In turn, this amounts to the capacity of separating and coordinatizing consistently results of measurement of quantum mechanical observables solely on the basis of the perspectival functioning of such a spectral sieve.
Consequently, the physical significance of the adjunction of a spectral sieve on $L$ is that it induces partial or local structural congruence relations between the Boolean and quantum levels of event structure in functorial terms, that is, without the invocation of any {\em ad hoc} choices of Boolean subalgebras of a quantum event algebra.

\subsection{Interrelating the Local Boolean and the Global Quantum Structural Level in Perspectivist Terms}
The crucial conceptual feature of the proposed categorical setting concerns the fact that the foregoing bi-directional correlation between the local Boolean and the global quantum structural level admits a rigorous formulation in terms of a pair of adjoint functors, thus giving rise to a categorical adjunction.
On this account, the consideration of the left adjoint functor $\mathbb L :  {{\mathcal {S} et}^{{\mathcal B}^{op}}}  \to {\mathcal L}$ 
to the Boolean realization functor ${\mathbb {R}}: { {\mathcal L}} \rightarrow {{\mathcal {S} et}^{{{\mathcal B}}^{op}}}$ paves the way for an explicit perspectivist representation of quantum event algebras via the appropriate adjunction of locally variable, interconnected families of Boolean frames. 
That is, if for a particular 
quantum event algebra $L$ in $\mathcal L$, the right adjoint functor $\mathbb R$ partitions $L$ in an orderly manner via the action of Boolean probing frames $\psi_B: {\mathbb M}(B) \rightarrow L$, comprising for variable $B$ in $\mathcal B$ spectral sieves on $L$, and thus functioning as suitable perspectives or contexts for measurement of observables, then, inversely,
the left adjoint functor $\mathbb L$ provides a perspectival synthesis of a quantum event algebra, in a structure preserving manner, by gluing compatibly together structured families or diagrams of variable local Boolean frames.

Specifically, it has been recently proved that there exists a categorical adjunction between the category of presheaves ${{\mathcal {S} et}^{{{\mathcal B}}^{op}}}$ on Boolean event algebras and the category of quantum event algebras $\mathcal L$, called the Boolean frames-quantum adjunction, established by the pair of adjoint functors ${\mathbb L}$ and ${\mathbb R}$,
\begin{equation}
\mathbb L : {{\mathcal {S} et}^{{\mathcal B}^{op}}} \adjar {\mathcal L} : \mathbb R \\,
\end{equation}
where the right adjoint,
\begin{equation}
\mathbb R: { {\mathcal L}} \rightarrow {{\mathcal {S} et}^{{{\mathcal B}}^{op}}} \\,
\end{equation}
is the Boolean realization functor of a quantum categorical event structure ${\mathcal L}$ in ${{\mathcal {S} et}^{{\mathcal B}^{op}}}$, whereas, the left adjoint,
\begin{equation}
\mathbb L : {{\mathcal {S} et}^{{{\mathcal B}}^{op}}} \rightarrow \mathcal L \\,
\end{equation}
is the colimit-preserving functor providing the perspectival synthesis of a quantum categorical event structure by means of diagrams of Boolean frames.\renewcommand{\baselinestretch}{1}\footnote{The Boolean frames-quantum adjunction is proved in [39]; see also [18, 40] for a more detailed investigation including the involved logical and semantic aspects, and [6] in relation to the establishment of the perspectivist character of the stated adjunction. The latter categorical adjunction has also been applied to the complementarity concept in quantum mechanics, providing a suitable extended interpretation of the complementarity principle [41]. In view of the latter, complementarity is not only understood as a relation between mutually exclusive experimental arrangements, as envisaged by the original conception, but it is primarily comprehended as a reciprocal relation concerning information transfer between two hierarchically different structural kinds of event structure---the Boolean and quantum kinds of structure---that can be brought into local or partial structural congruence.} \renewcommand{\baselinestretch}{1.15}

The Boolean frames-quantum adjunction consists, therefore, of the pair of adjoint functors ${\mathbb L} \dashv {\mathbb R}$, called left and right adjoints, as well as the following bijection, 
which is natural in both $\mathbb P$ in ${{\mathcal {S} et}^{{\mathcal B}^{op}}}$ and $L$ in ${\mathcal L}$,
\begin{equation}
{Hom}_{{{\mathcal {S} et}^{{\mathcal B}^{op}}}} ({\mathbb P},{\mathbb R}(L)) \cong {{Hom}_{\mathcal
L}({\mathbb L \mathbb P}, L)} \\,
\end{equation}
abbreviated as
\begin{equation}
Nat({\mathbb P},{\mathbb R}(L)) \cong {Hom}_{\mathcal L}({\mathbb L \mathbb P}, L) \\.
\end{equation}
The established bijective correspondence assures that the Boolean realization functor of ${\mathcal L}$, realized for each quantum event algebra $L$ in ${\mathcal L}$ by its functor of Boolean probing frames or perspectives,
\begin{equation}
{\mathbb R}(L) : B \mapsto {{Hom}_{\mathcal L}({\mathbb M}(B), L)} \\,
\end{equation}
has a left adjoint functor $\mathbb L :  {{{\mathcal {S} et}}^{{\mathcal B}^{op}}}  \to {\mathcal L}$,
which is defined for each presheaf $\mathbb P$ of Boolean algebras in ${{{\mathcal {S} et}}^{{\mathcal B}^{op}}}$ as the colimit ${\mathbb L}({\mathbb P})$,
\begin{equation}
{\mathbb
L}({\mathbb P})= {\it Colim} \{ \bf{{\int}}({\mathbb P},{\mathcal
B}) \Ar {{\mathbf {\int}}_{\mathbb P}} {\mathcal B} \Ar {\mathbb
M} {\mathcal L} \} \;,
\end{equation}
where $\bf{{\int}}({\mathbb P},{\mathcal B})$ defines the category of elements of a presheaf ${\mathbb P}$.\renewcommand{\baselinestretch}{1}\footnote{For readers not familiar with the categorical construction of colimits we note that their existence expresses in category-theoretic language the basic intuition that a complex object may be conceived as arising from the interconnection of partially or locally defined informational units within a category. In a nutshell, colimits may be viewed as binding factors `gluing' parts together.} \renewcommand{\baselinestretch}{1.15} 

It is now important to stress that due to the existence of the Boolean frames--quantum adjunction, $\mathbb L :{{{\mathcal {S} et}}^{{\mathcal B}^{op}}} \adjar {\mathcal L} : \mathbb R$,
every probing relation from a Boolean event algebra $B$ in $\mathcal B$ to a quantum event algebra $L$ in $\mathcal L$ shaped by the functor $\mathbb M: \mathcal B \rightarrow \mathcal L$ ---or, equivalently, every Boolean frame-perspective on $L$---factors uniquely through the category of presheaves of sets ${{{\mathcal {S} et}}^{{\mathcal B}^{op}}}$, as revealed by the following commutative diagram:
\begin{diagram}
€{\mathcal B}€€ €\Smono {\mathbb y} €\Sear {\mathbb M}€€ €{\mathcal
Set}^{{\mathcal B}^{op}}€\Edotar {\mathbb L}  \wdotaR {\mathbb R}€{\mathcal L}€€
\end{diagram}
This encapsulates the fact that there exists an exact solution to the problem of specifying a quantum event algebra $L$ in perspectivist terms, by means of structured interconnected families of Boolean probing frames acting on it, which is provided by the left adjoint colimit functor $\mathbb L: {{{\mathcal {S} et}}^{{\mathcal B}^{op}}} \rightarrow \mathcal L$ of the Boolean frames-quantum adjunction.

It is instructive to point out in this respect that in the proposed perspectivist approach to quantum mechanics, a global quantum event structure is not contemplated as an already given entity, but it is constituted in a continuous process of extension from the local to the global level by actualization of new potential facts with respect to local Boolean frames. 
For, each quantum event actualized relative to a particular probing frame serves as a datum for subsequent potential actualizations, thus instantiating a bundle of potential relations referring to this frame. Importantly, all these potential relations---namely, all relations among observables at the local level---are captured by the internal relations among their underlined Boolean probing frames and are extended to the global quantum level through suitable colimit-gluing conditions of structured families of partially ordered Boolean frames.
The exact formulation of the gluing conditions of any pair of Boolean probing frames on their overlap, targeting a quantum event structure, is given in [6, pp. 837-838].
Consequently, the specification of a quantum event algebra $L$ in perspectivist terms is synthesized in the limit of the joint compatible action of all Boolean frames-perspectives acting on it.

The fundamental functioning of the Boolean frames--quantum adjunction, specified by the pair of adjoint functors ${\mathbb L} \dashv {\mathbb R}$,
is made transparent if we consider that it provides a bi-directional mechanism of encoding and decoding information between diagrams of Boolean event algebras $B$ and quantum event algebras $L$ via the action of Boolean probing frames or perspectives $\psi_B: {\mathbb M}(B) \rightarrow L$.
Thus, if we think of ${{\mathcal {S} et}^{{\mathcal B}^{op}}}$ as the categorical universe of variable local Boolean frames modeled in $\mathcal {S} et$, and of ${\mathcal L}$ as the categorical universe of quantum event structures, then the left adjoint functor $\mathbb L :  {{\mathcal {S} et}^{{\mathcal B}^{op}}}  \to {\mathcal L}$ signifies an encoding process of information from the level of local Boolean algebras to the level of global quantum event algebras, whereas the Boolean realization functor ${\mathbb {R}}: { {\mathcal L}} \rightarrow {{\mathcal {S} et}^{{{\mathcal B}}^{op}}}$ signifies a decoding process in the inverse direction.
In general, the content of the information cannot remain completely invariant with respect to translating from one categorical universe to another, and conversely.
Note, in this respect, that the functors $\mathbb L$ and $\mathbb R$ are not inverses, since neither ${\mathbb L} {\mathbb R}$ nor ${\mathbb R} {\mathbb L}$ need be isomorphic to an identity functor.
However, as implied by the existence of the Boolean frames-quantum adjunction,
there remain two alternatives for a variable set $\mathbb P$ over local Boolean frames, standing for a presheaf functor $\mathbb P$ in
${{\mathcal {S} et}^{{\mathcal B}^{op}}}$, to exchange information with a quantum algebra $L$.
Either the content of information is transferred in quantum terms with the colimit in the category of elements of $\mathbb P$ translating, represented as the quantum morphism
${\mathbb L \mathbb P} \rightarrow {L}$, or the content of information is transferred in Boolean terms with the functor of Boolean frames of $L$ translating, represented correspondingly as the natural transformation ${\mathbb P} \rightarrow {{\mathbb R}({L})}$.
In the first case, from the setting of $L$, information is being received in quantum terms, while in the second, from the setting of $\mathbb P$, information is being sent in Boolean terms. Then, the natural bijection of relation (5.7) corresponds to the assertion that these two distinct ways of information transfer are equivalent. Consequently, the fact that these two functors are adjoint underlines a bi-directional dependent variation, safeguarding that the global information encoded in a quantum kind of event structure is retrievable in a structure-preserving manner by all possible partial structural congruences with the Boolean kind of event structure.

Thus, by virtue of the proposed category-theoretic perspectivist approach to quantum mechanics, a quantum
event structure can only be unfolded through structured interconnected families of
Boolean probing frames capable of carrying all the information encoded in the former.
And inversely, the non-directly accessible quantum event structure is uniquely
constituted (up to equivalence) by a multiplicity of intertwined local perspectives
directed towards it and covering the object of inquiry entirely under their joint action
via the colimit-gluing process.

\section{Perspectivist Semantics in Quantum Mechanics}
The logic of a physical theory, in addition to its syntactic aspect, incorporates the task of providing a semantics, focusing on the endeavor of establishing a criterion of truth or, in a strict sense, defining an account of truth.
In relation to the latter, the correspondence theory of truth has frequently been regarded within physical science as the most eminent.
Indeed, the elementary or atomic propositions, constituting the elements of the logical structures of fundamental theories of physics, concern the values of physical quantities pertaining to a system and are of the form ``the value of the physical quantity $A$ lies in a measurable subset $\it\Delta$ of the real numbers'', abbreviated usually by the shorthand notation $``A \in \it\Delta"$.
Propositions of the type $``A \in \it\Delta"$ refer to the world in the most direct feasible manner.
They can only be ascertained as true or false by corresponding, or not corresponding, to `facts' or, more generally, to `actual states of affairs' or, in contemporary physical terms, to `events' conceived in the physicist's language as actual measurement outcomes, as concrete elements of empirical reality.
In this respect, elementary propositions are posits in the process of probing the natural world.
They can be used to indicate properties of physical systems in the following sense: a system $S$ is characterized by a property $a$, namely a particular value of the physical quantity $A$, if and only if the corresponding proposition ${P}_A$ is true, that is, if and only if $``A = a"$ or $``A \in \it\Delta"$.
Physics is fundamentally concerned about the truth values of such propositions $``A \in \it\Delta"$, where $\it\Delta$ varies over the Borel subsets of the real line and $A$ varies over the physical quantities of system $S$, when $S$ is in a given state. Of course, the state may change in time, and hence the truth values also will change in time.
It is instructive to note in this connection that the consistent assignment of truth values to elementary propositions presupposes to take into account the overall logical structure of the relevant physical theory, especially in relation to quantum mechanics due to its non-Boolean character.

Although the notion of correspondence truth admits various different formulations, the core of any correspondence theory is the idea that a proposition is true if and only if it corresponds to or matches reality. The classical version of the theory describes this relationship as a correspondence to the facts about the world (e.g. [42, pp. 70-72]). Then, adopting a correspondence theory of truth amounts to endorsing instances of the following scheme:
\begin{quote}
{\it Correspondence to Facts} [CF]: The proposition that $P$ is true if and only if $P$ corresponds to a fact.
\end{quote}

Alternatively, if one construes the notion of a ``fact" in terms of the weaker notion of an obtaining ``state of affairs", then, [CF] is re-expressed as follows:
\begin{quote}
{\it Correspondence to States of Affairs} [CS]: The proposition that $P$ is true if and only if there is a state of affairs $X$ such that $P$ corresponds to $X$ and $X$ obtains.
\end{quote}
\noindent {The useful feature of states of affairs is that they refer to something that can be said to obtain or not to obtain, to be a fact or fail to be a fact, that is, they exist even when they are not concretely manifested or realized, thus overcoming conceptual difficulties related to ``negative facts".}

A noticeable criticism exerted to the traditional conception of correspondence truth is that the correspondence relation between propositional terms and corresponding facts (or states of affairs) is neither specified nor explained; it is simply presupposed. For, ``corresponds to the facts" functions merely as a synonym, as an alternative extended way of saying ``is true" [43].
Using a popular example towards this direction, the proposition ``snow is white" is true, simply because it corresponds to the fact that snow is white. Thus, the fact that makes a proposition true is a restatement of the proposition itself. Facts are merely re-expressions of the propositions they make true.
A full-fledged correspondence theory, however, must articulate an explanation of the correspondence relation that is more complex, and thus not amenable to the immediate restatement reply. In addition, there must provide a genuine account of facts as special kinds of entities that can be candidates for the relationship of truth-making.
Certainly, the connection between propositions (or, in general, truth-bearers) and the world need not be of the simplistic relation of mirror image that is attributed to traditional correspondence truth (e.g. [44]). Nor for a proposition to be true, it has to be necessarily tied to the world in an unmediated, context-independent manner, as assumed by the traditional alethic scheme.

Specifically, when examining the functioning of correspondence truth within the propositional structures of fundamental theories of physics, it should be clear that it requires an understanding not just of the logical form of correspondence in abstracto, but of the specific field of knowledge in which the correspondence relation is realized. It requires, for instance, a systematic analysis of the nature of properties and relations that may be instantiated by objects within the domain of the relevant theory.
Then, truth in terms of correspondence may be appropriately understood as property instantiation, in the following sense: if $P$ is a true proposition, then $P$ attributes a specific property to an object of the relevant domain.

In the quantum domain, due to the essentially probabilistic non-Boolean logical structure of the theory, for any quantum system of dimension higher than two, the totality of elementary propositions represented by projection operators or closed subspaces of Hilbert space are not partitioned into two mutually exclusive and collectively exhaustive sets representing either true or false propositions. 
For instance, propositions represented by subspaces that are at some non-zero or non-orthogonal angle to the unit vector $|\psi\rangle$---or, more appropriately, to the ray representing the quantum state---are not assigned any truth value in $|\psi\rangle$. These propositions are neither true nor false; they are assigned by $|\psi\rangle$ a probability value which is neither $0$ nor $1$, but between $0$ and $1$, and the law of bivalence is violated. Thus, they are undecidable or indeterminate for the system in state $|\psi\rangle$ and the corresponding properties are taken as indefinite.
In fact, such value indefiniteness resulting to semantic indeterminacy is a global property of quantum logical structures, as formally expressed by Kochen-Specker's theorem. 
The latter, however, does not exclude local two-valued truth-functional assignments with respect to complete Boolean algebras of projection operators on the Hilbert space of a quantum system. 
More explicitly, each self-adjoint operator representing an observable has associated with it a Boolean subalgebra which is identified with the Boolean algebra of projection operators belonging to its spectral decomposition.
Hence, given a set of observables of a quantum system, there always exists a complete Boolean algebra of projection operators, that is, a local Boolean subalgebra of the global quantum event algebra of the system with respect to which a local bivalent truth-functional assignment is meaningful, if and only if the given observables are compatible and, thus, simultaneously measurable.
Consequently, the existence of local two-valued truth-functional assignments, against a global non-Boolean background of indefinite possibilities in Hilbert space, points to the natural assumption, formulated in Sect. 3, that complete Boolean algebras play the role of local probing frames for attributing true/false values to propositions within a perspective or context of discourse.
The categorical perspectivist framework, developed in Sect. 5, implements this idea in a universal way by consistently representing the global structure of a quantum algebra of events in terms of a category of presheaves of variable interconnected families of Boolean probing frames, physically realized as suitable perspectives or contexts for measuring physical quantities.

In particular, truth-makers of quantum mechanical propositions, namely facts or actual states of affairs, are not pre-determined, pre-fixed; they are not `out there' wholly unrestrictedly.
As established by virtue of Kochen-Specker's theorem, they are context-dependent in an essential sense. This is in complete harmony with the categorical perspectivist framework put forward, according to which, truth valuation is localized and subsequently contextualized with respect to distinct Boolean probing frames or perspectives targeting the quantum system of interest.
In view of the preceding considerations, therefore, we propose a perspectivist/contextual account of truth that is compatible with the propositional structure of quantum theory by conforming to the following instance of the correspondence scheme:
\begin{quote}
{\it Perspectivist/Conetxtual Correspondence} [PCC]: The proposition that $P$-in-$C$ is true if and only if there is a state of affairs $X$ such that $(1)$ $P$ expresses $X$ in $C$ and $(2)$ $X$ obtains,
\end{quote}
\noindent where $C$ denotes, in general, the context of discourse and, specifically, in relation to the aforementioned quantum mechanical considerations, the experimental context $C_{A}({\mathbb M})$, linked to the proposition $P$ that is associated with a particular Boolean frame ${\mathbb M({B}_A)}$, defining a local perspective on the quantum system, and $A$ indicates the physical quantity under investigation.
As will be argued in the sequel, in the light of the alethic scheme [PCC], perspectivist truth is perspective or context-bound correspondence of a de re nature.
Given the determination of a perspective, or the specification of a local context of reference, the attribution of truth values to propositions designates an objectively existing state of affairs that does not depend upon the epistemic system of a knower's cognizance.
Agents sharing the same perspective reproduce empirical facts of phenomena intersubjectively.

Let us initially note that the proposed perspectivist/contextual account of truth satisfies Tarski's [45, p. 188] criterion of material adequacy, known as ``convention T" or ``schema T", for a theory of truth:
\begin{quote}
(T) The proposition that $``P"$ is true if, and only if, $P$
\end{quote}
where the symbol $``P"$ in (T) represents the name of the proposition which $P$ stands for. To this purpose, let us consider a particular proposition $P$: ``system $S$ has the property $P(A)$". Assume context-dependence with regard to $P(A)$, that is, the latter property of $S$ holds within context $C$. Then, proposition $P$ is concisely stated as: ``system $S$ has the property $P(A)$-in-$C$". Suppose now that this proposition is true. If so, the following instance of the T-schema must be true a priori:
\begin{quote}
The proposition that ``system $S$ has the property $P(A)$-in-$C$" is true if, and only if, system $S$ has the property $P(A)$-in-$C$.
\end{quote}

If, however, the property $P(A)$ of $S$ is context-dependent upon $C$, then the proposition that system $S$ has the property $P(A)$ must also be context-dependent upon $C$. Thus, in conformity with the propositional status entering into the scheme [PCC], the preceding instance of the T-schema can be written equivalently in succinct form as:
\begin{quote}
\noindent The proposition that $``P$-in-$C$" is true if, and only if, $P$-in-$C$.
\end{quote}

Evidently, the logical operation of the bi-conditional in the preceding T-sentence is governed again by the T-schema, or the truism, that the content of a proposition determines the necessary and sufficient conditions under which it is true. Thus for any given true proposition which is context-dependent upon $C$, the fact that makes it true is the context-dependent fact (or state of affairs) upon $C$ that the proposition expresses. Truth contextuality follows naturally from the contextuality of makers of propositional truths. If the latter are context-dependent, then whatever truths may be expressed about them must also be contextual.

The proposed account of truth, as encapsulated by the scheme of perspectivist/con\-textual correspondence [PCC], incorporates explicitly a context-dependence texture of the `world-word' relation, if the world in its microphysical dimension is to be correctly describable.
The truth-making relationship is now established, not in terms of a raw unconceptualized reality, as envisaged by the traditional scheme of correspondence truth, but between a well-defined portion of reality as carved out by the experimental context and the propositional content that refers to the selected context.
Such interdependence of propositional content and referential context is not by virtue of some kind of philosophical predilection, but by virtue of the microphysical nature of physical reality displaying a context-dependence of facts.
For, in view of Kochen-Specker's theorem, there simply does not exist, within a quantum mechanical discourse, a consistent global binary assignment of determinately true or determinately false propositions independently of the appeal to a context; propositional content ought to be linked to a context.
This connection between referential context and propositional content means that a descriptive elementary proposition in the domain of quantum mechanics is, in a sense, incomplete unless it is accompanied by the specified conditions of an experimental context under which the proposition becomes effectively truth-valued. In view of our approach, the latter observation underlines the fact that the conceptual significance of a logic of propositions referring to the description of a quantum system lies at the level of contextual propositions holding in distinct Boolean probing frames. The proposed categorical perspectivist framework reveals precisely that the logic of a quantum event structure is to be sought not in its internal constitution as a set-theoretical entity endowed with unrestricted primary qualities, but rather, in the form of its relationship with the Boolean kind of structure through the established network of adjoint functors between the category of presheaves of variable overlapping Boolean frames and the category of quantum event algebras. This conception enlightens yet further the connection between a quantum algebra of events and its underlying building blocks of Boolean algebras by clarifying the contextual character of quantum theory.
It also points out the significant role of Boolean probing frames in providing empirical access to the non-Boolean quantum world.

It is worthy to note in this respect that the proposed account of truth of perspectivist/contextual correspondence [PCC] essentially denies that there can be a God's eye view or an absolute Archimedean standpoint from which to logically evaluate the totality of facts of nature.
Hence, the reference to a Boolean preparatory experimental context $C_{A}({\mathbb M})$ should not be affiliated with practices of instrumentalism, operationalism and the like; it does not aim to reduce theoretical terms to products of operational procedures. It provides, in fact, the appropriate conditions under which it is possible for a quantum mechanical proposition to receive consistently a truth value.
In other words, the specification of the context is part and parcel of the truth-conditions that should obtain for a proposition in order the latter to be invested with a determinate (albeit unknown) truth value. Otherwise, the proposition is, in general, semantically undecidable.
In the quantum description, therefore, the introduction of the experimental context is to select at any particular time $t$ a specific complete Boolean subalgebra ${\mathbb M({B}_A)}$ of compatible propositions
in the global non-Boolean lattice ${L}_{\mathcal H}$ of a quantum system as co-definite; that is, each proposition in ${\mathbb M({B}_A)}$  is assigned at time $t$ a definite truth value, ``true" or ``false". Equivalently, each corresponding property of the system either obtains or does not obtain.
Importantly, the truth of a proposition in a Boolean frame remains invariant with respect to truth valuations of all other compatible propositions in all Boolean frames ${\mathbb M(B)}$, belonging to a Boolean probing functor of a quantum event algebra $L$.
Thus, in our approach, the specification of a particular Boolean perspective, conceptualized materially as a concrete experimental context, provides in a consistent manner the necessary conditions whereby bivalent assignment of truth values to quantum mechanical propositions is in principle applicable.
This marks the fundamental difference between conditions for well-defined attribution of truth values to propositions and mere operational conditions.

This element also signifies the transition from the uncritical qualification of truth values to propositions by acknowledging them as being true or false {\it simpliciter}, as in traditional correspondence theory of truth, to the demarcation of the limits of possible experience or to the establishment of pre-conditions which make possible the attribution of truth values to propositions.
From the standpoint of our interpretative framework, the proposed account of truth [PCC] is associated with a perspectivist method of inquiry that seeks to investigate---from within the sphere of our worldly conditions, scientific principles and practices---the presuppositions and limits of experience and knowledge, thus opposing a non-perspectival, metaphysically fixed point of reference.
In quantum semantics, therefore, the consideration of a particular Boolean probing frame ${\mathbb M({B}_A)}$, defining a Boolean perspective on the quantum system under investigation, forms a {\it pre-condition} of quantum physical experience, which is necessary if quantum mechanics is to grasp empirical reality at all.
Any microphysical event that occurs is raised at the empirical level only in conjunction with the specification of an experimental context that conforms to a set of observables co-measurable by that context.
The introduction of the experimental context furnishes thus the status of a presupposition of any empirical access to the quantum level of reality, of any possible empirical inquiry on the microscopic scale, and hence of any possible cognizance of microphysical objects as scientific objects of experience. In this respect, the specification of the context constitutes a methodological act preceding any empirical truth in the quantum domain and making it possible.

We note in this respect that the proposed account of truth of perspectivist/contextual correspondence [PCC] is not a relative notion itself; the propositions to which it applies are relative.
They are relative to a specific Boolean subalgebra of compatible propositions which are determinately true or false of a system at any given time.
For, as already argued, a quantum mechanical proposition is not true or false {\it simpliciter}, but acquires a determinate truth value with respect to a well-defined context of discourse as specified by the state of the quantum system concerned and a particular observable to be measured.
Thus, the conditions under which a proposition is true are jointly determined by the context in which the proposition is expressed and the actual microphysical state of affairs as projected into the specified context.
What makes a proposition true, therefore, is not that is relative to the context---as an alethic relativist must hold---but whether the conditions in
question obtain.
The obtainment of the conditions implies that it is possible for us to make, in an overall consistent manner, meaningful statements that the properties attributed to quantum objects are part of physical reality.
In our approach, therefore, a proposition is true because it designates an objectively existing state of affairs, albeit of a perspectivist/contextual nature.
In relation to the latter, we stress the fact that, in contrast to a panoptical view from nowhere of the classical paradigm, the general epistemological implication of quantum theory acknowledges in an essential way a perspectivist/contextual character of knowledge.
The classical doctrine that one can reasonably talk about `all entities', as if the attribution of properties to an `entity' had a unique, fixed meaning, independently of the appeal to a particular context of reference, is inadequate in the microphysical level of discourse.
Nonetheless, the traditional conception of correspondence truth, involving a direct context-independent relation between singular terms of propositions and definite
autonomous facts of an external reality, may be viewed as a limit case of the more generic alethic scheme of perspectivist/contextual correspondence [PCC], when the latter is applied in straightforward unproblematic circumstances where the non-explicit specification of a perspective or a context of discourse poses no further consequences.
This is immediately revealed if as context $C$ in [PCC] is taken the whole of reality and thus any particular reference to the conditions of a local context is silently removed.

\section{Concluding Remarks}
The primary objective of the proposed endo-theoretic approach to perspectivism is to actively engage with the science of today as a means of analyzing the origin, generation and systematization of scientific knowledge of nature by focusing on the conditions under which such knowledge may arise in perspectivist terms. Specifically,  it aims to elevate scientific perspectivism from a general philosophical position to a methodologically robust, formally rigorous, and conceptually fruitful framework of scientific inquiry in quantum mechanics, by establishing how locally variable contextual perspectives can be endo-theoretically integrated to grasp a complex, non-classical physical reality. This integration is achieved in the present study not by adding extrinsic postulates to quantum formalism, but by maximizing the inherent mathematical structures of quantum mechanics through the lens of category theory.

Our perspectivist framework in relation to quantum mechanics has been developed from the ground up, using initially the Kochen-Specker theorem to motivate the necessity of Boolean probing frames as perspectives applied on a quantum system, and then employ category theory to demonstrate that the global structure of a quantum algebra of events can be consistently represented (up to equivalence) through a multilevel interconnected structure of locally variable perspectives. The technically distinctive feature that underpins the foregoing representation is precisely the Boolean frames-quantum adjunction. It provides a universal, bi-directional mechanism of encoding and decoding information between the local Boolean and the global quantum level under the action of Boolean probing frames or perspectives, thereby formalizing how perspectival information transfer can be synthesized into a coherent whole. 

In this framework, truth valuation is contextualized with respect to distinct Boolean perspectives targeting the quantum system of interest, thus leading naturally to a perspectivist/contextual account of truth. In light of the latter, perspectivist truth conforms to perspective or context-bound correspondence, designating locally an objectively existing state of affairs as delineated by the considered context $C$. Such an account derives by virtue of the microphysical nature of physical reality in displaying a context-dependence of facts. It opposes, therefore, a non-perspectival, metaphysically fixed point of reference or a panoptical standpoint from which to state all facts of nature.

Intersubjective agreement within a perspective is an essential internal feature of our theoretical framework, reflecting a fundamental characteristic of how perspectival knowledge is structured, while consistency across perspectives results from the mathematical coherence of the overall framework, guided by the Boolean frames-quantum adjunction. The mechanism for comparing and interrelating different perspectives is embodied in the network of the mathematical relations (morphisms, functors, adjunctions) within the category of Boolean probing frames covering the quantum event algebra itself. As already stated, this allows for a systematic extension and global synthesis of local perspectival information.

In closing, we note that the endo-theoretic perspectivist approach to quantum mechanics provides the appropriate formal basis for developing a post-classical, structured view of scientific theorizing in the sense of comprehending a theory not just as a class of empirical models simpliciter, as a structureless set of ``models of the data", but also establishing mappings between these models allowing thereby their coherent embedding in a global theoretical structure. The aforementioned proposition refers to the actual scientific theorizing and practice of contemporary science and thus the suggested approach, due to its general character, may acquire the form of a theoretical pattern of scientific inquiry in the natural sciences, especially when dealing with complex trans-perspectival phenomena, the analysis of which requires the use of information of a multi-scale variety, resulting from more than one perspective.

\newpage

\begin{center}{ {\bf \large Appendix

}

\vspace{.5mm} }\end{center}
It is instructive to attempt a concise comparative assessment between our perspectivist framework of quantum theory and Rovelli's relational interpretation of quantum mechanics (RQM) focusing on their essential differences, thus revealing the profoundly different apprehension of the quantum world's architecture between the two proposals. The appendix is partly based on comments and clarificatory questions raised by the reviewer, giving us the opportunity to explicate on certain issues ranging from the concept of perspective to the way different perspectives may relate to each other and from the meaning of perspectivist truth to issues concerning the ontology of quantum events and their impact on the comprehension of reality.

\paragraph{[A] Methodology and Mechanism for Relating Perspectives -- Intersubjectivity} We initially point out that both our proposed endo-theoretic perspectivist framework and
Rovelli's RQM, particularly in its recent development involving the postulate of ``Cross-Perspective Links" (CPL), affirm the intersubjective agreement on empirical facts among agents under certain conditions. However, the underlying conceptual and formal mechanisms by which this agreement is established, and indeed the very notion of ``perspective", differ significantly.
For brevity, we shall call henceforth the updated version of RQM, as modified by the addition of the CPL postulate, RQM+.

\begin{enumerate}
\item {\bf Rovelli's Relational Quantum Mechanics and the CPL Postulate}

\begin{itemize}
\item[\textendash] {\it RQM}: In Rovelli's [24] initial formulation of RQM, the state of a system $S$ is always relative to an observer $O$. Facts---namely, values of physical quantities---are established only during interactions and are relative to the interacting systems.
    Comparisons of facts relative to different observers (e.g., Alice's measurement outcome relative to Alice, and Bob's measurement outcome relative to Bob) are only meaningful from the perspective of a third observer $W$ who interacts with both Alice and Bob [46]. This led to concerns about ``island universes" of information [47], where an observer cannot, in principle, know anything beyond their immediate present experience or confirm empirical facts relative to other observers. The implied lack of intersubjectivity regarding measurement outcomes seemed to challenge the character of the scientific practice or even undermine scientific objectivity itself.
\item[\textendash] {\it Introducing CPL}: To address the aforementioned intersubjectivity issues, Adlam \& Rovelli [48, p. 7] proposed a new postulate, the ``cross-perspective links" (CPL), formulated as follows:
``In a scenario where some observer Alice measures a variable $V$ of a system $S$, then provided that Alice does not undergo any interactions that destroy the information about $V$ stored in Alice's physical variables, if Bob subsequently measures the physical variable representing Alice's information about the variable $V$, then Bob's measurement result will match Alice's measurement result".
\begin{itemize}
\item[\textbullet] Under the CPL process, point-like quantum events---that is, the actualization of a value in an interaction---become absolute, observer-independent features of reality [48, p. 12]. However, quantum states remain relational, describing the relationship between systems.
\item[\textbullet] The value of a variable $V$ obtained by Alice is still initially relativized to Alice, since, at the beginning, she is the only one with that information, yet CPL establishes that this information, being physical and ``stored in Alice's physical variables", is accessible in principle through measurements to any other observers. It is a specific mechanism that has been added on an already existing structure for linking different observers' information but now about absolute quantum events.
\item[\textbullet] The CPL modification, while confronting at least theoretically the intersubjectivity problem, leads to the de-relativization of facts, thereby potentially undermining the distinctively relational aspects of the classical formulation of RQM.
\end{itemize}
\end{itemize}

\item {\bf Karakostas \& Zafiris' Endo-Theoretic Perspectivist Framework: The K\&Z Approach}

\begin{itemize}
\item[\textendash] {\it Nature of Perspective}: In our framework, a ``perspective" is not primarily an individual observer's interaction history but is formally defined as a Boolean probing frame. This is a complete Boolean algebra of projection operators associated with a set of compatible and thus comeasurable observables. It represents a particular experimental context by partitioning the ``space" of possible outcomes. Furthermore, it constitutes a local structural invariant within the global non-Boolean quantum event algebra, satisfying the general desiderata put forward in Sect. 1 concerning the suitability of a structural unit to be qualified as a perspective.
\item[\textendash] {\it Intersubjectivity}: Intersubjective agreement on the determinateness and repeatability of empirical facts, among agents sharing the same perspective, is an explicit consequence of the logical status characterizing a Boolean frame.
\begin{itemize}
\item[\textbullet] If agents Alice and Bob share the same perspective, it means they are operating within or have adopted the same Boolean frame ${\mathbb M({B}_A)}$ (as defined in Sect. 6).
\item[\textbullet] Within such a definite Boolean frame, propositions are assigned bivalent truth values. Thus, each proposition in ${\mathbb M({B}_A)}$ is assigned at any given time $t$ a definite truth value, ``true" or ``false". The Kochen-Specker theorem, although globally forbids a bivalent truth-value assignment to {\it all} quantum mechanical propositions, nonetheless, does not prohibit such local assignments within a selected Boolean subalgebra.
\item[\textbullet] Therefore, if Alice and Bob are considering propositions and performing measurements within the same explicitly defined Boolean frame, the empirical facts established---namely, the outcomes of measurements corresponding to projection operators within that frame---will be determinate and repeatable for both, simply because they are operating under identical logical conditions provided by that shared frame.
\end{itemize}
\end{itemize}
\begin{itemize}
\item[\textendash] {\it The Interrelation Mechanism Codified -- From Local Perspectives to a Global Structured Whole}: 
These local Boolean frames are not isolated ``island universes". They are systematically related to each other as homomorphic images via the modeling functor ${\mathbb M}$ within the overarching, non-Boolean quantum event algebra $L$. The K\&Z categorical perspectivist framework provides the precise syntax of perspectives for how they interconnect:
\begin{itemize}
\item[\textbullet] {\it Functor of Boolean Frames on $L$}: For any quantum event algebra $L$, we define a functor ${\mathbb {R}}(L)$ whose action on any Boolean algebra $B$ provides the set of all possible Boolean probing frames of the type $B$ that can be adjoined to $L$. This functor, ${\mathbb {R}}(L)={Hom_{\mathcal L}}({\mathbb M}(-),L)$, captures the totality of all possible local perspectives on $L$ (Sect. 5.2).
\item[\textbullet] {\it Compatibility and Gluing}: The framework establishes rigorous compatibility conditions for how these frames overlap consistently. When two Boolean frames, say, 
$\psi_B: {\mathbb M} (B) \rightarrow L$ and $\psi_C: {\mathbb M} (C) \rightarrow L$, are not disjoint, their relationship is governed by their {\it pullback}. This ensures that any information they share is consistent. The {\it cocycle conditions} for the gluing isomorphisms guarantee that these local perspectives can be ``pasted" together seamlessly into a coherent global structure [6, Sect. 4.2]. This mechanism explicitly prevents fragmentation.
\item[\textbullet] {\it Synthesis -- The Colimit Construction}: The actual assembly of the global structure from these coherently interconnected local Boolean frames is performed by a universal mathematical construction known as the {\it colimit}. Our central result is that the global quantum event algebra $L$ is precisely the colimit of the diagram formed by all of its local Boolean probing frames:
    \[L= Colim({\mathbb {R}}(L))\,.\]
    This means $L$ is the unique object that synthesizes all the information contained in the local perspectives composing ${\mathbb {R}}(L)$ in a way that respects all their interrelations. It is the ``universal solution" that arises from gluing all the locally variable perspectives together (Sect. 5.3).
\end{itemize}
\end{itemize}
\begin{itemize}
\item[\textendash] {\it The Boolean frames-quantum adjunction}: The preceding solution is based on the Boolean frames-quantum adjunction, which ensures that local information can be globally synthesized (Sect. 5.3). This adjunction provides a bi-directional mechanism for encoding (from Boolean to quantum) and decoding (from quantum to Boolean) information in perspectivist terms. Instead of being an added postulate, it is a fundamental consequence of the category-theoretic representation of quantum event structures via the action of Boolean probing frames, conceived as perspectives applied on a quantum system. It establishes that the global structure of a quantum algebra of events can be consistently comprehended through a multilevel structure of locally variable Boolean perspectives, interconnected in a category-theoretic environment, yielding jointly all the information encoded in the former. Thus, inter-perspectival consistency is an essential feature of the proposed framework.
\begin{itemize}
\item[\textbullet] The structural concept of ``perspective", our methodology on the conditions under which knowledge is possible, as analyzed in Sect. 2,  the explicit use of category theory, and the Boolean frames-quantum adjunction, are unique to our endo-theoretic approach in providing a consistent perspectivist framework of quantum mechanics.
\end{itemize}
\end{itemize}
\end{enumerate}

\paragraph{[B] Explicating the Meaning of the Perspectivist/Contextual Correspondence Theory of Truth} We are addressing the following clarificatory questions raised by the reviewer in relation to our proposed account of truth: (1) Is a given measurement valid only for those agents sharing the context $C$? (2) What happens if somebody is outside $C$? Can the measurement result be different or indeterminate? (3) According to the proposed account of truth, once an experimental context is set up and a measurement is performed, its result depends on the context. In what sense this result may be considered observer independent? (4) Furthermore, such contextualization criterion is very different from saying that scientific truths depend on what an epistemic community agrees upon, as usually said in perspectival considerations. 

\begin{enumerate}
\item {\bf Agents Sharing Context $C$}

\begin{itemize}
\item[\textendash] In quantum mechanics, a global Boolean context of measurement does not exist, and by virtue of Kochen-Specker's theorem, only single local contexts, as defined by our concept of a perspective, can be considered as being definite, thus capturing Boole's ``conditions of possible experience" (Sect. 2.2). Henceforth, in the endo-theoretic approach to perspectivism, truth valuation is localized and subsequently contextualized with respect to distinct Boolean probing frames conceived as  perspectives targeting the quantum system of interest. 
    That is, bivalent truth-value assignments hold consistently only within a perspective and its associated measurement context $C$. The proposition ``$P$-in-$C$" is only well-defined and capable of possessing a determinate truth value with respect to $C$. Consequently, the truth of ``$P$-in-$C$" characterizes an objectively existing state of affairs, as delineated by $C$. Any agent, who wishes to evaluate $P$ in the context $C$, would arrive at the same truth value, assuming access to the relevant empirical facts.
\end{itemize}

\item {\bf Agent ``Outside" Context $C$}

\begin{itemize}
\item[\textendash] If an agent is ``outside" $C$, it means they are operating within a different context $C^{\prime}$ (a different Boolean frame), or perhaps they have not adopted any specific Boolean frame relevant to $P$.
\item[\textendash] If $C$ and $C^{\prime}$ are incompatible contexts, corresponding to non-commuting sets of observables, then a proposition $P$ meaningful in $C$ might not even be well-defined in $C^{\prime}$, or a related proposition $P^{\prime}$  in $C^{\prime}$  might yield a different outcome. This is a standard feature of quantum contextuality as revealed by Kochen-Specker's theorem. For example, if $C$ involves measuring spin-$z$ and $C^{\prime}$ involves measuring spin-$x$, the outcomes correspond to different complementary aspects of reality, which, in principle, cannot be realized simultaneously.
\item[\textendash] From the standpoint of an agent in $C^{\prime}$, the truth value of ``$P$-in-$C$"---a proposition tied to a different, potentially incompatible context---might be considered indeterminate from $C^{\prime}$'s perspective. However, the truth value of ``$P$-in-$C$" itself is determinate within its own defining context $C$. Our framework aims to show how these determinate truth-valuations from different perspectives can be systematically interrelated---to the degree authorized by quantum mechanics---via the use of the functorial apparatus of category theory (Sect. 5.2).
\end{itemize}

\item {\bf Perspectivist/Contextual Truth is an Empirical Fact}

\begin{itemize}
\item[\textendash] Let us initially remind that in the endo-theoretic approach to perspectivism, the kind of the experimental context to be utilized corresponds to the specification of a particular Boolean probing frame $C$. Once $C$ is fixed, the measurement outcome for a proposition $P$ defined within $C$ yields a truth value for ``$P$-in-$C$". The latter truth value is objective and observer-independent given that context $C$; it corresponds to an objective, yet contextually delimited, state of affairs. Any observer or agent correctly applying the rules of quantum mechanics within that specific context $C$ will ascertain the same result. Since the truth of $P$ is context-dependent upon $C$, if a different context $C^{\prime}$, incompatible with $C$, were chosen, a different set of propositions would become determinate. In the quantum domain, perspectivist/contextual truth is an empirical fact of reality. For, within any well-defined experimental context (or Boolean probing frame), measurements yield definite, repeatable, and intersubjectively confirmable outcomes.
\end{itemize}

\item {\bf Contextualization Criterion -- Epistemic Community Agreement}
\begin{itemize}

\item[\textendash] Our contextualization criterion should be distinguished from typical perspectival theses in the general philosophy of science where truth usually depends on an epistemic community's agreement. In the endo-theoretic approach to perspectivism, truth is context-bound correspondence to an objective state of affairs. When applied to quantum mechanics, the ``context" is a structural feature of the physical theory---a Boolean algebra of compatible observables---not an epistemic community. The agreement of agents within that context is then a consequence of this objective, context-bound truth, yielding empirically confirmed facts. This is a contemporary form of perspectival realism, associated to the philosophy and conceptual foundations of quantum mechanics, where the consideration of a perspective forms an essential feature of the theoretical landscape for describing consistently microphysical reality (Sect. 6).
\end{itemize}
\end{enumerate}

\paragraph{[C] The K\&Z Approach Compared to Rovelli's RQM} We proceed to a concise comparison between our perspectivist framework of quantum theory and Rovelli's relational interpretation of quantum mechanics giving emphasis on certain features that reveal the essentially different accounts of the quantum world's architecture between the two proposals.

\begin{enumerate}
\item {\bf Notion of Perspective}

\begin{itemize}
\item[\textendash] {\it RQM}: A ``perspective" is tied to an observer and their interaction history, including the probabilistic predictions that the ensemble of events relative to the observer entails [49].
\item[\textendash] {\it K}\,\&{\it Z Approach}: A ``perspective" is an endo-theoretic structural unit, a Boolean probing frame serving as a window on physical reality. It constitutes a local structural invariant within quantum theory itself, characterizing a maximal set of comeasurable observables, or, equivalently, a complete Boolean algebra of commuting projection operators generated by this set. Such a kind of perspective constitutes the principal structural unit for investigating reality at her microphysical dimension. It provides experimental accessibility on the investigated system, functions locally as a probing frame for the individuation of events, resolves a targeted object of inquiry, generally, probes consistently the physical world.
\end{itemize}

\item {\bf Role of the Observing or Measuring System}

\begin{itemize}
\item[\textendash] {\it RQM}: It is a core thesis of the relational interpretation that any system can acquire the status of an ``observer" (or ``observing system") in a physical interaction. Accordingly, a measurement is considered as the obtainment of mutual information between the physical variables of two interacting systems, in which either system can undertake the role of the ``observer" while the other stands in the role of the ``observed system". This equivalence relation between different systems in their status of possible observers reflects the radically relational character of RQM.
\item[\textendash] {\it K}\,\&{\it Z Approach}: In the endo-theoretic approach to perspectivism, given a quantum system under investigation, the observer, apart from selecting the observable to be measured and thus specifying the perspective applied on the system, plays no special role. During the measurement process, the observer or the experimentalist is naturalized in terms of a measuring apparatus. Every reference is exclusively related to the possible results of interactions between the measuring apparatus and the measured system as parts of the experimental set-up. Values of quantum mechanical observables or outcomes of measurement processes are perspective-dependent since they are context-dependent. Their relational character is with respect to the context of measurement, not between any two interacting systems. Hence, in our approach, the measuring system cannot be any physical system, as in relational quantum mechanics, but a suitable measuring apparatus by means of which outcomes of measurements are irreversibly recorded. This is reminiscent of Wheeler's [50, p. 189] expression: ``Until the act of detection the phenomenon-to-be is not yet a phenomenon". In the quantum domain, the individual character of the quantum phenomenon removes the possibility for comparing consistently {\it what is} with {\it what might have been}.
\end{itemize}

\item {\bf Nature of Truth}

\begin{itemize}
\item[\textendash] {\it RQM}: In the classical formulation of RQM, facts are observer-depen\-dent. The value of a variable for system $S$ becomes definite relative to observer $O$ upon interaction. Another observer $O^{\prime}$ might not have this fact actualized relative to them. Since a comparison of measurement results between any pair of observers is only meaningful in RQM by invoking a third observer $W$, truth is radically relative to an observer's perspective.
\item[\textendash] {\it RQM}+: Quantum events become now absolute. If Alice measures variable $V$ on $S$ and gets v, this event, relative to Alice, is an absolute fact. The CPL postulate ensures that Bob, as any other observer, can later access this information about Alice's past relative fact. While the initial value ascription is to Alice's perspective, the event itself has an absolute status. To the extent, therefore, that the information gained by Alice when a variable becomes definite relative to her is in principle accessible to any observer who measures Alice's variable in the right basis, then the existence of an empirical fact is absolute, as absolute or observer-independent is in the CPL scenario the instantaneously realizable set of point-like quantum events [48, p. 11].
\item[\textendash] {\it K}\,\&{\it Z Approach}: Truth is perspectivist/contextual correspondence of a de re nature. Truth assignment is contextual since it depends on the selected Boolean frame $C$. Once $C$ is specified, the truth of ``$P$-in-$C$" is an objective matter determined by correspondence to the state of affairs carved out by $C$. Different Boolean frames $C^{\prime}$, $C^{\prime\prime}$, \ldots, provide locally possible perspectives onto the quantum world. The category-theoretic formalism establishes how the various perspectival truth assignments across perspectives are globally coherent. 
    In the relational interpretation, the cross-perspective links is a specific postulate added on an already existing structure for achieving inter-observer agreement about past events. 
    In contradistinction, our framework's consistency and interrelation of perspectives arise from the mathematical coherence of the categorical perspectivist representation of a quantum event structure via Boolean probing frames, which is of sheaf-theoretic nature. The quantum event algebra itself is comprehended functorially as a sheaf over the base category of Boolean probing frames, meaning that it is ``constituted" or ``glued" together from these locally variable perspectives in a consistent way [6].
\end{itemize}

\item {\bf Determining the Existence of a Preferred Basis}

\begin{itemize}
\item[\textendash] {\it RQM (pre-} \& {\it post-CPL)}: Unlike the relational interpretation, we take the prevalent view that while every measurement presupposes an interaction between two systems---the measuring system and the system under measurement---nonethe\-less, every interaction does not qualify as a measurement. RQM, functioning under pure unitary quantum dynamics, cannot single out a unique variable that should take on definite values during an interaction that leads to a quantum event [51]. Furthermore, two quantum systems, forming an entangled state, admit correlations of a different kind depending on the selected basis to represent the state, as in a Bell-type entangled state of two spin-half particles. Consequently, lacking an exact method for determining a preferred definite basis, there exists an inherent ambiguity in RQM associated to the theoretical feasibility that an ``observer" may assign multiple distinct relative states to the same ``observed system".\renewcommand{\baselinestretch}{1}\footnote{As a response to the preferred basis problem, it is argued in [48, p. 17] that under the action of the CPL postulate an observer can eventually observe a definite value in one particular basis, say, the basis of position, by assuming that ``my conscious experience emerges from the unified perspectives of the particles in my brain, the definite value that I will become aware of is the one on which a significant number of particles in my brain agree---so I will have the experience of seeing a point in a particular coarse-grained position on the detector screen. Note that this account would not work at all without CPL---if we do not have cross-perspective links, then it will not be the case that the particles in my brain come to agree on certain values via decoherence".} \renewcommand{\baselinestretch}{1.15}
\item[\textendash] {\it K}\,\&{\it Z Approach}: In the endo-theoretic approach to perspectivism there is no such ambiguity since the perspective, materialized as an experimental context, defines locally the preferred basis, that is, the eigenbasis of the set of observables that are compatible with the observable to be measured.
\end{itemize}

\item {\bf Global Event Ontology}

\begin{itemize}
\item[\textendash] {\it RQM}+: As a direct implication of the CPL postulate, quantum events, once radically relational in the classical version of RQM, are considered now as absolute, observer-independent facts of reality. The CPL process achieves intersubjective agreement among observers at the expense of inducing a nonlocal dependence of quantum events on one another, in the sense that the choice of which variable becomes definite in a quantum event at a given time will generally be dependent on the values of variables of previous events. In order to avoid potential compatibility issues with relativity theory, Adlam [52, p. 33] suggests that ``we must allow that the network of events is determined in an `all-at-once' manner, rather than being generated in some temporal order". Accordingly, this introduces in the context of RQM+ an ontology of a mosaic of absolute events whose distribution is obtained in an instantaneous, structureless and fragmented manner, impossible to be calculated or characterized as a whole [48, p.12].
\item[\textendash] {\it K}\,\&{\it Z Approach}: Our framework provides a constructive and structural account resulting in an ontology that is fundamentally different from a primitive, instantaneous, fragmented collection of events.
\begin{itemize}
\item[\textbullet] {\it Holistic and Structured}: Now, the global event ontology is not analogous to a mosaic of disparate event-flashes. It is a unified, coherent, and highly structured algebraic object, the quantum event algebra $L$, forming a complete, atomic, orthomodular lattice. The relations between events---such as inclusion, orthogonality, compatibility, and so on---are intrinsic to this algebraic structure. The structure is neither imposed externally nor is it realized in an ``all-at-once" manner.
\item[\textbullet] {\it Calculable and Characterizable as a Whole}: Far from being impossible for a distribution of quantum events to be calculated or characterized as a whole, the critical point of the K\,\&Z framework is to demonstrate that the global object $L$  is the characterization of the whole, and that it is uniquely determined by the synthesis of its local perspectival manifestations. The colimit construction provides the universal categorical solution for this characterization. The guiding principle, therefore, is not that the world is a mere collection of fragmented facts but that the world is a structured whole which is knowable through a multiplicity of partial, internally consistent perspectives.
\end{itemize}
\end{itemize}

\item {\bf The Nature of Events and Nonlocality}

\begin{itemize}
\item[\textendash] In our framework, as in standard quantum mechanics, an ``event" corresponds to an element of the global quantum algebra $L$; it is represented by a projection operator.
\item[\textendash] This event does not acquire a pre-given, absolute, context-independent truth value. The global algebra $L$  is inherently value-indefinite, which is the logical-structural meaning of the Kochen-Specker theorem.
\item[\textendash] An event obtains a definite truth value and thus it becomes an actual fact, only perspectivally, that is, within a Boolean probing frame or perspective from which the quantum system is considered (Sect. 6).
\item[\textendash] Quantum nonlocality and constraints on probabilistic correlations among quantum physical observables are not the result of peculiar nonlocal influences between absolute, fragmented events. They are a direct consequence of the non-Boolean, holistic structure of the global quantum algebra $L$. The entangled correlations of an EPR pair, for instance, are encoded in the global state vector of the compound system, which defines the relations within the tensor product algebra $L_{A} \otimes L_{B}$. A local measurement that involves the adoption of a Boolean frame, say, at system $B$ projects out a specific slice of this global structure, revealing information that is constrained by the whole. The correlation is thus structural, not causal.
\end{itemize}

\item {\bf ``All-at-Once" Realization vs. Structural Constitution}

\begin{itemize}
\item[\textendash] In RQM, including the CPL postulate, the strongly nonlocal dependencies of quantum events on one another force an ``all-at-once", atemporal explanatory stance for the entire mosaic of facts. Within our framework, such an ``all-at-once" realization of events is redundant. The non-Boolean quantum algebra $L$, equipped with the projective geometry of Hilbert space, encodes generic kinematic (pre-dynamic) constraints on probabilistic correlations between measurement outcomes (events), ensuring the consistent incorporation of probabilities in a physically indefinite world. The relationship between the global structure and local measurements is explicated by the Boolean frames-quantum adjunction, which formalizes a bi-directional, structure-preserving information flow between these two hierarchically distinct structural levels. This provides a structural explanation for how local contexts, within a generic non-Boolean quantum framework, reveal global properties, surpassing the need for an ``all-at-once" hypothesis concerning the realization of disparate events.
\end{itemize}
\end{enumerate}

In conclusion, the K\,\&Z approach offers in comparison to RQM+ a profoundly different picture of the quantum world's architecture. Instead of postulating a distribution of absolute but fragmented events and then seeking an external hypothesis of an ``all-at-once" realization that induces a generic form of nonlocality, we demonstrate how a holistic indefinite quantum reality, conceptualized as a global quantum event algebra, is mathematically constituted by the coherent synthesis of all possible, locally variable Boolean perspectives. The distribution of events is neither instantaneous nor fragmented; it is an expression of the intricate non-Boolean probabilistic correlations within a single unified quantum event structure.

\vspace{1cm}

\begin{center}{ {\bf \large References

}

\vspace{.5mm} }\end{center}

\begin{enumerate}
\renewcommand{\baselinestretch}{0}
{\small
\item {Giere, R.N.: Scientific Perspectivism. The University of Chicago Press, Chicago (2006)}
\item {Teller, P.: What Is Perspectivism, and Does It Count as Realism? In: Massimi, M., McCoy, C. (eds.) Understanding Perspectivism, pp. 49-64. Routledge, New York (2019)}
\item {Massimi, M.: Perspectival Realism. Oxford University Press, New York (2022)}
\item {Rueger, A.: Perspectival Models and Theory Unification. Br. J. Philos. Sci. {\bf 56}, 579-594 (2005)}
\item {Chang, H.: Realism for Realistic People. Cambridge University Press, Cambridge (2022)}
\item {Karakostas, V., Zafiris, E.: On the Structure and Function of Scientific Perspectivism in Categorical Quantum Mechanics. Br. J. Philos. Sci. {\bf 73}, 811-848 (2022)}
\item {Karakostas, V.: Forms of Quantum Nonseparability and Related Philosophical Consequences. J. Gen. Philos. Sci. {\bf 35}, 283-312 (2004)}
\item {Devitt, M.: Realism and Truth, 2nd edn.  Princeton University Press, Princeton (1997)}
\item {Karakostas, V.: Realism and Objectivism in Quantum Mechanics. J. Gen. Philos. Sci. {\bf 43}, 45-65 (2012)}
\item {Pauli, W.: Probability and Physics. In: Enz, C., Meyenn, K. (eds.), Wolfgang Pauli, Writings on Physics and Philosophy, pp. 43-48. Springer-Verlag, New York (1994)}
\item {Putnam, H.: Reason, Truth, and History. Cambridge University Press, Cambridge (1981)}
\item {Kochen, S., Specker, E.P.: The Problem of Hidden Variables in Quantum Mechanics. J. Math. Mech. {\bf 17}, 59-87 (1967)}
\item {Boole, G.: An Investigation of the Laws of Thought on Which are Founded the Mathematical Theories of Logic and Probabilities. Dover, New York (1958)}
\item {Kant, I.: Critique of Pure Reason, trans. Guyer, P., Wood, A. Cambridge University Press, New York (1998)}
\item {Varadarajan, V.S.: Geometry of Quantum Theory, 2nd edn. Springer, New York (2007)}
\item {Abramsky, S., Brandenburger, A.: The Sheaf-Theoretic Structure of Non-Locality and Contextuality. New J. Phys. {\bf 13}, 113036 (2011)}
\item {Cabello, A., Severini, S., Winter, A.: Graph-Theoretic Approach to Quantum Correlations. Phys. Rev. Lett. {\bf 112}, 04040 (2014)}
\item {Karakostas, V., Zafiris, E.: Contextual Semantics in Quantum Mechanics from a Categorical Point of View. Synthese {\bf 194}, 847-886 (2017)}
\item {Gleason, A.M.: Measures on the Closed Sub-Spaces of Hilbert Spaces. J. Math. Mech. {\bf 6}, 885-893 (1957)}
\item {Lison\^{e}k, P., Badziag, P., Portillo, J.R., Cabello, A.: Kochen-Specker Set with Seven Contexts. Phys. Rev. A {\bf 89}, 042101 (2014)}
\item {Wang, P. et al.: Significant Loophole-Free Test of Kochen-Specker Contextuality Using Two Species of Atomic Ions. Sci. Advan. {\bf 8}, 1660 (2022)}
\item {Howard, M., Wallman, J., Veitch, V., Emerson, J.: Contextuality Supplies the `Magic' for Quantum Computation. Nature {\bf 510}, 351-355 (2014)}
\item {Svozil, K.: Contexts in Quantum, Classical and Partition Logic. In: Engesser, K., Gabbay, D., Lehmann, D. (eds.), Quantum Logic, pp. 551-586. Elsevier, Amsterdam (2009)}
\item {Rovelli, C.: Relational quantum mechanics. Int. J. Theor. Phys. {\bf 35}, 1637-1678 (1996)}
\item {van Fraassen, B.: Rovelli's World. Found. Phys. {\bf 40}, 390-417 (2010)}
\item {Bene, G., Dieks, D.: A Perspectivist Version of the Modal Interpretation of Quantum Mechanics and the Origin of Macroscopic Behavior. Found. Phys. {\bf 32}, 645-671 (2002)}
\item {French, S.: A Phenomenological Approach to Quantum Mechanics: Cutting the Chain of Correlations. Oxford University Press, Oxford (2023)}
\item {Fuchs, C.A., Schack, R.: Quantum-Bayesian Coherence. Rev. Mod. Phys. {\bf 85}, 1693 (2013)}
\item {Healey, R.: Quantum Bayesian and Pragmatist Views of Quantum Theory. In: E. N. Zalta (ed.) The Stanford Encyclopedia of Philosophy, Winter Edition (2023)}
\item {Berghofer, P., Wiltsche, H. (eds.): Phenomenology and QBism. Routledge, New York (2024)}
\item {Bell, J.L.: Toposes and Local Set Theories, 2nd edn. Dover, New York (2008)}
\item {Goldblatt, R.: Topoi: The Categorial Analysis of Logic, 2nd edn. Dover, New York (2006)}
\item {Mac Lane, S., Moerdijk, I.: Sheaves in Geometry and Logic. Springer, New York (1992)}
\item {Borceux, F.: Categories of Sheaves, 2nd edn. Cambridge University Press, Cambridge (2008)}
\item {Johnstone, P.T.: Sketches of an Elephant: A Topos Theory Compendium, Vol. 2. Clarendon Press, London (2002)}
\item {Chiara, M.D., Giuntini, R., Greechie, R.: Reasoning in Quantum Theory. Kluwer, Dordrecht (2004)}
\item {Awodey, S.: Category Theory, 2nd edn. Oxford University Press, Oxford (2010)}
\item {Eddington, A.S.: The Theory of Groups. In: Newman, J.R. (ed.) The World of Mathematics, Vol. 3, pp. 1558-1575. Dover, New York (2003)}
\item {Zafiris, E.: Sheaf-Theoretic Representation of Quantum Measure Algebras. J. Math. Phys. {\bf 47}, 092103 (2006)}
\item {Zafiris, E., Karakostas, V.: A Categorial Semantic Representation of Quantum Event Structures. Found. Phys. {\bf 43}, 1090-1123 (2013)}
\item {Zafiris, E., Karakostas, V.: Category-Theoretic Interpretative Framework of the Complementarity Principle in Quantum Mechanics. Int. J. Theor. Phys. {\bf 58}, 4208-4234 (2019)}
\item {Burgess, A., Burgess, J.P.: Truth. Princeton University Press, Princeton (2011)}
\item {Lewis, D.: Forget about the Correspondence Theory of Truth. Analysis {\bf 61}, 275-280 (2001)}
\item {Sher, G.: Correspondence Pluralism. Synthese {\bf 202}, 169 (2023)}
\item {Tarski, A.: The Concept of Truth in Formalized Languages. In: Tarski, A. (ed.) Logic, Semantics, Metamathematics, pp. 152-278. Clarendon Press, Oxford (1956)}
\item {Di Biagio, A., Rovelli, C.: Relational Quantum Mechanics is about Facts, not States: A Reply to Pienaar and Brukner. Found. Phys. {\bf 52}(3), 62 (2022)}
\item {Pienaar, J.L.: A Quintet of Quandaries: Five No-Go Theorems for Relational Quantum Mechanics. Found. Phys. {\bf 51}, 97 (2021)}
\item {Adlam, E., Rovelli, C.: Information is Physical: Cross-perspective Links in Relational Quantum Mechanics. Phil. Phys. {\bf 1}, 1-19 (2023)}
\item {Laudisa, F., Rovelli, C.: Relational Quantum Mechanics. In: E. N. Zalta (ed.) The Stanford Encyclopedia of Philosophy, Spring Edition (2021)}
\item {Wheeler, J. A., Zurek, W.H. (eds.): Quantum Theory and Measurement. Princeton University Press, Princeton (1983)}
\item {Brukner, \v{C}.: Qubits are not Observers -- A No-Go Theorem, arXiv:2107.03513 (2021)}
\item {Adlam, E.: What Does `(Non)-Absoluteness of Observed Events' Mean? Found. Phys. {\bf 54}, 13 (2024)}

}

\end{enumerate}

\end{document}